\documentclass[12pt,letterpaper]{article}
\usepackage{jheppub}

\usepackage{graphicx}
\usepackage{bbm}
\usepackage{amsmath}
\usepackage{amssymb}
\usepackage{mathtools}
\usepackage{amsthm}
\usepackage{mathrsfs}
\usepackage{marvosym}
\usepackage{dsfont}
\usepackage[labelformat=simple]{subcaption}
\usepackage{xcolor}
\usepackage{braket}
\usepackage{cleveref}
\usepackage{comment}
\usepackage{float}
\usepackage{fancybox}
\usepackage[skins,theorems]{tcolorbox}
\tcbset{highlight math style={enhanced,
		colframe=black,colback=white,arc=0pt,boxrule=1pt}}

\newtheorem{theorem}{Claim}

\allowdisplaybreaks

\definecolor{dark-gray}{gray}{0.20}
\definecolor{gray}{gray}{0.30}
\definecolor{light-gray}{gray}{0.80}
\definecolor{dark-red}{rgb}{0.7,0,0}
\definecolor{dark-green}{rgb}{0.1,0.4,0}
\definecolor{dark-blue}{rgb}{0.3,0.3,0.7}
\definecolor{light-blue}{rgb}{0.8,0.8,1}
\definecolor{swamp}{RGB}{240, 199, 197}

\newcommand{\be}{\begin{equation}}
	\newcommand{\ee}{\end{equation}}

\def\be{\begin{equation}}
	\def\ee{\end{equation}}
\def\bea{\begin{eqnarray}}
	\def\eea{\end{eqnarray}}

\newcommand{\beq}{\begin{equation}}  \newcommand{\eeq}{\end{equation}}
\newcommand{\bal}{\begin{aligned}}   \newcommand{\eal}{\end{aligned}}
\def\beqa{\begin{eqnarray}}
	\def\eeqa{\end{eqnarray}}

\newcommand{\cK}{\mathcal{K}}

\newcommand{\dd}{\mathrm{d}}

\newcommand{\volume}{\text{{\small} vol}\, }

\def\Mpf{M_{\text{Pl;}\, 4}}

\def\LSP{\Lambda_{\text{sp}}}

\numberwithin{equation}{section}

\def\simleq{\; \raise0.3ex\hbox{$<$\kern-0.75em
		\raise-1.1ex\hbox{$\sim$}}\; }
\def\simgeq{\; \raise0.3ex\hbox{$>$\kern-0.75em
		\raise-1.1ex\hbox{$\sim$}}\; }

\numberwithin{equation}{section}

\hypersetup{
	colorlinks=true,
	linkcolor=dark-blue,
	citecolor=dark-red,
	urlcolor=dark-green,
	linktoc=page
}

\theoremstyle{remark}

\newtheoremstyle{named}{}{}{\itshape}{}{\bfseries}{.}{.5em}{#3}
\theoremstyle{named}

\newtheorem{condition}{Condition}

\title{\centering Stringy Evidence for a Universal Pattern at Infinite Distance}

\author{Alberto Castellano$^1$,} 
\author{Ignacio Ruiz$^1$,}
\author{Irene Valenzuela$^{1,2}$}
\affiliation{$^1$Instituto de F\'{i}sica Te\'{o}rica UAM-CSIC and Departamento de F\'{i}sica Te\'{o}rica, Universidad Aut\'{o}noma de Madrid, Cantoblanco, 28049 Madrid, Spain}
\affiliation{$^2$CERN, Theoretical Physics Department, 1211 Meyrin, Switzerland}

\emailAdd{alberto.castellano@csic.es, ignacio.ruiz@uam.es, irene.valenzuela@cern.ch}

\abstract{Infinite distance limits in the moduli space of a quantum gravity theory are characterized by having infinite towers of states becoming light, as dictated by the Distance Conjecture in the Swampland program. These towers imply a drastic breakdown in the perturbative regimes of the effective field theory at a quantum gravity cut-off scale known as the species scale. In this work, we find a universal pattern satisfied in all known infinite distance limits of string theory compactifications, which relates the variation in field space of the mass of the tower and the species scale: $\frac{\vec\nabla m}{m}  \cdot\frac{\vec\nabla  \Lambda_{\rm sp}}{ \Lambda_{\rm sp}}=\frac{1}{d-2}$ in $d$ spacetime dimensions. This implies a more precise definition of the Distance conjecture and sharp bounds for the exponential decay rates. We provide plethora of evidence in string theory in diverse dimensions and with different levels of supersymmetry, and  identify some sufficient conditions that  allow the pattern to hold from a bottom-up perspective.}

\begin{document}
	\makeatletter
	\let\old@fpheader\@fpheader
	\renewcommand{\@fpheader}{\old@fpheader  \vspace*{-0.1cm} \hfill CERN-TH-2023-204\\ \vspace*{-0.1cm} \hfill IFT-UAM/CSIC-23-134}
	\makeatother
	
	\maketitle
	\setcounter{page}{1}
	\pagenumbering{roman}

	\hypersetup{
		pdftitle={A cool title},
		pdfauthor={.. Irene Valenzuela},
		pdfsubject={}
	}

	\newcommand{\remove}[1]{\textcolor{red}{\sout{#1}}}

	\newpage
	\pagenumbering{arabic} 
	\section{Introduction}
	\label{s:intro}
	
	A long-standing dream for  an effective field theorist is to determine the cut-off scale at which the effective field theory (EFT) must break down and new physics should arise. The logic of naturalness has served this purpose in the history of High Energy Physics, but we might be living times of change, where this logic is failing for the first time when applied to the mass of the Higgs boson or to the cosmological constant. Instead, a more modern perspective  suggests to use guiding principles arising from requiring ultra-violet (UV) consistency of the EFT; in particular, consistency with a UV completion in a Quantum Gravity (QG) theory --- since our universe of course includes gravitational interactions. 
	The power of UV consistency to constrain low energy physics is supported by results of the Swampland \cite{Brennan:2017rbf,Palti:2019pca,vanBeest:2021lhn,Grana:2021zvf,Harlow:2022gzl,Agmon:2022thq,VanRiet:2023pnx} and the S-matrix bootstrap \cite{kruczenski2022snowmass,Mizera:2023tfe} programs. Moreover, they can provide  information about the scale at which the EFT (weakly coupled to Einstein gravity) drastically breaks down and must be replaced by a quantum gravity description. Interestingly, in certain regimes of the space of parameters of the EFT, this cut-off scale can be much lower than the Planck scale.
	
	The Distance Conjecture \cite{Ooguri:2006in} in the Swampland program provides the concrete mechanism by which the EFT drastically breaks down: The existence of an infinite tower of states becoming light in the perturbative regimes. The presence of a tower signals the breakdown of semiclassical Einstein gravity at a cut-off scale known as the species scale \cite{Arkani-Hamed:2005zuc, Distler:2005hi, Dimopoulos:2005ac, Dvali:2007hz, Dvali:2007wp}. In string theory, all continuous parameters are given by the vacuum expectation value of some scalar fields, so scanning over different values of the EFT parameters is tantamount to moving within the scalar field space of the theory (known as the moduli space if they happen to be exactly massless). From this perspective, perturbative regimes correspond to infinite distance limits in field space, where one typically recovers some approximate global symmetry \cite{Grimm:2018ohb, Gendler:2020dfp, Corvilain:2018lgw, Heidenreich_2021,Lanza:2020qmt,Lanza:2021udy}. Since global symmetries are not allowed in quantum gravity, it is precisely in these perturbative corners where the quantum gravity cut-off (i.e., the species scale) can get much lower than the Planck scale, possibly leading to observable quantum gravity effects at low energies.
	
	However, the Distance Conjecture does not specify the rate at which the tower becomes light, only that it should do so exponentially in terms of the traversed geodesic field space distance. Therefore, it is not possible to give a quantitative bound on the EFT cut-off unless we specify the lowest possible value for this exponential rate. Moreover, to derive the species scale one needs to know in principle about all towers of states becoming light, and not only the leading (i.e. lightest) one, which complicates the story considerably. In recent years, a lot of work has been dedicated to sharpen this conjecture and constrain the nature of the tower of states in the context of string theory compactifications, with the goal of finding perhaps some \emph{universal} bound for the exponential rate of the mass of the tower, and as a byproduct, the cut-off scale. 
	
	In this work, we have found that all (up to now explored) string theory examples seem to follow a very simple and sharp pattern relating the characteristic mass of the leading tower of states $m_{\rm t}$, and the species scale $\LSP$, which is given asymptotically by
	\beq \label{eq:patternmass}
	\frac{\vec\nabla m_{\text{t}}}{m_{\text{t}}} \cdot\frac{\vec\nabla \LSP}{\LSP}= \frac{1}{d-2}\, ,
	\eeq
	where the product is taken using the metric in the moduli space and $d$ is the spacetime dimension of our theory. This is quite surprising given the rich casuistics that typically arise when checking diverse string theory compactifications. In a companion paper \cite{patternPRL}, we present the pattern and its consequences, while here we will mainly discuss the string theory evidence. Notice that when written in terms of the number of light species (i.e. the number of weakly coupled fields whose mass falls at or below the species scale), \eqref{eq:patternmass} reduces to
	\beq
	\label{patternN}
	\frac{\vec\nabla m_{\text{t}}}{m_{\text{t}}} \cdot\frac{\vec\nabla N}{N}=-1\, ,
	\eeq
	since $\LSP=M_{\text{Pl};\, d}\, N^{-1/(d-2)}$. The universality of the pattern, which becomes independent of the number of spacetime dimensions or the nature of the infinite distance/perturbative limit, is at the very least tantalizing, and suggests that there might be an underlying reason constraining the structure of the tower. Notice that \eqref{patternN} puts constraints on the variation on the density of states below the species scale and the rate at which they are becoming light. Roughly speaking, the more dense the spectrum becomes, the faster the species scale goes to zero and therefore the slower the tower should become light.
	
	Since by definition $m_{\text{t}}\leq \LSP$, eq. \eqref{eq:patternmass} implies a definite bound on how slow the tower mass can go to zero asymptotically in comparison to the species scale. Upon using our pattern above, we obtain a lower bound for the exponential rate of the tower given by $\frac{1}{\sqrt{d-2}}$, which reproduces precisely the bound proposed in the sharpened Distance Conjecture \cite{Etheredge:2022opl}. This is closely related to the Emergent String Conjecture \cite{Lee:2019wij}, as the bound is saturated by a tower of oscillator modes of a fundamental string, while Kaluza-Klein modes usually have larger exponential rates. Hence, understanding the pattern \eqref{patternN} from the bottom-up opens a new avenue to test the Emergent String Conjecture independently of string theory.
	
	In this paper, we provide evidence for the pattern by checking multiple string theory constructions in different number of spacetime dimensions and with different amounts of supersymmetry. This includes setups with maximal supergravity, theories with sixteen or eight supercharges and 4d $\mathcal{N}=1$ settings arising from diverse string theory compactifications. For each different level of supersymmetry, we have selected a few representative examples to illustrate the realization of the pattern. In certain moduli spaces, we can even derive the pattern in full generality. However, for the moment, it should be taken purely as an observation, since we do not have a clear-cut argument that allows us to discern whether it is a lamppost effect or a general feature of quantum gravity. We believe, though, that it is interesting either way. In the former case, it provides at the very least an elegant and universal constraint that summarizes the casuistics of infinite distance limits observed in known string theory compactifications. In the latter case, it could be the definite criterium that characterizes the tower of the Distance Conjecture and constrains its exponential mass decay rate, providing therefore information about the quantum gravity cut-off of an EFT from the bottom-up perspective. The aim of this work is to bring the attention to this pattern so as to invite everyone to test it in more examples and look for a bottom-up rationale. On the quest for such a bottom-up explanation we have also investigated whether the pattern could arise from the Emergence Proposal \cite{Palti:2019pca,vanBeest:2021lhn}, and identified some sufficient conditions that the structure of the towers of states should satisfy to allow eq. \eqref{eq:patternmass} to hold, which can be motivated by Swampland or string theory considerations.
	
	The outline of the paper is as follows. We start with an explanation of the pattern and its consequences in Section \ref{s:pattern}, and provide evidence for it in large classes of string theory compactifications in the rest of the paper. Section \ref{s:maxsugra} is dedicated to setups of maximal supergravity, Sections \ref{s:16supercharges} and \ref{s:8supercharges} to theories with sixteen and eight supercharges, respectively, whilst Section \ref{s:N=1} analyzes diverse 4d $\mathcal{N}=1$ string theory compactifications. In Section \ref{s:bottomup}, we give the first steps towards providing a bottom-up rationale and identify some underlying sufficient conditions. We conclude in Section \ref{s:conclusions} with some final remarks.

	\subsection*{Guide to read this paper}	
	
	Since this is a long paper, we provide here a guide to help the reader navigate through it, according to their interest. Thus, a minimal read would include, apart from the introduction and conclusions, Sections \ref{s:pattern} and \ref{s:bottomup}. The first one describes in detail how and why the pattern works, even in the multi-moduli case (see also Section \ref{ss:summary}), as well as its most immediate consequences. The latter explains how it can be motivated from a bottom-up perspective, emphasizing what are the \emph{sufficient conditions} for it to hold, based both on Swampland considerations and examples taken from the rest of the text. With this, one can get a general understanding of the pattern. In a companion paper \cite{patternPRL}, we summarize the main results of these two sections and present the pattern for a more general audience, including some further remarks. On the other hand, the reader interested in the details behind its realization in concrete string theory constructions is encouraged to go through Sections \ref{s:maxsugra} to \ref{s:N=1}, which constitute the bulk of this paper. In these sections, several string theory examples in different dimensions and with different levels of supersymmetry are thoroughly discussed. The different setups can be read independently, and can also serve as a review of the string theory tests of the Distance conjecture along different types of infinite distance limits that have been performed in the literature in the past years. 	
	\section{The Pattern and its consequences}
	\label{s:pattern}

	Consider a $d$-dimensional theory containing a set of massless scalars (moduli), weakly coupled to Einstein gravity as follows
	\begin{equation}
		\mathcal{L}_{\text{scalar}} = \frac{1}{2\kappa_d^2} \mathsf{G}_{i j} (\phi)\, \partial \phi^i \cdot \partial \phi^j\, ,
	\end{equation}
	where $\mathsf{G}_{ij}(\phi)$ is the field space metric in the moduli space $\mathcal{M}_{\phi}$ spanned by the vacuum expectation value (vev) of the massless scalars. According to the Distance Conjecture \cite{Ooguri:2006in}, we should have an infinite tower of states becoming exponentially light along every infinite distance geodesic within this moduli space. In other words, along any such limit there should exist a tower with scaling $m\sim e^{-\lambda \Delta_{\phi}}$ as $\Delta_{\phi}\rightarrow \infty$, where $\lambda$ is an order one coefficient and $\Delta_{\phi}$ denotes the traversed geodesic distance.
	
	Following \cite{Calderon-Infante:2020dhm,Etheredge:2022opl,Etheredge:2023odp}, let us define the $\zeta$-vectors of the towers --- also referred to as scalar charge-to-mass ratios\footnote{The name originates from the Scalar Weak Gravity Conjecture \cite{Palti_2017}, as these vectors measure the strength of the scalar force induced by the moduli in comparison to the gravitational one (see also \cite{Andriot:2020lea,Lee:2018spm,Gonzalo:2019gjp,DallAgata:2020ino, Benakli:2020pkm}).} --- by
	\begin{equation}\label{eq:chargetomass}
		\zeta^i \equiv -\partial^i \log m\, ,
	\end{equation}
	where the index is raised with the (inverse) field metric $\mathsf{G}^{ij}(\phi)$. These $\zeta$-vectors provide information about how fast the tower becomes light. More concretely, the exponential rate of the mass of the tower is given by the projection of $\vec{\zeta}$ along the asymptotic geodesic direction, i.e. $\lambda=\vec{\zeta} \cdot \hat T$, where $\hat T$ is the normalized tangent vector of the geodesic approaching the infinite distance limit. Note that $\lambda$ depends on the trajectory taken, so that is not an intrinsic property of the tower $m$. Given the set of all possible towers becoming light, we will denote by $m_{\rm t}$ the one that does so at the fastest rate (i.e. $\lambda_{\rm t}$ is the largest exponent).
	
	Due to the presence of the infinite towers of light states, the EFT dramatically breaks down at some cut-off scale known as the species scale $\LSP$ \cite{Arkani-Hamed:2005zuc, Distler:2005hi, Dimopoulos:2005ac, Dvali:2007hz, Dvali:2007wp}. Above this scale, it is not possible to have a semiclassical Einstein gravity description anymore. The value of the species scale will depend on the nature and mass of the towers becoming light. In general, it is given by
	\beq
	\label{LSP}
	\LSP \simeq \frac{M_{\text{Pl};\, d}}{N^{\frac{1}{d-2}}}\, ,
	\eeq
	with $N$ being essentially the number of species --- i.e. weakly coupled light fields --- at or below the species scale itself. In other words, it is given by
	\beq
	\label{NSP}
	N=\int^{\LSP}_0\rho(m)\dd m\, ,
	\eeq
	where $\rho(m)$ is the density of species per unit mass. Hence, not only the leading but all light towers of states indeed matter when computing $\LSP$. Given the masses and structure of the towers, the species scale can be computed using the above two equations. This cut-off can be motivated both by perturbative arguments of renormalizing the graviton propagator \cite{Donoghue:1994dn,Aydemir:2012nz,Anber:2011ut,Calmet:2017omb,Han:2004wt} as well as black hole entropy arguments \cite{Dvali:2007hz, Dvali:2007wp}. Equivalently, the species scale also determines the scale at which higher derivative gravitational terms become of the same order as the Einstein tree-level one, which has been proven to be a more powerful technique to identify the species scale in string theory setups \cite{vandeHeisteeg:2022btw,vandeHeisteeg:2023ubh,vandeHeisteeg:2023dlw,Castellano:2023aum} (see also \cite{Caron-Huot:2022ugt} for a derivation in the context of S-matrix bootstrap).
	
	Since the towers become massless asymptotically, the species scale will also vanish in the infinite distance limit, but this typically happens at a different rate than that of the towers. Analogously, following \cite{Calderon-Infante:2023ler}, we can define the $\mathcal{Z} $-vectors as follows
	\begin{equation}\label{eq:speciescalechargetomass}
		\mathcal{Z}^i_{\text{sp}} \equiv -\partial^i \log \LSP\, ,
	\end{equation}
	providing the rate at which the species scale goes to zero asymptotically. 
	
	Depending on the infinite distance limit under consideration, we will have a different microscopic interpretation of the leading tower and the species scale, which is tied to the value of their exponential rates. In principle, the relation between $m_{\rm t}$ and $\LSP$ is independent of their exponential decay rates as we move in the moduli space $\mathcal{M}_{\phi}$.
	
	However, interestingly, by exploring a plethora of string theory compactifications, we find that the variation of the mass of the leading tower and the species scale seem to be always related by the following simple constraint that is satisfied asymptotically: 
	\begin{equation}\label{eq:pattern}
		\tcbhighmath[boxrule=1pt,drop fuzzy shadow=black]{
			\vec{\zeta}_{\rm t} \cdot \vec{\mathcal{Z}}_{\text{sp}} = \mathsf{G}^{ij} \left(\partial_i \log m_{\rm t}\right) \left(\partial_j \log \LSP\right)= \frac{1}{d-2}\; ,}
	\end{equation}
	where $d$ is the spacetime dimension of our theory. This pattern holds universally in all the string theory examples that we present in this paper, regardless of the nature of the infinite distance limit and the microscopic interpretation of the light towers. 
	Even more interestingly, using \eqref{LSP}, we can re-write the pattern as
	\beq\label{eq:patternN}
	\tcbhighmath[boxrule=1pt,drop fuzzy shadow=black]{
		\mathsf{G}^{ij} \left(\partial_i \log m_{\rm t}\right) \left(\partial_j \log N\right)=-1\; ,}
	\eeq
	which is moreover independent of the number of dimensions. This hints toward a universal relation between the density of states becoming light and their characteristic mass. The faster they become light as we approach the infinite distance limit, the less dense the towers can get, and viceversa. In some sense (that we will make more concrete later),  the variation of the mass and the number of states in the moduli space act as `dual variables'.
	
	\subsection*{Implied bounds on the exponential decay rates}
	
	Notice that a relation like \eqref{eq:pattern} implies a lower bound for the scalar charge-to-mass ratio of the leading tower asymptotically, since the latter should be always lighter than the species scale, i.e. $m_{\rm t}\leq \LSP$. This consistency condition implies $|\vec{\zeta}_{\rm t} \cdot \vec{\mathcal{Z}}_{\text{sp}}| \leq |\vec{\zeta}_{\rm t}|^2$ and, therefore,
	\begin{equation}\label{eq:Rudelius}
		|\vec{\zeta}_{\rm t}|^2 \geq \frac{1}{d-2}\, ,
	\end{equation}
	which leads to the lower bound for the exponential rate of the leading tower
	\beq
	\lambda_{\text{t}} = |\vec{\zeta}_{\rm t}| \geq \frac{1}{\sqrt{d-2}}\, ,
	\eeq
	recently proposed in the sharpened Distance Conjecture \cite{Etheredge:2022opl}. Analogously, in those cases (as it happens in all known examples) in which there exists a tower $\vec{\zeta}\propto\vec{\mathcal{Z}}_{\rm sp}$ satisfying the pattern \eqref{eq:pattern}, then one gets an upper bound on the exponential rate of the species scale since $|\vec{\mathcal{Z}}_{\rm sp}|\leq |\vec{\zeta}| $, yielding
	\beq
	\lambda_{\text{sp}} = |\vec{\mathcal{Z}}_{\text{sp}}| \leq \frac{1}{\sqrt{d-2}}\, ,
	\eeq
	which matches the recently proposed bound in \cite{vandeHeisteeg:2023ubh}\footnote{The pattern \eqref{eq:pattern} is only valid asymptotically, and this is why it is consistent that the constant in the upper bound for the species scale is fixed to $\frac{1}{\sqrt{d-2}}$. This might get modified when moving to the interior of the moduli space.} based both on EFT arguments and string theory evidence.

	Notice that the above bounds are always saturated by the oscillator modes of a fundamental string. Hence, if we assume that Kaluza-Klein (KK) towers always have a larger exponential rate $\lambda_{\text{t}}$ (as indeed happens in all examples known so far), we are essentially recovering the Emergent String Conjecture (ESC) \cite{Lee:2019wij} as well, assuming that membranes decay at a slower rate than particles and strings, as happens in all known string theory examples (see also \cite{Alvarez-Garcia:2021pxo}). It would be interesting, though, to show that the only possible towers of states satisfying the pattern are indeed KK towers or oscillator string modes (as implied by the ESC) from a purely bottom-up perspective.
	
	We want to remark that the pattern \eqref{eq:pattern} is much more concrete than previous analyses as it provides a sharp \emph{equality} relating the asymptotic behavior of the species scale and the leading tower of states, instead of just some bound on their respective decay rates. We expect that, upon further exploration, this may highly constrain the nature of the possible towers of states predicted by the Distance Conjecture.
	
	Furthermore, we can also recover the recently proposed lower bound  for the exponential rate of the species scale (named the Species Scale Distance Conjecture \cite{Calderon-Infante:2023ler}),
	\begin{equation}\label{eq:lambdaspmin}
		\lambda_{\text{sp}} \geq \lambda_{\text{sp, min}} = \frac{1}{\sqrt{(d-1)(d-2)}}\, ,
	\end{equation}
	if we \emph{assume} (based on string theory evidence) that the maximum possible value for the exponential rate of the leading tower is given by that of a KK tower decompactifying one (unwarped) extra dimension, i.e. $\lambda_{\text{t, max}} = \sqrt{\frac{d-1}{d-2}}$. In this regard, all the examples analyzed in the present paper can be equivalently seen to provide further evidence in favor of the bound \eqref{eq:lambdaspmin}.
	
	\subsection*{First steps towards decoding the pattern}
	
	Before getting into more complicated examples, let us first show how the pattern is satisfied for the case of a single modulus and a single tower of states becoming light. Let us consider two cases; either the leading tower is a KK tower or a tower of string oscillator modes, as dictated by the Emergent String Conjecture and as observed in all string theory examples so far.
	The species scale associated to a KK tower decompactifying $n$ (unwarped) extra  dimensions  is given by the higher dimensional Planck mass
	\beq
	\label{Mp}
	\LSP\equiv M_{\text{Pl};\, d+n}=M_{\text{Pl};\, d}\, \left(\frac{m_{{\rm KK},\, n}}{M_{\text{Pl};\, d}}\right)^{\frac{n}{d+n-2}},
	\eeq
	as can be derived from applying \eqref{LSP} and \eqref{NSP} to an equi-spaced tower with $m_k=k^{1/n} m_{\text{KK},\, n}$, where $k=1,\ldots,\infty$. By dimensional reduction of the theory, it is also well-known that the exponential rates of the KK tower and the species scale read
	\beq\label{eq:zeta&speciesveconemodulus}
	\zeta_{{\rm KK},\, n} = \sqrt{\frac{d+n-2}{n (d-2)}}\, , \qquad \mathcal{Z}_{{\rm KK}, \, n}=\sqrt{\frac{n}{(d+n-2) (d-2)}}\, ,
	\eeq
	where $\mathcal{Z}_{{\rm KK}, \, n}$ can be obtained from $\zeta_{{\rm KK},\, n}$ upon using \eqref{Mp}.
	It can be easily checked that this always reproduces the pattern \eqref{eq:pattern} independently of the number of dimensions that get decompactified,
	\beq\label{eq:patternKKn}
	\zeta_{{\rm KK},\, n} \cdot \mathcal{Z}_{{\rm KK},\, n}= \frac{1}{d-2}\, .
	\eeq
	Let us remark, though, that the above expressions for the exponential rates are valid when decompactifying to a higher dimensional \emph{vacuum}, since the story is more complicated when the theory decompactifies to a running solution instead, as recently shown in \cite{Etheredge:2023odp}. We will comment more on this in Section \ref{s:16supercharges}. 
	
	The other relevant case is that of a tower of string oscillator modes. If these states arise from a fundamental string, we have
	\beq\label{eq:zeta&speciesvecstring}
	\zeta_{\rm osc}= \frac{1}{\sqrt{d-2}}=\mathcal{Z}_{\rm osc}\, ,
	\eeq
	since the species scale coincides with the string scale (up to maybe logarithmic corrections that will not be relevant here) due to the exponential degeneracy of states at the string scale. It is then automatic that 
	\beq\label{eq:patternstringsinglemodulus}
	\zeta_{\rm osc} \cdot \mathcal{Z}_{\rm osc} = \frac{1}{d-2}\, .
	\eeq
	In summary, for a single modulus, the pattern implies that the exponential rate of the species scale verifies $\lambda_{\text{sp}}=\left((d-2)\lambda_{\text{t}} \right)^{-1}$ or, in other words, $\Lambda_{\text{sp}}\sim m_{\rm t}^{1/(d-2)\lambda_{\text{t}}^2 }$, which holds regardless of whether we consider KK or stringy towers. In the multi-moduli case, though, these vectors are  not parallel to each other in general. Thus, the pattern is not giving a direct relation between the exponential rates along a given trajectory, but rather between the scalar charge-to-mass vectors $\vec\zeta_{\text{t}}$ and $\vec{\mathcal{Z}}_{\text{sp}}$ as we take an asymptotic limit. This is essential for the pattern to hold in a multi-moduli setup.
	
	At this moment, the claimed universality of the pattern should surprise you for two reasons:
	\begin{itemize}
		\item The structure of the tower fixes the relation between $m_{\text{t}}$ and $\Lambda_{\text{sp}}$ at a given point of the moduli space. However, a priori, this relation is independent of the exponential decay rate of $m_{\text{t}}$ and $\Lambda_{\text{sp}}$ as we move in moduli space. The pattern implies that they are not independent but can be derived from each other, leading to a universal relation satisfied both for KK and string towers.
		\item The pattern is satisfied even in the presence of multiple towers, when the species scale is not simply determined by the leading tower. For instance, we will see that there can be regions of the moduli space where e.g., the leading tower is a KK tower while the species scale corresponds to some string scale. Even then, the pattern is still satisfied as the angle between the vectors precisely compensates for the change in the magnitude, such that \eqref{eq:pattern} holds in a non-trivial manner. The same occurs when decompactifying to a larger number of dimensions than those associated to the leading tower, due to the presence of other subleading KK towers that change the value of the species scale.
	\end{itemize}

	Sometimes, it gets useful to define the convex hull of the $\zeta$-vectors of all light towers in a given asymptotic regime \cite{Calderon-Infante:2020dhm}, since this provides us information about which tower is dominating along each direction. Analogously, one can define the convex hull of $\mathcal{Z}$-vectors of the species scale as in \cite{Calderon-Infante:2023ler}, thus informing us about the nature of the infinite distance limit, namely the quantum gravity theory above $\LSP$. Notice, though, that these convex hulls can only be defined if there is a region of moduli space in which the hull of the scalar charge-to-mass vectors does not change. In such a case, it follows from \eqref{eq:pattern} that both polytopes are dual to each other, as hinted in \cite{Calderon-Infante:2023ler}. This implies, in particular, that given any one of them one can simply retrieve the other upon imposing the aforementioned relation as a constraint. Therefore, both convex hulls contain  the same information. It is then equivalent to keep track of all towers becoming light along a given trajectory (which allows one to compute the species scale), than to focus just on the leading tower along all asymptotic geodesics of a given asymptotic regime. Starting from a tower in some particular limit, we can then use the pattern to predict the nature of the towers in other asymptotic limits, and even reconstruct global information about how different limits (and different duality frames) glue together in moduli space.\footnote{This will be explored in more detail in \cite{taxonomyTBA} where we will present some rules about how to glue different asymptotic limits together, which can be equivalently derived from maximal supergravity string theory setups, or from assuming the pattern \eqref{eq:pattern}.}
	
	In the upcoming sections we will test this pattern in the multi-moduli case within several familiar string theory vacua, differing in the number of spacetime dimensions, the amount of supersymmetry preserved, etc. We will see that the pattern is always satisfied, independently of how complicated the tower structure may look like a priori.
	
	\section{Derivation in string theory setups with 32 supercharges}
	\label{s:maxsugra}
	
	We begin by deriving the pattern in string theory compactifications with 32 supercharges, i.e. maximal supergravity setups arising from toroidal compactifications of M-theory. The advantage of these setups is that the $\zeta$-vectors associated to the leading towers of states take some very specific values that remain \emph{fixed} as we move within the moduli space. This will allow us, in turn, to show that the pattern \eqref{eq:pattern} is verified in full generality at every infinite distance limit of the moduli space.
	
	Due to the simplicity of these setups, we can basically summarize the results in two main scenarios that highlight the key features underlying the realization of the pattern. Hence, we will first explain these main features, and then exemplify them in concrete examples of M-theory toroidal compactifications down to $d=9,8$ later on. We finish the section by generalizing the discussion to any number of spacetime dimensions for the sake of completeness. 
	
	\subsection{Summary of underlying key features}
	\label{ss:summary}
	
	Consider a $D$-dimensional theory compactified down to $d=D-n$ spacetime dimensions, both preserving maximal supersymmetry in flat space. As shown in \cite{Etheredge:2022opl}, such setups in Minkowski space satisfy the Emergent String Conjecture \cite{Lee:2019wij}, in the sense that every infinite distance limit corresponds either to an emergent string limit or to some decompactification. Hence, there are essentially two main scenarios, depending on whether the species scale associated to a given asymptotic regime corresponds to a higher dimensional Planck mass or to the fundamental string scale. In the following, we explain the underlying key features that make a relation like \eqref{eq:pattern} to be satisfied in these two cases, which we will later exemplify in some concrete examples. For a detailed derivation of the relevant formulae involved see Appendix \ref{ap:generalities}.
	
	%%%%%%%%%%%%%%%%%%%%%%%%%%%%%%%%%%%%
	\begin{figure}
		\begin{center}
			\begin{subfigure}{0.475\textwidth}
				\center
				\includegraphics[width=65mm]{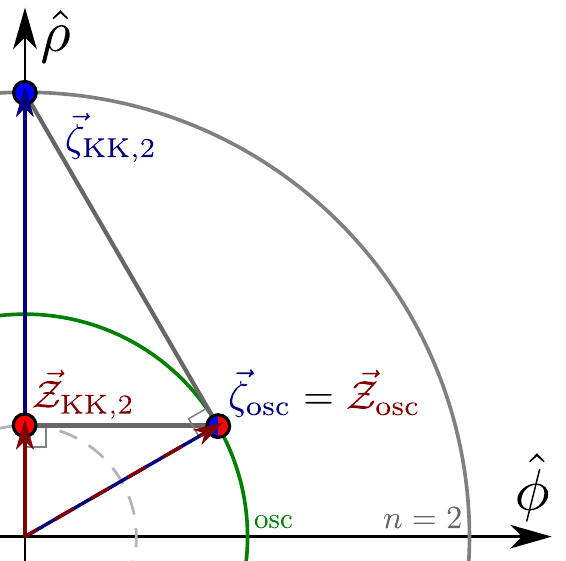}
				\caption{KK and emergent string limit} \label{sfig:KKstring}
			\end{subfigure}
			\begin{subfigure}{0.475\textwidth}
				\center
				\includegraphics[width=70mm]{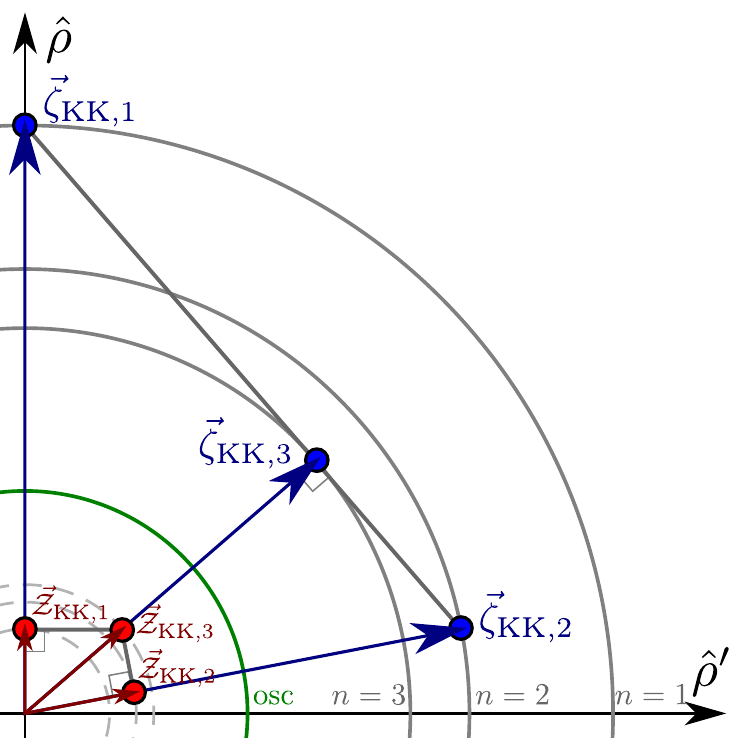}
				\caption{Two KK limits} \label{sfig:twoKK}
			\end{subfigure}
			\caption{\small Sketches depicting two possibilities in multi-field limits for maximal supergravity, both in $d=8$, with vectors associated to light towers in blue and to the species scale in red. \textbf{(a)} Decompactification of two internal dimensions and an emergent string limit. The species scale is controlled by the oscillator modes of the string unless we move along the decompactification direction, where it coincides with the ten-dimensional Planck mass. In this case, $\hat{\rho}$ and $\hat{\phi}$ denote the normalized radion and the ten-dimensional dilaton. \textbf{(b)} Two decompactification limits, of one and two internal dimensions, with towers $\vec{\zeta}_{{\rm KK},\, 1}$ and $\vec{\zeta}_{{\rm KK},\, 2}$ (as well as the total volume, $\vec{\zeta}_{{\rm KK},\, 3}$). Note that unless we decompactify a single cycle, the species scale is controlled by the eleven-dimensional Planck mass. The axes $\hat{\rho}$ and $\hat{\rho}'$ correspond to the normalized radions associated to decompactifying the 1- and 2-cycles, respectively.\label{fig:MT2radions} }
		\end{center}
	\end{figure}
	%%%%%%%%%%%%%%%%%%%%%%%%%%%%%%%%%%%%

	\subsubsection*{Perturbative string limit}
	
	This first scenario is characterized by having the species scale equal to the string scale. Hence, the $\mathcal{Z}$-vector of the species scale is the same than the $\zeta$-vector associated to the tower of string oscillator modes. However, this does not necessarily mean that the tower of string modes is the leading one. As noted in \cite{Castellano:2022bvr}, if we have both a KK and a string tower becoming light, the species scale will indeed correspond to the string scale (even if the KK tower is parametrically lighter) as long as the string scale remains below the species scale associated to the KK tower (i.e. the higher dimensional Planck mass). Hence, the most general scenario with $\LSP \simeq m_s$ can contain both KK and string modes below the species scale. For the sake of concreteness, let us focus on the KK tower associated to the overall volume of the compactification space and the oscillator modes arising from a fundamental string already existing in the higher dimensional theory. We can then restrict to a slice of the tangent space of the moduli space spanned by the overall volume modulus $\hat{\rho}$ and the string dilaton $\hat{\phi}$. The relevant $\zeta$-vectors for such towers within this subspace are (in the flat frame $\{\hat{\phi},\hat{\rho}\}$, c.f. eqs. \eqref{eq:Ddim}-\eqref{eq:chargetomasstower}) \cite{Etheredge:2022opl,Calderon-Infante:2023ler}: 
	\begin{equation}\label{eq:zetasstringlimit}
		\begin{split} 
			\vec{\zeta}_{{\rm KK},\, n} &= \left( 0 , \sqrt{\frac{d+n-2}{n (d-2)}} \right)\, ,\quad \vec{\mathcal{Z}}_{{\rm KK},\, n} = \left( 0 , \sqrt{\frac{n}{(d+n-2) (d-2)}} \right)\, ,\\
			\vec{\zeta}_{\text{osc}}=\vec{\mathcal{Z}}_{\text{osc}} &= \left( \frac{1}{\sqrt{d+n-2}} , \sqrt{\frac{n}{(d+n-2)(d-2)}} \right)\, .
		\end{split}
	\end{equation}
	These vectors are plotted in Figure \ref{sfig:KKstring}. The tangent vectors of asymptotic geodesics in this slice of the moduli space are radial vectors (i.e straight lines passing through the origin) \cite{Etheredge:2022opl}. As explained in Section \ref{s:pattern}, to obtain the exponential rate $\lambda$ of a tower (or the species scale) along a given geodesic, we just need to compute the projection of the associated $\zeta$-vector (resp. $\mathcal{Z}$-vector) along such direction. The larger this projection is, the fastest the mass (or the species scale) goes to zero asymptotically. The leading (i.e. the lightest) tower of states is therefore the one with the largest projection of $\vec\zeta$ over such direction; and the same applies to the species scale, which will be the one with the largest projection of $\vec{\mathcal{Z}}$.
	
	If we move parallel to $\vec{\zeta}_{{\rm KK},\, n} $, both the Planck scale and the string scale decay at the same rate, so we can simply take the species scale vector as $\vec{\mathcal{Z}}_{{\rm KK},\, n}$. Otherwise, for any other intermediate direction, $\LSP$ will be given by the string scale, as it always remains below the Planck scale, so we should take instead $\vec{\mathcal{Z}}_{\text{osc}}$. On the other hand, the leading tower is always the KK one, except if we move parallel to $\vec{\zeta}_{\text{osc}}$, where both towers present the same exponential rate.\footnote{Note that precisely in this case the limit qualifies as equi-dimensional, in the notation defined in \cite{Lee:2019wij}. Such limits probe gravitational theories in the same number of spacetime dimensions as the starting point of the (infinite distance) trajectory. The fact that there is a KK tower decaying at the same rate than the string tower along this direction is also expected from the Emergent String Conjecture \cite{Lee:2019wij}.} It is clear from Section \ref{s:pattern} that $\vec{\zeta}_{\text{KK,}\, n} \cdot \vec{\mathcal{Z}}_{\text{KK},\, n}=\frac1{d-2}$ and $\vec{\zeta}_{\text{osc}} \cdot \vec{\mathcal{Z}}_{\text{osc}}=\frac1{d-2}$ for each tower independently, but it is less obvious that the pattern will continue working when considering both towers simultaneously. We find here that even in the case in which the species scale is the string scale and the leading tower corresponds to the KK tower, the pattern still holds:
	\beq\label{eq:KKstring}
	\vec{\zeta}_{{\rm KK},\, n} \cdot \vec{\mathcal{Z}}_{\text{osc}}=\frac1{d-2}\, .
	\eeq
	This can be easily understood geometrically from Figure \ref{sfig:KKstring} as follows. Since $\vec{\mathcal{Z}}_{\text{osc}}$ is perpendicular to the convex hull generated by $\vec{\zeta}_{{\rm KK},\, n}$ and $\vec{\zeta}_{\rm osc}$, it turns out that $\vec{\zeta}_{\rm osc}$ is the projection of $\vec{\zeta}_{{\rm KK},\, n}$ along the direction associated to $\vec{\mathcal{Z}}_{\text{osc}}$, so that the pattern holds in general. Alternatively, the projection of $\vec{\mathcal{Z}}_{\text{osc}}$ along the direction determined by $\vec{\zeta}_{{\rm KK},\, n}$ coincides with $\vec{\mathcal{Z}}_{\text{KK,}\, n}$ since the radial component of $\vec{\mathcal{Z}}_{\text{osc}}$ arises from changing the masses to lower dimensional Planck units and it is therefore equal to the radial component of $\vec{\mathcal{Z}}_{\text{KK},\, n}$  as can be seen in \eqref{eq:zetasstringlimit}. 
	
	\subsubsection*{Decompactification limit}
	
	The second scenario occurs when all the light towers below the species scale are KK modes (possibly decompactifying to different number of dimensions), and we do not find any additional tower of string modes before reaching the lightest higher dimensional Planck mass. Hence, the species scale is a Planck scale in higher dimensions. For concreteness, let us focus on a two-dimensional slice spanned by two KK towers decompactifying to $d+n$ and $d+n'$ dimensions, respectively, with associated volume moduli $\hat{\rho}$ and $\hat{\rho}'$. The $\zeta$-vectors are given by \cite{Etheredge:2022opl} 
	\begin{equation}\label{eq:n&n'zetas}
		\begin{split} 
			\vec{\zeta}_{{\rm KK},\, n} &= \left( 0 , \sqrt{\frac{d+n-2}{n (d-2)}} \right)\, ,\\
			\vec{\zeta}_{{\rm KK},\, n'} &= \left( \sqrt{\frac{d+n+n'-2}{n' (d+n-2)}} ,\, \sqrt{\frac{n}{(d+n-2)(d-2)}} \right)\, .
		\end{split}
	\end{equation}
	Depending on the infinite distance trajectory that we explore, the species scale will correspond to the Planck scale of decompactifying $n$, $n'$ or $n+n'$ extra dimensions. The associated $\mathcal{Z}$-vectors are \cite{Calderon-Infante:2023ler}
	\begin{equation}
		\begin{split} 
			\vec{\mathcal{Z}}_{{\rm KK},\, n} &= \left( 0 , \sqrt{\frac{n}{(d+n-2) (d-2)}} \right) \, ,\\
			\vec{\mathcal{Z}}_{{\rm KK},\, n'} &= \left( \sqrt{\frac{n'(d+n+n'-2)}{(d+n'-2)^2 (d+n-2)}} ,\, \frac{n'}{d-2+n'} \sqrt{\frac{n}{(d+n-2) (d-2)}} \right) \, ,\\
			\vec{\mathcal{Z}}_{{\rm KK},\, n+ n'} &= \left( \sqrt{\frac{n'}{(d+n-2) (d+n+n'-2)}},\, \sqrt{\frac{n}{(d+n-2) (d-2)}} \right) \, \label{eq:combinedZ}.
		\end{split}
	\end{equation}
	All these vectors are represented in Figure \ref{sfig:twoKK}. The species scale corresponds to the lightest Planck scale along any chosen infinite distance trajectory. Hence, it will always be given by $\vec{\mathcal{Z}}_{{\rm KK},\,  n+ n'}$ in the entire asymptotic regime unless we move parallel to either $\vec{\zeta}_{{\rm KK},\, n}$ or $\vec{\zeta}_{{\rm KK},\, n'}\,$, in which case it reduces to $\vec{\mathcal{Z}}_{{\rm KK},\, n}$ or $\vec{\mathcal{Z}}_{{\rm KK},\, n'}\,$, respectively. However, the leading tower corresponds to decompactifying only $n$ or $n'$ extra dimensions unless we move precisely parallel to $\vec{\mathcal{Z}}_{{\rm KK},\,  n+ n'}$. The latter case would physically correspond to an isotropic decompactification of both $n$- and $n'$-dimensional internal cycles, with an effective KK tower of charge-to-mass vector given by (c.f. eq. \eqref{eq:effectivezeta})
	\beq
	\vec{\zeta}_{{\rm KK},\, n+ n'} = \left( \sqrt{\frac{n' (d+n+n'-2)}{(d+n-2) (n+n')^2}}, \sqrt{\frac{n (d+n+n'-2)^2}{(n+n')^2(d+n-2) (d-2)}} \right) \, .
	\eeq
	Again, the pattern is clearly satisfied whenever we move along the asymptotic trajectories determined by any of the individual KK towers (due to \eqref{eq:patternKKn}), but it also nicely holds for intermediate directions within the asymptotic regime, since 
	\beq\label{eq:doubleKK}
	\vec{\zeta}_{{\rm KK},\, n} \cdot \vec{\mathcal{Z}}_{{\rm KK},\, n+ n'}= \vec{\zeta}_{{\rm KK},\, n'} \cdot \vec{\mathcal{Z}}_{{\rm KK},\,n+ n'} = \frac1{d-2}\, .
	\eeq
	Notice that such relation may be easily understood from geometrical considerations as follows. The species vector $\vec{\mathcal{Z}}_{{\rm KK},\, n+n'}$ appears to be always perpendicular to the convex hull generated by $\vec{\zeta}_{{\rm KK},\, n}$ and $\vec{\zeta}_{{\rm KK},\, n'}$ (see Figure \ref{sfig:twoKK}), such that they both project to $\vec{\zeta}_{{\rm KK},\,  n+n'}$ along the direction determined by the former. Alternatively, $\vec{\mathcal{Z}}_{{\rm KK},\,  n+n'}$ projects to $\vec{\mathcal{Z}}_{{\rm KK},\, n}$ (analogously $\vec{\mathcal{Z}}_{{\rm KK},\, n'}$) along the direction determined by $\vec{\zeta}_{{\rm KK},\, n}$ (respectively $\vec{\zeta}_{{\rm KK},\, n'}$), which may be understood again as coming from a change of Planck units in both cases, given the commutativity of the compactification process (see Appendix \ref{ap:generalities}).
	
	\subsubsection*{Summary}
	
	What can be learned from the two scenarios above? The species scale vector $\vec{\mathcal{Z}}$ always happens to be perpendicular to the convex hull of the light towers of states. Conversely, the leading scalar charge-to-mass vector $\vec{\zeta}_{\rm t}$ is orthogonal to the convex hull generated by the species vectors. This is a feature that will hold in general for M-theory toroidal compactifications, as we show below. In fact, such constraints are restrictive enough so as to ensure that, once we assume that the pattern \eqref{eq:pattern} is verified by any pair of collinear vectors $\vec{\zeta}$ and $\vec{\mathcal{Z}}$ (i.e. when both are associated to the same tower of states), then the pattern extends automatically to any other asymptotic limit of the moduli space.\footnote{For instance, if $\vec{\zeta}_{\rm t}$ is orthogonal to the convex hull generated by $\vec{\mathcal{Z}}_{\rm sp}$ (the total specie scale) and $\vec{\mathcal{Z}}_{\rm t}$ (the one obtained only from considering the leading tower), then satisfying $\vec{\zeta}_{\rm t}\cdot \vec{\mathcal{Z}}_{\rm t}=\frac1{\sqrt{d-2}}$ guarantees that $\vec{\zeta}_{\rm t}\cdot \vec{\mathcal{Z}}_{\rm sp}=\frac1{\sqrt{d-2}}$, as the difference between the two species scales vectors is given by a vector which is orthogonal to $\vec{\zeta}_{\rm t}$.}
	
	Notice, however, that the same story does not apply immediately when the amount of supersymmetry of our theory is reduced, since then the charge-to-mass and species vectors can `slide' (or jump) non-trivially depending on where we sit in moduli space, see Section \ref{s:16supercharges}. In any event, most of our efforts in the upcoming sections will be dedicated to show that, even in such cases, the pattern is still verified at any infinite distance boundary, and it does so in a way that can be easily understood from pictures similar to those shown in Figure \ref{fig:MT2radions} above.

	\subsection{Maximal supergravity in 9d}
	\label{ss:9d}
	
	Next, we will illustrate the above general scenarios in concrete examples, starting with the unique 9d $\mathcal{N}=2$ supergravity theory arising from compactifying M-theory on a two-dimensional torus. The $\zeta$- and $\mathcal{Z}$-vectors of the towers of states and the species scale for this particular setup have been recently analyzed in \cite{Etheredge:2022opl} and \cite{Calderon-Infante:2023ler} respectively. Here we will build upon these results and simply check if the pattern \eqref{eq:pattern} is verified, paying special attention to the way in which this happens.
	
	Consider M-theory compactified on a $\mathbb{T}^2$ with a metric parametrized as 
	\begin{equation}\label{eq:T2metric}
		g_{mn}= \frac{e^U}{\tau_2} \left(
		\begin{array}{cc}
			1 & \tau_1  \\
			\tau_1 & |\tau|^2  \\
		\end{array}
		\right) \, ,
	\end{equation}
	where $\tau=\tau_1+{\rm i}\tau_2$ denotes the complex structure of the torus and $U$ controls its overall volume (in M-theory units). The scalar and gravitational sectors in the 9d Einstein frame read \cite{Etheredge:2022opl}
	\begin{equation}\label{eq:9d}
		S_{\text{9d}} \supset \frac{1}{2\kappa_9^2} \int \dd^{9}x\, \sqrt{-g}\,  \left[ \mathcal{R} - \frac{9}{14} \left( \partial U \right)^2 -\frac{\partial \tau \partial \bar \tau}{2 \tau_2^2} \right]\, .
	\end{equation}
	This theory has a moduli space which is classically exact and parameterizes the manifold $\mathcal{M}_{\text{mod}}=SL(2, \mathbb{Z})\backslash SL(2, \mathbb{R})/U(1) \times \mathbb{R}$, where we have taken into account the $SL(2, \mathbb{Z})$ U-duality symmetry associated to the full quantum theory \cite{Schwarz:1995dk,Aspinwall:1995fw}. 
	
	In the following, we will effectively forget about the axion $\tau_1$,\footnote{In other words, we restrict ourselves to explore geodesic paths that leave the axionic component of $\tau$ fixed to a constant value, since any other geodesic reaching infinite distance within $SL(2, \mathbb{R})/U(1)$ can be mapped to the former via some modular transformation.} since it plays no role in our discussion \cite{Calderon-Infante:2023ler}, and we moreover define canonically normalized fields $\hat U$ and $\hat \tau$ as follows
	\begin{equation} \label{eq:canonicalnormalization}
		U =  \kappa_9 \sqrt{\frac{14}{9}}\, \hat U \, , \quad \tau_2 = \kappa_9\, e^{\sqrt{2} \, \hat\tau} \, .
	\end{equation}
	As discussed in \cite{Etheredge:2022opl}, the relevant towers of states becoming light at the infinite distance limits of this moduli space are $\frac{1}{2}$-BPS particles. For this particular example, the convex hull determined by the (asymptotic) scalar charge-to-mass vectors of all light towers is spanned by Kaluza-Klein modes with the following $\zeta$-vectors:
	\begin{equation}\label{eq:KKzetas9d}
		\vec{\zeta}_{\text{KK},\, 1} = \left( \frac{3}{\sqrt{14}},\frac{1}{\sqrt{2}} \right) \, , \qquad
		\vec{\zeta}_{\text{KK},\, 1'} = \left( \frac{3}{\sqrt{14}},-\frac{1}{\sqrt{2}}\right) \, ,
	\end{equation}
	as well as M2-branes wrapping the compactification manifold, with
	\begin{equation}\label{eq:M2zeta9d}
		\vec{\zeta}_{\text{M2}} = \left( -\sqrt{\frac{8}{7}},0 \right) \, .
	\end{equation}
	We have adopted the notation $\vec{\zeta}=\left(\zeta^{\hat U}, \zeta^{\hat \tau} \right)$. Notice that they all satisfy the relation $|\vec{\zeta}|^2=8/7$, in accordance with \eqref{eq:zeta&speciesveconemodulus} above for $d=9$ and $n=1$.
	
	On the other hand, their associated species scale vectors, $\vec{\mathcal{Z}}$, are seen to be given by \cite{Calderon-Infante:2023ler}
	\begin{equation} \label{eq:SSvec9d}
		\begin{split} 
			\vec{\mathcal{Z}}_{\text{KK},\, 1} &= \left( \frac{3\sqrt{14}}{112},\frac{\sqrt{2}}{16} \right) \, , \qquad \vec{\mathcal{Z}}_{\text{KK},\, 1'} = \left( \frac{3\sqrt{14}}{112},-\frac{\sqrt{2}}{16}\right) \, ,\\
			\vec{\mathcal{Z}}_{\text{M2}} &= \left( -\frac{1}{2\sqrt{14}} , 0\right) \, ,
		\end{split}
	\end{equation}
	corresponding to the appropriate 10d Planck mass of the decompactified (dual) theories. In particular, since $|\vec{\mathcal{Z}}|^2=\frac{1}{(d-1)(d-2)}=\frac{1}{56}$, these saturate the lower bound proposed in \cite{Calderon-Infante:2023ler} for the decay parameter, $\lambda_{\text{sp}}$, of the species scale.
	
	As explained in \cite{Calderon-Infante:2023ler}, a crucial ingredient when determining the set of all possible species scales is the concept of effective tower \cite{Castellano:2021mmx}. Indeed, for intermediate directions between $\vec{\zeta}_{\text{KK},\, 1}$ and $\vec{\zeta}_{\text{KK},\, 1'}$, despite one KK tower being (in general) parametrically lighter than the other, one still needs to account for bound states thereof in order to properly compute the species scale in that asymptotic regime. Upon doing so, one arrives at the following species scale vector \cite{Calderon-Infante:2023ler}
	\begin{equation}\label{eq:effectivespecies}
		\vec{\mathcal{Z}}_{{\rm KK},\, 2}=\frac{1}{d} \left( \vec{\zeta}_{{\rm KK},\, 1} +\vec{\zeta}_{{\rm KK},\, 1'} \right) = \left( \frac{\sqrt{14}}{21} , 0\right)\, ,
	\end{equation}
	to which we can associate an effective (averaged) mass scale and charge-to-mass vector as follows
	\begin{equation}\label{eq:effectivemass}
		\vec{\zeta}_{{\rm KK},\,2}=\frac{1}{2} \left( \vec{\zeta}_{{\rm KK},\, 1} +\vec{\zeta}_{{\rm KK},\, 1'} \right) = \left( \frac{3}{\sqrt{14}} , 0\right)\, .
	\end{equation}
	The physical interpretation for \eqref{eq:effectivespecies} is clear. It simply corresponds to the 11d Planck scale, signalling full decompactification of the internal torus. On the other hand, the charge-to-mass vector \eqref{eq:effectivemass} is a meaningful quantity only when one takes the decompactification limit in an isotropic way, namely for an asymptotic vector $\hat T = \partial_{\hat U}$. Still it may be useful to think in terms of `averaged' geometric quantities when computing the species scale vectors and checking the pattern \eqref{eq:pattern} explicitly, as we discuss later on in this section.
	
	Apart from these, there is also another set of $\frac{1}{2}$-BPS states comprised by critical type IIA strings arising from M2-branes wrapped on a non-trivial 1-cycle. These can be seen to lead to the following charge-to-mass vectors
	\begin{equation}\label{eq:9dstrings}
		\vec{\zeta}_{\text{osc}} = \left( -\frac{1}{2 \sqrt{14}},\frac{1}{2\sqrt{2}} \right) \, , \quad 
		\vec{\zeta}_{\text{osc'}} = \left( -\frac{1}{2 \sqrt{14}},- \frac{1}{2\sqrt{2}} \right) \, ,
	\end{equation}
	which coincide with those of their associated species scale \cite{Castellano:2022bvr} and moreover satisfy $|\vec{\mathcal{Z}}_{\text{osc}}|^2=\frac{1}{d-2}=\frac{1}{7}$ (c.f. \eqref{eq:zeta&speciesvecstring}).
	
	%%%%%%%%%%%%
	\begin{figure}[htb]
		\begin{center}
			\includegraphics[scale=.65]{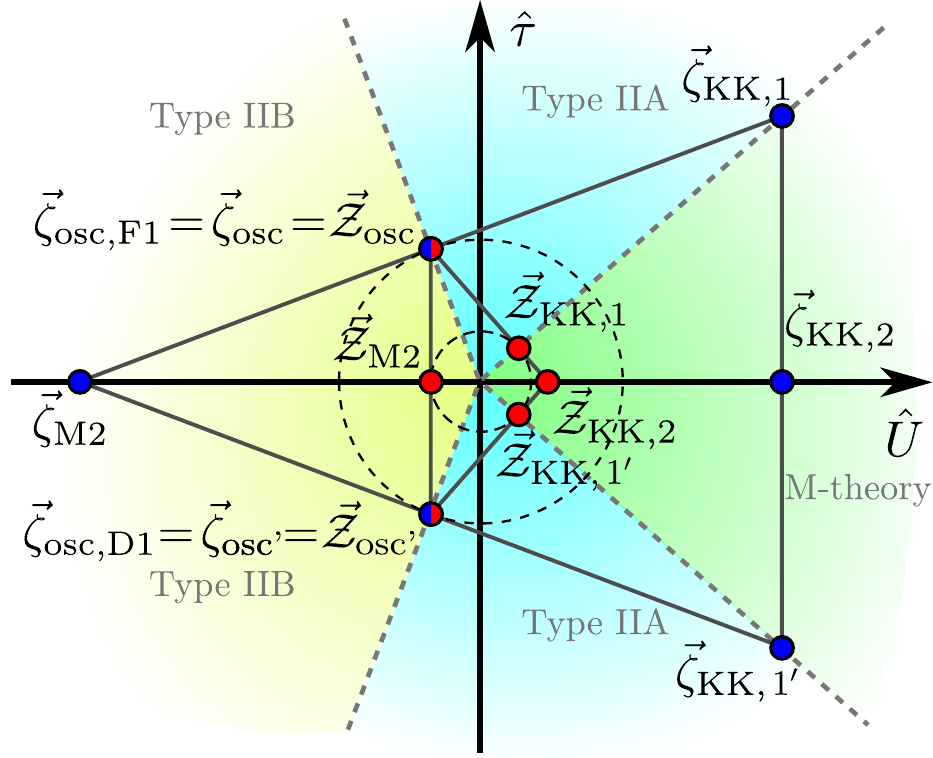}
			\caption{\small Convex hulls spanned by the species scale (red) and mass scales of the leading towers (blue) in nine-dimensional maximal supergravity. The 1-spheres of radii $\frac{1}{\sqrt{d-2}}=\frac{1}{\sqrt{7}}$ and $\frac{1}{\sqrt{(d-1)(d-2)}}=\frac{1}{\sqrt{56}}$ are plotted in dashed lines. Also plotted in distinct shades are the different duality frames of the theory. Notice that both type IIA and type IIB string theory have two different duality frames, while there is a single one for M-theory.} 
			\label{fig:T2SDC&SSDC}
		\end{center}
	\end{figure}
	%%%%%%%%%%%%
	
	In Figure \ref{fig:T2SDC&SSDC} we depict the convex hulls associated to the towers of states along with their species scale vectors \cite{Calderon-Infante:2023ler}, which are constructed from the expressions \eqref{eq:KKzetas9d}-\eqref{eq:9dstrings}. Notice that there is a $\mathbb{Z}_2$ symmetry with respect to the $\hat \tau$-axis, which may be thought of as a discrete remnant of the U-duality group of the theory, more specifically its associated Weyl group (see footnote \ref{fnote.Weyl}). Therefore, it is enough to focus just on the upper-half plane in order to check the pattern \eqref{eq:pattern}.
	
	First, notice that for those directions in which both $\vec{\zeta}_{\text{t}}$ and $\vec{\mathcal{Z}}_{\text{sp}}$ are aligned --- i.e. for $\hat T$ parallel to the $\vec{\zeta}_I$ associated to any leading tower $m_I$ --- the condition $\vec{\zeta}_{\text{t}} \cdot \vec{\mathcal{Z}}_{\text{sp}}= \frac{1}{d-2} =\frac{1}{7}$ is satisfied. Moreover, this turns out to be sufficient for the pattern to hold also along intermediate directions. The reason behind is a duality between both convex hull diagrams. In fact, as one can see from Figure \ref{fig:T2SDC&SSDC}, the vertices from one correspond to edges of the other and viceversa, the latter being orthogonal to the former. Therefore, it follows that whenever we take $\vec{\zeta}_{\text{t}}$ (analogously $\vec{\mathcal{Z}}_{\text{sp}}$) to be given by any of the two vertices generating an edge of its corresponding diagram, its inner product with the dual $\vec{\mathcal{Z}}_{\text{sp}}$ (analogously $\vec{\zeta}_{\text{t}}$) orthogonal to such edge reduces to that of the previous `parallel' cases and thus satisfies the pattern \eqref{eq:pattern}.

	\subsection{Maximal supergravity in 8d}
	\label{ss:8d}
	
	As our second example, we now take M-theory compactified on a $\mathbb{T}^3$, leading to 8d $\mathcal N=2$ supergravity, whose bosonic action reads as follows \cite{Obers:1998fb}
	\begin{equation}\label{eq:8d}
		S_{\text{8d}} = \frac{1}{2\kappa_8^2} \int \dd^{8}x\, \sqrt{-g}\,  \left[ \mathcal{R} - \frac{1}{4} \left(g^{m m'} g^{n n'} + \frac{1}{6} g^{m n} g^{m' n'}\right) \partial g_{m n} \partial g_{m' n'} - \frac{1}{2 \mathcal{V}_3^2} \left(\partial c \right)^2 \right] + \ldots\, ,
	\end{equation}
	where $g_{mn}$ is the internal metric, $\mathcal{V}_3$ its overall volume in M-theory units and the scalar $c$ arises by reducing the 11d 3-form field along the 3-cycle. The dots in \eqref{eq:8d} above indicate higher $p$-form fields also present in the gravity supermultiplet. 
	
	As is well-known, the Narain moduli space (see e.g., \cite{Ibanez:2012zz}) of such toroidal compactification --- which is again exact at the classical level --- is enhanced thanks to the additional compact field, $c(x)$, to a coset space of the form $\mathcal{M}_{\text{mod}}=SL(2, \mathbb{Z})\backslash SL(2, \mathbb{R})/U(1) \times SL(3, \mathbb{Z})\backslash SL(3, \mathbb{R})/SO(3)$, where the discrete piece corresponds to the U-duality group of the eight-dimensional theory \cite{Hull:1994ys}. Instead of choosing a parametrization which makes the U-duality group manifest, we take here the same approach as in \cite{Calderon-Infante:2023ler} and simply dimensionally reduce the 9d theory described in \eqref{eq:9d} on a circle. Upon doing so one arrives at
	\begin{equation}\label{eq:8dcanonical}
		S_{\text{8d}} = \frac{1}{2\kappa_8^2} \int \dd^{8}x\, \sqrt{-g}\,  \left[ \mathcal{R} - \left( \partial \hat U \right)^2 - \left(\partial \hat \tau\right)^2 - \left( \partial \hat \rho \right)^2 \right] + \ldots\, ,
	\end{equation}
	with $\hat \rho$ denoting the canonically normalized radion associated to the extra circle within $\mathbb{T}^3$ and the ellipsis indicate further compact scalar fields and higher $p$-forms in the theory.
	
\begin{figure}
		\begin{center}
			\begin{subfigure}{0.475\textwidth}
				\center
				\includegraphics[width=70mm]{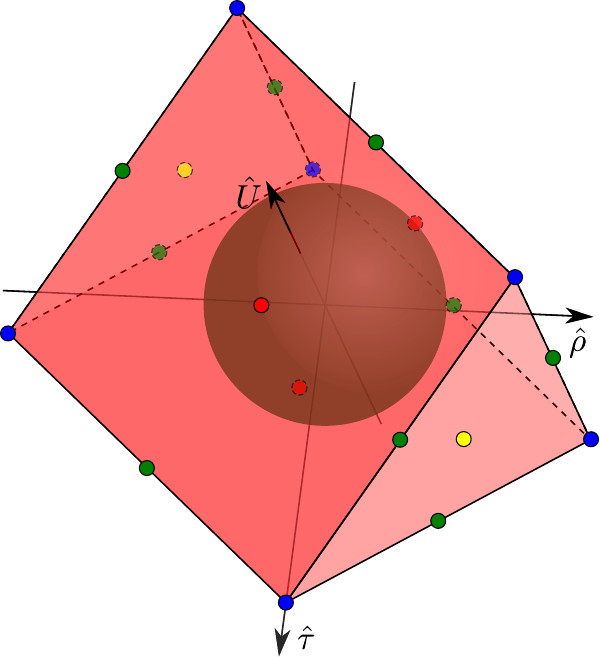}
				\caption{Towers $\{\vec{\zeta}_I\}_I$} \label{sfig:mass8d}
			\end{subfigure}
			\begin{subfigure}{0.475\textwidth}
				\center
				\includegraphics[width=70mm]{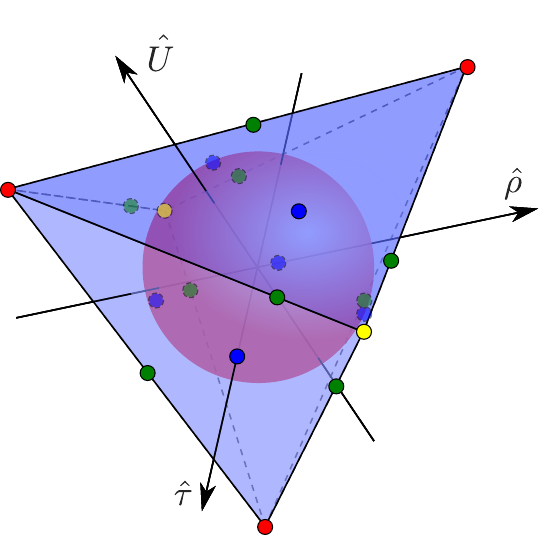}
				\caption{Species scales $\{\vec{\mathcal{Z}}_J\}_J$} \label{sfig:species8d}
			\end{subfigure}
			\caption{\small Convex hull conditions for the masses $\{\vec{\zeta}_I\}_I$ and species scales $\{\vec{\mathcal{Z}}_J\}_J$ of the leading towers in eight-dimensional maximal supergravity, containing the `extremal balls' of radii $\frac{1}{\sqrt{d-2}}=\frac{1}{\sqrt{6}}$ and $\frac{1}{\sqrt{(d-1)(d-2)}}=\frac{1}{\sqrt{42}}$, respectively. The string towers are depicted in red \fcolorbox{black}{red}{\rule{0pt}{6pt}\rule{6pt}{0pt}}, whilst KK towers associated to decompactification of one, two and three dimensions appear in blue \fcolorbox{black}{blue}{\rule{0pt}{6pt}\rule{6pt}{0pt}}, green \fcolorbox{black}{dark-green}{\rule{0pt}{6pt}\rule{6pt}{0pt}} and yellow \fcolorbox{black}{yellow}{\rule{0pt}{6pt}\rule{6pt}{0pt}}, respectively. Note that the string vectors coincide in the two diagrams.}\label{fig:convexhulls8d}
		\end{center}
	\end{figure}

	Similarly to the previous 9d example, the convex hull spanned by the different towers is generated (saturated) here by $\frac{1}{2}$-BPS particles (strings) \cite{Etheredge:2022opl}.\footnote{A complete list of the relevant towers of $\frac{1}{2}$-BPS states in the present eight-dimensional setup can be found e.g., in Table 1 from ref. \cite{Calderon-Infante:2023ler}.} Let us start with the BPS particles. The advantage of choosing the parametrization in \eqref{eq:8dcanonical} is that one can essentially read off most of the scalar charge-to-mass vectors characterizing the infinite towers of states from the previous 9d example, by simply dimensionally reducing those. Therefore, for the KK towers one obtains
	\begin{equation} \label{eq:KKvectors}
		\begin{split} 
			\vec{\zeta}_{\text{KK},\, 1} &= \left( \frac{1}{\sqrt{2}} , \frac{1}{\sqrt{42}}, \frac{3}{\sqrt{14}} \right) \, , \qquad \vec{\zeta}_{\text{KK},\, 1'} = \left( -\frac{1}{\sqrt{2}} , \frac{1}{\sqrt{42}}, \frac{3}{\sqrt{14}} \right) \, ,\\
			\vec{\zeta}_{\text{KK},\, 1''} &= \left( 0 , \sqrt{\frac{7}{6}}, 0 \right) \, ,
		\end{split}
	\end{equation}
	where the last $\zeta$-vector arises from the extra $\mathbb{S}^1$ and the notation is $\vec{\zeta}=\left(\zeta^{\hat \tau}, \zeta^{\hat \rho}, \zeta^{\hat U}\right)$. Analogously, one finds a triplet of towers comprised by M2-branes wrapping different 2-cycles within $\mathbb{T}^3$, with the following charge-to-mass vectors
	\begin{equation} \label{eq:M2vectors}
		\begin{split} 
			\vec{\zeta}_{\text{M},\, 1} &= \left( \frac{1}{\sqrt{2}} , -\frac{5}{\sqrt{42}}, -\frac{1}{\sqrt{14}} \right) \, , \qquad \vec{\zeta}_{\text{M},\, 1'} = \left( -\frac{1}{\sqrt{2}} , -\frac{5}{\sqrt{42}}, -\frac{1}{\sqrt{14}} \right) \, ,\\
			\vec{\zeta}_{\text{M},\, 1''} &= \left( 0, \frac{1}{\sqrt{42}}, -\sqrt{\frac{8}{7}} \right)  \, ,
		\end{split}
	\end{equation}
	where the last one is inherited from the 9d setup, whilst the first two are new \cite{Calderon-Infante:2023ler}. Notice that these vectors already generate the convex hull of light towers, see Figure \ref{sfig:mass8d}. However, there also exist additional towers of states associated to the oscillation modes of critical (type IIA) strings, whose $\zeta$-vectors read as
	\begin{equation} \label{eq:stringvectors}
		\begin{split} 
			\vec{\zeta}_{\text{osc}} &= \left( \frac{1}{2\sqrt{2}} , \frac{1}{\sqrt{42}}, -\frac{1}{2 \sqrt{14}} \right) \, , \qquad \vec{\zeta}_{\text{osc}'} = \left( -\frac{1}{2\sqrt{2}} , \frac{1}{\sqrt{42}}, -\frac{1}{2 \sqrt{14}} \right) \, ,\\
			\vec{\zeta}_{\text{osc}''} &= \left( 0 , -\sqrt{\frac{2}{21}}, \frac{1}{\sqrt{14}} \right)  \, ,
		\end{split}
	\end{equation}
	and which happen to lie at the extremal ball, thus saturating the sharpened Distance Conjecture \cite{Etheredge:2022opl}. The first two are inherited from the 9d example above (c.f. \eqref{eq:9dstrings}), whilst the third one arises from the M2-brane of 11d supergravity wrapped along the additional circle.
	
	On a next step, one can analogously compute the would-be species scale vectors within each asymptotic direction of the 8d moduli space. This was done in \cite{Calderon-Infante:2023ler}, and we simply state here the results highlighted there. First, for the triplet of type IIA critical strings one finds $\mathcal{Z}$-vectors which coincide with those of their associated charge-to-mass ratios, namely \eqref{eq:stringvectors}. Additionally, one can extract a total of six species scale vectors from the $n=1$ towers --- thus signaling decompactification of one extra dimension --- arising either from Kaluza-Klein modes or the M2-particles (see eqs. \eqref{eq:KKvectors} and \eqref{eq:M2vectors}) via the relation \cite{Calderon-Infante:2023ler}\footnote{\label{fnote:lambdafnofmass}Relation \eqref{eq:masterformula} arises from the usual dependence of the species scale on the characteristic mass of the infinite tower of states, namely $\Lambda_{\text{sp}}\, \sim\, m_{\text{tower}}^{\frac{n}{d-2+n}}$, with $n$ denoting the density parameter of the tower \cite{Castellano:2021mmx, Castellano:2022bvr}.}
	\begin{equation}\label{eq:masterformula}
		\vec{\mathcal{Z}}_{\text{sp, t}} = \frac{n}{d-2+n} \, \vec{\zeta}_{\text{t}}\, ,
	\end{equation}
	with $n\in \mathbb{N}$ being an effective density parameter \cite{Castellano:2021mmx} capturing the number of spacetime dimensions that decompactify upon taking the asymptotic limit (see also \eqref{Mp}).
	
	However, this turns out not being enough so as to fully generate the convex hull diagram for the species scale vectors. Indeed, as discussed in \cite{Calderon-Infante:2023ler}, the role of generating/saturating towers gets exchanged between the two hulls, and it is now crucial to take also into account the combined effective towers. In particular, one can easily construct $\frac{1}{2}$-BPS particles from bound states of the aforementioned $n=1$ towers \cite{Obers:1998fb}, resulting in the following $n=2$ triplets \cite{Calderon-Infante:2023ler}
	\begin{equation} \label{eq:n=2KKvectors}
		\begin{split} 
			\vec{\mathcal{Z}}_{\text{KK},\,2} &= \left( 0, \frac{1}{4 \sqrt{42}}, \frac{3}{4 \sqrt{14}} \right) \, , \qquad \vec{\mathcal{Z}}_{\text{KK},\,2'} = \left( \frac{1}{8 \sqrt{2}}, \frac{1}{ \sqrt{42}}, \frac{3}{8 \sqrt{14}} \right) \, ,\\
			\vec{\mathcal{Z}}_{\text{KK},\, 2''} &= \left( -\frac{1}{8 \sqrt{2}}, \frac{1}{ \sqrt{42}}, \frac{3}{8 \sqrt{14}} \right) \, ,
		\end{split}
	\end{equation}
	for Kaluza-Klein bound states, where the notation follows that of \eqref{eq:effectivespecies}. Analogously, one finds
	\begin{equation} \label{eq:n=2M2vectors}
		\begin{split} 
			\vec{\mathcal{Z}}_{\text{M},\, 2} &= \left( 0, -\frac{5}{4 \sqrt{42}}, -\frac{1}{4 \sqrt{14}} \right) \, , \qquad \vec{\mathcal{Z}}_{\text{M},\, 2'} = \left( \frac{1}{8 \sqrt{2}}, -\frac{1}{ 2 \sqrt{42}}, -\frac{5}{8 \sqrt{14}} \right) \, ,\\
			\vec{\mathcal{Z}}_{\text{M},\, 2''} &= \left( -\frac{1}{8 \sqrt{2}}, -\frac{1}{ 2 \sqrt{42}}, -\frac{5}{8 \sqrt{14}} \right) \, ,
		\end{split}
	\end{equation}
	for bound states (with $n=2$ again) of M2-particles and also
	\begin{equation} \label{eq:n=2KKM2vectors}
		\begin{split} 
			\vec{\mathcal{Z}}_{\text{KK-M},\, 2} &= \left( 0, \frac{1}{\sqrt{42}}, -\frac{1}{2 \sqrt{14}} \right) \, , \qquad \vec{\mathcal{Z}}_{\text{KK-M},\, 2'} = \left( -\frac{1}{4 \sqrt{2}}, -\frac{1}{2 \sqrt{42}}, \frac{1}{4 \sqrt{14}} \right) \, ,\\
			\vec{\mathcal{Z}}_{\text{KK-M},\, 2''} &= \left( \frac{1}{4 \sqrt{2}}, -\frac{1}{2 \sqrt{42}}, \frac{1}{4 \sqrt{14}} \right) \, ,
		\end{split}
	\end{equation}
	arise from BPS bound states between wrapped M2-branes and KK replica \cite{Calderon-Infante:2023ler}. These come along with their corresponding `effective' $\zeta$-vectors, which are parallel to the $\mathcal{Z}$ ones and may be defined as in \eqref{eq:effectivemass} above. All these species scales correspond to the Planck scale of the possible two-higher dimensional theories (i.e. 10d theories) that arise in the diverse decompactification limits.
	
	Additionally, there is an $SL(2, \mathbb{Z})$-doublet of $n=3$ towers, signalling full decompactification of the 3-torus back to 11d M-theory, whose species scale vectors read
	\begin{equation}\label{eq:n=3vectors}
		\vec{\mathcal{Z}}_{\text{KK},\,3} = \left( 0, \frac{1}{ \sqrt{42}}, \frac{2}{3 \sqrt{14}} \right) \, , \qquad
		\vec{\mathcal{Z}}_{\text{M},\, 3} = \left( 0, -\frac{1}{ \sqrt{42}}, -\frac{2}{3 \sqrt{14}} \right) \, .
	\end{equation}
	%
	
	%%%%%%%%%%%%%%%%%%%%%%%%%%%%%%%%%%%%
	%%%%%%%%%%%%%%%%%%%%%%%%%%%%%%%%%%%%
	
\begin{figure}[htb]
		\begin{center}
			\includegraphics[scale=.8]{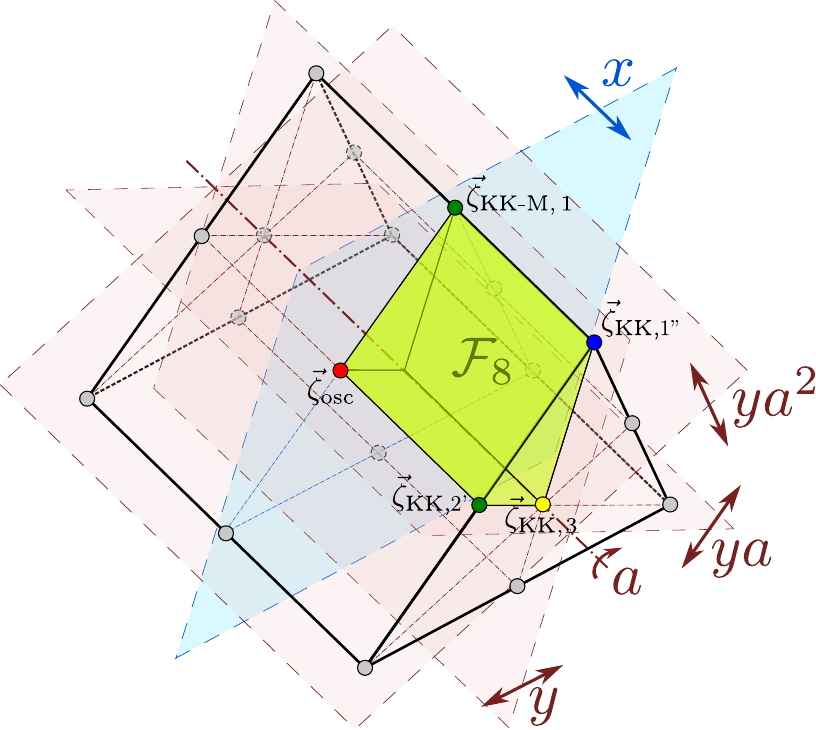}
			\caption{\small Sketch of the fundamental domain $\mathcal{F}_8$ of the $S_3\times S_2=\left \langle y,\,a: y^2=a^3=e,\, yay=a^{-1} \right\rangle\times\left\langle x: x^2=e\right\rangle$ symmetry group acting on the scalar charge-to-mass vectors associated to the relevant towers in 8d maximal supergravity. The figure also shows the towers spanning the fundamental domain as well as the individual actions of the symmetry group. An analogous fundamental domain for the species scale polytope from Figure \ref{sfig:species8d} could be built, as both $\{\vec{\zeta}_I\}_I$ and $\{\vec{\mathcal{Z}}_J\}_J$ present the same symmetries.} 
			\label{fig:sym}
		\end{center}
	\end{figure}

	With this at hand, one can draw the corresponding species scale convex hull diagram, which is depicted in Figure \ref{sfig:species8d}. In order to check \eqref{eq:pattern} one can do as in the 9d example above and focus --- thanks to the U-duality group of the theory --- on a strictly smaller polyhedron. Indeed, since the symmetry group of the convex polytope is $S_2\times S_1$, it is enough for our purposes to take 1/12 of the full diagram, namely the one containing e.g., the set $\lbrace \vec{\zeta}_{\text{KK},\, 1''}, \vec{\zeta}_{\text{osc}}, \vec{\zeta}_{\text{KK-M},\,2}, \vec{\zeta}_{{\rm KK},\, 2'}, \vec{\zeta}_{{\rm KK},\, 3}\rbrace$. Figure \ref{fig:sym} depicts the aforementioned vertices and the fundamental domain they span, as well as the discrete symmetries associated to the diagram. As can be easily verified, along these directions the product $\vec{\zeta}_{\text{t}} \cdot \vec{\mathcal{Z}}_{\text{sp}}= \frac{1}{d-2} =\frac{1}{6}$ is verified since the species scale and the charge-to-mass vectors are aligned. As also happened with our previous example, this is all we need to check in order to get convinced that the pattern holds along every asymptotic (intermediate) direction as well, since from Figure \ref{fig:convexhulls8d} it becomes clear that the vertices spanning one convex hull are orthogonal to the faces of the other, and viceversa.	
	
	%%%%%%%%%%%%
	%%%%%%%%%%%%
	
	\subsection{Maximal supergravity in $d<8$}
	\label{ss:generaldim}
	
	After the previous concrete examples, we will argue in what follows that the results discussed there hold more generally in the context of maximal supergravity. The strategy will be to isolate the key ingredients from the nine- and eight-dimensional setups and translate them into the more general case in $d$ spacetime dimensions. This is done in Section \ref{sss:sketch}, whilst the computational details are relegated to Section \ref{sss:generalcomputations}.
	
	\subsubsection{A sketch of the proof}
	\label{sss:sketch}
	
	The argument proceeds in a recursive manner, relying essentially on the duality properties of the theory as well as the uniqueness of maximal supergravity for $d\leq9$ \cite{Hull:1994ys}. 
	
	Let us start by noticing from the examples above that the charge-to-mass vectors associated to towers with density parameter $n$ lie always along a facet\footnote{Actually, they are located at the point of the facet closest to the origin.} of the convex hull polytope with \emph{dimension} equal to $n-1$ (see Figures \ref{fig:T2SDC&SSDC} and \ref{sfig:mass8d}), whilst those vectors controlling the species scale belong to a facet of \emph{codimension} $n$ (c.f. Figures \ref{fig:T2SDC&SSDC} and \ref{sfig:species8d}).\footnote{The vectors associated to string towers, having density $n\to\infty$, appear at facets of maximal (co-)dimension for the charge-to-mass (resp. species) diagram.} This holds in lower spacetime dimensions as well, since the length of the vectors is fully determined once $d$ and $n$ are specified (c.f. \eqref{eq:zeta&speciesveconemodulus}), and it is indeed a  clear manifestation of the duality between both convex hulls in the sense that the facets of one correspond to the vertices of the other, and viceversa. 
	
	One also notices that the diagrams present some symmetry properties that reflect the U-duality group of the quantum theory (see Table \ref{tab:irreps} below). This, in turn, allows us to restrict ourselves to some \emph{fundamental domain} (i.e. a subset of the original convex hull) containing all the relevant information for the diagram, whilst the remaining parts of the hull appear to be mere copies of the former, obtained upon acting with the different elements of the symmetry group.
	In fact, one may view such fundamental domain as the region whose boundaries precisely arise as fixed submanifolds under some element(s) of the symmetry group, which moreover coincides with the Weyl subgroup associated to the U-duality group (see Figure \ref{fig:sym}).\footnote{\label{fnote.Weyl}Consider some EFT with a $\mathsf{n}$-dimensional moduli space $\mathcal{M}_{\varphi}$ parametrized by the scalars $\{\varphi^i\}_{i=1}^{\mathsf{n}}$. The U-duality group $G$ of said theory transforms said scalars in a way such that the different states of the EFT are mapped to one another. However, if we are interested only in non-compact scalars (thus ignoring compact axionic fields, since they play no role for our considerations, see \cite{Calderon-Infante:2023ler}), some of the transformations of $G$ might affect only the compact scalars, which we left fixed. These transformations are the elements of a \emph{maximal torus of $G$, $T_G\hookrightarrow G$}, which is the maximal Abelian, connected and compact subgroup of $G$. As in general $T_G$ is not a normal subgroup of $G$, in order to properly quotient $G$ by $T_G$, the normalizer $N_G (T_G)=\{g\in G:gT_G=T_Gg\}$ is introduced, corresponding to the largest subgroup of $G$ such that $T_G$ is a normal subgroup. Then the \emph{Weyl group} of $G$ is defined as ${\rm W(G)}:=N_G(T_G)/T_G$, and it will correspond with the symmetries of the non-compact scalars (and thus of the different vectors under consideration). It is finite (there are only so many ways of exchanging points) and a subgroup of $GL(\mathbb{R}^{\mathsf{m}})$, where $\mathsf{m}\leq \mathsf{n}$ is the number of unbounded moduli.}
	
	Therefore, what we need first to know is how to select a fundamental domain $\mathcal{F}_d$, in practice. For this, we note that the towers of states with $n=1$ arrange themselves into a \emph{single} irreducible representation of the U-duality group for $d<9$, as shown in the second column of Table \ref{tab:irreps}. These include perturbative (i.e. KK, winding modes, etc.) as well as non-  perturbative states (wrapped branes, KK-monopoles, etc.), and for us it will be enough to focus on just one of them, which we take to be of perturbative nature, namely a Kaluza-Klein vector. Hence, we work inductively, starting from M-theory compactified on $\mathbb{T}^k$ down to $d+1=11-k$ dimensions, where we assume the pattern \eqref{eq:pattern} to hold. Then, we dimensionally reduce on an extra circle, leading to M-theory on $\mathbb{T}^k \times \mathbb{S}^1 \simeq \mathbb{T}^{k+1}$, and we consider the `cone' of asymptotic directions comprised by the large radius direction (of the additional $\mathbb{S}^1$) and the KK replica of the vectors determining some fundamental domain, $\mathcal{F}_{d+1}$, in the parent $(d+1)$-dimensional theory.
	Upon doing so, one can easily check (see Section \ref{sss:generalcomputations} below) that eq. \eqref{eq:pattern} is verified along any asymptotic trajectory within $\mathcal{F}_d$. Finally, since the pattern has already been shown to hold for $k=1,2,3$ (corresponding to maximal supergravity in ten, nine and eight dimensions, respectively), one concludes that it extends to all lower dimensional cases as well.
	
	%%%%%%%%%%%%%%%%%%%%%%%%%%%%%%%
	\begin{table}[h!!]\begin{center}
			\renewcommand{\arraystretch}{1.00}
			\begin{tabular}{|c||c|c|c|c|}
				\hline
				$d$ &  U-duality group  & Irrep. &  $\{\vec{\zeta}_I\}_I$ sym. group & Order\\
				\hline 
				10A & 1  & $\mathbf{1}$ & 1 & 1 \\
				10B & $SL(2,\mathbb{Z})$  & $\mathbf{2}$ & $\mathbb{Z}_2 \simeq S_2$ & 2 \\
				9 & $SL(2,\mathbb{Z})$   & $\mathbf{2} \oplus \mathbf{1}$ & $\mathbb{Z}_2 \simeq S_2$ & 2 \\
				8 & $SL(2,\mathbb{Z}) \times SL(3,\mathbb{Z})$  & $(\mathbf{2},\mathbf{3} )$ & $S_2\times S_3$ & 12 \\
				7 & $SL(5,\mathbb{Z})$  & $\mathbf{10}$ & $S_5$ & 120 \\
				6 & $SO(5, 5, \mathbb{Z})$  & $\mathbf{16}$ & $\text{W}\, ({\rm Spin}(5,5))$ & $1\,920$ \\
				5 & $E_{6 (6)} (\mathbb{Z})$  & $\mathbf{27}$ & $\text{W}\, (E_6)$ & $51\,840$ \\
				4 & $E_{7 (7)} (\mathbb{Z})$  & $\mathbf{56}$ & $\text{W}\, (E_7)$ & $2\,903\,040$ \\
				3 & $E_{8 (8)} (\mathbb{Z})$  & $\mathbf{248}$ & $\text{W}\, (E_8)$ & $719\,953\,920$ \\
				\hline
			\end{tabular}
			\caption{\small U-duality representations of the particle multiplets in M-theory on $\mathbb{T}^{k}$ \cite{Obers:1998fb} for $10\geq d\geq 3$. Note that there are two possibilities for $d=10$, corresponding to ten-dimensional type IIA and type IIB supergravities. The second column shows the vector and charge representations for $n=1$ BPS towers, which for $d<9$ arrange into a single irrep. Additionally, the symmetry group acting on the $\zeta$ (equivalently $\mathcal{Z}$)-vectors is displayed, which corresponds to the Weyl subgroup of the associated U-duality group, as well as its finite order \cite{wilson2009finite}. The latter controls the number of copies of $\mathcal{F}_d$ that comprise the convex hull of $\zeta$- or $\mathcal{Z}$-vectors.}
			\label{tab:irreps}
		\end{center}
	\end{table}
	%%%%%%%%%%%%%%%%%%%%%%%%%%%%%%%%%
	
	\subsubsection{Relevant computations}
	\label{sss:generalcomputations}
	The aim of this subsection is to provide some of the details that corroborate our claims before regarding the analysis of the pattern \eqref{eq:pattern} in $d<8$ maximal supergravity. Let us assume that we have already fixed a fundamental domain $\mathcal{F}_d$, as outlined in Section \ref{sss:sketch}. Such polytope is thus generated by the reference $n=1$ tower, with charge-to-mass vector $\vec{\zeta}_{\rm KK, 1}$, together with the KK replica of those vectors determining the fundamental domain of the theory in one dimension higher (see Figure \ref{fig:sym}). In the following, we will denote the latter as $\lbrace \vec{\zeta}_{{\rm KK,}\,n+1} \rbrace$, with $n\in\{1,\ldots,10-d,\infty\}$. 
	First, we notice that whenever we focus on a given direction determined by some $\vec{\zeta}$ within $\mathcal{F}_d$, the pattern is automatically satisfied, since both the species and charge-to-mass vectors are associated to one and the same tower and thus parallel to each other (c.f. \eqref{eq:patternKKn}). The non-trivial task is to show that eq. \eqref{eq:pattern} is still satisfied along intermediate directions as well, where the vectors $\lbrace \vec{\zeta}_{\text{t}}, \vec{\mathcal{Z}}_{\text{sp}}\rbrace$ are no longer aligned. To do so, we first prove the following claim:
	\begin{theorem}\label{claim1}
		The leading tower of states within $\mathcal{F}_d$ always corresponds to $\vec{\zeta}_{\rm KK,\, 1}$. Additional towers $m_I$ can become light at the same rate along certain asymptotically geodesic trajectories, characterized by some normalized tangent vector $\hat{T}$.
	\end{theorem}
	This can be easily shown upon computing the inner product between $\vec{\zeta}_{\rm KK,\, 1}$ and any other charge-to-mass vector belonging to the set $\lbrace \vec{\zeta}_{{\rm KK,}\,n+1} \rbrace$. One finds
	\begin{equation}\label{eq:proofKK1dominant}
		\vec{\zeta}_{\rm KK,\, 1} \cdot \vec{\zeta}_{\text{KK},\, n+1} = \vec{\zeta}_{\rm KK, 1} \cdot \left[\frac{1}{n+1} \left( \vec{\zeta}_{\rm KK,\, 1} + n\, \vec{\zeta}_{{\rm KK,}\, n} \right) \right] = \frac{d+n-1}{(d-2) (n+1)} = |\vec{\zeta}_{{\rm KK},\, n+1}|^2\, ,
	\end{equation}
	where we have used eq. \eqref{eq:effectivezeta} in the second equality. The fact that it coincides with $|\vec{\zeta}_{{\rm KK},\, n+1}|^2$ implies, geometrically, that the segment of the hull determined by both vectors is indeed orthogonal to $\vec{\zeta}_{{\rm KK},\, n+1}$ itself (see e.g., Figure \ref{fig:MT2radions}). Now, given any normalized tangent vector $\hat{T}$, we can split it into parallel and perpendicular components with respect to the plane spanned by $\vec{\zeta}_{\rm KK,\, 1}$ and $\vec{\zeta}_{\text{KK,}\, n+1}$, such that $\hat{T}=\hat{T}^\parallel+\hat{T}^\perp$, where $\hat{T}^{\parallel}=a\,\vec{\zeta}_{\rm KK,\, 1}+b\,\vec{\zeta}_{\text{KK,}\, n+1}$, and with $a,\,b\geq 0$. Therefore, we have
	\begin{align}
		\hat{T}\cdot(\vec{\zeta}_{\rm KK,\, 1} - \vec{\zeta}_{\text{KK},\, n+1})&=\hat{T}^\parallel\cdot(\vec{\zeta}_{\rm KK,\, 1} - \vec{\zeta}_{{\rm KK},\, n+1})\notag\\
		&=a\, \vec{\zeta}_{\rm KK,\, 1}\cdot(\vec{\zeta}_{\rm KK,\, 1} - \vec{\zeta}_{{\rm KK},\, n+1})+ b\,\vec{\zeta}_{{\rm KK},\, n+1}\cdot(\vec{\zeta}_{\rm KK,\, 1} - \vec{\zeta}_{{\rm KK},\, n+1})\notag\\
		&=a\,\underbrace{(|\vec{\zeta}_{{\rm KK},\, 1}|^2-|\vec{\zeta}_{{\rm KK},\, n+1}|^2)}_{>0}\geq 0,
	\end{align}
	so that the $\vec{\zeta}_{\rm KK,\, 1}$ tower always becomes light faster than $\vec{\zeta}_{\text{KK},\, n+1}$ except for $a=0$, namely when $\hat{T}^{\parallel} \propto \vec{\zeta}_{\text{KK},\, n+1}$, in which case they do so at the same rate. This ends our proof of Claim \ref{claim1} above.
	On the other hand, the species scale strongly depends on the chosen asymptotic trajectory (see e.g., Figure \ref{fig:convexhulls8d}). Hence, in order to check the pattern \eqref{eq:pattern}, one needs to demonstrate the following statement:
	\begin{theorem}\label{claim2}
		For any possible species scale vector spanning $\mathcal{F}_d$, that we collectively denote $\lbrace \vec{\mathcal{Z}}_{\rm{KK,}\,n+1} \rbrace$ with $n\in\{1,\ldots,10-d,\infty\}$, we find:
		\begin{subequations}
			\begin{equation}\label{eq:claim2a}
				\vec{\zeta}_{\rm{KK,}\, 1} \cdot \vec{\mathcal{Z}}_{\rm{KK},\, n+1} = \frac{1}{d-2}\, ,
			\end{equation}
			\begin{equation}\label{eq:claim2b}
				\vec{\zeta}_{\rm{KK},\, n'+1} \cdot \vec{\mathcal{Z}}_{\rm{KK},\, n+1}=\frac{1}{d-2}\, .
			\end{equation}
		\end{subequations}
		In particular, the second equality holds provided the parent vectors satisfy the pattern in the higher $(d+1)$-dimensional theory. 
	\end{theorem}
	Note that the first part of the claim above trivially follows from eqs. \eqref{eq:masterformula} and \eqref{eq:proofKK1dominant}. The second statement, however, requires a bit more work. Intuitively, it means that the condition \eqref{eq:pattern} is consistent (or preserved) under dimensional reduction. 
	Thus, we take, without loss of generality, some vector $\vec{\mathcal{Z}}_{\text{KK},\, n+1}$ as the one dominating certain asymptotic region of moduli space within the fundamental domain, and we consider the inner product \eqref{eq:claim2b}. Here, $\vec{\zeta}_{\text{KK},\, n'+1}$ is taken to be any other charge-to-mass vector within $\mathcal{F}_d$ such that it verifies the pattern with respect to $\vec{\mathcal{Z}}_{\text{KK},\, n+1}$ in the parent $(d+1)$-dimensional theory. Recall that, upon dimensionally reducing some vectors $\vec{\zeta}^{\,(d+1)}_{\text{KK},\, n'}$ and $\vec{\mathcal{Z}}^{\,(d+1)}_{\text{KK},\, n}$ on a circle, one gets \cite{Calderon-Infante:2023ler}
	\begin{align}
		\vec{\zeta}_{\text{KK},\, n'} = \left( \vec{\zeta}^{\,(d+1)}_{\text{KK},\, n'}\, ,\, \frac{1}{\sqrt{(d-1)(d-2)}}\right)\, , \quad  \vec{\mathcal{Z}}_{\text{KK},\, n+1} = \left( \vec{\mathcal{Z}}^{\,(d+1)}_{\text{KK},\, n}\, ,\, \frac{1}{\sqrt{(d-1)(d-2)}}\right)\, ,
	\end{align}
	where the first components of both vectors are directly inherited from the ones of the theory in $d+1$ dimensions, whilst the last entry corresponds to the $\mathbb{S}^1$ radion direction (see also Appendix \ref{ap:generalities}). Hence, requiring $\vec{\zeta}_{\text{KK},\, n'+1}$ to verify the pattern in the higher-dimensional theory translates into the following statement
	\begin{equation}\label{eq:patternd+1}
		\vec{\zeta}^{\,(d+1)}_{\text{KK},\, n'} \cdot \vec{\mathcal{Z}}^{\,(d+1)}_{\text{KK},\, n} = \frac{1}{d-1}\, ,
	\end{equation}
	such that we finally obtain
	\begin{align}
		\vec{\zeta}_{\text{KK},\, n'+1} \cdot \vec{\mathcal{Z}}_{\text{KK},\, n+1} &= \left[\frac{1}{n'+1} \left( \vec{\zeta}_{\text{KK},\, 1} + n'\, \vec{\zeta}_{\text{KK},\, n'} \right) \right] \cdot \vec{\mathcal{Z}}_{\text{KK},\, n+1} \notag\\
		& = \frac{1}{n'+1} \vec{\zeta}_{\text{KK},\, 1} \cdot \vec{\mathcal{Z}}_{\text{KK},\, n+1} + \frac{n'}{n'+1} \left[ \vec{\zeta}^{\,(d+1)}_{\text{KK},\, n'} \cdot \vec{\mathcal{Z}}^{\,(d+1)}_{\text{KK},\, n} + \frac{1}{(d-1)(d-2)}\right] \notag\\
		&=\frac{1}{d-2}\, ,
	\end{align}
	where in order to arrive at the last equality one needs to use eqs. \eqref{eq:claim2a} and \eqref{eq:patternd+1} above. This completes the proof of Claim \ref{claim2}, which ensures that both convex hull diagrams, namely that associated to the $\zeta$-vectors and the species one, are completely dual to each other (with respect to a sphere of radius $\frac{1}{\sqrt{d-2}}$), as also happened for the 9d and 8d case. This proves that the pattern \eqref{eq:pattern} holds in complete generality in flat space compactifications with maximal supergravity.
	For completeness, let us mention that this property holds as well between vectors in- and outside the selected fundamental region (see e.g., Figures \ref{fig:T2SDC&SSDC} and \ref{fig:convexhulls8d}). Notice that this follows immediately from the analysis restricted to $\mathcal{F}_d$ just performed, since any vector outside the fundamental domain can be reached from another one within the latter via the action of some element $g\in G$ of the finite symmetry group $G$ of the diagram. However, since $G$ is a subgroup of the U-duality group of the theory (c.f. Table \ref{tab:irreps}), and this itself is a subgroup of the coset which parameterizes the moduli space (see e.g., \cite{Cecotti:2015wqa}), the scalar product defined with respect to the bi-invariant metric $\mathsf{G}_{i j}$ is automatically preserved.

	\section{Examples in setups with 16 supercharges}
	\label{s:16supercharges}
	
	As we lower the level of supersymmetry, Kaluza-Klein replica are not necessarily BPS anymore, and the vectors generating the convex hull of the towers and the species scale can change upon exploring different regions of the moduli space. Satisfying the pattern in those cases becomes less trivial and provides strong evidence for it beyond maximal supergravity. In this section, we will discuss certain slices of the moduli space of heterotic string theory on a circle, for which all asymptotic corners (as well as how they fit together) are well-known \cite{Aharony:2007du}. In this case, it is still possible to define a globally flat metric\footnote{When referring to a `flat' frame in a certain moduli space we always ignore the compact (axionic) directions, since taking them into account usually introduces a non-vanishing curvature, thus obstructing the definition of a global flat chart.} which will allow us to draw the convex hull in a global fashion \cite{Etheredge:2023odp}, and discuss how it changes as we move in moduli space. For completeness, we will also briefly discuss the case of M-theory on $K3$, before turning in the rest of the paper to more complicated lower-supersymmetric 4d setups.
	
	\subsection{Heterotic string theory in 9d}
	\label{ss:het s1}
	
\begin{figure}
		\begin{center}
			\begin{subfigure}{0.475\textwidth}
				\center
				\includegraphics[width=70mm]{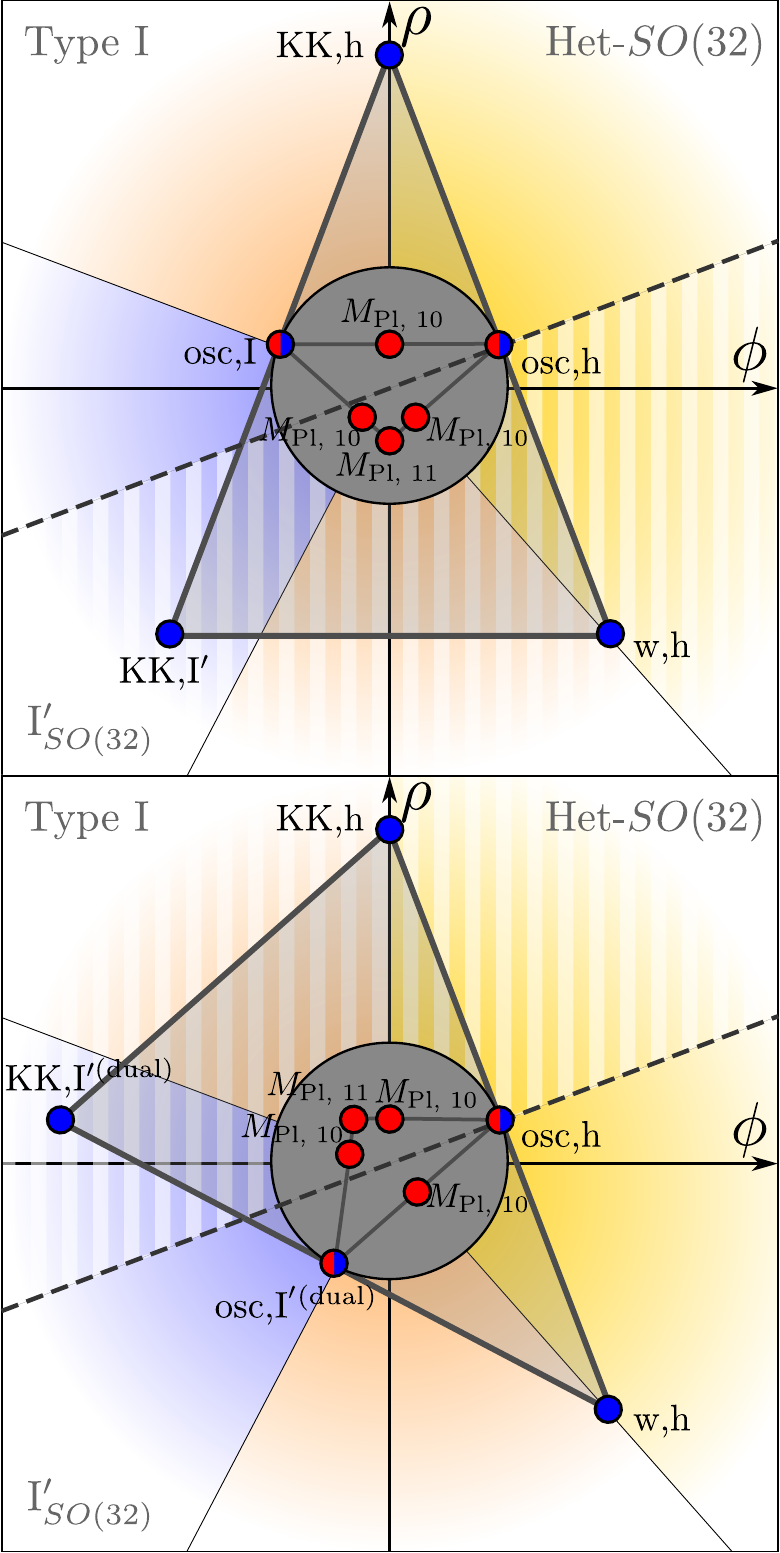}
				\caption{$SO(32)$} \label{sfig:so32}
			\end{subfigure}
			\begin{subfigure}{0.475\textwidth}
				\center
				\includegraphics[width=70mm]{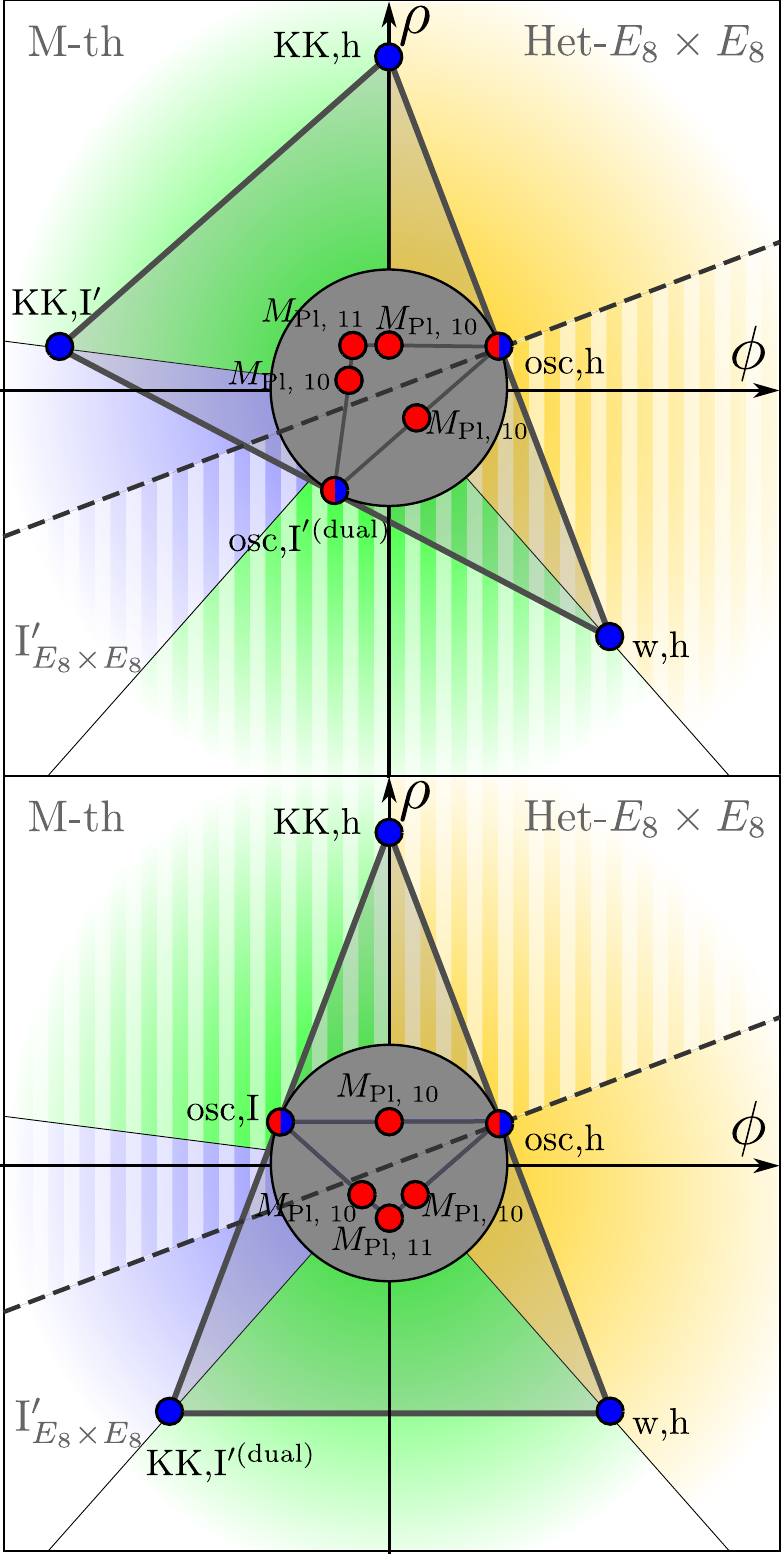}
				\caption{$E_8\times E_8$} \label{sfig:e8e8}
			\end{subfigure}
			\caption{\small Scalar charge-to-mass vectors for the towers (blue) and species scales (red) observed for the $SO(32)$ and $E_8\times E_8$ slices of the heterotic on $\mathbb{S}^1$ moduli spaces, depending on whether the infinite distance limits (along the \emph{non}-dashed regions) are above or below the self-dual line (dashed), following the convention for the canonically normalized moduli as in \cite{Etheredge:2023odp}.}\label{fig:hets1}
		\end{center}
	\end{figure}

\begin{figure}[htb]
		\begin{center}
			\includegraphics[scale=1]{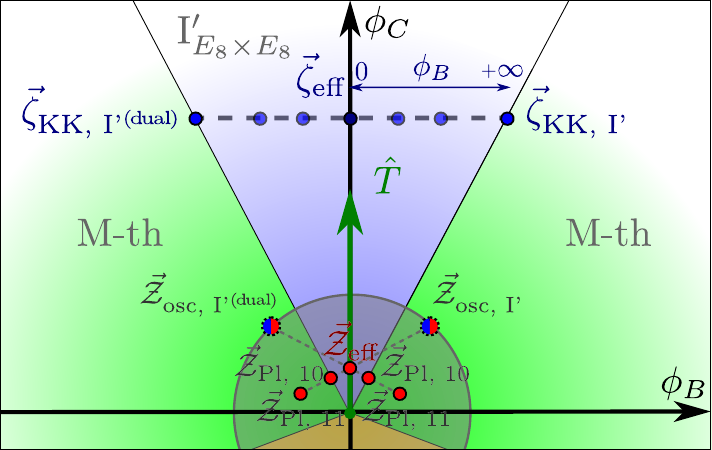}
			\caption{\small Details of the $E_8\times E_8$ slice of $\mathcal{M}_{\rm het}$, parameterized in terms of $\{\phi_B,\phi_C \}$. When moving with $\hat{T}=(0,1)$, thus parallel to the self-dual line, the type I$'$ KK tower (and its dual) has a scalar charge-to-mass vector $\vec{\zeta}_{\rm KK,\; I'}$ whose expression depends on the distance $\phi_B$ of the trajectory to the self-dual line (c.f. eq. \eqref{eq: sliding}), coalescing for $\phi_B\to 0$ to $(0,\frac{5}{2\sqrt{7}})$. The fixed $\vec{\mathcal{Z}}_{\rm eff}=\left(0,\frac{2}{5\sqrt{7}} \right)$ vector satisfying the pattern is also depicted. Additional $\mathcal{Z}$-vectors associated to the obstructed emergent string towers as well as the heavier Planck masses are also presented. The $SO(32)$ slice has an analogous behavior, with $\vec{\zeta}_{\rm KK,\;I'}$ located on the other side of the self-dual line, see \cite{Etheredge:2023odp}.} 
			\label{fig:sliding}
		\end{center}
	\end{figure}

	A typical example of a theory with 16 supercharges is that obtained by the compactification of the heterotic string on $\mathbb{S}^1$. This results in an 18-dimensional moduli space $\mathcal{M}_{\text{het}}=\mathbb{R}\times SO(17,1;\mathbb{Z})\backslash SO(17,1;\mathbb{R})/SO(17)$, parametrizing the 10d dilaton $\phi$, radion $\rho$ and the 16 Wilson lines. We can then study two-dimensional $\{\phi,\rho\}$ slices of $\mathcal{M}_{\text{het}}$ with fixed Wilson line moduli. In particular, we will be interested in two concrete slices of the moduli space of rank 16 (for the gauge group), which can be obtained by compactifying the $SO(32)$ and $E_8\times E_8$ 10d heterotic string theories on a circle, with all Wilson lines turned off. We expect equivalent results for the disconnected components of the moduli space with lower rank \cite{Aharony:2007du, Etheredge:2023odp}. Depending on the values taken by the dilaton and radion vevs, the theory can be better presented in terms of a different dual description, resulting in a finite chain of duality frames, as shown in Figure \ref{fig:hets1} and described in more detail in \cite{Aharony:2007du, Etheredge:2023odp}. Both slices present a self-dual line at $\rho=\frac{1}{\sqrt{7}}\phi$ (depicted in red in Figure \ref{fig:hets1} below) splitting each diagram in two mirrored regions.
	%%%%%%%%%%%%%%%%%%%%%%%%%%%%%%%%%%%%
	%%%%%%%%%%%%%%%%%%%%%%%%%%%%%%%%%%%%
	The most interesting duality frame is that corresponding to type I$'$ string theory, which is an orientifolded version of type IIA on a circle, with two $O8^-$ planes at the endpoints of the interval and 16 D8-branes, whose location determines the gauge group (16 of then stacked on one orientifold for $SO(32)$ and a symmetric pair of 8 D8-stacks for $E_8\times E_8$), with the dilaton running between the $O8^-$ planes and the branes \cite{Polchinski:1995df}. As a result, the large radius limit of type I$'$ leads to decompactification to a running solution of massive type IIA in 10 dimensions (rather than a higher dimensional vacuum). This makes the scalar charge-to-mass vector of the type I$'$ KK tower (which is non-BPS) to change non-trivially as we move in moduli space. The main result of \cite{Etheredge:2023odp} shows that warping effects make this vector to \emph{slide} perpendicularly to the self-dual line as we move along a trajectory parallel to self-dual line and change the distance to the latter (see Figure \ref{fig:sliding}). As a function of the asymptotic direction, though, it is simply seen as a \emph{jumping} of the KK vector from one unwarped value to the other as we cross the self-dual line. This jump occurs in opposite directions for the $SO(32)$ or $E_8\times E_8$ theories. This implies that, in each duality frame, the location of the $\zeta$-vectors of the towers is the same as in the above moduli spaces of 9d maximal supergravity (with 32 supercharges). This is clear upon comparing Figure \ref{fig:hets1} with Figure \ref{fig:T2SDC&SSDC} of Section \ref{ss:9d}. The lower level of supersymmetry plays only an important role when determining how to `glue' the different patches altogether, which occurs in a very non-trivial way.
	
	Hence, as long as we do not move parallel to the self-dual line in the type I$'$ region, it is then clear that the pattern \eqref{eq:pattern} is satisfied, since the distribution of the towers and the species scale vectors is locally the same as in maximal supergravity. Each region will be characterized by a different realization of the species scale (either the 10d string scale or the 11d M-theory Planck scale), such that the convex hulls of the towers and species scale are dual to each other and the pattern is thus realized. The tower vectors were already computed in \cite{Etheredge:2023odp}, so we are simply computing the species vectors as well here in order to represent everything together in Figure \ref{fig:hets1} below.
	
	It remains to be seen, though, whether the pattern will also hold if moving parallel to the self-dual line in the type I$'$ region. As explained, this limit decompactifies to a running solution in massive type IIA with a non-trivial spatial dependence of the dilaton. In particular, this changes the exponential rate of the KK tower in comparison to the unwarped result \eqref{eq:zeta&speciesveconemodulus}, as computed in \cite{Etheredge:2023odp}.  For the $E_8\times E_8$ slice\footnote{The $SO(32)$ is analogous but with slightly more cumbersome expressions, see Section 3 in \cite{Etheredge:2023odp}.} one has
	\begin{equation}\label{eq: sliding}
		\frac{m_{\rm KK,\, I'}}{M_{\rm Pl;\, 9}}\sim e^{-\frac{5}{2\sqrt{7}}\phi_C+\frac{3}{2}\phi_B}(1+3e^{2\phi_B})^{-1}\Longrightarrow \vec{\zeta}_{\rm KK,\, I'}=\left(\frac{1}{2}-\frac{2}{1+3e^{2\phi_B}},\frac{5}{2\sqrt{7}}\right)
	\end{equation}
	which is written in a basis of flat coordinates $\{\phi_B,\phi_C\}$.\footnote{This amounts to a clockwise $\frac{\pi }{2}+\arctan\left(\frac{1}{\sqrt{7}}\right)$ turn from the $\{\phi,\rho\}$ coordinates shown in Figure \ref{fig:hets1}.} Each of these coordinates measures, respectively, the moduli space distance perpendicular and parallel to the self-dual line in the type I$'$ frame. As already mentioned, this implies that the type I$'$ KK modes move orthogonal to the self-dual line as a function of $\phi_B$ (see Figure \ref{fig:sliding}).
	At each side of the self-dual line (i.e. in each of the type I$'$ frames) we seem to have a different tower of KK states, whose scalar charge-to-mass ratio coincides when moving exactly along the interface. We expect that these towers actually correspond to different sets of states that are mapped to each other upon performing the duality. If that is the case, they should both contribute to $\LSP$, yielding a lower value for the species scale (i.e. a larger value of the exponential rate) than what each tower alone would yield. The type I$'$ string oscillator modes, though, are not expected to contribute since the string perturbative limit is obstructed. Computing this species scale from top-down string theory would be a project by itself, so we leave it for future work. Here, we will simply determine what should be the value for the species scale along the self-dual line such that the pattern holds even for these decompactifications to running solutions. We hope that this can be useful to elucidate the fate of the pattern in these special cases.
	
	Along the self-dual line, the scalar charge-to-mass vector of the KK towers is given by $\vec{\zeta}_{\rm eff}=\left(0,\frac{5}{2\sqrt{7}}\right)$, with an associated species scale $\vec{\mathcal{Z}}_{\rm eff}$ that should also point towards this direction. For decompactification limits, the species scale can be computed \cite{Castellano:2021mmx,Castellano:2022bvr} (see also \eqref{eq:multspecies}) in terms of an \emph{effective tower} $m_{\rm eff,\, n}\sim n^{1/p_{\rm eff}}m_{\rm eff,\, 0}$ with  $p_{\rm eff}=\sum_i p_i$ and $m_{i,\, n}\sim n^{1/p_{i}}m_{i,\, 0}$. We do not expect $p_{\rm eff}=1$ since this would correspond to having a single KK tower decompactifying one dimension, nor $p_{\rm eff}=2$ since it would rather indicate a double decompactification. In fact, for the pattern to hold, we can check that the required value for the density parameter is something in between, namely $p_{\rm eff}=\frac{4}{3}$, which can be obtained upon identifying
	\begin{equation}
		\frac{\Lambda_{\rm eff}}{M_{\text{Pl;}\, 9}}=\left(\frac{m_{\rm eff}}{M_{\text{Pl;}\, 9}}\right)^{\frac{p_{\rm eff}}{9-2+p_{\rm eff}}}=e^{-\frac{2}{5\sqrt{7}}\phi_C}\Longrightarrow \vec{\mathcal{Z}}_{\rm eff}=\left(0,\frac{2}{5\sqrt{7}} \right)\, .
	\end{equation}
	This value would imply $\vec{\zeta}_{\rm eff}\cdot\vec{\mathcal{Z}}_{\rm eff}=\vec{\zeta}_{\rm KK,\, I'}\cdot\vec{\mathcal{Z}}_{\rm eff}=\vec{\zeta}_{\rm KK,\, I'^{\,(dual)}}\cdot\vec{\mathcal{Z}}_{\rm eff}=\frac{1}{7}$, satisfying the pattern for any $\phi_B\geq 0$. Along the self-dual line, the type I$'$ radion (in 10d Planck units) and string coupling scale as $R_{\rm I'}M_{\rm Pl; 10}=g_{\rm I'}^{-5/4}\sim e^{\frac{5\sqrt{7}}{16}\phi_C}$. This implies that the species cut-off should scale as $\frac{\LSP}{M_{\rm Pl;\, 10}}\sim (R_{\rm I'}M_{\rm Pl;\, 10})^{-\frac{32}{175}}\sim g_{\rm I'}^{\frac{8}{35}}$, although it is not possible for us to elucidate the separate dependence on the radion and the dilaton. It would be interesting to check, from string theory, whether this behaviour of the species scale is indeed realized and the structure of the KK towers (taking into account the large warping associated to decompactifying to a running solution) is such that effectively implies  $p_{\rm eff}=\frac{4}{3}$. Hence, whether the pattern is fulfilled in this particular asymptotic direction remains open and is left for future investigation.

	\subsection{M-theory in 7d}
	
	For completeness, let us consider M-theory compactified on a $K3$ surface, leading to a supersymmetric setup in 7d with 16 supercharges as well. This example will resemble many features that will be explained in more detail when discussing 4d theories arising from Calabi--Yau compactifications. Furthermore, our analysis here nicely complements the work performed in \cite{Lee:2019xtm}, where the emphasis was placed on the leading tower of states rather than the species scale.
	
	For simplicity, we will focus on \emph{attractive} $K3$ manifolds, namely those $K3$ spaces where the rank of the Picard group is maximal.\footnote{The Picard group is defined as $\text{Pic}(K3)= H^{1,1}(K3) \cap H^2(K3,\mathbb{Z})$, such that it corresponds to (dual) curve classes which have some holomorphic representative \cite{Aspinwall:1996mn}. For attractive $K3$s, $\text{rk}(\text{Pic}(K3))=20$.} For such manifolds, the complex structure is completely fixed (see e.g., \cite{Moore:1998zu} for details on this), so that both the 7d lagrangian as well as the mass of the different (non-)perturbative states depend only on the K\"ahler moduli $\{t^a\}$. The latter arise, as usual, as expansion parameters of the K\"ahler 2-form $J= t^a\omega_a$, where $\{\omega_a\}$ constitutes a basis of $H^{1,1}(X_3,\mathbb{Z})$ . 
	
	All in all, the relevant piece of the 7d (bosonic) reads as follows (see e.g., \cite{Townsend:1995de})
	\begin{equation}\label{eq:7dMthy}
		\begin{aligned}
			\ S_{\text{7d}} \supset & \frac{1}{2\kappa^2_7} \int \dd^{7}x\, \sqrt{-g}\,  \left[ \mathcal{R} - \frac{9}{20} \left( \partial \log \mathcal{V}_{K3} \right)^2 - \mathsf{G}_{a b}\, \left( \partial \tilde{t}^a \cdot \partial \tilde{t}^b \right) \right]\, ,
		\end{aligned}
	\end{equation}
	where we have defined $\mathcal{V}_{K3} = \frac{1}{2} \eta_{ab} t^a t^b$ with $\eta_{ab}= \omega_a \cdot \omega_b$ denoting the intersection form of the $K3$ surface, and $\tilde{t}^a= t^a/\mathcal{V}_{K3}^{1/2}$ are rescaled moduli subject to the constraint $ \frac{1}{2} \eta_{ab} \tilde{t}^a \tilde{t}^b \stackrel{!}{=} 1$. The moduli space (which is classically exact \cite{Witten:1995ex}) can be seen to be isomorphic to $\mathcal{M}_{\text{7d}}= O(3,19;\mathbb{Z})\backslash O(3,19;\mathbb{R})/O(19) \times \mathbb{R}_{+}$ \cite{Aspinwall:1996mn}, where the overall volume parameterizes the $\mathbb{R}_{+}$ factor with a metric of the form $\mathsf{G}_{\mathcal{V}_{K3} \mathcal{V}_{K3}}= \frac{9}{20 \mathcal{V}_{K3}^2}$. The coset piece admits a natural metric as well, which is given by \cite{Lee:2019xtm}
	\begin{equation}\label{eq:7dmodspacemetric}
		\begin{aligned}
			\mathsf{G}_{a b} = \int_{K3} \omega_a \wedge \star \omega_b = \frac{t_a t_b}{\mathcal{V}_{K3}}- \eta_{a b} = \tilde{t}_a \tilde{t}_b -\eta_{a b}\, ,
		\end{aligned}
	\end{equation}
	where the indices are lowered with the intersection form $\eta_{a b}$. 
	
	Regarding the infinite distance boundaries of such moduli space, there are several of them, according to which moduli are sent to infinity: the large volume point, the small `radius' limit, a \emph{unique} type of infinite distance degeneration at constant $\mathcal{V}_{K3}$ \cite{Lee:2019xtm} and combinations thereof. We discuss each of them in the following.
	
	\subsubsection*{The Large/Small Volume Limits}
	
	Let us start with the large volume singularity $\mathcal{V}_{K3} \to \infty$, which of course lies at infinite distance in the field space metric defined from the action \eqref{eq:7dMthy} above. It corresponds to the decompactification limit, where the $K3$ manifold grows large and we come back effectively to 11d supergravity. Thus, the infinite tower of asymptotically light states is given by the KK tower, whose mass is
	\begin{equation}
		\frac{m_{\text{KK, K3}}}{M_{\text{Pl};\, 7}} = \mathcal{V}_{K3}^{-9/20} \quad \Longrightarrow \quad \vec{\zeta}_{\text{KK, K3}} = \left( \frac{9}{20} \frac{1}{\mathcal{V}_{K3}}, 0, \ldots, 0 \right)\, ,
	\end{equation}
	where we have used that the 7d and 11d Planck scales are related by $M_{\text{Pl};\, 7}^5= M_{\text{Pl};\, 11}^5 \mathcal{V}_{K3}$. The associated species scale corresponds to the 11d Planck mass, such that upon taking the inner product between $\vec{\zeta}_{\text{KK, K3}}$ and $\vec{\mathcal{Z}}_{\text{sp}} = \frac{4}{9} \vec{\zeta}_{\text{KK, K3}}$ (c.f. eq. \eqref{eq:masterformula}) we find that $\vec{\zeta}_{\text{KK, K3}} \cdot \vec{\mathcal{Z}}_{\text{sp}}= \frac{1}{5}$, in agreement with \eqref{eq:pattern}.
	
	The small `radius' limit, namely $\mathcal{V}_{K3} \to 0$, is of different physical nature. One can argue that it corresponds to an emergent string limit \cite{Lee:2019wij}, where an asymptotically tensionless and weakly coupled heterotic string emerges at infinite distance. Indeed, it is possible to construct an heterotic-like string by wrapping the M5-brane on the whole $K3$ surface \cite{Cherkis:1997bx,Park_2009}, with a tension in 7d Planck units which reads as follows
	\begin{equation}
		\frac{T_{\text{M5}}}{M_{\text{Pl};\, 7}^2} = \mathcal{V}_{K3}^{3/5} \quad \Longrightarrow \quad \vec{\zeta}_{\text{osc, M5}} = \left( -\frac{3}{10} \frac{1}{\mathcal{V}_{K3}}, 0, \ldots, 0 \right)\, .
	\end{equation}
	Moreover, there are additional $\frac{1}{2}$-BPS states arising from wrapped M2-branes on certain holomorphic curves within the $K3$, which moreover correspond to perturbative winding modes of the dual heterotic string on $\mathbb{T}^3$.\footnote{Note that since $H^2(K3, \mathbb{Z})$ defines a lattice of signature $(3,19)$ there are precisely 3 non-equivalent holomorphic curves with non-negative self-intersection, and thus non-contractible. These should correspond to the 3 winding modes sectors of the dual heterotic string on $\mathbb{T}^3$.} Their mass dependence can be deduced from the DBI action, and yields \cite{Castellano:2022bvr}
	\begin{equation}\label{eq:M2mass&vectors}
		\frac{m_{\text{M2}}^{(a)}}{M_{\text{Pl};\, 7}} = t^a\, \mathcal{V}_{K3}^{-1/5} = \tilde{t}^a\, \mathcal{V}_{K3}^{3/10} \quad \Longrightarrow \quad \vec{\zeta}^{\,(a)}_{\text{M2}} = \left( -\frac{3}{10} \frac{1}{\mathcal{V}_{K3}}, 0, \ldots, -\frac{1}{\tilde{t}^a}, \ldots, 0 \right)\, ,
	\end{equation}
	where the non-zero entries correspond to the overall volume component and the one associated to the rescaled $\tilde{t}^a$ modulus (see discussion after eq. \eqref{eq:7dMthy}). It is therefore clear that, upon contracting $\vec{\zeta}_{\text{t}}= \lbrace \vec{\zeta}_{\text{osc, NS5}}, \vec{\zeta}^{\,(a)}_{\text{M2}} \rbrace$ with $\vec{\mathcal{Z}}_{\text{sp}}=\vec{\zeta}_{\text{osc, NS5}}$, one obtains $\vec{\zeta}_{\text{t}} \cdot \vec{\mathcal{Z}}_{\text{sp}}= \frac{1}{5}$, thus fulfilling the pattern.
	
	\subsubsection*{Infinite Distance at fixed (overall) Volume}
	
	Let us consider now infinite distance limits with the overall volume kept fixed and constant. In fact, as demonstrated in \cite{Lee:2019xtm} (see also earlier related works in \cite{Lee:2018urn,Lee:2018spm}), for such a limit to exist it must be possible to select some $\omega_0 =\sum_a c^a \omega_a \in H^{1,1}(X_3,\mathbb{Z})$ (with $c^a \geq 0$) such that\footnote{The fact that the limit \eqref{eq:T2limit7d} leads to an infinite distance with respect to the metric \eqref{eq:7dmodspacemetric} follows from the asymptotic dependence of $\mathsf{G}_{a b}$:
		\begin{equation}
			\notag \Delta= \int_1^{\infty} \dd \sigma \sqrt{\mathsf{G}_{a b} \frac{\dd \tilde{t}^a}{\dd \sigma} \frac{\dd \tilde{t}^b}{\dd \sigma}}\, \sim\, \int_1^{\infty} \dd \log \left(\tilde{t}^0 \right)\, \to\, \infty\, ,
		\end{equation}
		where we have used that $\mathsf{G}_{i j}=\eta_{0i} \eta_{0j} \left( \tilde{t}^0\right)^2 + \mathcal{O} (\sigma^0)$, $\mathsf{G}_{0 j}=\eta_{0j} \eta_{i 0} \tilde{t}^j \tilde{t}^0 - \eta_{0i} + \mathcal{O} (1/\sigma^2)$ and $\mathsf{G}_{0 0}= \eta_{0j} \eta_{i 0} \tilde{t}^i \tilde{t}^j$.}
	\begin{equation}\label{eq:T2limit7d}
		J= t^0 \omega_0 + t^i \omega_i %= \sigma\, \omega_0 + \frac{t^i}{\sigma}\, \omega_i
		\, , \qquad \text{with}\ \ t^0 = \sigma,\ t^i = \frac{a^i}{\sigma}, \quad \sigma \to \infty\, ,
	\end{equation}
	where $i=1, \ldots, 19$ and the basis $\lbrace \omega_0, \omega_i\rbrace$ verifies that $\omega_0 \cdot \omega_0=0$ and $\sum_i a^i\, \omega_0 \cdot \omega_i= \mathcal{V}_{K3} + \mathcal{O}(1/\sigma^2)$. Geometrically, the very existence of such a limit enforces the attractive $K3$ to admit some elliptic fibration over a $\mathbb{P}^1$-base, with the genus-one fibre $\mathcal{C}_0$ being Poincaré dual to the K\"ahler cone generator $\omega_0$. Such holomorphic curve shrinks upon taking the limit \eqref{eq:T2limit7d}, whilst the base grows at the same rate so as to keep the overall $\mathcal{V}_{K3}$ fixed and finite.
	
\begin{figure}[htb]
		\begin{center}
			\includegraphics[scale=.35]{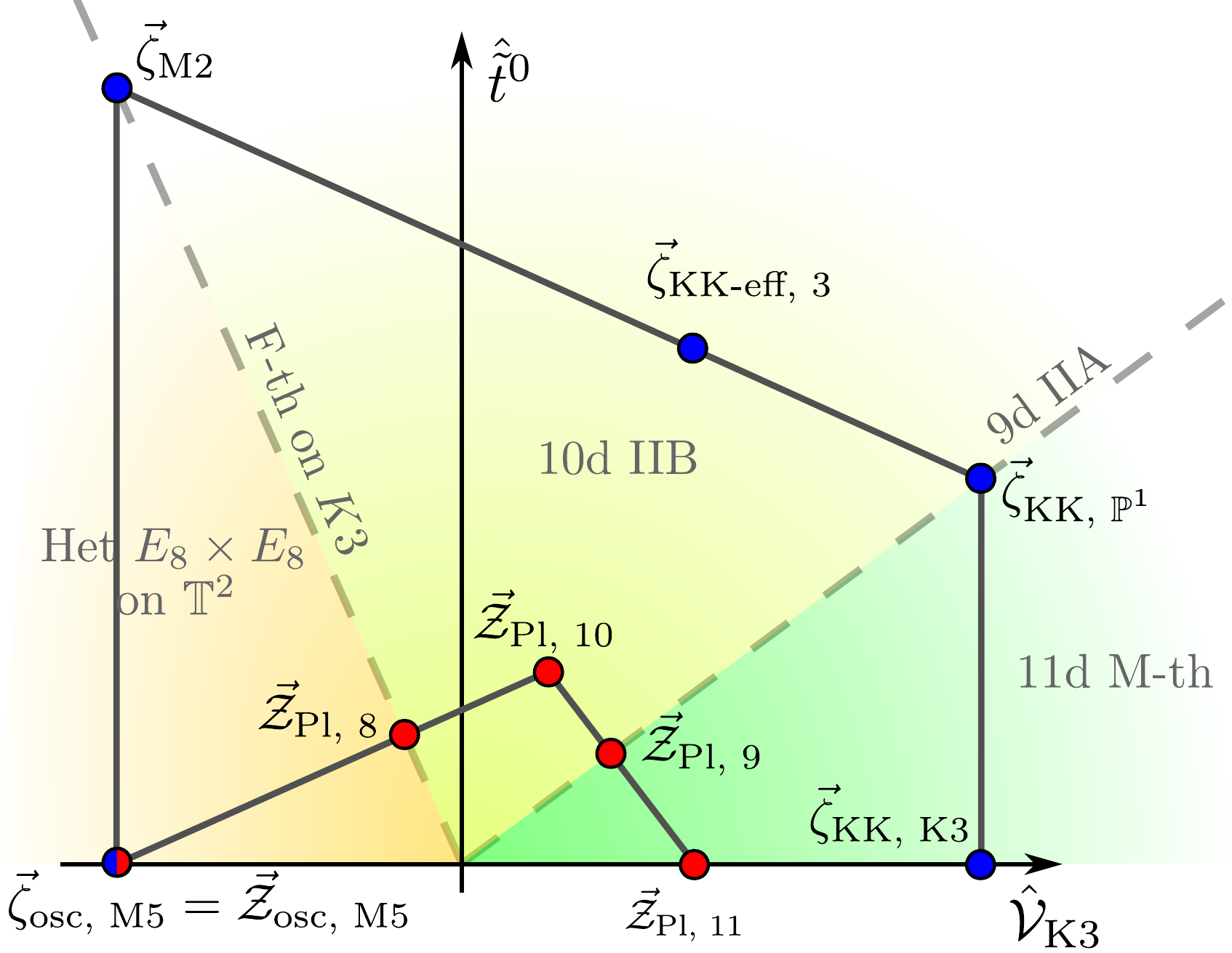}
			\caption{\small Convex hulls for the lightest towers (blue) and species scale (red) in M-theory compactified on an attractive $K3$ surface, using a flat frame $\lbrace \hat{\tilde{t}}^0, \hat{\mathcal{V}}_{K3}\rbrace$, in which the expressions of the different vectors are $\vec{\zeta}_{\rm osc,\; M5}=\vec{\mathcal{Z}}_{\rm osc,\; M5}=\left(-\frac{1}{\sqrt{5}},0\right)$, $\vec{\zeta}_{\rm M2}=\left(-\frac{1}{\sqrt{5}},1\right)$, $\vec{\zeta}_{\rm KK-eff,\; 3}=\left(\frac{2}{3\sqrt{5}},\frac{2}{3}\right)$, $\vec{\zeta}_{{\rm KK},\; \mathbb{P}^1}=\left(\frac{3}{2\sqrt{5}},\frac{1}{2}\right)$, $\vec{\zeta}_{{\rm KK},\; K3}=\left(\frac{3}{2\sqrt{5}},0\right)$, $\vec{\mathcal{Z}}_{\rm Pl,\; 8}=\left(-\frac{1}{6\sqrt{5}},\frac{1}{6}\right)$, $\vec{\mathcal{Z}}_{\rm Pl,\; 10}=\left(\frac{1}{4\sqrt{5}},\frac{1}{4}\right)$, $\vec{\mathcal{Z}}_{\rm Pl,\; 9}=\left(\frac{3}{7\sqrt{5}},\frac{1}{7}\right)$ and $\vec{\mathcal{Z}}_{\rm Pl,\; 11}=\left(\frac{2}{3\sqrt{5}},0\right)$. It is easy to see that both polytopes are dual to each other (with respect to the 1-sphere of radius $\frac{1}{\sqrt{d-2}}=\frac{1}{\sqrt{5}}$), and thus the pattern is satisfied. The different limiting theories, which can be deduced by looking at the dominant species scale in each region of the moduli space, are also shown for completeness.} 
			\label{fig:CHMthy7d}
		\end{center}
	\end{figure}

	Given the behavior of the different 2-cycles along the limit \eqref{eq:T2limit7d}, there are potentially two kinds of infinite towers of states. First, there are the supergravity KK modes associated to the $\mathbb{P}^1$-base, whose volume grows asymptotically. The mass scale of such tower behaves as follows
	\begin{equation}\label{eq:massP1}
		\frac{m_{\text{KK},\, \mathbb{P}^1}}{M_{\text{Pl};\, 7}}=\frac{1}{\left(\tilde{t}^0\right)^{1/2}\, \mathcal{V}_{K3}^{9/20}} \quad \Longrightarrow \quad \vec{\zeta}_{\text{KK},\, \mathbb{P}^1} = \left(\frac{9}{20} \frac{1}{\mathcal{V}_{K3}}, \frac{1}{2 \tilde{t}^0}, 0, \ldots, 0 \right)\, ,
	\end{equation}
	so that it becomes (exponentially) light upon probing the $\tilde{t}^0 \to \infty$ limit. In addition, there is a second infinite set of states becoming light even faster, which arise from M2-branes wrapping the genus-one fibre. Their mass is controlled by the volume of the latter
	\begin{equation}\label{eq:M2massFthy}
		\frac{m_{\text{M2}}}{M_{\text{Pl};\, 7}} = \mathcal{V}_{\mathcal{C}_0}\, \mathcal{V}_{K3}^{-1/5}=\frac{\mathcal{V}_{K3}^{3/10}}{\tilde{t}^0} \quad \Longrightarrow \quad \vec{\zeta}_{\text{M2}} = \left( -\frac{3}{10} \frac{1}{\mathcal{V}_{K3}}, \frac{1}{\tilde{t}^0}, 0, \ldots, 0 \right)\, ,
	\end{equation}
	and they can be seen to correspond to the dual KK replica implementing the duality between M-theory on $K3$ and F-theory on $K3\times \mathbb{S}^1$ \cite{Vafa:1996xn,Lee:2019xtm}. However, in order to correctly interpret what is the resolution of the singularity in QG, we need to study the behavior of the species scale. One can thus associate two such scales, one for each tower, as follows (c.f. \eqref{eq:masterformula})
	\begin{align}\label{eq:species7d}
		\frac{\Lambda_{\rm M2}}{M_{\text{Pl};\, 7}} &\simeq \left(m_{\text{M2}}\right)^{1/6} = \frac{\mathcal{V}_{K3}^{1/20}}{\left( \tilde{t}^0\right)^{1/6}}\quad \Longrightarrow \quad \vec{\mathcal{Z}}_{\text{Pl},\, 8} = \left( -\frac{1}{20} \frac{1}{\mathcal{V}_{K3}}, \frac{1}{6\, \tilde{t}^0}, 0,\ldots, 0 \right)\, ,\\
		\frac{\Lambda_{\text{Pl},\, 9}}{M_{\text{Pl};\, 7}} &\simeq \left(m_{\text{KK},\, \mathbb{P}^1}\right)^{2/7} = \frac{1}{\left(\tilde{t}^0\right)^{1/7}\, \mathcal{V}_{K3}^{9/70}} \quad \Longrightarrow \quad \vec{\mathcal{Z}}_{\text{Pl},\, 9} = \left(\frac{9}{70} \frac{1}{\mathcal{V}_{K3}}, \frac{1}{7 \tilde{t}^0}, 0, \ldots, 0 \right)\, , 
	\end{align}
	which coincide with the 8d Planck scale\footnote{This can be easily checked upon identifying $R_8 = \frac{(\tilde{t}^0)^{5/6}}{\mathcal{V}_{K3}^{1/4}}$, where $R_8$ denotes the radius (in 8d Planck units) of the F-theory circle, as well as the relation between the 8d and 7d Planck scales, namely $M_{\text{Pl};\, 7}^5= M_{\text{Pl};\, 8}^5 2\pi R_8$.} (in the F-theory frame) and the 9d Planck scale, respectively. We are not done yet though, since both sets of states can be combined together forming bound states, namely the wrapped M2-branes may have non-trivial momentum along the $\mathbb{P}^1$-base. Furthermore, such `mixed' states contribute to the computation of a third candidate for the species scale, whose $\mathcal{Z}$-vector reads (see eq. \eqref{eq:effectiveKKspeciesvector})
	\begin{align}\label{eq:speciesKK10d}
		\vec{\mathcal{Z}}_{\text{Pl},\, 10} = \frac{1}{8} \left( \vec{\zeta}_{\text{M2}} + 2 \vec{\zeta}_{\text{KK},\, \mathbb{P}^1}\right) = \left(\frac{3}{40} \frac{1}{\mathcal{V}_{K3}}, \frac{1}{4 \tilde{t}^0}, 0, \ldots, 0 \right)\, , 
	\end{align}
	thus signalling towards decompactification to 10d type IIB string theory. In Figure \ref{fig:CHMthy7d} all these vectors are plotted, both for the scalar charge-to-mass and species scale are shown, including those relevant in the large/small $K3$ volume limit, as previously discussed. 
	
	With this, we are now ready to check what is the minimum $\LSP$ dominating the asymptotic physics along the limit \eqref{eq:T2limit7d}. Indeed, it is easy to see either from the formulae above or the diagram in Figure \ref{fig:CHMthy7d}, that this becomes the 10d Planck scale. Therefore, such limit may be interpreted as some `nested' decompactification, first from 7d M-theory to 8d F-theory (as remarked in \cite{Lee:2019xtm}), and then up to ten dimensions, effectively sending all supersymmetry breaking defects (i.e. D7-branes and O7-planes) to infinity and thus restoring maximal (chiral) supergravity in 10d. Hence, a quick computation reveals that the pattern $\vec{\zeta}_{\text{M2}} \cdot \vec{\mathcal{Z}}_{\text{Pl},\, 10}= \frac{1}{5}$ is also verified in this limit (to leading order in $1/\tilde{t}^0$). 
	
	%%%%%%%%%%%%
	%%%%%%%%%%%%
	
	\subsubsection*{Mixed Limits}
	
	To conclude, let us briefly comment on the possibility of superimposing any of the previous limits, thus sending both the overall $K3$ volume and the $\tilde{t}^0$ K\"ahler modulus to infinity at different rates, a priori. In fact, upon comparing the different species scale that can arise (and even compete) at distinct corners of the moduli space, one can indeed separate these asymptotic regions into different sectors, depending on which specific scale dominates (see Figure \ref{fig:CHMthy7d}). In any event, one can still verify that the pattern is respected in all such cases, due to the non-trivial gluing conditions between the different patches.
	
	\section{Examples in 4d $\mathcal{N}=2$ EFTs}
	\label{s:8supercharges}
	
	We now turn to theories with 8 supercharges. In particular, we will focus on 4d $\mathcal{N}=2$ setups arising upon compactifying type II string theory on Calabi--Yau threefolds. The singularity structure of the moduli space of these theories is very rich and has been thoroughly studied in the past, providing for different types of infinite distance limits. We will first introduce the basic notation in Section \ref{ss:preliminary} and then study different concrete examples in later subsections, as well as presenting general arguments in favour of satisfying the pattern in full generality within the vector multiplet moduli space. Section \ref{ss:hypers} analyzes the effect of (towers of) instanton corrections on singularities located classically at infinite distance, which are nevertheless excised and deflected within the true quantum hypermultiplet moduli space.
	
	\subsection{Setting the stage: the vector multiplet moduli space}
	\label{ss:preliminary}
	
	Let us start by reviewing the main ingredients that will be necessary in what follows so as to check the pattern \eqref{eq:pattern} in 4d $\mathcal{N}=2$ setups. Such theories arise upon compactifying e.g., type IIA string theory on a Calabi--Yau threefold, $X_3$, and in the low energy regime they may be described by the following (bosonic) action \cite{Bodner:1990zm}
	\begin{equation}\label{eq:IIAaction4d}
		\begin{aligned}
			\ S_{\text{4d}} = & \frac{1}{2\kappa^2_4} \int \mathcal{R} \star 1 - \frac{1}{2} \text{Re}\, \mathcal{N}_{AB} F^A \wedge F^B-\frac{1}{2} \text{Im}\, \mathcal{N}_{AB} F^A \wedge \star F^B \\
			& - \frac{1}{\kappa^2_4} \int \mathsf{G}_{a\bar b}\, \dd z^a\wedge \star \dd\bar z^b + h_{pq}\, \dd q^p \wedge \star \dd q^q\, ,
		\end{aligned}
	\end{equation}
	where $F^B=dA^B$, with $B=0, \ldots, h^{1,1}(X_3)$, denote the field strengths associated to the Abelian $U(1)$ gauge bosons belonging to the vector multiplets as well as the graviphoton. On the other hand, the complex scalars $z^a=b^a+{\rm i}t^a$, $a=1,\ldots, h^{1,1}$, describe the (complexified) K\"ahler sector of the theory and determine altogether the vector multiplet moduli space \cite{Candelas:1990pi}, whereas the scalars in the various hypermultiplets (including e.g., the complex structure moduli) are denoted collectively by $q^p$.
	
	Since we will only be interested in the computation of the relevant scalar charge-to-mass vectors as well as the corresponding species scale, we will focus on the scalar-tensor sector of the action \eqref{eq:IIAaction4d} and effectively forget about the vector fields. In particular, we will need the explicit expression for the moduli space metrics. The K\"ahler moduli $\{t^a\}_{a=1}^{h^{1,1}}$ can be used to expand the K\"ahler form $J= t^a\omega_a$, where $\{\omega_a\}_{a=1}^{h^{1,1}}$ is a basis of integral 2-forms dual to a basis of Mori cone generators in $H_2(X_3,\mathbb{Z})$ \cite{Lanza:2021udy}. With this choice, one finds the following metric for the scalars within the vector multiplets \cite{Candelas:1990pi,Strominger:1985ks}:
	\beq\label{eq:kahlersectormetric}
	\mathsf{G}_{a\bar b}=\partial_a \partial_{\overline{b}} K_{\text{ks}}=\partial_a \partial_{\overline{b}} \left(- \log \frac{4}{3} \mathcal{K} \right)\, .
	\eeq
	where $\frac{\mathcal{K}}{6}=\frac{1}{6} \kappa_{abc}t^a t^b t^c =\frac{1}{6}\int_{X_3}J\wedge J\wedge J= \mathcal{V}(X_3)$ denotes the volume of the threefold in string units, $K_{\text{ks}}$ is the K\"ahler potential \cite{Grimm:2005fa} and $\kappa_{abc}$ are the triple intersection numbers of the Calabi--Yau $X_3$, given by
	\begin{equation}
		\kappa_{abc}=\int_{X_3}\omega_a\wedge\omega_b\wedge \omega_c\;.
	\end{equation}
	
	The analysis of the hypermultiplet sector will be postponed until Section \ref{ss:hypers}. By Mirror Symmetry (see e.g., \cite{Hori:2003ic}), this effective theory can be equivalently described as arising from compactifying type IIB on the mirror Calabi--Yau threefold, such that the role of K\"ahler and complex structure moduli get exchanged. The different types of infinite distance limits in the vector multiplet sector can then be nicely classified using the theory of Mixed Hodge Structures within the complex structure moduli space of type IIB \cite{Grimm:2018ohb, Grimm:2018cpv}. However, in the present work, we will analyze each of these limits using the language of type IIA compactifications, since the microscopic interpretation of the corresponding asymptotic limit (either decompactification or emergent string limit \cite{Lee:2019wij}) becomes more apparent from this point of view.
	
	\subsubsection*{Classification of infinite distance limits at large volume}
	\label{sss:largevolume}
	
	From the perspective of type IIA string theory, we need to particularize to the large volume patch, where one can safely ignore both $\alpha'$ and worldsheet instanton contributions which further correct the form of the metric displayed in \eqref{eq:kahlersectormetric}. Still, the structure of possible infinite distance singularities is very rich as we review in what follows. Following \cite{Corvilain:2018lgw,Grimm:2018cpv,Grimm:2018ohb}, we can parametrize infinite distance limits within the K\"ahler cone, in terms of trajectories of the form
	\beq \label{eq:singlefieldlim}
	\lbrace t^i\rbrace = t^1,\ldots , t^{n}\to \infty\, ,\qquad n\leq h^{1,1}(X_3)\, ,
	\eeq
	with $b^i$ approaching finite values. The several distinct types of infinite distance limits have been thoroughly studied and classified by different means in \cite{Corvilain:2018lgw,Lee:2019wij}, and can be divided into three classes shown in Table \ref{tab:intersN=2} below, according to the behavior of the intersection numbers $\kappa_{abc}$ in terms of the asymptotic direction taken. More details about the notation in terms of Roman numerals can be found in \cite{Grimm:2018ohb} whilst the notation in terms of $J$-class A/B can be found in \cite{Lee:2019wij} (see also \cite{Lee:2019tst}). Geometrically, these three classes correspond to different fibration structures: the \emph{unique} limit in which the overall volume of $X_3$ blows up uniformly, thus corresponding to the large volume point; the ones in which the CY$_3$ possesses an elliptic fibration over some K\"ahler two-fold; and those in which the threefold develops either some $K3$ or $\mathbb{T}^4$ fibration over a $\mathbb{P}^1$-base. We will consider in the upcoming subsections specific examples of each representative class of limit followed by a general analysis of each singularity type, providing all of them more evidence in favour of the pattern here described. 
	
	In Table \ref{tab:limitsN=2}, we summarize the microscopic interpretation of the leading tower of states becoming light at each type of infinite distance limit, as well as the physical realization of the species scale for each case. Recall that, in this section, we consider infinite distance limits lying purely within the vector multiplet moduli space, while all hypermultiplet scalars (including the 4d dilaton) remain fixed. To achieve this, sometimes we will need to co-scale properly certain ten-dimensional variables \cite{Lee:2019wij}. For instance, if we want to keep the 4d dilaton, $\varphi_4 =\phi-\frac{1}{2} \log \mathcal{V}$, fixed and \emph{finite}, one needs to rescale accordingly the 10d dilaton $\phi$, which will bring us to the strong coupling regime of type IIA as we will see below in more detail. For other limits involving also the hypermultiplets, see Section \ref{ss:hypers}.
	%%%%%%%%%%%%%%%%%%%%%%%%%%%%%%%
	\begin{table}[h!!]\begin{center}
			\renewcommand{\arraystretch}{0.80}
			\begin{tabular}{|c|c|c|}
				\hline
				Type \cite{Corvilain:2018lgw} & Type \cite{Lee:2019wij} & Intersection numbers \\
				\hline 
				IV$_d$ & --- & $\text{rk}(\boldsymbol{\kappa}^{(n)}) = \text{rk}(\boldsymbol{\kappa}^{(n)}_a)= 1$ and $\text{rk}(\boldsymbol{\kappa}^{(n)}_{a b})=d$   \\
				\hline
				III$_c$ & $J$-class A & $\text{rk}(\boldsymbol{\kappa}^{(n)}) = 0$, $\text{rk}(\boldsymbol{\kappa}^{(n)}_a)= 1$ and $\text{rk}(\boldsymbol{\kappa}^{(n)}_{a b})=c+2$ \\
				\hline
				II$_b$ & $J$-class B & $\text{rk}(\boldsymbol{\kappa}^{(n)}) = \text{rk}(\boldsymbol{\kappa}^{(n)}_a)= 0$ and $\text{rk}(\boldsymbol{\kappa}^{(n)}_{a b})=b$ \\
				\hline
			\end{tabular}
			\caption{\small Infinite distance limits in the large volume regime within the vector multiplet moduli space of type IIA compactified on a CY$_3$. They can be classified in terms of the behavior of the triple intersection numbers $\kappa_{abc}$ via Mixed Hodge Theory \cite{Corvilain:2018lgw}, or using a purely geometrical analysis \cite{Lee:2019wij}. The following notation has been introduced: $\boldsymbol{\kappa}^{(n)}_{a b} = \sum_{i=1}^n \kappa_{i a b}$, $\boldsymbol{\kappa}^{(n)}_{a} = \sum_{i,j=1}^n \kappa_{i j a}$, $\boldsymbol{\kappa}^{(n)} = \sum_{i,j,k=1}^n \kappa_{i j k}$ and $\text{rk}(\cdot)$ denotes the rank of the corresponding matrix.
			}
			\label{tab:intersN=2}
		\end{center}
	\end{table}
	%%%%%%%%%%%%%%%%%%%%%%%%%%%%%%%
	\begin{table}[h!!]\begin{center}
			\renewcommand{\arraystretch}{1.00}
			\begin{tabular}{|c|c|c|c|c|}
				\hline
				Type \cite{Corvilain:2018lgw} & Type \cite{Lee:2019wij} &  Fibration structure  &  Dominant Tower & $\LSP$\\
				\hline 
				IV$_d$ & --- & Unspecified  & D0 & $M_{\text{Pl};\, 5}$\\
				\hline
				III$_c$  & $J$-class A & Elliptic Fibration & D0 and D2 on $\mathbb{T}^2$ & $M_{\text{Pl};\, 6}$\\
				\hline
				II$_a$ & $J$-class B & $K3$ or $\mathbb{T}^4$ Fibration & NS5 on $K3/\mathbb{T}^4$ & $\sqrt{T_{\rm NS5}}$\\
				\hline
			\end{tabular}
			\caption{\small Infinite distance limit classification according to refs. \cite{Corvilain:2018lgw} and \cite{Lee:2019wij}. We also show the kind of asymptotic fibration structure exhibited by the threefold as well as the dominant tower(s) of states controlling the species scale for each case.}
			\label{tab:limitsN=2}
		\end{center}
	\end{table}
	%%%%%%%%%%%%%%%%%%%%%%%%%%%%%%%%%

	\subsection{Type IV limits: M-theory circle decompactification}
	\label{s:typeIVlimits}
	
	\subsubsection{A simple example: the Quintic}
	\label{ss:ExampleI}
	
	As our first example, we consider a one-modulus case and we explore the large volume point, which is always present within the vector multiplet moduli space \cite{Corvilain:2018lgw}. For concreteness, we will particularize to the quintic threefold studied in \cite{Candelas:1987se,Candelas:1990rm}, which may be defined as a family of degree 5 hypersurfaces in $\mathbb{P}^4$. Such threefold presents 101 complex parameters (appearing in the quintic polynomial) associated to complex structure deformations, as well as a single (complexified) K\"ahler structure modulus that we denote by $z=b+{\rm i}t$. Within the vector multiplet moduli space one finds three singularities: the large volume point at $z\to {\rm i}\infty$, the conifold locus, that is located at $z \simeq 1.21\, {\rm i}$, and the Landau-Ginzburg orbifold point, which happens for $z = \frac{1}{2} \left( 1+ {\rm i} \cot{\pi/5}\right)$ \cite{Blumenhagen:2018nts}. 
	
	Close to the large volume point, the K\"ahler potential behaves as \cite{Candelas:1990rm}
	\beq\label{eq:KahlerpotLV}
	e^{-K_{\text{ks}}} = \frac{256 \pi^6 }{9375}t^3 + \mathcal{O}\left(t^0 \right)\, ,
	\eeq
	with $t= \text{Im}\, z$, such that the moduli space metric can be approximated by
	\beq\label{eq:quinticmetric}
	\mathsf{G}_{z\bar z}= \frac{3}{4} \frac{1}{(\text{Im}\, z)^2} + \mathcal{O}\left(1/t^5 \right)\, .
	\eeq
	Next, we need to compute the scalar charge-to-mass vector associated to the leading infinite tower of states, as well as the corresponding species scale. Regarding the former point, there is indeed a plethora of perturbative (e.g., KK modes) and non-perturbative states becoming light upon exploring the large volume singularity (see e.g., \cite{Font:2019cxq,Corvilain:2018lgw,Lee:2019wij}). The former can be easily seen to be subleading (contrary to what happens in the 4d $\mathcal{N}=1$ heterotic example from Section \ref{ss:heteroticCY3}), whilst the latter arise as $\frac{1}{2}$-BPS bound states of D0- and D2-branes wrapping minimal 2-cycles of the CY$_3$, whose mass is controlled by the normalized $\mathcal{N}=2$ central charge\footnote{We do not consider here magnetically charged states corresponding to wrapped D4- or D6-particle states \cite{Ceresole:1995ca}, since they do not become massless in the limit of interest (see e.g., \cite{Font:2019cxq}).}
	\beq\label{eq:centralcharge}
	\frac{m_{n_{2p}}}{\Mpf} = \sqrt{8\pi } e^{K_{\text{ks}}/2} |Z_{\text{IIA}}| = \sqrt{\frac{\pi}{ \mathcal{V}(X_3)}} |n_0+n_{2,a} z^a|\, ,
	\eeq
	where $n_0, n_{2,a}\in \mathbb{Z}$ correspond to D0- and D2-brane charges, respectively, and the subscript $a$ indicates the holomorphic 2-cycle wrapped by the 2-brane. For the quintic, given that $h^{1,1}=1$, the previous mass formula reduces to
	\beq\label{eq:centralchargequintic}
	\frac{m_{n_{2p}}}{\Mpf} \sim \frac{|n_0+n_{2} z|}{ t^{3/2}}\, .
	\eeq
	Any state with D2-brane charge (i.e. $n_2\neq 0$) will scale as $m_{\rm D2}\sim t^{-1/2}M_{\rm Pl;\, 4}$, while for $n_2=0$ we have instead $m_{\rm D0}\sim t^{-3/2}M_{\rm Pl;\, 4}$. This means, in particular, that the leading tower becomes that comprised by D0-branes alone, whilst there is another one (which is additive, in the sense of \cite{Castellano:2021mmx, Castellano:2022bvr})  made out of bound states of D0- and D2-particles \cite{Corvilain:2018lgw}.\footnote{\label{fnote:stabilityBPS}In general, it becomes difficult to properly argue for the existence of an \emph{infinite} tower of states which become asympotically stable \cite{Grimm:2018ohb,Palti:2021ubp}. This is why in the original work \cite{Grimm:2018ohb}, the monodromy transformations characterizing the infinite distance singularities were exploited, in order to argue at least for the existence of the \emph{monodromy tower}, which may or may not be the leading tower depending on the case. In certain circumstances, however, one may instead use dualities to support the existence of the tower, as happens in the present case, where the D0-tower corresponds to the KK replicas of the 5d fields along the M-theory circle, see Section \ref{sss:IIA/Mthy}.}
	
	Therefore, from eq. \eqref{eq:centralchargequintic}, we obtain
	\begin{equation}\label{eq:D0zeta4d}
		\vec{\zeta}_{\text{D0}} = -\partial_t \log m_{\text{D0}} = \frac{3}{2t}\, \quad \Longrightarrow \quad |\vec{\zeta}_{\text{D0}}| = \sqrt{\frac{3}{2}}\, ,
	\end{equation}
	where  we have used the field space metric \eqref{eq:quinticmetric} to compute the norm of the charge-to-mass vector, namely $|\vec{\zeta}_{\text{D0}}|= 2\mathsf{G}^{z\bar z} \partial_z \log m_{\text{D0}}\, \partial_{\bar z} \log m_{\text{D0}}$. Note that this precisely matches that of a KK decompactification of one extra dimension, c.f. \eqref{eq:zeta&speciesveconemodulus}. This is of course no coincidence since the D0-branes correspond to the KK tower of the M-theory circle, so that the large volume limit induces a circle decompactification to a 5d $\mathcal{N}=1$ theory described in terms of M-theory on the same threefold $X_3$ (see Section \ref{sss:IIA/Mthy} below). 	
	
	The species scale can then be computed as usual for a single KK tower \cite{Castellano:2022bvr}, leading to
	\beq \label{eq:D0tower}
	\frac{\LSP}{\Mpf}\, \simeq\, \left( \frac{m_{\text{D0}}}{\Mpf} \right)^{1/3}\, \sim\, \frac{1}{\mathcal{V}(X_3)^{1/6}}\, \sim\, \frac{1}{t^{1/2}}\, ,
	\eeq
	which goes to zero upon exploring the $t \to \infty$ limit, as expected. It moreover matches with the 5d Planck scale, as we show later on. Hence, from eq. \eqref{eq:D0tower} one obtains
	\begin{equation}\label{eq:D0Zeta4d}
		\vec{\mathcal{Z}}_{\text{sp}} = -\partial_t \log \LSP = \frac{1}{2t}\, ,
	\end{equation}
	such that upon contracting with \eqref{eq:D0zeta4d} using the moduli space metric \eqref{eq:quinticmetric} we find
	\begin{equation}
		\vec{\zeta}_{\text{D0}} \cdot \vec{\mathcal{Z}}_{\text{sp}} = \frac{1}{2}\, ,
	\end{equation}
	thus fulfilling the pattern in the present $d=4$ setup.
	
	\subsubsection{General story}
	\label{sss:IIA/Mthy}
	
	The above large volume singularity is always present within the vector multiplet moduli space of any type IIA CY$_3$ compactification, such that the results found for the quintic can be easily extended to the more general case, as we argue in the following. 
	
	Recall from \eqref{eq:IIAaction4d} that the relevant piece of the 4d lagrangian obtained from type IIA compactified on a generic threefold is \cite{Bodner:1990zm}
	\begin{equation}\label{eq:IIAlagrangian4d}
		\mathcal{L}_{\text{IIA, 4d}} \supset \frac{1}{2} \mathcal{R} - \frac{1}{2} \mathsf{G}_{a b}(\tilde{t})\, \partial \tilde{t}^a \cdot \partial \tilde{t}^b - \frac{1}{12} \left( \partial \log \mathcal{V} \right)^2 - \left( \partial \varphi_4 \right)^2\, ,
	\end{equation}
	where we defined $\mathsf{G}_{a b} = 2 \mathsf{G}_{a \bar b}$ (c.f. \eqref{eq:kahlersectormetric}) and we split the K\"ahler coordinates into unimodular ones, $\tilde{t}^a = t^a/\mathcal{V}^{1/3}$ --- which satisfy the constraint $\kappa_{abc}\tilde{t}^a \tilde{t}^b \tilde{t}^c = 6$ --- and the overall volume modulus $\mathcal{V}$. Now, since we take a limit here where $\mathcal{V} \to \infty$ with the 4d dilaton fixed and finite, the 10d dilaton needs to be co-scaled, such that we end up probing the strong $g_s$ regime, i.e. $\phi \to \infty$, which can be better described by M-theory. Comparing then the lagrangian \eqref{eq:IIAlagrangian4d} with the one obtained by dimensionally reducing M-theory on the same manifold times a circle of radius $R_5$ (in 5d Planck units), which reads \cite{Cadavid:1995bk}
	\begin{equation}
		\mathcal{L}_{\text{M-th, 4d}} \supset \frac{1}{2} \mathcal{R} - \frac{1}{2} \mathsf{G}_{a b}(\tilde{t})\, \partial \tilde{t}^a \cdot \partial \tilde{t}^b - \frac{3}{4} \left( \partial \log R_5 \right)^2 - \frac{1}{4} \left( \partial \log \mathcal{V}_5 \right)^2\, ,
	\end{equation}
	we arrive at the following moduli identifications (taking also into account quantum corrections \cite{Gopakumar:1998ii,Gopakumar:1998jq,Lawrence:1997jr})
	\beq
	\label{eq:modulimatching}
	R_5^3 = \mathcal{V}(X_3)\, , \qquad \mathcal{V}_5(X_3)=e^{-2\varphi_4}\, ,
	\eeq
	where $\mathcal{V}_5(X_3)$ denotes the volume of the threefold measured in 11d Planck units.
	
	With these identifications at hand it is now easy to see how the masses of the D0- and D2-particles in 4d Planck units are translated into 5d quantities \cite{Castellano:2022bvr}
	\begin{equation}\label{eq:massesD0D2}
		\begin{aligned}
			\frac{m_{\text{D0}}}{\Mpf} & =\sqrt{8\pi } e^{K_{\text{ks}}/2} = \sqrt{\frac{\pi}{\mathcal{V}}} = \frac{m_{\text{KK},\, 5}}{\Mpf}\, ,\\
			\frac{m_{\text{D2}}}{\Mpf} & =\sqrt{8\pi } e^{K_{\text{ks}}/2} |t^a| = \frac{\sqrt{\pi}\, \tilde t^a}{\mathcal{V}^{1/6}} = \frac{m_{\text{M2}}}{\Mpf}\, ,
		\end{aligned}
	\end{equation}
	where in the last expression we have considered a single D2-brane wrapping some 2-cycle once and for simplicity we have set the corresponding axion vev $b^a$ to zero. Proceeding similarly to what we did in the quintic example, and taking the limit $\mathcal{V} \to \infty$ (whilst keeping the $\tilde t^a$ fixed and non-degenerate) we obtain the following charge-to-mass and species vectors
	\begin{equation}
		\left(\zeta_{\text{D0}}\right)_{\mathcal{V}} = - \partial_{\mathcal{V}} \log (m_{\text{D0}}) = \frac{1}{2 \mathcal{V}}\, , \qquad
		\left(\mathcal{Z}_{\text{sp}}\right)_{\mathcal{V}} = - \partial_{\mathcal{V}} \log(\LSP) = \frac{1}{6 \mathcal{V}}\, ,
	\end{equation}
	where the remaining components of the vectors, namely those arising from log-derivatives with respect to the $\tilde t^a$ fields, are vanishing. Note that the species scale, as computed from \eqref{eq:D0tower}, coincides asymptotically with the 5d Planck mass, which can be related to the 4d one by $M_{\text{Pl};\, 5}^2 2\pi R_5 = \Mpf^2$. Therefore, upon contracting them using the moduli space metric in \eqref{eq:IIAlagrangian4d}, i.e. $\mathsf{G}_{\mathcal{V}\mathcal{V}}=\frac{1}{6\mathcal{V}^2}$, we find that $\vec{\zeta}_{\text{D0}} \cdot \vec{\mathcal{Z}}_{\text{sp}} = \frac{1}{2}$ is again fulfilled.	
	
	Interestingly, there is an alternative very simple way to show the realization of the pattern in general for this type of limit. The leading tower of states is made of D0-branes, so that we can write $\zeta_{\text{D0}}^a=\frac{\mathsf{G}^{a b}}{2}\frac{\partial K_{\text{ks}}}{\partial t^{b}}$. Furthermore, since we decompactify a single extra dimension, the species scale vector is given by $\vec{\mathcal{Z}}_{\text{sp}}=\frac{1}{3} \vec{\zeta}_{\text{D0}}$ (c.f. eq. \eqref{eq:masterformula}). Therefore, we may have alternatively computed the inner product as follows
	\beq 
	\label{noscale}
	\vec{\zeta}_{\text{D0}} \cdot \vec{\mathcal{Z}}_{\text{sp}} = \frac{1}{12} \frac{\partial K_{\text{ks}}}{\partial t^{a}} \mathsf{G}^{a b} \frac{\partial K_{\text{ks}}}{\partial t^b} = \frac{1}{2}\, ,
	\eeq
	where in order to arrive at the RHS, one needs to use the no-scale property of the vector-multiplet metric \eqref{eq:kahlersectormetric}, namely $K_a \mathsf{G}^{a b} K_b=6$.

	\subsection{Type III limit: Partial decompactification}
	\label{s:typeIIIlimits}
	
	\subsubsection{Example: Type IIA on $\mathbb{P}^{(1,1,1,6,9)} [18]$}
	\label{ss:ExampleII}
	
	As a second example, we consider type IIA string theory compactified on the threefold $X_3=\mathbb{P}^{(1,1,1,6,9)}$ [18], which may be seen as an elliptic fibration over a $\mathbb{P}^2$-base, with $h^{1,1}=2$ \cite{Candelas:1994hw}. We denote the (real-valued) K\"ahler moduli by $\lbrace t^1, t^2 \rbrace$, which at large volume control the $\mathcal{N}=2$ K\"ahler potential
	\begin{equation}\label{eq:kahlerpotP11169}
		e^{-K_{\text{ks}}} = \frac{4}{3}\kappa_{abc}t^a t^b t^c = 12(t^1)^3 + 12(t^1)^2 t^2 + 4 t^1(t^2)^2 + \ldots\, ,
	\end{equation}
	with $\kappa_{abc}$ being the triple intersection numbers in an integral basis of $H^2(X_3)$ and the ellipsis denote further perturbative and non-perturbative $\alpha'$-corrections. From this one may already compute the (inverse) moduli space metric, whose entries read
	\begin{equation}\label{eq:fullinversemetric}
		\mathsf{G}^{-1}\, = \, \left(
		\begin{array}{cc}
			2(t^1)^2 +\frac{3(t^1)^4}{3t^1t^2+(t^2)^2}& - \frac{3(t^1)^2 \left(t^1+t^2\right)}{t^2}  \\
			- \frac{3(t^1)^2 \left(t^1+t^2\right)}{t^2} & 9(t^1)^2+ \frac{9(t^1)^3}{t^2} +3t^1t^2 +(t^2)^2 \\
		\end{array}
		\right) \, .
	\end{equation}
	On the other hand, the infinite distance boundaries present in this example were analyzed from the MHS point of view in \cite{Grimm:2018cpv}, and three types of infinite distance limits were found therein: \emph{(i)} $t^1 \to \infty$ with $t^2$ finite (a Type IV$_1$ singularity), \emph{(ii)} $t^2 \to \infty$ with $t^1$ finite (a Type III$_0$ singularity) and \emph{(iii)} $t^1, t^2 \to \infty$ (a Type IV$_2$ singularity). The asymptotic regime in the latter case can be divided into two subregions (i.e., growth sectors) depending on whether $t^1\gg t^2 $ or $t^2\gg t^1 $ as $t^1, t^2 \to \infty$. 	
	
\begin{figure}[t!]
		\begin{center}
			\begin{subfigure}[t]{0.375\textwidth}
				\centering
				\includegraphics[width=\textwidth]{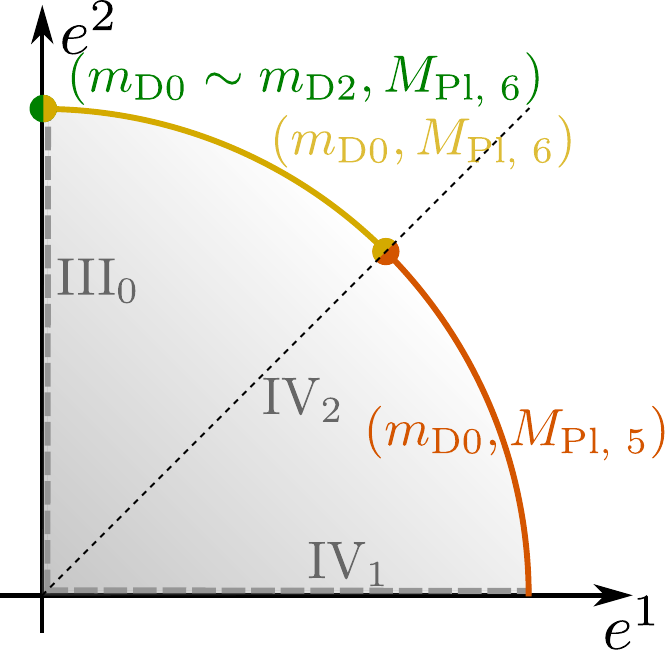}
				\caption{
					\label{fig:asympt lim  IIAP11169-lim}}
			\end{subfigure}
			~ 
			\begin{subfigure}[t]{0.495\textwidth}
				\centering
				\includegraphics[width=\textwidth]{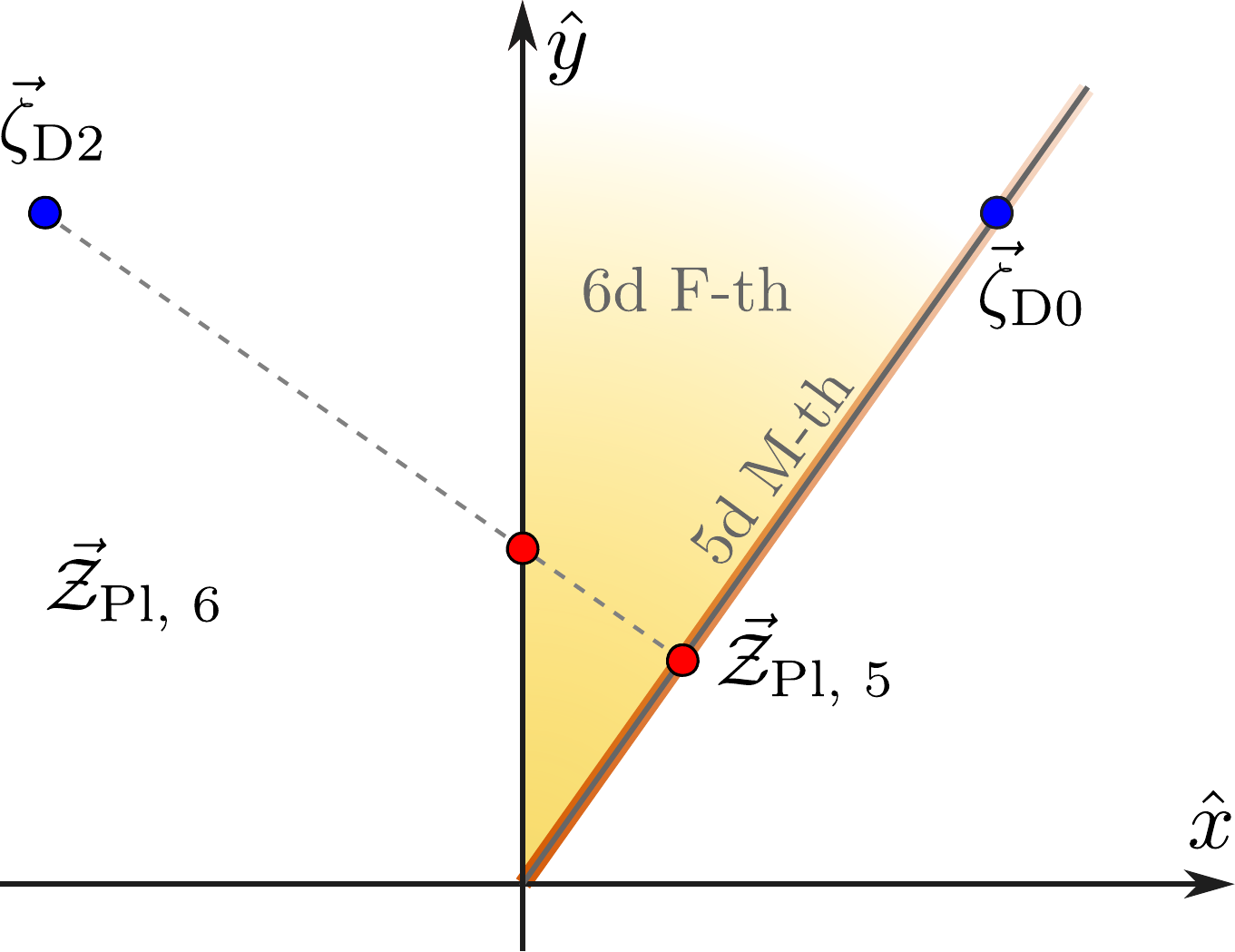}
				\caption{%\small Scalar charge-to-mass and species vectors.
					\label{fig:asympt lim  IIAP11169-vec}}
			\end{subfigure}
			\caption{\small For type IIA string theory on $\mathbb{P}^{(1,1,1,6,9)}[18]$, the infinite distance limits are classified by their singularity type according to \cite{Corvilain:2018lgw}, as well as their leading tower and species scales, as depicted in Figure \ref{fig:asympt lim  IIAP11169-lim}. Figure \ref{fig:asympt lim  IIAP11169-vec} shows the relevant (scalar) charge-to-mass vectors for towers (blue) and species scales (red) in a flat frame $\{\hat{x},\hat{y}\}$ (where $\{\hat{y}\geq \sqrt{2}\hat{x}\geq 0\}\equiv\{t^1\geq 0,\,t^2\geq 0\}$), with $\vec{\zeta}_{\rm D2}=\left(-\frac{1}{\sqrt{2}},1\right)$, $\vec{\zeta}_{\rm D0}=\left(\frac{1}{\sqrt{2}},1\right)$, $\vec{\mathcal{Z}}_{\rm Pl,\; 6}=\left(0, \frac{1}{2}\right)$ and $\vec{\mathcal{Z}}_{\rm Pl,\; 5}=\left(\frac{1}{3\sqrt{2}}, \frac{1}{3}\right)$. As argued in the text, along the $t^1\gtrsim t^2$ regime (orange), where the leading tower of states is given by $m_{\rm D0}$ whilst the species scale is set by $M_{\rm Pl;\, 5}$, asymptotically the moduli space becomes effectively one-dimensional. The other two relevant scales, namely $m_{\rm D2}$ and $M_{\rm Pl;\, 6}$ also become parametrically light at the same rate as the 5d Planck scale for these directions. The different duality frames are identified,}
			\label{fig:asympt lim  IIAP11169}
		\end{center}
	\end{figure}

	In what follows, we will study each of them in turn, to show that the pattern
	\begin{equation}
		\left.\vec{\zeta}_{\text{t}}\cdot\vec{\mathcal{Z}}_{\rm sp}\right|_{\mathbf{t}(\sigma)}=\left.\left(\mathsf{G}^{a b}\partial_{a}\log m_{\text{tower}}\,\partial_{b}\log \Lambda_{\rm sp}\right)\right|_{\mathbf{t}(\sigma)} = \frac{1}{2}\, ,
	\end{equation}
	indeed holds for any trajectory $\mathbf{t}(\sigma)$ within each region, upon replacing $\LSP$ with the properly identified species scale in each growth sector. This is summarized in Figure \ref{fig:asympt lim  IIAP11169-lim}, where the leading towers of states and species scales are shown.
	
	Notice that in this example, unlike the situation in simple toroidal compactifications where the $\zeta$-vectors remain fixed (see examples from Sections \ref{s:maxsugra} and \ref{s:N=1}), both the mass formulae and the metric $\mathsf{G}_{a b}$ vary non-trivially across the moduli space. For the latter, and using \eqref{eq:kahlerpotP11169}, one finds
	\begin{equation}
		\mathsf{G}|_{t^1\gg t^2}=
		\begin{pmatrix}
			\frac{3}{2(t^1)^2}& \frac{1}{2(t^1)^2}\\
			\frac{1}{2(t^1)^2}&\frac{1}{6(t^1)^2}
		\end{pmatrix}
		+\mathcal{O}\left(\frac{t^2}{(t^1)^3}\right)\,,\quad
		\mathsf{G}|_{t^2\gg t^1}=
		\begin{pmatrix}
			\frac{1}{2(t^1)^2}& \frac{3}{2(t^2)^2}\\
			\frac{3}{2(t^1)^2}&\frac{1}{(t^2)^2}
		\end{pmatrix}+\mathcal{O}\left(\frac{t^1}{(t^2)^3}\right)\,.
	\end{equation}
	Consider first limits with $t^1\gg t^2$. As one can see, in the asymptotic regime, $t^2$ becomes irrelevant and its growth rate does not affect the expression for the metric $\mathsf{G}_{a b}$. One can check that indeed this happens as well for any mass scale or species cut-off along this limit. Therefore, in such cases $\mathcal{M}_{\text{VM}}$ becomes effectively one-dimensional, with any original dependence on $t^2$ effectively lost. However, this is not the case when $t^2\gg t^1$, where subleading $t^1$ terms enter in the expression of the metric, as well as towers and species cut-offs. More generally, a simple computation reveals that the slice of the moduli space parametrized by $\{t^1,t^2\}$ is Riemann flat, such that global flat coordinates may be defined as follows
	\begin{equation}
		\bigg \{ \hat{x}=\frac{1}{\sqrt{2}}\log t^1+\mathcal{O}\left(\frac{t_1}{t_2}\right),\, \hat{y}=\log t^2\left(1+\mathcal{O}\left(\frac{t^1}{t^2}\right) \right) \bigg \}\, , \quad \text{with}\ \ \hat{y}> \sqrt{2}\hat{x}\geq 0\, ,
	\end{equation}
	which can then be used to depict the different $\zeta$- and $\mathcal{Z}$-vectors, see Figure \ref{fig:asympt lim  IIAP11169-vec} below. Finally, the $t^1\gtrsim t^2$ region is asymptotically mapped to the one-dimensional `boundary', namely to the line $\{\hat{y}=\sqrt{2}\hat{x}\}$.\footnote{Notice that the fact that the moduli space in flat coordinates ends precisely along the line aligned with the vector $\vec{\mathcal{Z}}_{\text{Pl},\, 5}$ prevents the Species Scale Distance Conjecture, namely the lower bound $\lambda_{\rm sp} \geq \frac{1}{\sqrt{(d-1)(d-2)}}$ on the decay rate for the species scale recently proposed in \cite{Calderon-Infante:2023ler}, from being immediately violated. This provides further (strong) evidence for the latter, which was only shown to hold in maximal supergravity setups.}

	%%%%%%%%%%%%%%%%
	
	%%%%%%%%%%%%%%%%

	\subsubsection*{Growth sector $t^1\gg t^2$ with $t^1, t^2 \to \infty$}
	
	This includes the particular case of sending $t^1 \to \infty$ with $t^2$ finite (a Type IV$_1$ singularity), since it shares the same leading behavior of the mass of the towers as well as the species scale. The asymptotic behavior for the volume reads
	\begin{equation}\label{eq:volP11169n=3}
		\mathcal{V} = \frac{3}{2} (t^1)^3 \left(1 + \mathcal{O}\left(t^2/t^1\right) \right)\, .
	\end{equation}
	
	Following the discussion of the previous section, this limit corresponds again to decompactifying to 5d M-theory, as expected from being a Type IV singularity together with the above behavior of the volume modulus. Thus, it is clear that the pattern will hold in this limit due to the general argument around eq. \eqref{noscale}, but let us show it explicitly here for illustrative purposes. By repeating the same kind of computations as in the previous example we find 
	\begin{equation}\label{eq:massesD0D2P11169}
		\frac{m_{\text{D0}}}{\Mpf} =\sqrt{8\pi } e^{K_{\text{ks}}/2} \sim \frac{1}{(t^1)^{3/2}}\, , \qquad \frac{m_{\text{D2}}}{\Mpf} =\sqrt{8\pi } e^{K_{\text{ks}}/2} t^1 \sim \frac{1}{(t^1)^{1/2}}\, ,
	\end{equation}
	for the mass scale of the D0- and D2-particles, respectively. Strictly speaking, there are two possibilities for obtaining four-dimensional BPS states from wrapped D2-branes, since there exist two different non-trivial classes of holomorphic curves within $\mathbb{P}^{(1,1,1,6,9)}[18]$. The one corresponding to the mass scale computed in \eqref{eq:massesD0D2P11169} is associated to the `horizontal' class \cite{Candelas:1994hw}, namely the fibre of the elliptic fibration. For the other `vertical' class, since the supersymmetric cycle wrapped by the 2-brane is topologically equivalent to a rational --- i.e. a $\mathbb{P}^{1}$ --- curve that is moreover contractible, there is only a finite number of associated Gopakumar-Vafa (GV) invariants \cite{Gopakumar:1998ii, Gopakumar:1998jq} which are non-zero (see e.g., \cite{Candelas:1994hw, Hosono:1993qy}). This means, incidentally, that such D2-particles do not give rise to an infinite tower of states becoming massless upon exploring the $t^1 \to \infty$ limit, such that we can safely ignore them for our purposes here. 
	
	As discussed in the previous subsection, the species scale can be computed through D0-brane state counting, arriving at the following result
	\beq\label{eq:speciesn=3} 
	\frac{\LSP}{\Mpf}\, \simeq\, \left( \frac{m_{\text{D0}}}{\Mpf} \right)^{1/3} \sim\, \frac{1}{(t^1)^{1/2}}\, ,
	\eeq
	which is nothing but the 5d Planck scale.
	
	With this, we now have all the necessary information so as to check whether the pattern \eqref{eq:pattern} is satisfied. Thus, let us first compute the charge-to-mass vectors of the leading set of states, namely the D0-brane tower, as well as the species vector obtained from eq. \eqref{eq:speciesn=3} above. The former is given by
	\beq\label{eq:zetaD0n=3} 
	\vec{\zeta}_{\text{D0}} = \left( \frac{\left(3t^1+t^2\right)^2}{6 (t^1)^3 + 6 (t^1)^2 t^2 + 2t^1(t^2)^2}\, , \, \frac{3t^1+2t^2}{6 (t^1)^2 + 6 t^1 t^2 + 2(t^2)^2}\right) \, ,
	\eeq
	where the notation is $\vec{\zeta}=\left( \zeta_{t^1}, \zeta_{t^2} \right)$, and which at leading order becomes just $\vec{\zeta}_{\text{D0}} \simeq \left( \frac{3}{2t^1}, \frac{1}{2t^1}\right)$. The latter, on the other hand, is simply proportional to the charge-to-mass vector associated to the D0-branes, such that $\vec{\mathcal{Z}}_{\text{sp}}=\frac{1}{3} \vec{\zeta}_{\text{D0}}$. Hence, upon contracting both vectors using the inverse moduli space metric \eqref{eq:fullinversemetric}, one confirms that indeed $\vec{\zeta}_{\text{D0}} \cdot \vec{\mathcal{Z}}_{\text{sp}}= \frac{1}{d-2}=\frac{1}{2}$ is verified \emph{exactly}, namely even before performing the expansion in $t^2/t^1$.

	\subsubsection*{Growth sector $t^2\gg t^1$ with $t^1, t^2 \to \infty$}
	
	For this growth sector, the situation turns out to be quite different. First, note that it includes the particular limit of sending $t^2 \to \infty$ with $t^1$ finite (a Type III$_0$ singularity \cite{Grimm:2018cpv}) and, as can be easily checked, the volume \eqref{eq:kahlerpotP11169} is dominated by the last term in the RHS
	\begin{equation}\label{eq:volumen=2limit}
		\mathcal{V} = \frac{1}{2} t^1 (t^2)^2 \left(1 + \mathcal{O}\left(t^1/t^2\right) \right)\, ,
	\end{equation}
	which implies the following asymptotic dependence for the inverse metric components (to leading order in $t^2$)
	\begin{equation}\label{eq:modspacemetricP11169n=2}
		\mathsf{G}^{-1}\, \simeq \, \left(
		\begin{array}{cc}
			2(t^1)^2 & -3 (t^1)^2  \\
			-3 (t^1)^2 & (t^2)^2 \\
		\end{array}
		\right) \, .
	\end{equation}
	The QG resolution of the singularity requires from a double decompactification to 6d F-theory on the same elliptic threefold $X_3$ \cite{Lee:2019wij,Castellano:2022bvr, Marchesano:2022axe}. This may be intuitively understood by looking again at the asymptotic behavior of the mass scales of the infinite towers of light states\footnote{For the D2-branes wrapping the elliptic fibre $k \in \mathbb{Z} \setminus \lbrace 0 \rbrace$ times one obtains \cite{Klemm:2012sx, Klemm:1996hh}
		\beq \label{eq:GVinvariantsT2limit}
		n_{\textbf{k}} = \chi_E(X_3) = 2 \left ( h^{1,1} (X_3) -  h^{2,1} (X_3) \right)\, ,
		\eeq
		thus behaving like a KK spectrum associated to a circle reduction from 5d to 4d.}
	\begin{equation}\label{eq:massesD0D2P11169n=2}
		\frac{m_{\text{D0}}}{\Mpf}  \sim \frac{1}{\sqrt{t^1} t^2}\, , \qquad \frac{m_{\text{D2}}}{\Mpf} \sim \frac{\sqrt{t^1}}{t^2}\, ,
	\end{equation}
	which present both the same dependence, contrary to the previous case (c.f. \eqref{eq:massesD0D2P11169}). Additionally, one can form $\frac{1}{2}$-BPS bound states of D0- and D2-particles upon turning on some (quantized) worldvolume flux $F$ for the wrapped D2-brane \cite{Polchinski:1998rr}. As a consequence, and following the algorithmic computation of the species scale proposed in \cite{Castellano:2021mmx}, one arrives at a cut-off of the form
	\beq 
	\frac{\LSP}{\Mpf} \simeq \left(N_{\text{D0}}\, N_{\text{D2}}\right)^{1/2} \sim \frac{1}{\sqrt{t^2}}\, ,
	\eeq
	where $N_{\text{D0, D2}}$ counts the number of D-brane states of the specified kind falling below the species scale. This moduli dependence of the species scale indeed matches with the result of the 6d Planck scale (see discussion around \eqref{eq:6dPlanckscale} below). We can then compute the scalar charge-to-mass vectors for the two towers of states, which to leading order in $1/t^2$, read as
	\begin{equation}\label{eq:zetavectorsD0D2n=2}
		\vec{\zeta}_{\text{D0}} \simeq \left( \frac{1}{2 t^1}, \frac{1}{t^2} \right)\, , \qquad \vec{\zeta}_{\text{D2}} \simeq \left( -\frac{1}{2 t^1}, \frac{1}{t^2} \right)\, .
	\end{equation}
	Analogously, one finds for the species vector
	\begin{equation}\label{eq:speciesvectorIIA}
		\vec{\mathcal{Z}}_{\text{sp}} \simeq \left( \frac{3}{4 t^2}, \frac{1}{2t^2} \right)\, ,
	\end{equation}
	such that upon taking the product with respect to the (inverse) moduli space metric \eqref{eq:modspacemetricP11169n=2}, the pattern \eqref{eq:pattern} still holds for \emph{both} towers. Notice that in order to arrive at this result it is crucial to take into account that $t^1/t^2 \to 0$ asymptotically along the limit of interest.

	\subsubsection{General story}
	\label{sss:IIA/Fthy}
	
	The previous example contained two types of limits, one belonging to the category of Section \ref{s:typeIIlimits} and a new one: a partial decompactification to 6d F-theory. Let us elaborate a bit more on this second case, which corresponds to the regime where $t^2$ grows faster than $t^1$. From \eqref{eq:modspacemetricP11169n=2}, one can check that the length of the tower of bound states charge-to-mass and species vectors behave as follows
	\begin{equation}
		|\vec{\zeta}_{(\text{D0, D2})}|= \left|\frac{1}{2} (\vec{\zeta}_{\text{D0}} + \vec{\zeta}_{\text{D2}}) \right|=\left|\left(\frac{3}{2 t^2},\frac{1}{t^2}\right)\right| \simeq 1\, , \qquad
		|\vec{\mathcal{Z}}_{\text{sp}}| \simeq \frac{1}{2}\, ,
	\end{equation}
	to leading order in the expansion parameter $t^1/t^2$. These indeed coincide with the typical values for Kaluza-Klein vectors associated to a two-dimensional compact manifold, matching with the microscopic interpretation of the singularity as a decompactification from 4d to 6d. Our aim here will be to stress that this will be generically the case whenever we explore a type $\mathbb{T}^2$ limit in the language of \cite{Lee:2019wij} (or a Type III singularity in the language of \cite{Grimm:2018ohb}). Subsequently, this will allow us to show that the pattern $\eqref{eq:pattern}$ holds in general for such kind of infinite distance limits. 
	
	Let us consider an infinite distance limit in which the curve associated to the fastest growing modulus has the intersection numbers of a Type III singularity (see second row in Tables \ref{tab:intersN=2} and \ref{tab:limitsN=2}). It was shown in \cite{Lee:2019wij} that, geometrically, this corresponds to a limit in which the Calabi--Yau threefold exhibits an elliptic fibration  over a K\"ahler surface $B_2$, and the volume of the latter grows faster than the fiber (i.e.  belongs to the type $\mathbb{T}^2$ class).	After resolving any Kodaira-Néron type singularity that may be present \cite{Weigand:2018rez}, we can then divide the K\"ahler moduli into two sets: those parametrizing fibral curves, $\lbrace t^a_f \rbrace$, and the ones inherited from the base, $\lbrace t^{\alpha}_b \rbrace$.

	The limit at hand then corresponds to
	\begin{equation}\label{eq:n=2limit}
		t^a_f = \text{const.}\, , \qquad t^{\alpha}_b = \xi^{\alpha} \sigma\, , \qquad \text{with}\, \, \sigma\to \infty\, , 
	\end{equation}
	accompanied by a suitable co-scaling of the 10d dilaton --- so as to keep fixed all moduli in the hypermultiplet sector. Microscopically, the Quantum Gravity resolution of the singularity requires from a double decompactification to F-theory on the same elliptic threefold $X_3$, as we review in the following.
	
	On the one hand, at the level of the spectrum, one finds --- at least in the simplest instances --- only two infinite sets of asymptotically light states: those comprised by D0-branes and D2-branes wrapping the elliptic fibre class. Notice that, since the volume of the latter 2-cycle, which we denote by $\mathcal{V}_{\mathbb{T}^2}$, does not diverge in the limit \eqref{eq:n=2limit}, the central charges associated to both towers of states are controlled by the same quantity, namely the (square root of the) overall threefold volume:
	\begin{equation}\label{eq:massesD0D2Fthylimit}
		\frac{m_{\text{D0}}}{\Mpf}  = \sqrt{\frac{\pi}{\mathcal{V}}}\, , \qquad \frac{m_{\text{D2}}}{\Mpf} = \sqrt{\frac{\pi}{\mathcal{V}}}\, \mathcal{V}_{\mathbb{T}^2}\, .
	\end{equation}
	and indeed they furnish the Kaluza-Klein replica along the torus of the 6d F-theory massless fields. 
	
	From this set of (asymptotically light) towers, one can easily compute the species scale dominating the infinite distance limit. In fact, upon using type IIA/F-theory duality \cite{Lee:2019wij}, we conclude that the species scale should be controlled parametrically by the six-dimensional Planck mass, namely\footnote{The second equality in \eqref{eq:6dPlanckmass} follows from the moduli identifications $R_5 = \mathcal{V}^{1/3}$ (c.f. \eqref{eq:modulimatching}) as well as $R_6^{-4/3}=\frac{\mathcal{V}_{\mathbb{T}^2}}{\mathcal{V}^{1/3}}$ \cite{Corvilain:2018lgw}.}
	\beq \label{eq:6dPlanckmass}
	M_{\text{Pl; 6}}^2 \simeq\frac{\Mpf^2}{R_6^{2/3}R_5} = \frac{\Mpf^2 \mathcal{V}_{\mathbb{T}^2}^{1/2}}{\mathcal{V}^{1/2}} \sim \frac{1}{\sigma^{1/2}}\, ,
	\eeq
	with $R_6$ and $R_5$ being the corresponding radii of the 6d-to-5d and 5d-to-4d circle compactifications (measured in the 6d and 5d Planck units respectively). Indeed, one can check that (c.f. eq. \eqref{eq:massesD0D2Fthylimit})
	\beq\label{eq:6dPlanckscale}
	\left(\frac{M_{\text{Pl; 6}}}{\Mpf}\right)^2 \simeq \left( \frac{m_{\text{D0}}\, m_{\text{D2}}}{\Mpf^2}\right)^{\frac{1}{2}} \sim \left(\frac{\LSP}{\Mpf}\right)^2\, ,
	\eeq
	in agreement with the usual species counting computation \cite{Castellano:2021mmx, Castellano:2022bvr}. 
	
	On the other hand, for the K\"ahler potential one finds the following leading asymptotic behavior \cite{Lee:2019wij,Cota:2022maf}
	\begin{equation}\label{eq:kahlerpotn=2}
		K_{\text{ks}}= - \log \left(\frac{\sigma^2}{2} \left( c_a t^a_f\right)\eta_{\alpha \beta} \xi^{\alpha} \xi^{\beta} + \mathcal{O} (\sigma)\right)\, ,
	\end{equation}
	where $c_a$ are some positive coefficients\footnote{\label{fnote:ellipticclass}The coefficients $c_a$ in eq. \eqref{eq:kahlerpotn=2} determine the generic elliptic fibre class $[C_{\mathbb{T}^2}]$. Hence, in terms of a basis $\lbrace \mathcal{C}_f^a \rbrace$ of generators of the relative Mori cone $\mathsf{M}(X_3/B_2)$, the former becomes $C_{\mathbb{T}^2}= \sum_a c_a \mathcal{C}_f^a$, where the notation follows that of \cite{Cota:2022maf}.} determined by the particular fibration structure of the threefold and $\eta_{\alpha \beta}$ denote the intersection numbers of the two-fold base \cite{Corvilain:2018lgw}. From eq. \eqref{eq:kahlerpotn=2} one can compute the moduli space metric, which reads as follows
	%
	%\begin{equation}\label{eq:metricn=2}
	\begin{align}\label{eq:metricn=2}
		\mathsf{G}_{\alpha \beta} &= \frac{1}{2} \frac{\partial^2 K_{\text{ks}}}{\partial t^{\alpha}_b \partial t^{\beta}_b} = \frac{1}{\sigma^2} \left[\mathsf{G}^{(\rm const.)}_{\alpha \beta} + \mathcal{O}(1/\sigma) \right]\, , \quad \mathsf{G}_{\alpha b} = \frac{1}{2} \frac{\partial^2 K_{\text{ks}}}{\partial t^{\alpha}_b \partial t^{b}_f} = \frac{1}{\sigma} \left[\mathsf{G}^{(\rm const.)}_{\alpha b} + \mathcal{O}(1/\sigma) \right]\, ,\notag\\
		\mathsf{G}_{a b} &= \frac{1}{2} \frac{\partial^2 K_{\text{ks}}}{\partial t^a_f \partial t^b_f} = \mathsf{G}^{(\rm const.)}_{a b} + \mathcal{O}(1/\sigma)\, ,
	\end{align}
	%\end{equation}
	%
	where all the constant matrices above have full rank except for $\mathsf{G}^{(\rm const.)}_{a b}$, which has rank one. 
	
	Armed with all this, one can readily check upon using the no-scale property of the (leading order) metric $\mathsf{G}_{\alpha \beta}$, namely 
	\begin{equation}\label{eq:noscale}
		\frac{\partial K_{\text{ks}}}{\partial t^{\alpha}_b} \mathsf{G}^{\alpha \beta} \frac{\partial K_{\text{ks}}}{\partial t^{\beta}_b} = 4\, ,
	\end{equation}
	that the product
	\begin{equation}
		\vec{\zeta}_{\text{D0, D2}}\cdot\vec{\mathcal{Z}}_{\rm sp}=\left(\mathsf{G}^{A B}\partial_{A}\log m_{\text{tower}}\,\partial_{B}\log \Lambda_{\rm sp}\right) \stackrel{~\eqref{eq:n=2limit}~}{=} \frac{1}{2}\, , \qquad A, B= \lbrace a, \beta \rbrace\, ,
	\end{equation}
	is indeed satisfied for any trajectory of the form specified in \eqref{eq:n=2limit} above. To see this, it is important to realize that any term involving derivatives with respect to the fibral moduli, $t^a_f$, provides ultimately a contribution to the scalar product $\vec{\zeta}_{\text{t}}\cdot\vec{\mathcal{Z}}_{\rm sp}$ which is of $\mathcal{O}\left(1/\sigma\right)$ or higher, such that it goes away upon taking the infinite distance limit. For this same reason, the result also applies to more general limits in which the fiber volume is also sent to infinity, but at a slower rate than that of the base.
	
	\subsection{Type II limits: Emergent String Limits}
	\label{s:typeIIlimits}
	
	\subsubsection{Type IIA on $\mathbb{P}^{(1,1,2,2,6)} [12]$}
	\label{ss:ExampleIII}
	
	As a final example, we consider type IIA string theory compactified on the threefold $X_3=\mathbb{P}^{(1,1,2,2,6)}$ [12], which was originally introduced and studied in \cite{Candelas:1993dm, Hosono:1994ax}. Such two-parameter CY$_3$ can be seen as a $K3$ fibration over a $\mathbb{P}^1$-base \cite{Lee:2019wij} (see also \cite{Blumenhagen:2018nts}), whose K\"ahler moduli $\lbrace t^1, t^2 \rbrace$ appear in the K\"ahler potential as follows\footnote{Here $t^2$ measures the classical volume of the $\mathbb{P}^1$-base, whilst $t^1$ parameterizes the volume of a second $\mathbb{P}^1$ corresponding to a rational curve (of non-negative self-intersection) inside the $K3$-fibre \cite{Candelas:1993dm}.}
	\begin{equation}\label{eq:kahlerpotP11226}
		e^{-K_{\text{ks}}} = \frac{32}{3}(t^1)^3 + 16(t^1)^2 t^2 + \ldots\, ,
	\end{equation}
	where the ellipsis denote further $\alpha'$ as well as worldsheet instanton corrections which are exponentially suppressed in the asymptotic limit. The (inverse) moduli space metric that derives from such K\"ahler potential reads 
	\begin{equation}\label{eq:fullinversemetricP11226}
		\mathsf{G}^{-1}\, = \, \left(
		\begin{array}{cc}
			(t^1)^2 & - \frac{2}{3} (t^1)^2  \\
			- \frac{2}{3} (t^1)^2 & \frac{4}{3}(t^1)^2+\frac{8}{3} t^1 t^2 +2 (t^2)^2 \\
		\end{array}
		\right) \, .
	\end{equation}
	Using the nomenclature of MHS, we have the following infinite distance limits (see e.g., \cite{kerr2019polarized, Bastian:2021eom}): \emph{(i)} $t^1 \to \infty$ with $t^2$ finite (a Type IV$_2$ singularity), \emph{(ii)} $t^2 \to \infty$ with $t^1$ finite (a Type II$_1$ singularity), and \emph{(iii)} $t^1,t^2 \to \infty$ (a Type IV$_2$ singularity). The Type IV singularities will again correspond to M-theory circle decompactifications, so the analysis of Sections \ref{sss:IIA/Mthy} and \ref{ss:ExampleII} carries over. In fact, notice that all $t^2$-dependence essentially disappears when taking the limit $t^1\gg t^2 \gg 1$. As a consequence, the moduli space becomes effectively one-dimensional within such regime, as also happened in the example from Section \ref{ss:ExampleII} above. Moreover, the fact that the $\lbrace t^1, t^2\rbrace$ slice of the moduli space is Riemann flat allows us to define some global flat coordinates,
	\begin{equation}\label{e:mod 11226}
		\bigg \{ \hat{x}=\log t^1+\mathcal{O}\left(\frac{t_1}{t_2}\right),\, \hat{y}=\frac{1}{\sqrt{2}}\log t^2\left(1+\mathcal{O}\left(\frac{t^1}{t^2}\right) \right) \bigg \}\, , \quad \text{with}\ \ \hat{y}> \frac{1}{\sqrt{2}}\hat{x}\geq 0\, ,
	\end{equation}
	which we will use to depict the different relevant $\zeta$- and $\mathcal{Z}$-vectors in the present setup. Let us also mention that the $t^1\gtrsim t^2$ region is again mapped to the line $\{\hat{y}=\frac{1}{\sqrt{2}}\hat{x}\}$. For clarity reasons, however, we will use the $\{t^1, t^2\}$ coordinates for our subsequent analysis.
	
	On the other hand, things get more interesting upon probing the limit $t^2 \to \infty$ (either with $t^1$ fixed or growing at a smaller rate), since the QG resolution of the corresponding Type II singularity is of a different type than the ones discussed so far. The purely geometric analysis of \cite{Lee:2019wij} shows that it  corresponds to an emergent string limit in which a critical heterotic string becomes asymptotically tensionless at the infinite distance boundary. This string can be seen to arise from an NS5-brane wrapping the $K3$-fibre \cite{Harvey:1995rn}, whose \emph{quantum} volume remains constant along the limit. It is clear then that the pattern \eqref{eq:pattern} is going to be satisfied in this case, since the species scale is equal to the string scale, whose exponential rate should simply be $\frac{1}{\sqrt{d-2}}$ if corresponding to the fundamental string propagating in $d$ dimensions. Nevertheless, let us check it explicitly by computing the relevant vectors in this regime. We first calculate the leading contribution to the threefold volume in \eqref{eq:kahlerpotP11226}, which scales asymptotically as follows 
	\begin{equation}
		\mathcal{V} = 2 (t^1)^2\, t^2 \left(1 + \mathcal{O}\left(t^1/t^2\right) \right)\, .
	\end{equation}
	Next we need to determine both the charge-to-mass vectors associated to the leading tower of states and the appropriate species scale. There is indeed a plethora of potentially light towers, both of perturbative and non-perturbative nature. First of all, one finds a critical string --- of heterotic nature --- arising from wrapping an NS5-brane on the $K3$ surface, whose tension behaves as
	\begin{equation}
		T_{\text{NS5}} = \frac{2\pi}{ \ell_s^2 g_s^2} \mathcal{V}_{K3}\, ,
	\end{equation}
	with $\ell_s = 2\pi \sqrt{\alpha'}$ the fundamental string length and $\mathcal{V}_{K3}$ the volume of the $K3$-fibre. Notice that along the $t^2$-limit that we consider here, the volume of the fibre either remains constant or grows at a smaller rate,	since $\mathcal{V}_{K3} \propto (t^1)^2$. Hence, upon properly co-scaling the 10d dilaton so as to keep the 4d one fixed and finite (see discussion at the end of Section \ref{ss:preliminary}) one arrives at
	\begin{equation}
		\frac{T_{\text{NS5}}}{\Mpf^2} = \frac{\mathcal{V}_{K3}}{ 2 \mathcal{V}} \sim \frac{1}{t^2} \quad \Longrightarrow \quad \vec{\zeta}_{\text{osc, NS5}} = \vec{\mathcal{Z}}_{\text{osc, NS5}} \simeq \left( \frac{1}{3t^2}, \frac{1}{2 t^2} \right)\, ,
	\end{equation}
	which holds at leading order in $t^1/t^2$. Apart from these, there are also additional infinite towers of states which become asymptotically massless in the limit of interest. These can be seen to correspond to Kaluza-Klein modes associated to the diverging $\mathbb{P}^1$-base, with mass
	\begin{equation}\label{eq:KKP1scale}
		\frac{m^2_{\text{KK}, \, \mathbb{P}^1}}{\Mpf^2} = \frac{e^{2\varphi_4}}{4 \pi \mathcal{V}_{\mathbb{P}^1}} \sim \frac{1}{t^2}\, ,
	\end{equation}
	as well as non-perturbative states arising from D0- and D2-branes wrapping the rational curve within the $K3$-fibre,
	whose masses scale as follows
	\begin{equation}\label{eq:D0D2emergenthet}
		\frac{m_{\text{D0}}}{\Mpf} = \frac{\sqrt{\pi}}{\mathcal{V}^{1/2}}\sim \frac{1}{t^1 (t^2)^{1/2}}\, , \qquad \frac{m_{\text{D2}}}{\Mpf} = \frac{\sqrt{\pi} t^1}{\mathcal{V}^{1/2}}\sim \frac{1}{(t^2)^{1/2}}\, .
	\end{equation}
	The latter infinite set of D2-branes are mapped through type IIA/Heterotic duality \cite{Kachru:1995wm} to winding modes of the dual heterotic string on $\widehat{K3} \times \mathbb{T}^2$ \cite{Harvey:1995fq, Kawai:1996te}. Note that all these towers decay at the same rate than the string along the infinite distance limit under consideration.
	
	From the above mass formulae one may readily compute the associated charge-to-mass vectors upon taking derivatives with respect to the non-compact K\"ahler fields,\footnote{Strictly speaking, the vector $\vec{\zeta}_{\text{KK},\, \mathbb{P}^1}$ presents an additional non-trivial component due to the dependence of the KK scale on the 4d dilaton in \eqref{eq:KKP1scale}. Such component, despite not contributing to the inner product \eqref{eq:patternemergentheterotic} below, must be taken into account when computing the length of the charge-to-mass vector, thus leading to a perfect matching with \eqref{eq:zeta&speciesveconemodulus} for $d=4$ and $n=2$.} yielding (to leading order):
	\begin{align}
		\vec{\zeta}_{\text{KK},\, \mathbb{P}^1} &= \left( 0, \frac{1}{2 t^2} \right)\, , \qquad \vec{\zeta}_{\text{D0}} \simeq \left( \frac{1}{t^1}, \frac{1}{2 t^2} \right)\, ,\qquad
		\vec{\zeta}_{\text{D2}} \simeq \left( \frac{1}{3t^2}, \frac{1}{2 t^2} \right)\, .
	\end{align}
	%\end{equation}
	%
	Therefore, taking into account that the species counting is dominated by the excitation modes of the dual heterotic string, such that $\vec{\mathcal{Z}}_{\text{sp}}=\vec{\mathcal{Z}}_{\text{osc, NS5}}$, one can explicitly check that
	\begin{equation}\label{eq:patternemergentheterotic}
		\vec{\zeta}_{\text{t}} \cdot \vec{\mathcal{Z}}_{\text{osc, NS5}} = \frac{1}{2}\, ,
	\end{equation}
	where $\text{t}\in \lbrace \text{KK, D0, D2, NS5} \rbrace $ includes all the light leading towers, and we have made use of the inverse metric shown in eq. \eqref{eq:fullinversemetricP11226}. In fact, the above inner product holds exactly (i.e. even before taking the infinite distance limit) for all charge-to-mass vectors except for $\vec{\zeta}_{\text{KK},\, \mathbb{P}^1}$, in which case \eqref{eq:pattern} is satisfied at leading order in $t^1/t^2$.
	
	%%%%%%%%%%%%%%%%
	\begin{figure}[htb]
		\begin{center}
			\begin{subfigure}{0.425\textwidth}
				\includegraphics[width=0.9\textwidth]{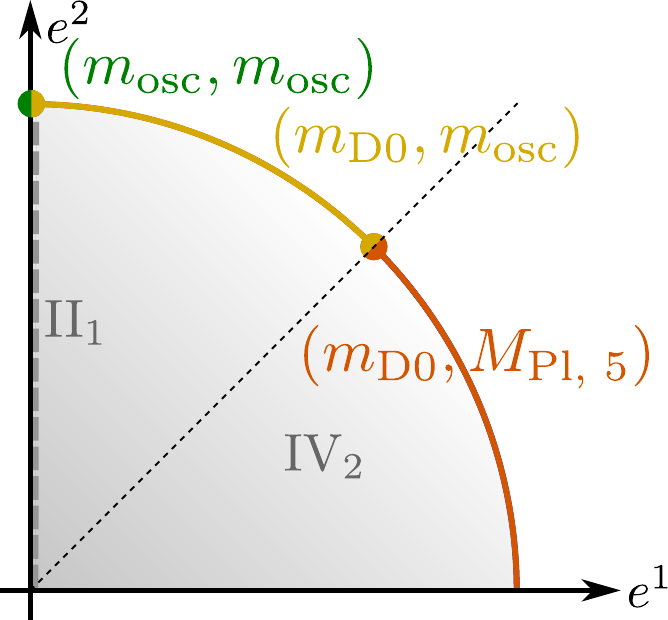}
				\caption{
					\label{fig:asympt lim  IIAP1126-lim}}
			\end{subfigure}
			\begin{subfigure}{0.455\textwidth}
				\includegraphics[width=\textwidth]{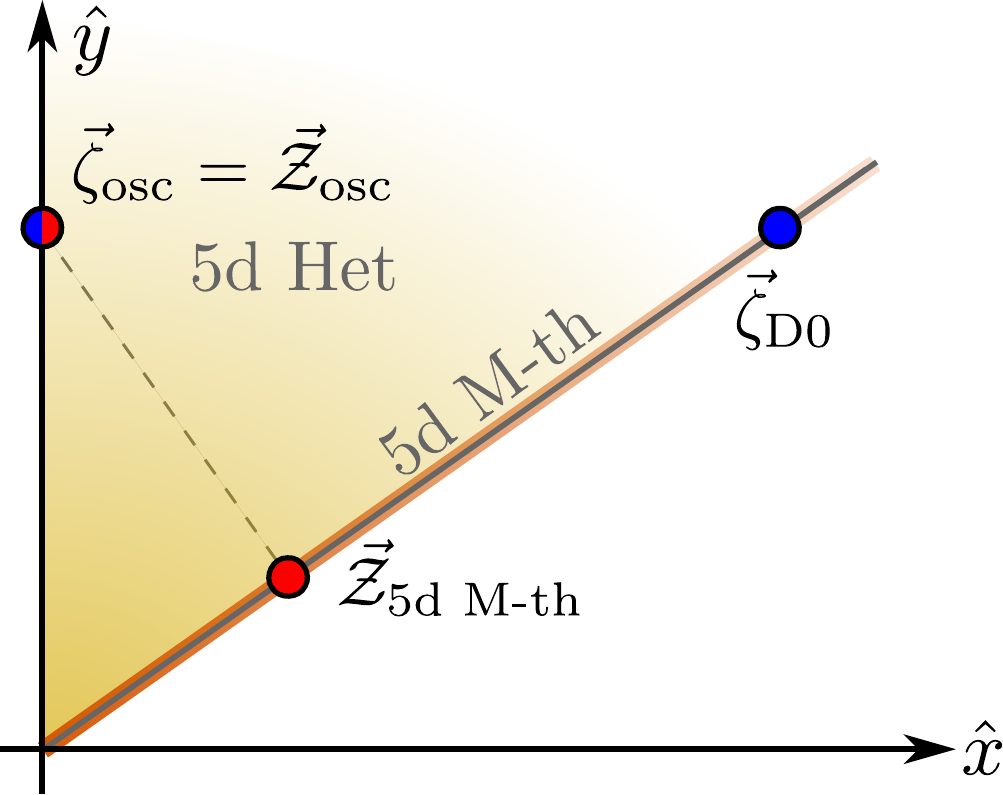}
				\caption{
					\label{fig:asympt lim  IIAP1126-vec}}
			\end{subfigure}
			\caption{\small For type IIA string theory on $\mathbb{P}^{(1,1,2,2,6)}[12]$, the infinite distance limits are classified by their singularity type according to \cite{Corvilain:2018lgw}, as well as their leading tower (blue) and species scales (red), as depicted in Figure \ref{fig:asympt lim  IIAP1126-lim}. Figure \ref{fig:asympt lim  IIAP1126-vec} shows the relevant (scalar) charge-to-mass and species vectors in the flat frame $\{\hat{y}\geq\frac{1}{\sqrt{2}}\hat{x}\geq 0\}$ given in \eqref{e:mod 11226}, with the $t^1\gtrsim t^2$ asymptotically corresponding to the one-dimensional region $\{\hat{x}=\sqrt{2}\hat{y}\}$. For this direction the string tower falls as fast as the 5d Planck mass, which sets the species scale. The vectors are given by $\vec{\zeta}_{\rm osc}=\vec{\mathcal{Z}}_{\rm osc}=\left(0,\frac{1}{\sqrt{2}}\right)$, $\vec{\zeta}_{\rm D0}=\left(1,\frac{1}{\sqrt{2}}\right)$ and $\vec{\mathcal{Z}}_{\rm Pl,\; 5}=\left(\frac{1}{3}, \frac{1}{3\sqrt{2}}\right)$. The different duality frames are identified.}
			\label{fig:asympt lim  IIAP1126}
		\end{center}
	\end{figure}

	To summarize, consider some limit of the form $\mathbf{t}(\sigma) = \left( t^1, t^2 \right) = \left(\sigma^{e^1}, \sigma^{e^2}\right)$, with the vector $\mathbf{e}$ belonging to the first quadrant of $\mathbb{S}^1$. If  $e^2 > e^1$, we get an emergent string limit and the analysis above applies. If $e^1 \geq e^2$, we rather decompactify to 5d M-theory and the general argument of Section \ref{ss:ExampleI} carries over so that the pattern also holds. In Figure \ref{fig:asympt lim  IIAP1126-lim} these limits, as well as the leading towers and species, are depicted, while in \ref{fig:asympt lim  IIAP1126-vec} the associated scalar charge-to-mass vectors (which in this case are constant in flat coordinates) are represented. Hence, the pattern
	\begin{equation}
		\left.\vec{\zeta}_{\text{t}}\cdot\vec{\mathcal{Z}}_{\rm sp}\right|_{\mathbf{t}(\sigma)}=\left.\left(\mathsf{G}^{a b}\partial_{a}\log m_{\text{tower}}\,\partial_{b}\log \Lambda_{\rm sp}\right)\right|_{\mathbf{t}(\sigma)} = \frac{1}{2}\, ,
	\end{equation}
	is verified for any asymptotic trajectory $\mathbf{t}(\sigma)$. 
	
	Finally, let us remark here that some of the towers of particles in the present setup, such as the one associated to $m_{\text{KK}, \, \mathbb{P}^1}$, suffer from the \emph{sliding} phenomenon first described within the heterotic string theory context in ref. \cite{Etheredge:2023odp} (c.f. Section \ref{ss:het s1}). Moreover, note that the charge-to-mass and species vectors arrangement in Figure \ref{fig:asympt lim  IIAP1126-vec} corresponds to a rotated version of that shown in Figure \ref{sfig:KKstring}, thus making manifest that they both share the same physical interpretation.
	
	\subsubsection{General story}
	\label{sss:IIA/heterotic}	
	
	Here we want to generalize our previous discussion so as to systematically check the pattern for any Type II singularity (in the MHS nomenclature) located within the large volume patch. The following analysis builds on the intuition gained from the previous $\mathbb{P}^{(1,1,2,2,6)}$ [12] example and it parallels that from Section \ref{sss:IIA/Fthy} above.   
	
	First, notice that this class of limits can be equivalently described in a purely geometrical way as exhibiting some kind of asymptotic surface fibration \cite{Lee:2019wij}, where the fibre is isomorphic to either a $K3$ or a $\mathbb{T}^4$ two-fold (see Table \ref{tab:limitsN=2}). This allows us to separate the K\"ahler moduli into $t^0$, which measures the volume of the $\mathbb{P}^1$-base, and $\lbrace t^{\alpha}_f\rbrace_{\alpha=1}^{h^{1,1}-1}$ which are instead associated to the $K3/\mathbb{T}^4$ fibre. Let us then consider the particular infinite distance limit described by 
	\begin{equation}\label{eq:n=1limit}
		t_f^{\alpha}  = \text{const.}\, , \qquad t^0= \sigma\, , \qquad \text{with}\, \, \sigma\to \infty\, , 
	\end{equation}
	which indeed belongs to the $K3/\mathbb{T}^4$ class. Microscopically, such a limit is believed to correspond to an emergent heterotic (or type II) string limit, where the critical string arises from compactifying an NS5-brane on the generic $K3$ (respectively $\mathbb{T}^4$) fibre.\footnote{This is difficult to prove in general, since one would need to study the excitation spectrum of the world-volume theory of the wrapped NS5-brane and match it (at all mass levels) with that of the fundamental dual string, which is of course a very non-trivial task.} Mirroring our discussion in Section \ref{sss:IIA/Fthy}, we both look at the relevant light spectrum and the moduli space metric. Regarding the former, one finds a $\frac{1}{2}$-BPS string obtained by wrapping the NS5-brane on the generic fibre (that is assumed to be \emph{fundamental}), D0-branes as well as D2-branes wrapped on 2-cycles within the fibre class, and a (double) KK tower associated to the base of the fibration. Their mass/tension read as
	\begin{align}
		\notag \frac{T_{\text{NS5}}}{\Mpf^2} &= \frac{\mathcal{V}_{K3}}{ 2 \mathcal{V}}\, , \qquad \frac{m_{\text{D0}}}{\Mpf} = \frac{\sqrt{\pi}}{\mathcal{V}^{1/2}}\, ,\\
		\frac{m_{\text{D2}}^{(\alpha)}}{\Mpf} &= \frac{\sqrt{\pi} t_f^{\alpha}}{\mathcal{V}^{1/2}}\, , \qquad \frac{m_{\text{KK}, \, \mathbb{P}^1}}{\Mpf} = \frac{e^{\varphi_4}}{\sqrt{4 \pi \mathcal{V}_{\mathbb{P}^1}}}\, ,
	\end{align}
	with $\mathcal{V}$ the overall threefold volume, $\mathcal{V}_{K3}= \frac{1}{2} \int_{K3} J\wedge J$ that of the fibre and $\mathcal{V}_{\mathbb{P}^1} = t^0$ controls the volume of the $\mathbb{P}^1$-base. 
	
	On the other hand, for the K\"ahler potential one finds the following leading asymptotic behavior \cite{Lee:2019wij}
	\begin{equation}\label{eq:kahlerpotn=1}
		K_{\text{ks}}= - \log \left(\sigma\, \eta_{\alpha \beta} t_f^{\alpha} t_f^{\beta} + \mathcal{O} (\sigma^0)\right)\, ,
	\end{equation}
	where $\eta_{\alpha \beta} = \kappa_{0 \alpha \beta}$ denotes the intersection form associated to the $K3/\mathbb{T}^4$-fibre. From this one can compute the moduli space metric, which can be expanded as a power series in $1/t^0$, similarly to the case of partial decompactification in F-theory (c.f. eq. \eqref{eq:metricn=2}).
	
	With this, we can finally prove that the pattern \eqref{eq:pattern} holds for the present Type II degenerations. Indeed, using the fact that (to leading order in $1/t^0$)
	\begin{equation}\label{eq:noscalen=1}
		\frac{\partial K_{\text{ks}}}{\partial t^0} \mathsf{G}^{0 0} \frac{\partial K_{\text{ks}}}{\partial t^0} = 2\, ,
	\end{equation}
	which can be seen as a sort of no-scale property of the metric $\mathsf{G}_{0 0}=\frac{1}{2} \frac{\partial^2 K_{\text{ks}}}{\partial t^0 \partial t^0}$, the product
	\begin{equation}\label{eq:patternn=1limits}
		\vec{\zeta}_{\text{t}} \cdot\vec{\mathcal{Z}}_{\text{osc, NS5}}=\left(\mathsf{G}^{A B}\partial_{A}\log m_{\text{tower}}\,\partial_{B}\log \Lambda_{\rm sp}\right) \stackrel{~\eqref{eq:n=1limit}~}{=} \frac{1}{2}\, , \qquad A, B= \lbrace a, \beta \rbrace\, ,
	\end{equation}
	is indeed satisfied for all $\text{t}= \lbrace \text{KK, D0, D2, NS5} \rbrace$. We stress the fact that eq. \eqref{eq:patternn=1limits} holds to leading order in $1/t^0$, since it can be seen upon using the metric derived from \eqref{eq:kahlerpotn=1} that any term involving derivatives with respect to the fibral moduli, $t_f^{\alpha}$, contributes at order $\mathcal{O}\left(1/\sigma\right)$ or higher. Once again, this is the reason why the result also applies to more general limits in which the fiber volume is also sent to infinity, but at a slower rate than that of (any curve in) the base.
	
	\subsection{Comments about the complex structure moduli space of type IIB}
	
	Let us briefly mention here how the previous analysis extends to the vector multiplet moduli space of type IIB string theory compactified on the (mirror) threefold $Y_3$. In principle, via Mirror Symmetry, a similar story should hold also for the complex structure moduli space of type IIB on $Y_3$, where the charge-to-mass and species vectors must behave in the same fashion as in its type IIA counterpart. In practice, however, the microscopic physics is oftentimes lurked, preventing us from performing a clean geometrical analysis as in type IIA. The reason for this is two-fold: First, it is difficult to argue for the existence of infinite towers of BPS bound states, since not every BPS charge may be actually populated due to the possible presence of walls of marginal stability (see footnote \ref{fnote:stabilityBPS}).\footnote{Notice that the results of ref. \cite{Palti:2021ubp} suggest that a tower of electric BPS states would always exist as long as we also have the corresponding BPS extremal black hole solution for large charges.} Therefore, it is usually not at all clear which is the lightest tower, whose $\zeta$-vector we would need to compute. Related to this, the fact that we cannot determine all towers of states becoming light for each limit means that the species scale can be hard to calculate, in general.
	
	Our aim here will be to comment on how some of these difficulties can be sidestepped, using both techniques from the Mixed Hodge Structure  literature (see e.g., \cite{Grimm:2018ohb, Grimm:2018cpv}) as well as building on our previous type IIA analysis. Thus, regarding the leading tower of states, we will assume that there is \emph{at least} one tower given by D3-branes wrapping the fastest shrinking 3-cycle. This can be motivated from the examples of Sections \ref{ss:ExampleII} and \ref{ss:ExampleIII}, where there was always some D0 or D2-brane tower becoming light at the fastest rate (even in the emergent string limits, c.f. \eqref{eq:D0D2emergenthet}). These states are all mapped through Mirror Symmetry to certain D3-branes wrapping special Lagrangian 3-cycles. From this, one can deduce at least one co-leading scalar charge-to-mass vector $\vec{\zeta}_{\rm t}$, whose components read
	\begin{equation}
		\left(\zeta_{\text{D3}}\right)_i = - \frac{1}{2} \frac{\partial K_{\text{cs}}}{\partial \text{Im}\, z^i}\, ,
	\end{equation}
	where $\lbrace z^i\rbrace$ denote the complex structure moduli and $K_{\text{cs}}$ is the associated K\"ahler potential (see Section \ref{ss:hypers} for details on this).
	
	To compute the species scale, on the other hand, one needs to know not only how many towers there are but also their microscopic degeneracy. However, we will avoid having to deal with these subtleties by looking instead at certain moduli dependent functions that correct the 4d $\mathcal{N}=2$ two-derivative lagrangian, which have been recently argued to capture the (global) behavior of the species scale within the vector multiplet moduli space \cite{vandeHeisteeg:2022btw,vandeHeisteeg:2023ubh} (see also \cite{vandeHeisteeg:2023dlw,Castellano:2023aum,Cribiori:2023sch} for related works). Following these works, we will take the topological genus-one partition function $\mathcal{F}_1$, which reads as
	\beq
	\mathcal{F}_1 =   \frac {1}{2}\left( 3+h^{2,1}-\frac {\chi (Y_3)}{12}\right)K_{\text{cs}} + \frac {1}{2}\log \det \mathsf{G}_{i\bar j} + \log|f|^2\, ,
	\label{eq:F1}
	\eeq
	to give a proxy for the number of species in the vector multiplet sector. Here, $h^{2,1}$ is the (complex) dimension of the moduli space, $\chi$ denotes the Euler characteristic of the threefold $Y_3$, $\mathsf{G}_{i\bar j}$ is the metric on the complex structure sector derived from the K\"ahler potential $K_{\text{cs}}$ (c.f. eq. \eqref{eq:CSmetric}) and $f(z^i)$ is an holomorphic anomaly which can be generically fixed upon comparing with the known asymptotic behavior of $\mathcal{F}_1$ \cite{Cecotti:1992vy,Bershadsky:1993ta}.
	
	For concreteness, we particularize in what follows to the large complex structure (LCS) regime, where a plethora of infinite distance degenerations may occur. Let us note in passing that the argument works equally well for any other such singularity, not necessarily belonging to the LCS patch. We will thus need the leading order behavior of $\mathcal{F}_1$, which is given by \cite{Bershadsky:1993ta} 
	\beq
	\mathcal{F}_1 = \frac{2\pi}{12} \int_{X_3} J\wedge c_2 + \ldots = \frac{2\pi}{12} c_{2,\, i}\, \text{Im}\, z^i + \ldots\, ,
	\label{eq:F1LCS}
	\eeq
	where $X_3$ is the mirror threefold with associated K\"ahler 2-form $J$, $c_2$ denotes its second Chern class and the ellipsis indicate further subleading contributions when $\text{Im}\, z^i \gg 1$. From this, one obtains \cite{vandeHeisteeg:2023ubh}
	\begin{equation}
		\left(\mathcal{Z}_{\text{sp}}\right)_i = - \partial_i \log \LSP = \frac{1}{2} \partial_i \log \mathcal{F}_1 = \frac{c_{2,\, i}}{2 \int J\wedge c_2} + \mathcal{O} \left( \frac{\log \text{Im}\, z^i}{\text{Im}\, z^i} \right)\, ,
	\end{equation}
	where we have used that $\LSP= \Mpf\, N^{-1/2}$, with $N= \mathcal{F}_1$. Therefore, what we want to show here is that the product
	\begin{equation}\label{eq:IIBproduct}
		\vec{\zeta}_{\text{t}} \cdot \vec{\mathcal{Z}}_{\text{sp}} =-\frac{1}{4} K_i\, \mathsf{G}^{i j}\, \frac{c_{2,\, j}}{\int J\wedge c_2} = -\frac{1}{2} K_i\, K^{i j}\, \frac{c_{2,\, j}}{\int J\wedge c_2}\, ,
	\end{equation}
	gives $\frac{1}{2}$ regardless of the kind of limit that we explore. Note that in the previous expression we have substituted the metric element along the saxionic directions $\mathsf{G}_{i j}$ in favour of $ K_{i j} = \partial_i \partial_j K_{\rm cs}$. 
	
	In a nutshell, this follows from the homogeneous dependence of the quantities $\exp (K_{\rm cs})$ and $\int J\wedge c_2$ with respect to the complex structure moduli $z^i$. Indeed, for Type II, III and IV degenerations in the complex structure moduli space, Mixed Hodge Theory tells us that the K\"ahler potential behaves (to leading order) as follows (see e.g., \cite{Grimm:2018cpv})
	\begin{equation}
		K_{\text{cs}} \to K_{\text{cs}} - \omega \log \sigma\, , \qquad \text{as}\ \ \text{Im}\, z^a \to \sigma\, \text{Im}\, z^a\, \quad \text{with }\sigma\rightarrow \infty\, ,
	\end{equation}
	with $\omega=1,2,3$ respectively, and where the set $\lbrace z^a \rbrace \subseteq \lbrace z^i \rbrace$ denotes those moduli which are sent to infinity upon approaching the infinite distance singularity. From the above relation one can prove a number of useful identities. In particular, one finds asymptotically
	\begin{align}
		\text{Im}\, z^a \partial_a K_{cs} &= -\omega\, , \qquad \text{Im}\, z^a\, \text{Im}\, z^b\, \partial_a \partial_b K_{cs} = \omega\, ,
	\end{align}
	which can then be used to show that
	\begin{subequations}
		\begin{align}
			K_a K^{a b} K_b &= \omega+\dots \, ,\\
			K^{ab} K_b &= -\text{Im}\, z^a+\dots \, .
		\end{align}
	\end{subequations}
	where the corrections vanish asymptotically. The first relation is nothing but the familiar no-scale condition of the metric $\mathsf{G}_{i j}$, whilst upon plugging the second one into eq. \eqref{eq:IIBproduct} we obtain
	\begin{equation}
		\vec{\zeta}_{\text{t}} \cdot \vec{\mathcal{Z}}_{\text{sp}} = \frac{\text{Im}\, z^a\, c_{2,\, a}}{2\int J\wedge c_2} = \frac{1}{2}\, ,
	\end{equation}
	where one needs to use that $\text{Im}\, z^a\, \partial_a \log \left( \int J\wedge c_2\right) =1$ in order to arrive at the final result, which again follows from the asymptotic homogeneity of the integrated second Chern class.

	\subsection{The Hypermultiplet moduli space}
	\label{ss:hypers}

	Up to now we have restricted ourselves to a purely classical analysis, where quantum effects can be safely neglected. The purpose of this subsection is to study the fate of the pattern \eqref{eq:pattern} within heavily quantum corrected moduli spaces, thus providing strong evidence for its robustness. We will still restrict ourselves to 4d $\mathcal{N}=2$ setups, now focusing on the hypermultiplet sector, which locally decouples --- at the two-derivative level --- from its vector multiplet counterpart \cite{Seiberg:1996ns}
	\begin{equation}\label{eq:N=2modspace}
		\mathcal{M} = \mathcal{M}_{\rm VM} \times \mathcal{M}_{\rm HM}\, .
	\end{equation}
	In type IIA CY$_3$ compactifications, the first factor in \eqref{eq:N=2modspace} is described as a projective special K\"ahler manifold of complex dimension $h^{1,1} (X_3)$ \cite{deWit:1984wbb} (c.f. discussion around \eqref{eq:kahlersectormetric}), and the second piece is a quaternionic-K\"ahler space parametrized by $4(h^{2,1} (X_3)+1)$ real scalars \cite{Bagger:1983tt}. The former includes the K\"ahler structure deformations of the compactification space $X_3$, whilst the latter contains the complex structure moduli, which are parametrized by complex coordinates, $z^I= \left(1, z^i \right)$, arising from the periods of the holomorphic $(3,0)$-form $\Omega$ as follows
	\begin{equation}\label{eq:CSmoduli}
		z^I = \frac{Z^I}{Z^0}\, , \qquad \text{with}\ \ Z^I = \int_{A_I} \Omega\, , \quad \mathcal{F}_J (z^i) = \int_{B^J} \Omega\, .
	\end{equation}
	The $A_I$\,- and $B^J$-cycles introduced above define an integral symplectic basis of $H_3(X_3)$, such that
	\begin{equation}\label{eq:symplecticpairing}
		A_I \cdot B^J = - B^J \cdot A_I= \delta^J_I\, ,
	\end{equation}
	where $I,J =0, \ldots, h^{2,1}(X_3)$. Apart from these, the hypermultiplet sector includes the 4d dilaton $\varphi_4$, which controls the Planck-to-string mass ratio $\Mpf^2/m_s^2 = 4 \pi e^{-2\varphi_4}$, a compact scalar field field $\varrho$ which is dual to the Neveu-Schwarz 2-form $B_2$, and a total of $2h^{2,1}+2$ additional axions arising from the periods of the Ramond-Ramond 3-form field
	\begin{equation}
		\zeta^I = \int_{A_I} C_3\, , \qquad \tilde{\zeta}_J = \int_{B^J} C_3\, .
	\end{equation}
	Classically, the sigma-model metric for this set of fields reads as follows \cite{Ferrara:1989ik,Cecotti:1988zx}
	\begin{align}\label{eq:classicalhypermetric}
		h_{p q}\, \dd q^p \dd q^q &= \left( \dd \varphi_4\right)^2 + \mathsf{G}_{i \bar j} \dd z^i \dd z^{\bar j} + \frac{e^{4\varphi_4}}{4} \left( \dd \varrho - \left( \tilde{\zeta}_J \dd\zeta^J-\zeta^J \dd\tilde{\zeta}_J \right)\right)^2 \notag\\
		& -\frac{e^{2\varphi_4}}{2} \left( \text{Im}\, \mathcal{B}\right)^{-1\ IJ} \left( \dd\tilde{\zeta}_I -\mathcal{B}_{IK} \dd\zeta^K\right) \left( \dd\tilde{\zeta}_J -\bar{\mathcal{B}}_{JL} \dd\zeta^L\right)\, ,
	\end{align}
	where $\mathsf{G}_{i \bar j}$ is the metric on the space of complex structures \cite{Candelas:1990pi}
	\begin{equation}\label{eq:CSmetric}
		\mathsf{G}_{i\bar j} = \partial_{z^i}\partial_{\bar z^j} K_{\text{cs}} \, ,\qquad \text{with}\qquad K_{\text{cs}}=-\log \left[{\rm i} \int\Omega \wedge \bar \Omega \right]\, ,
	\end{equation}
	and $\mathcal{B}_{IJ} (z^i)$ denotes a complex matrix whose precise form will not be needed (see e.g., \cite{Ceresole:1995ca} for more details on this). Quantum-mechanically, however, the above line element receives both perturbative and non-perturbative corrections, the latter due to e.g., Euclidean D2-brane instantons wrapping special Lagrangian (sLag) 3-cycles \cite{Becker:1995kb}. In general, such corrections are difficult to obtain (see Appendix \ref{ap:hypermetric} for details).
	
	Our discussion here will closely follow the general analysis presented in \cite{Marchesano:2019ifh,Baume:2019sry}, where the effect of the aforementioned instanton corrections on certain \emph{classical} infinite distance singularities was studied.

	\subsubsection{Classical infinite distance points}
	\label{sss:classivalvsquantum}
	
	In the following, we will focus on trajectories within $\mathcal{M}_{\rm HM}$ which lie entirely along the non-compact directions, namely we set the axion vevs to zero value. This allows us to compute the relevant metric components, i.e.
	\begin{equation}\label{eq:hypermetricIIA}
		\dd s^2_{\text{HM}}= 2(\dd\varphi_4)^2 + 2\mathsf{G}_{i\bar j}\dd z^i \dd \bar z^j + (\text{axions})\, ,
	\end{equation}
	even after taking into account perturbative and non-perturbative corrections \cite{Marchesano:2019ifh,Baume:2019sry} (see Appendix \ref{ss:exactmetric}). Here we will be interested in studying the realization of the pattern \eqref{eq:pattern} along a certain family of  trajectories, which can be parametrized as follows\footnote{Recall that since we focus on trajectories lying entirely in the hypermultiplet moduli space, the overall volume $\mathcal{V}$ of the CY$_3$ is assumed to be fixed. Hence, the 10d and 4d dilaton agree up to this constant volume factor.}
	\begin{equation}\label{eq:generictraj}
		\text{Im}\, z^i \sim \sigma^{e^1}\, , \qquad e^{-\phi}\sim \sigma^{e^2}\, , \qquad \sigma \to \infty\, ,
	\end{equation}
	with $e^1, e^2 \geq 0$. Note that such paths correspond to geodesic trajectories with respect to the classical hypermultiplet metric. 
	We now consider different scenarios depending on the precise values of $\mathbf{e}=(e^1, e^2)$.
	
	\subsubsection*{Weak String Coupling Point}
	
	For the case in which we take $\mathbf{e}=(0, e^2)$, the only contribution to the classical moduli space distance $\Delta_{\rm HM}$ arises from the 4d dilaton piece. The leading tower of states are the oscillation modes of the fundamental string, whose mass behaves asymptotically as follows (we set $e^2=1$ without loss of generality)
	\begin{equation}\label{eq:fundstringmass}
		\left(\frac{m_s}{\Mpf} \right)^2= \frac{ e^{2\varphi_4}}{4 \pi} \sim \frac{1}{\sigma^2}\, ,
	\end{equation}
	thus leading to a charge-to-mass vector whose only non-zero entry corresponds to the 4d dilaton field: 
	\begin{equation}
		\vec{\zeta}_{\text{osc}} = \left(\zeta_{\varphi_4}\, ,\, \ldots\right)= \left(-1, 0\, , \ldots, 0\right)\, .
	\end{equation}
	Notice that, since the volume of the CY threefold is kept fixed, the associated KK-scale also behaves like \eqref{eq:fundstringmass} asymptotically, namely $m_{\rm KK,\, 6} = m_s/\mathcal{V}^{1/6} \sim \sigma^{-1}$. Its charge-to-mass vector, $\vec{\zeta}_{\text{KK},\, 6}$, may be easily obtained and indeed coincides with that of the fundamental type IIA string except for an extra non-trivial component along the overall volume direction\footnote{Note that upon computing the norm of the vector \eqref{eq:KKCYzetavector} using the metrics in eqs. \eqref{eq:hypermetricIIA} and \eqref{eq:IIAlagrangian4d} one gets $|\vec{\zeta}_{\text{KK}}|=\sqrt{\frac{2}{3}}$, in agreement with \eqref{eq:zeta&speciesveconemodulus} for $d=4$ and $n=6$.}
	\begin{equation}\label{eq:KKCYzetavector}
		\vec{\zeta}_{\text{KK},\, 6} = \left(\zeta_{\varphi_4},\zeta_{\mathcal{V}}\, ,\, \ldots\right)= \left(-1,\, \frac{1}{6\mathcal{V}}\, ,0\, , \ldots, 0\right)\, .
	\end{equation}
	On the other hand, the species scale coincides with the string scale, such that upon taking the inner product between the previous vectors and $\vec{\mathcal{Z}}_{\text{sp}}$, one gets $\vec{\zeta}_{\text{t}} \cdot \vec{\mathcal{Z}}_{\text{osc}} = \frac{1}{2}$, in agreement with the pattern \eqref{eq:pattern}. To show this, one needs to use that $\mathsf{G}_{\varphi_4 \varphi_4}=2$ as well as the factorization of the vector multiplet and hypermultiplet metrics, c.f. \eqref{eq:N=2modspace}.
	
	For completeness, let us mention that even though the scaling of the 10d dilaton in \eqref{eq:generictraj} has been chosen so as to probe the weak coupling behavior of the fundamental type IIA string, one could also in principle consider trajectories with $e^2 \leq 0$, thus exploring the strong coupling regime. It turns out, however, that both kind of limits are related by $SL(2, \mathbb{Z})$ duality (see Appendix \ref{ss:SL2Z} for details), such that everything said so far trivially extends to this dual scenario as well. In particular, for such S-dual limit the dominant critical string becoming light corresponds to a D4-brane wrapping the reference sLag 3-cycle of the CY manifold with $\mathbb{T}^3$ topology \cite{Strominger:1996it},	which is precisely mapped via Mirror Symmetry to a D1-string in the type IIB dual picture \cite{Alvarez-Garcia:2021pxo}.
	
	\subsubsection*{LCS Point}
	
	Let us now turn to the other (more interesting) possibility, namely we consider the case $\mathbf{e}=(e^1, 0)$ in \eqref{eq:generictraj}, thus exploring the LCS point at fixed dilaton vev. Note that the string scale is now fixed in Planck units, such that it can no longer provide for the leading tower of states. Moreover, the overall threefold volume is kept constant, but the fact that we take a large complex structure limit means that the compact manifold develops a highly anisotropic behavior, as can be confirmed by looking at the volume of supersymmetric 3-cycles, $\Gamma=n^I A_I + n_J B^J$. The latter can be computed in string units as follows \cite{Lee:2019wij}
	\begin{equation}\label{eq:slagvolumes}
		\mathcal{V}_{\Gamma} = \int_{\Gamma} \sqrt{g} = \left(\frac{8 \mathcal{V}}{{\rm i} \int\Omega \wedge \bar \Omega } \right)^{1/2} \text{Im}\, \int_{\Gamma} e^{-{\rm i} \theta}\Omega \, ,
	\end{equation}
	where $\theta$ determines the appropriate calibration 3-form. For the limit of interest, such volumes are controlled by the period vector $\Pi(z^i) = \left(Z^0, Z^i, \mathcal{F}_j, \mathcal{F}_0\right)^{T}$, as well as the would-be K\"ahler potential $K_{\text{cs}} = {\rm i} \int_{X_3} \Omega \wedge \bar \Omega$, thus leading to the following schematic behavior
	\begin{equation}
		\mathcal{V}_{\Gamma} \sim \left \lbrace (z^i)^{-3/2}, (z^i)^{-1/2}, (z^i)^{1/2}, (z^i)^{3/2} \right\rbrace\, .
	\end{equation}
	Therefore, it becomes clear that the relevant set of asymptotically light states are linked to the fastest shrinking/growing 3-cycles, namely the one associated to the reference period (i.e. $A_0$) and its symplectic dual ($B^0$), respectively. These determine the KK scale, which behaves as follows (we henceforth set $e^1=1$ for simplicity)
	\begin{equation}\label{eq:KKSYZ}
		\left(\frac{m_{\text{KK},\, B^0}}{\Mpf} \right)^2= \frac{1}{\mathcal{V}_{B^0}^{2/3}} \left(\frac{m_s}{\Mpf} \right)^2 \sim \frac{1}{\sigma}\, ,
	\end{equation}
	and the tension of the dual type IIA string arising from a D4-brane wrapping the reference $A_0$-cycle (see discussion after \eqref{eq:KKCYzetavector})
	\begin{equation}\label{eq:D4SYZ}
		\left(\frac{T_{\text{D4}}}{\Mpf^2} \right)= \frac{\mathcal{V}_{A_0}}{g_s} \left(\frac{m_s}{\Mpf} \right)^2 \sim \frac{1}{\sigma^{3/2}}\, .
	\end{equation}
	Notice that since the KK tower \eqref{eq:KKSYZ} is parametrically heavier than the mass scale of the emergent dual type IIA string, the limit thus explored is \emph{pathological}, as defined in \cite{Lee:2019wij} (see also \cite{Baume:2019sry, Alvarez-Garcia:2021pxo}), in the sense that upon approaching the singularity it seems that one can in principle retrieve a fundamental string in less than ten spacetime dimensions. Despite this abnormal behavior, the pattern \eqref{eq:pattern} seems to be nevertheless satisfied, as one can readily confirm:
	\begin{align}\label{eq:patternviolation}
		\vec{\zeta}_{\rm{D4}} \cdot \vec{\mathcal{Z}}_{\text{D4}} &= \frac{\partial \log m_{\rm{D4}}}{\partial \text{Im}\, z^i}\, \mathsf{G}^{ij}\, \frac{\partial \log \LSP}{\partial \text{Im}\, z^j} + \frac{\partial \log m_{\rm{D4}}}{\partial \varphi_4}\, \mathsf{G}^{\varphi_4 \varphi_4}\, \frac{\partial \log \LSP}{\partial \varphi_4}\notag\\
		&=\frac{3}{8}+\frac{1}{8}=\frac{1}{2}\, ,
	\end{align}
	where we have defined $\mathsf{G}_{i j} =2 \mathsf{G}_{i \bar j}$ and we made use of the no-scale property of $K_{\text{cs}}$ close to the LCS point, which reads $K_i K^{i \bar j} K_{\bar j} = 3$. Similarly, for the scalar product between the lightest KK tower and the species scale one finds
	\begin{align}\label{eq:patternviolationII}
		\vec{\zeta}_{\text{KK},\, B^0} \cdot \vec{\mathcal{Z}}_{\text{D4}} &= \frac{\partial \log m_{\text{KK},\, B^0}}{\partial \text{Im}\, z^i}\, \mathsf{G}^{ij}\, \frac{\partial \log \LSP}{\partial \text{Im}\, z^j} + \frac{\partial \log m_{\text{KK},\, B^0}}{\partial \varphi_4}\, \mathsf{G}^{\varphi_4 \varphi_4} \frac{\partial \log \LSP}{\partial \varphi_4}\notag\\
		&=\frac{1}{4}+\frac{1}{4}=\frac{1}{2}\, .
	\end{align}
	At this point, one would be tempted to conclude that the pattern \eqref{eq:pattern} also seems to hold for the hypermultiplet sector in $\mathcal{N}=2$ theories. However, as already mentioned, such moduli space receives strong quantum corrections, such that it is not clear at all whether the conclusions drawn from the present classical analysis will survive after taking into account perturbative and non-perturbative $g_s$-\,corrections. In the following, we will argue (building on earlier works in the topic \cite{Marchesano:2019ifh, Baume:2019sry, Alvarez-Garcia:2021pxo}), that the effect of including such quantum corrections is to correct the pathological behavior exhibited in  eqs. \eqref{eq:KKSYZ} and \eqref{eq:D4SYZ}, while ensuring that the pattern is still fulfilled.

	\subsubsection{Instanton Corrections}
	\label{sss:instantons}
	
	As explained in \cite{Marchesano:2019ifh, Baume:2019sry}, the reason why the previous classical analysis is incomplete hinges on the presence of large quantum corrections which had been ignored so far. Such quantum effects arise from Euclidean D2- as well as NS5-brane instantons, and when taken into account, they may strongly modify the tree-level hypermultiplet metric displayed in eq. \eqref{eq:classicalhypermetric}. In fact, the classical LCS singularity above gets heavily corrected and is traded at the quantum level for another infinite distance degeneration, now at weak 4d string coupling. However, a careful analysis on this matters becomes rather intricate, requiring moreover from the introduction of several new tools. Therefore, in order to not complicate unnecessarily the main discussion in this section, we summarize here the upshot and the main intuition behind it, leaving the details for Appendix \ref{ap:hypermetric} (see in particular the discussion in Section \ref{ss:detailshyper}). 
	
	The argument goes as follows. One can indeed exploit the $SL(2,\mathbb{Z})$ symmetry that the hypermultiplet moduli space enjoys (even at the quantum level) to translate any limit of the form \eqref{eq:generictraj} into a \emph{dual} one at weak string coupling and fixed complex structure moduli. Hence, it gets sufficient to know how the weak coupling point is affected by the aforementioned quantum corrections. Fortunately, we do not expect neither perturbative nor non-perturbative effects to play any important role at weak coupling, since those should be suppressed along the limit $g_4=g_s \mathcal{V}^{-1/2} \to 0$. This can be confirmed upon looking at how the moduli space metric deviates from the tree-level one. Indeed, there appear additional terms which at leading order behave as follows \cite{Gunther:1998sc,Becker:1995kb} (see Appendices \ref{ss:exactmetric} and \ref{ss:detailshyper} for details) 
	\begin{equation}
		\delta  \dd s_{\rm HM}^2 =  \delta  \dd s_{\rm HM}^2\rvert_{\text{1-loop}} + \delta  \dd s_{\rm HM}^2\rvert_{\text{D-inst}}\, \sim\, g_4^2\, +\, \sum_{\gamma} \Omega_{\gamma}\, e^{-S_{m,\, k_I}} \to 0\, ,
	\end{equation}
	where the sum runs over all (towers of) D2-brane instantons with action denoted by $S_{m,\, k_I} \sim \frac{1}{g_4}$ (c.f. eq. \eqref{eq:D2instantonaction}). Hence, it is enough to use the classical approximation \eqref{eq:classicalhypermetric} for all practical purposes here, such that we conclude that the calculations performed after \eqref{eq:fundstringmass} remain valid, and the pattern is still verified for all trajectories of the form \eqref{eq:generictraj}.	
	
	\subsubsection{Intertwining the vector and hypermultiplet sectors}
	\label{sss:mixedlimits}
	
	Finally, let us briefly consider the possibility of taking limits which imply moving both in the vector and hypermultiplet moduli spaces. As a representative example, we analyze in what follows the large volume limit at fixed 10d string dilaton, corresponding to decompactification from 4d to 10d type IIA supergravity. In terms of the appropriate 4d variables, we send $\mathcal{V}\to\infty$ and, consequently, $\varphi_4=\phi-\frac{1}{2}\log\mathcal{V} \to -\infty$. This means, in particular, that the string mass becomes light in 4d Planck units
	\begin{equation}
		m_s = (4\pi)^{-1/2}\, M_{\rm Pl;\, 4}\, e^{\varphi_4} =  (4\pi)^{-1/2}\, M_{\rm Pl;\, 4}\, e^{\phi}\, \mathcal{V}^{-1/2}\to 0\;.
	\end{equation}
	Furthermore, for such a decompactification limit, the overall KK tower becomes asymptotically massless at a faster rate,
	\begin{equation}
		m_{\rm KK,\, 6} = m_s\, \mathcal{V}^{-1/6}= (4\pi)^{-1/2}\, M_{\rm Pl;\, 4}\, e^{\varphi_4} \,\mathcal{V}^{-1/6} = (4\pi)^{-1/2}\, M_{\rm Pl;\, 4}\, e^{\phi}\, \mathcal{V}^{-2/3}\to 0\,,
	\end{equation}
	so that it corresponds to the leading tower, since the D0/D2-brane towers are of course slightly/much heavier than $m_{s}$ for the limit at hand. Regarding $\LSP$, we note that the 10-dimensional Planck mass scales asymptotically like the string scale,
	\begin{equation}
		M_{\rm Pl;\, 10} = (4\pi)^{1/8}\, m_s\, e^{-\phi/4} = (4\pi)^{-3/8}\, M_{\rm Pl;\, 4}\, e^{\frac{3}{4}\varphi_4}\, \mathcal{V}^{-1/8} = (4\pi)^{-3/8}\, M_{\rm Pl;\, 4}\, e^{\frac{3}{4}\phi}\, \mathcal{V}^{-1/2}\, ,
		\label{Mpl104d}
	\end{equation}
	so that we conclude that the species scale is set by the string scale. 
	Therefore, sticking to the $\{\varphi_4,\mathcal{V}\}$ basis, one obtains
	\begin{equation}
		\vec{\zeta}_{\rm t}=\vec{\zeta}_{\rm KK,\, 6}=\left(-1,\frac{1}{6\mathcal{V}}, 0\, , \ldots, 0\right),\qquad \vec{\mathcal{Z}}_{\rm sp}=\vec{\mathcal{Z}}_{\rm osc}= \left(-1, 0\, , \ldots, 0\right)\, ,
	\end{equation}
	for the charge-to-mass and species vectors, such that upon using the relevant metric components $\mathsf{G}_{\varphi_4 \varphi_4}=2$ and $\mathsf{G}_{\mathcal{V}\mathcal{V}}=\frac{1}{6\mathcal{V}^2}$, 
	it can be readily checked that indeed $\vec{\zeta}_{\rm t}\cdot \vec{\mathcal{Z}}_{\rm sp}=\frac{1}{d-2}=\frac{1}{2}$, thus fulfilling the pattern.\footnote{This particular limit is analogous to the large volume limit of a toroidal decompactification. It is then also verified that $\vec{\zeta}_{\rm KK,\, 6}\cdot \vec{\mathcal{Z}}_{\rm Pl,\, 10}=\frac12$ with $\vec{\mathcal{Z}}_{\rm Pl,\, 10}= \left(-\frac{3}{4},\frac{1}{8\mathcal{V}}, 0\, , \ldots, 0\right)$, as derived from \eqref{Mpl104d}.}
	
	In general, one can take several combinations of limits involving moduli from both sectors of the 4d $\mathcal{N}=2$ moduli space, resulting in different microscopic interpretations of the singularities. Some of them will be subjected to strong quantum corrections, as previously discussed, but nonetheless we expect the pattern \eqref{eq:pattern} to be satisfied in all such cases, as they will simply correspond to combinations of the building blocks already discussed.
	
	\section{Examples in 4d $\mathcal{N}=1$ EFTs}
	\label{s:N=1}
	
	In this section we will check different examples of 4d $\mathcal{N}=1$ EFTs realized by compactifications of different string theories. The instances here considered were already studied in \cite{Lanza:2021udy} in the context of EFT strings, but no mention to the species scale was made in most of the cases. For these examples, we will first identify what are the leading towers and species scales, and later on we show that $\vec{\zeta}_{\rm t}\cdot\vec{\mathcal{Z}}_{\rm sp}=\frac{1}{d-2}=\frac{1}{2}$ is fulfilled in all infinite distance limits. While this is not a general proof that the observed pattern holds for any 4d $\mathcal{N}=1$ theory, it serves as a powerful indicator for the case.
	
	\subsection{M-theory on $G_2$-manifolds}
	\label{s:MthT7}
	
	We first consider a M-theory compactification on a 7-dimensional smooth $G_2$\,-manifold $X$ \cite{Beasley:2002db, Acharya:2004qe}, with a corresponding associative 3-form $\Phi$ (for which the $G_2$\,-holonomy condition requires $\dd \Phi,\,\dd\star\Phi=0$) and volume in $\ell_{11}$ units
	\begin{equation}\label{eq:vx 61}
		V_X=\frac{1}{7}\int_X\Phi\wedge\star\Phi\, .
	\end{equation}

	In the large volume regime, the $\mathcal{N}=1$ chiral coordinates can be written as $t^j=a^j+{\rm i}s^j$, upon expanding the 11d and associative 3-forms as follows
	\begin{equation}
		\mathbf{t}=\mathbf{a}+{\rm i}\mathbf{s}=[C_3+{\rm i}\Phi]=t^j[\Sigma_j]\in H^3(X,\mathbb{C})\, ,
	\end{equation}
	where $[\Sigma_i]$ are the Poincar\'e duals to a basis of 4-cycles spanning the torsion-free part of $H_4(X,\mathbb{Z})$. Up to irrelevant constants, the K\"ahler potential can be obtained from \eqref{eq:vx 61} as $K=-3\log V_X$, with $\Phi$ and $\star\Phi$ being functions only of the saxions \cite{Papadopoulos:1995da,Harvey:1999as,Gutowski:2001fm}. The 11d spacetime-metric has the form
	\begin{equation}
		\dd s^2=e^{2A}\dd s_4^2+\ell_{11}^2\dd s_X^2\,,
	\end{equation}
	where $\dd s_X^2$ is the dimensionless $X$ line element compatible with the $G_2$\,-structure, and the Weyl rescaling factor so as to obtain the 4d Einstein frame metric is given by
	\begin{equation}
		e^{2A}=\frac{\ell_{11}^2\Mpf^2}{4\pi V_X}\, .
	\end{equation}
	We will consider Joyce's compact model (see \cite{joyce1996a,joyce1996b,joyce2000compact} for more details on these manifolds), where $X$ is the resolution of $\mathbb{T}^7/\Gamma$, with the 7-torus parametrized by $\{y_a\sim y_a+1\}_{a=1}^7$. Here $\Gamma$ denotes a finite group preserving the associative three form
	\begin{equation}\label{eq: joyce 3-form}
		\Phi=\eta_{123}+\eta_{145}+\eta_{167}+\eta_{246}-\eta_{257}-\eta_{347}-\eta_{356}\, ,
	\end{equation}
	where $\eta_{abc}=\eta_a\wedge\eta_b\wedge\eta_c$ and $\{\eta_a=R_a\dd y_a\}_{a=1}^7$ denoting the $\mathbb{T}^7$ 7-bein, with the torus radii $\{R_a\}_{a=1}^7$ given in M-theory units. For our particular example, we will take $\Gamma=\mathbb{Z}_2\times \mathbb{Z}_2\times \mathbb{Z}_2$, whose generators $\{\alpha,\beta,\gamma\}$ act on the toroidal coordinates as follows:
	\begin{align}
		\alpha:\, &(y_1,\ldots,y_7)\mapsto(y_1,y_2,y_3,-y_4,-y_5,-y_6,-y_7)\, ,\notag\\
		\beta:\, &(y_1,\ldots,y_7)\mapsto(y_1,-y_2,-y_3,y_4,y_5,\frac{1}{2}-y_6,-y_7)\, ,\\
		\gamma:\, &(y_1,\ldots,y_7)\mapsto(-y_1,y_2,-y_3,y_4,\frac{1}{2}-y_5,y_6,\frac{1}{2}-y_7)\, ,\notag
	\end{align}
	which leave invariant the seven 3-forms $\eta_{abc}$ from \eqref{eq: joyce 3-form}. The latter moreover span $H^3(X,\mathbb{Z})$, so that they can be identified with seven 3-cycles $\{C^a\simeq\mathbb \mathbb{T}^3\}_{a=1}^7$, each of them parametrized by $(y_{a_I},y_{a_J},y_{a_K})$, with the same indices as the dual 3-form $\eta_{a_Ia_Ja_K}$, precisely as
	\begin{equation}\label{eq: table IJK}
		\begin{array}{c|ccccccc}
			a&1&2&3&4&5&6&7\\\hline
			a_I&1&1&1&2&2&3&3\\
			a_J&2&4&6&4&5&4&5\\
			a_K&3&5&7&6&7&7&6	
		\end{array}\, 
	\end{equation} 
	Surviving after the singularities are resolved,\footnote{In the resolution process of the $\mathbb{T}^7/\Gamma$ singularities, which are located at the disjoint union of twelve $\mathbb{T}^3$, $3\times 12=36$ extra 3-cycles $\{\tilde{C}^\alpha\}_{\alpha=1}^{36}$ are introduced, resulting in 36 additional \emph{twisted} saxions, $\tilde{s}^\alpha=\int_{\tilde{C}^\alpha}\Phi$. We will consider asymptotic limits in which these remain fixed, and thus are subleading for the computations we are interested in.} we can introduce seven \emph{untwisted} saxions
	\begin{equation}
		s^a=\int_{C^a}\Phi\, , \qquad a=1,\ldots,7\, ,
	\end{equation}
	from where we find $s^a=R_{a_I}R_{a_J}R_{a_K}$. It is then straightforward to show that in the large volume limit, the volume of the $\mathbb{T}^7/(\mathbb{Z}_2\times \mathbb{Z}_2\times \mathbb{Z}_2)$ manifold is given by $V_X=R_1 \ldots R_7=(s^1 \ldots s^7)^{1/3}$ in 11d units, modulo subleading corrections. The KK scale associated to the decompactification of any individual radius $R_a$ is thus given by
	\begin{equation}\label{eq: 4d N1 Mth KK}
		m_{{\rm KK},\, R_a}^2=\frac{M_{\rm Pl;\, 11}^2}{R_a^2}=\frac{\Mpf^2}{R_a^2V_X}=\frac{\Mpf^2}{s^{a_I}s^{a_J}s^{a_K}}\, ,
	\end{equation}
	whilst the K\"ahler potential and metric for the saxion moduli space read
	\begin{equation}\label{eq: mod sp metric saxions 4dN1}
		K_V=-\log(s^1 \ldots s^7)\ +\ \text{const.}\quad \Longrightarrow \quad \mathsf{G}_{s^I s^J}=\frac{\delta_{IJ}}{2(s^I)^2}\, .
	\end{equation}
	As an illustration, let us check the limit $s^1\to \infty$ while leaving the remaining saxions fixed and finite. From eqs. \eqref{eq: table IJK} and \eqref{eq: 4d N1 Mth KK} this is seen to correspond to decompactifying $R_1,\, R_2,\, R_3\to \infty$, resulting in three KK towers becoming light:
	\begin{equation}
		m_{\text{KK},\, R_1}=\frac{M_{\rm Pl; 4}}{\sqrt{s^1 s^2 s^3}}\, ,\quad 
		m_{\text{KK},\, R_2}=\frac{M_{\rm Pl; 4}}{\sqrt{s^1 s^4 s^5}}\, ,\quad 
		m_{\text{KK},\, R_3}=\frac{M_{\rm Pl; 4}}{\sqrt{s^1 s^6 s^7}}\, .
	\end{equation}
	As we are decompactifying three internal dimensions, the species scale will correspond to the 7d Planck mass,
	\begin{align}
		\Lambda_{\rm sp}=M_{\rm Pl;\, 7}&=\left(R_4 R_5 R_6 R_7 \right)^{1/5}M_{\rm Pl;\, 11}=(R_4 R_5 R_6 R_7)^{1/5} V_X^{-1/2}\Mpf \notag\\
		&=\left((s^1)^3 s^2\ldots s^7\right)^{-1/10}\Mpf\, .
	\end{align}
	From these, one can easily obtain the following charge-to-mass and species vectors (in a flat basis)
	\begin{align*}
		\vec{\zeta}_{\text{KK},\, R_1}&=\left(\frac{1}{\sqrt{2}},\frac{1}{\sqrt{2}},\frac{1}{\sqrt{2}},0,0,0,0\right)\,, \; 
		\vec{\zeta}_{\text{KK},\, R_2}=\left(\frac{1}{\sqrt{2}},0,0,\frac{1}{\sqrt{2}},\frac{1}{\sqrt{2}},0,0\right)\,,  \\
		\vec{\zeta}_{\text{KK},\, R_3}&=\left(\frac{1}{\sqrt{2}},0,0,0,0,\frac{1}{\sqrt{2}},\frac{1}{\sqrt{2}}\right)\,, \; 
		\vec{\mathcal{Z}}_{\rm sp}=\left(\frac{3}{5 \sqrt{2}},\frac{1}{5 \sqrt{2}},\frac{1}{5 \sqrt{2}},\frac{1}{5 \sqrt{2}},\frac{1}{5 \sqrt{2}},\frac{1}{5 \sqrt{2}},\frac{1}{5 \sqrt{2}}\right)\,,
	\end{align*}
	which indeed result in $\vec{\zeta}_{\text{KK},\, R_i}\cdot \vec{\mathcal{Z}}_{\rm sp}=\frac{1}{d-2}=\frac{1}{2}$ being satisfied for $i=1,\, 2\,, 3$. In the following, we will show that this is still the case for any other infinite distance limit that we may consider. Note that, since the transformation between $\{R_i\}_{i=1}^7$ and $\{s^I\}_{I=1}^7$ is bijective, it is equivalent to work with one parametrization or the other. However, in order to better identify the leading tower and species scale, we will henceforth sitck to the former.
	
	Let us first analyze the species scale. Hence, consider the decompactification of $n$ radii $\{\hat{R}_i\}_{i=1}^{n}\subseteq\{R_i\}_{i=1}^7$ from 4 to $d=4+n$ dimensions. In this limit, the species scale will be given by the $d$-dimensional Planck mass, which depends on $\volume_{11-d}$, namely the volume of the remaining compactified dimensions, i.e. $\volume_{11-d}=\frac{V_X}{\hat{R}_1 \ldots \hat{R}_n}$. Thus,
	\begin{align}\label{e:lsp t7}
		\LSP&\sim M_{{\rm Pl};\, 4+n}\sim M_{\rm Pl;\, 11}\volume_{11-d}^{\frac{1}{d-2}}\sim %\Mpf \volume_{11-d}^{\frac{1}{d-2}}V_X^{-1/2}\sim
		\Mpf V_X^{-\frac{n}{2(n+2)}}\left[\prod_{i=1}^n\hat{R}_i\right]^{-\frac{1}{n+2}}\notag\\
		&\sim \Mpf \prod_{i=1}^7R_i^{-\frac{1}{n+2}\left(\frac{n}{2}+\mathds{I}_{\rm dec}(i)\right)}\, ,
	\end{align}
	where $\mathds{I}_{\rm dec}(i)=1$ if $R_i$ is decompactified and $0$ otherwise. 
	
	On the other hand, for the leading tower, given the decompactifying radii $\{\hat{R}_i\}_{i=1}^{n}\subseteq\{R_i\}_{i=1}^7$, it will correspond to the KK modes associated to $\volume_{\rm t}\sim \tilde{R}_i \ldots \tilde{R}_{n_{\max}}$, where $\{\tilde{R}_i\}_{i=1}^{n_{\max}}\subseteq \{\hat{R}_i\}_{i=1}^{n}$ are the radii decompactifying fastest. Notice that $n_{\max}\leq n$. Therefore, we obtain
	\begin{equation}\label{e:mt t7}
		m_{\rm t}\sim\frac{M_{\rm Pl;\, 11}}{\volume_{\rm t}^{1/n_{\max}}}\sim\frac{M_{\rm Pl;\, 4}}{\volume_{\rm t}^{1/n_{\max}}V_X^{1/2}}\sim \Mpf \prod_{i=1}^7R_i^{-\frac{1}{2}-\frac{1}{n_{\rm max}}\mathds{I}_{\max}(i)}\,,
	\end{equation}
	where $\mathds{I}_{\max}(i)=1$ if $R_i\in \{\tilde{R}_i\}_{i=1}^{n_{\max}}$ and $0$ otherwise. With this, we can finally evaluate the inner product
	\begin{equation}\label{eq: Mth 4d N1 prod}
		\vec{\mathcal{Z}}_{\rm sp}\cdot\vec{\zeta}_{\rm t}=\mathsf{G}^{R_iR_j}\partial_{R_i}\log\LSP\partial_{R_j}\log m_{\rm t}\, ,
	\end{equation}
	where the main difficulty lies in the fact that the expression for the moduli space metric has been given in terms of the saxions and not the radii (see eq. \eqref{eq: mod sp metric saxions 4dN1} above). To solve this, it is easier to use an intermediate `logarithmic' parametrization, namely we define $R_i=\exp(\rho_i)$ and $s^I=\exp(\sigma^I)$, which yields the following (inverse) metric
	\begin{align}\label{eq: metric G2}
		\mathsf{G}^{R_iR_j}&=\frac{\partial R_i}{\partial s^I}\frac{\partial R_j}{\partial s^J}\mathsf{G}^{s^I s^J}=
		%\frac{\partial R_i}{\partial \rho_k}\frac{\partial R_j}{\partial \rho_l}\frac{\partial\sigma^K}{\partial s^I}\frac{\partial \sigma^L}{\partial s^J}\frac{\partial\rho_k}{\partial \sigma^K}\frac{\partial\rho_l}{\partial \sigma^L}\mathsf{G}^{s^I s^J}\notag\\&=
		2R_iR_j \frac{\partial\rho_i}{\partial \sigma^I}\frac{\partial\rho_j}{\partial \sigma^J}\delta^{IJ}\qquad\text{(no sum over $i,\, j$)}\, ,
	\end{align}
	where the diagonal nature of $\mathsf{G}_{s^I s^J}$ has been used so as to reach the RHS. Analogously, the relation between $\vec{\sigma}$ and $\vec{\rho}$ can be extracted using e.g., eq. \eqref{eq: 4d N1 Mth KK}, from where we find
	\begin{equation}
		2\rho_i=\sigma^{I_i}+\sigma^{J_i}+\sigma^{K_i}-\sum_{j=1}^7\rho_j\Longrightarrow\rho_i=\frac{1}{2}(\sigma^{I_i}+\sigma^{J_i}+\sigma^{K_i})-\frac{1}{6}\sum_{J=1}^7\sigma^J\,,
	\end{equation}
	or equivalently, using \eqref{eq: table IJK},
	\begin{equation}\label{eq:matrixM}
		\vec{\rho}=M\vec{\sigma}\,,\qquad\text{with }\ \ M=\frac{1}{6}\left(
		\begin{array}{rrrrrrr}
			2 & 2 & 2 & -1 & -1 & -1 & -1 \\
			2 & -1 & -1 & 2 & 2 & -1 & -1 \\
			2 & -1 & -1 & -1 & -1 & 2 & 2 \\
			-1 & 2 & -1 & 2 & -1 & 2 & -1 \\
			-1 & 2 & -1 & -1 & 2 & -1 & 2 \\
			-1 & -1 & 2 & 2 & -1 & -1 & 2 \\
			-1 & -1 & 2 & -1 & 2 & 2 & -1 \\
		\end{array}
		\right)\,.
	\end{equation}
	Notice that the matrix $M$ is symmetric and invertible. With this, we finally obtain
	\begin{equation}
		\mathsf{G}^{R_iR_j}=2R_iR_jM_{iI}M_{jJ}\delta^{IJ}=\left(\delta^{ij}-\frac{1}{9}\right)R_iR_j\qquad\text{(no sum over $i,\, j$)}\, ,
	\end{equation}
	which can be used to evaluate \eqref{eq: Mth 4d N1 prod} along any limit in which we decompactify $n$ directions, where $n_{\max}$ of which do so at the fastest rate:
	\begin{align}
		\vec{\mathcal{Z}}_{\rm sp}\cdot\vec{\zeta}_{\rm t}&=\frac{1}{n+2}\sum_{i,j=1}^7\left(\delta^{ij}-\frac{1}{9}\right)\left(\frac{1}{2}+\frac{\mathds{I}_{\max}(i)}{n_{\max}}\right)\left(\frac{n}{2}+\mathds{I}_{\rm dec}(j)\right)\notag\\
		&=\frac{1}{n+2}\sum_{i,j= 1}^7\left(\frac{n}{4}\delta^{ij}+\frac{1}{2}\delta^{ij}\mathds{I}_{\rm dec}(j)+\frac{n}{2n_{\rm max}}\delta^{ij}\mathds{I}_{\max}(i)+\frac{\delta^{ij}}{n_{\max}}\mathds{I}_{\max}(i)\mathds{I}_{\rm dec}(j)\right.\notag\\
		&\qquad\left.-\frac{n}{36}-\frac{\mathds{I}_{\rm dec}(j)}{18}-\frac{n}{18n_{\max}}\mathds{I}_{\max}(i)-\frac{\mathds{I}_{\max}(i)\mathds{I}_{\rm dec}(j)}{9n_{\max}}\right)=\frac{1}{2}\, ,
	\end{align}
	as we wanted to show. Note that from the above general expression it follows that the pattern also holds for the scalar product of the $\vec{\zeta}_{\rm t}$ associated to the decompactification of $n$ internal dimensions and $\vec{\mathcal{Z}}_{\rm sp}$ given by $M_{{\rm Pl;}\,4+n}$, as well as the species scale with any of the subleading towers, since $\{\tilde{R}_i\}_{i=1}^{n_{\max}}\subseteq \{\hat{R}_i\}_{i=1}^{n}$. From \eqref{e:lsp t7} and \eqref{e:mt t7} one can check that those towers are always lighter than (and as a result its light states contribute to) the species scale, since
	\begin{equation}\label{e: sub lighter}
		\frac{m_{\{\tilde{R}_i\}_{i=1}^{n_{\max}}}}{\LSP}=\exp\left[-\sum_{i=1}^{7}\left\{\begin{array}{lr}
			\frac{1}{n+2}&\text{if }R_i\not\in \{\tilde{R}_i\}_{i=1}^{n_{\max}},\, \{\hat{R}_i\}_{i=1}^{n}\\
			0& \text{if }R_i\in \{\hat{R}_i\}_{i=1}^{n}\backslash\{\tilde{R}_i\}_{i=1}^{n_{\max}}\\
			\frac{n+1-n_{\rm max}}{n_{\rm max}(n+2)}&\text{if }R_i\in\{\tilde{R}_i\}_{i=1}^{n_{\max}}
		\end{array}\right\}e_i\lambda\right]\ll 1
	\end{equation}
	for any $R_i\sim \exp\{e_i\lambda\}$, with $\lambda\to\infty$ and $e_i\geq 0$, trajectory. This is not necessarily the case, however, if $\{\tilde{R}_i\}_{i=1}^{n_{\max}}\not\subset\{\hat{R}_i\}_{i=1}^{n}$.

	Therefore, since decompactification and large saxion limits are mapped in a one-to-one fashion,\footnote{Note that since the matrix $M$ in \eqref{eq:matrixM} is invertible, it defines a bijection between the variables $\lbrace \sigma^J \rbrace$ and $\lbrace \rho^j \rbrace$. In addition, such a map can be seen to provide an automorphism of $\mathbb{R}^7_{\geq 0}$, such that the large saxion and supergravity regimes are respected.} all possible (large volume) infinite distance limits of M-theory compactified on $\mathbb{T}^7/(\mathbb{Z}_2\times \mathbb{Z}_2\times \mathbb{Z}_2)$ fulfill the pattern $\vec{\mathcal{Z}}_{\rm sp}\cdot\vec{\zeta}_{\rm t}=\frac{1}{d-2}=\frac{1}{2}$. This is expected by the general discussion of Section \ref{s:maxsugra}, since further quotienting by an orbifold action should not change the structure of infinite distance limits in the untwisted sector. However, it is useful to work out an explicit example in full glory here, so that we can check whether the pattern is also satisfied by the subleading towers. The conclusion is that, as long as the towers survive the projection, this is not affected by the orbifolding, and for any practical use only the volume scaling plays a role in the pattern. One would expect this to be also the case when working with more general product manifolds $\mathcal{X}'_n=(\mathcal{X}_{n_1}\times \ldots \times\mathcal{X}_{n_N})/\Gamma$, with $\Gamma$ some appropriate finite group, where the towers and the species scales can be associated with partial decompactifications of intermediate sub-manifolds, as argued in Appendix \ref{ap:generalities}.
	
	\subsection{Heterotic string theory on a CY$_3$}
	\label{ss:heteroticCY3}
	
	As a second example, we consider heterotic string theory compactified on a Calabi--Yau threefold, $X$. Following \cite{Lanza:2021udy}, we take $X=\mathbb{P}^{(1,1,1,6,9)} [18]$, which we recall can be regarded as an elliptic fibration over $\mathbb{P}^2$, so that $b_2(X_3)=2$. The geometric classification of infinite distance limits in the K\"ahler sector parallels that of Section \ref{ss:ExampleII}, but the microscopic interpretation of the towers and the species scale will be quite different, since there are no BPS particles in 4d $\mathcal{N}=1$. We focus here on infinite distance limits associated to the K\"ahler moduli as well as the 4d dilaton. Thus, let us consider two holomorphic curves, $\mathcal{C}_{\mathbb{T}^2}$ and $\mathcal{C}_{\mathbb{P}^1}$,\footnote{These curves can be identified with the generic elliptic fiber and a $\mathbb{P}^1$ within the $\mathbb{P}^2$-base \cite{Candelas:1994hw}.} whose volume in string units is measured by the K\"ahler saxions $s^1,\,s^2\in\mathbb{R}_{>0}$, respectively, whilst the overall volume of $X$ is given by (again in string units)
	\begin{equation}\label{eq: vol P11169}
		V_X=\kappa(\mathbf{s},\mathbf{s},\mathbf{s})=\frac{1}{6}\kappa_{abc}s^a s^b s^c=\frac{3}{2}(s^1)^3+\frac{3}{2}(s^1)^2s^2+\frac{1}{2}s^1(s^2)^2\, .
	\end{equation}
	Here $\kappa_{abc}$ denote the triple intersection numbers of the threefold $X$ written in some basis $\lbrace \omega_a \rbrace$ of $H^2(X, \mathbb{Z})$. % integral basis of the K\"ahler form dual to the divisor set. 
	Furthermore, we include the universal saxion $s^0$, which is defined as
	\begin{equation}\label{eq: univ saxion}
		s^0=e^{-2\phi}V_X\, ,
	\end{equation}
	with $\phi$ the 10d dilaton. In the perturbative regime, the leading contribution to the K\"ahler potential in the previous parametrization has the form
	\begin{equation}
		K=-\log s^0-\log V_X+\ldots\, ,
	\end{equation}
	from where the moduli space metric can be derived as follows
	\begin{equation}
		\mathsf{G}_{s^a s^b}=\frac{1}{2}\frac{\partial^2 K}{\partial s^a \partial s^b}\, .
	\end{equation}
	Note that since $V_X$ is not an homogeneous function of the K\"ahler saxions, then $\mathsf{G}_{s^a s^b}$ will not be diagonal, although it will simplify when taking certain limits (see Section \ref{ss:ExampleII} for details on this point).
	
	The 10d string frame metric obeys the following ansatz,
	\begin{equation}
		\dd s^2=e^{2A}\dd s_4^2+\ell_s^2\dd s^2_{X_3}\, ,
	\end{equation}
	where $\dd s_4^2$ is the 4d Einstein frame metric and $\ell_{\rm s}$ denotes the string length, such that
	\begin{equation}\label{eq: weyl P11169}
		e^{2A}=\frac{\ell_s^2 \Mpf^2e^{2\phi}}{4\pi V_X}=\frac{\ell_s^2\Mpf^2}{4\pi s^0}\, .
	\end{equation}
	Following \cite{Lanza:2021udy}, let us start by enumerating all possible towers of states that could become light in some asymptotic corner. First of all, the fundamental string tension in 4d Planck units reads as
	\begin{equation}
		T_{\rm F1}=\frac{2\pi e^{2A}}{\ell_s^2}=\frac{\Mpf^2}{2s^0}\, ,
	\end{equation}
	implying the following moduli dependence for the string scale
	\begin{equation}\label{eq: P11169 string scale}
		m_{s}\sim \Mpf\, (s^0)^{-1/2}\, .
	\end{equation}
	As for the KK scale, denoting by $R_\ast$ the largest decompactification `radius' measured in string units, we have
	\begin{equation}
		m_{{\rm KK}_\ast}=\frac{2\pi e^A}{R_{\ast}\ell_s}\sim \frac{\Mpf}{(s^0)^{1/2}R_\ast}\, .
	\end{equation}
	These are associated to several even-dimensional cycles of the internal geometry, for which an useful basis can be constructed as follows:
	\begin{align}\label{eq:miscellanea}
		\mathcal{V}_a &= \frac{1}{2} \int_{X_3} \omega_a \wedge J\wedge J = \frac{1}{2} \kappa_{a b c} s^b s^c\, , \qquad \mathcal{V}_{a b} = \int_{X_3} \omega_a \wedge \omega_b \wedge J = \kappa_{a b c} s^c\, ,
	\end{align}
	which indeed measure the volume of the divisor dual to $\omega_a$ as well as that of the intersection curve between (the duals of) $\omega_a$ and $\omega_b$, respectively.
	
\begin{figure}[htb]
		\begin{center}
			\includegraphics[width=0.55\textwidth]{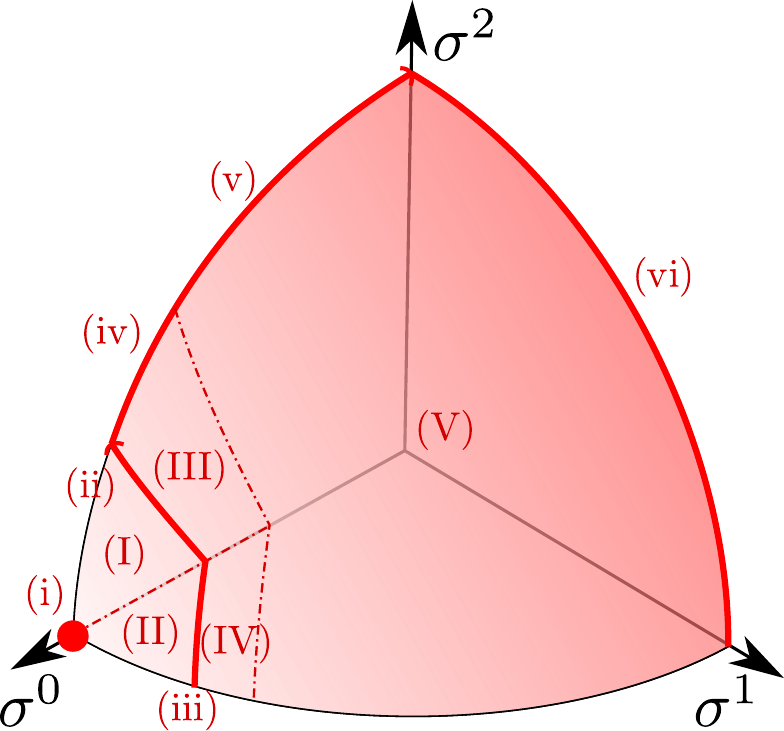}
			\caption{\small Classification of the different $(s^0,s^1,s^2)(\lambda)\sim (e^{\sigma^0 \lambda},e^{\sigma^1\lambda},e^{\sigma^2\lambda})$ limits in terms of their leading $m_{\rm t}$ and $\LSP$, as given in Table \ref{tab:limits hetCY}. In the interfaces between different regions their leading towers and species cut-off scale in the same way, respectively. Recall that measure-zero regions are depicted in lowercase.} 
			\label{fig:regions het 4dN=1}
		\end{center}
	\end{figure}

	Now, upon looking at the analytic expression of the threefold volume \eqref{eq: vol P11169} in terms of that of $s^1\sim\volume(\mathcal{C}_{\mathbb{T}^2})$ and $s^2\sim\volume(\mathcal{C}_{\mathbb{P}^1})$, we note that only two scales are allowed: If $s^2\gg s^1$, then $V_X\sim s^1(s^2)^2$ and the fastest growing cycle corresponds to the $\mathbb{P}^2$-base,\footnote{Of course, the volume of the $\mathbb{P}^1$ curve, $\mathcal{C}_{\mathbb{P}^1}$, inside the $\mathbb{P}^2$-base scales in the same way as the latter, namely $R_{\mathbb{P}^1} \sim R_{\mathbb{P}^2}\sim (s^2)^{1/2}$. However, the overall decompactified volume corresponds to the 4-dimensional divisor instead of just the complex curve $\mathcal{C}_{\mathbb{P}^1}$.} namely $R_\ast=R_{\mathbb{P}^2}\sim \left(\mathcal{V}_1\right)^{1/4} \sim (s^2)^{1/2}$; whilst any other limit results in full decompactification, with $R_\ast=R_{X}\sim V_X^{1/6}$. As a consequence, we only find the following two possibilities:
	\begin{equation}\label{eq:KKscales}
		m_{{\rm KK},\,X}\sim \Mpf\, (s^0)^{-1/2}\, V_X^{-1/6},\qquad m_{{\rm KK},\, \mathbb{P}^2}\sim \Mpf\, (s^0)^{-1/2}\, (s^2)^{-1/2}\, .
	\end{equation}
	Moreover, for limits in which $s^0$ is held fixed whilst the overall volume diverges, we enter into a strong coupling regime, since from \eqref{eq: univ saxion} we find that $e^{2\phi}=V_X/s^0 \sim V_X$. For such limits, new light degrees of freedom can appear. Take for example $E_8\times E_8$ heterotic string theory, where the strong coupling limit is given by Hořava-Witten's construction of M-theory compactified on $X_3\times\mathbb{S}^1/\mathbb{Z}_2$ \cite{Horava:1996ma, Horava:1995qa}. In this case, there is a new KK scale associated to the interval (which corresponds to the non-BPS D0-brane mass), whose mass scale is given by
	\begin{equation}
		m_{\text{KK, M-th}}=\frac{\sqrt{2\pi}e^{2A}e^{-2\phi}}{\ell_s}\sim \Mpf V_X^{-1/2}\, .
	\end{equation}
	Once the possible leading towers $m_{\rm t}$ have been laid out, we can start considering the species scale $\Lambda_{\rm sp}$. Apart from the already computed string scale \eqref{eq: P11169 string scale}, the species cut-off can be set by the higher dimensional Planck masses. Given  $m_{\rm s}\sim R_{10}^{1/2}M_{\rm Pl;\, 11}$, with $R_{10}$ being the length of the $\mathbb{S}^{1}/\mathbb{Z}_2$ interval, and using the fact that $e^{\phi}\equiv R_{10}^{3/2}$ \cite{Horava:1996ma}, we obtain 
	\begin{equation}
		M_{\rm Pl;\, 11}\sim \Mpf R_{10}^{-1/2}(s^0)^{-1/2}\sim \Mpf (s^0)^{-1/3}V_X^{-1/6}\, .
	\end{equation}
	On the other hand, $M_{\rm Pl;\, 10}\sim e^{-\phi/4}m_{\rm s}$, so that
	\begin{equation}\label{e:Mpl10}
		M_{\rm Pl;\, 10}\sim \Mpf(s^0)^{-3/8}V_X^{-1/8}\, .
	\end{equation}
	As for $M_{\text{Pl};\, d}$ with $4\leq d\leq 10$, we need to be more careful. In general we will have
	\begin{equation}
		M_{{\rm Pl};\, d}\sim M_{\rm Pl;\, 10}\volume_{10-d}|_{\rm Pl}^{\frac{1}{d-2}}=\Mpf(s^0)^{-3/8}V_X^{-1/8}\volume_{10-d}|_{\rm Pl}^{\frac{1}{d-2}}\, ,
	\end{equation}
	where $\volume_{10-d}$ is the volume of the remaining compactified dimensions, which we will need in Planck units. While our volumes are expressed in string units, we notice that from \eqref{e:Mpl10} $\ell|_{\rm Pl}=\ell|_{\rm string}(s^0)^{1/8}V_X^{-1/8}$. Hence, we finally arrive at
	\begin{equation}
		M_{\text{Pl};\, d}=\Mpf(s^0)^{\frac{4-d}{2(d-2)}}V_X^{-\frac{1}{d-2}}\volume_{10-d}^{\frac{1}{d-2}}\, ,
	\end{equation}
	where now $\volume_{10-d}$ is measured in string units, and which agree with the $d=4,\, 10$ cases. For our purposes, it will be enough to compute $M_{\rm Pl;\, 8}=M_{\rm Pl;\, 4}(s^0)^{-1/3}(s^2)^{-1/3}$, since as we commented before (see discussion around eq. \eqref{eq:KKscales}), there are no decompactification limits to arbitrary dimensions, but only to $d=5, 8, 10$ and 11 (M-theory). The first of these, on the other hand, is obtained by decompactifying only the M-theory interval without increasing the overall volume, so that the 5d Planck scale can be obtained through the same argument as in \eqref{eq:zetaD0n=3}, resulting in
	\begin{equation}
		M_{\rm Pl;\, 5}\sim \Mpf V_X^{-1/6}\, .
	\end{equation}
	Now, in order to test the pattern we will compute the decay rates of the different towers and scales along some trajectory $\mathbf{s}(\lambda)$,
	\begin{equation}
		\alpha_I=-\frac{\partial_\lambda\log m_I|_{\mathbf{s}(\lambda)}}{\|\dot{\mathbf{s}}(\lambda)\|}\, ,
	\end{equation}
	and then calculate for those dominating (this is, with larger $\alpha_I$)
	\begin{equation}\label{eq: patron check}
		\left.\vec{\zeta}_{\rm t}\cdot\vec{\mathcal{Z}}_{\rm sp}\right|_{\mathbf{s}(\lambda)}=\left.\left(\mathsf{G}^{s^i s^j}\partial_{s^i}\log m_{\rm t}\,\partial_{s^j}\log \LSP\right)\right|_{\mathbf{s}(\lambda)}\, .
	\end{equation}
	We can scan the space of possible asymptotic limits by considering trajectories of the form $\mathbf{s}(\lambda)\sim(e^{\sigma^0 \lambda},e^{\sigma^1\lambda},e^{\sigma^2\lambda})$, with $\mathbf{e}$ belonging to the first octant of $\mathbb{S}^2$, and indeed check that the $\vec{\zeta}_{\rm t}\cdot\vec{\mathcal{Z}}_{\rm sp}=\frac{1}{2}$ pattern is fulfilled.\footnote{In the cases for which $s^i\sim s^j$, one could wonder whether $s^i=a s^j$ could give rise to different limits with different finite $a$, but indeed they result in the same dominating towers and species scales.} In Figure \ref{fig:regions het 4dN=1} the different asymptotic limits, classified according to which scales $m_{\rm t}$ and $\LSP$ dominate, are depicted in terms of $\mathbf{\sigma}=(\sigma^0,\sigma^1,\sigma^2)$. These regions are also shown in Table \ref{tab:limits hetCY}. In the interfaces between two or more regions, the lightest towers and species scales from each of them scale in the same way, and indeed the product $\vec{\zeta}_{\rm t}\cdot\vec{\mathcal{Z}}_{\rm sp}=\frac{1}{2}$ is fulfilled.
	\begin{table}[h!!]
		\begin{center}
			\begin{tabular}{c|c|cc}
				Region&Limit&$m_{\rm t}$&$\LSP$\\\hline
				(I)&$s^0\gg s^1(s^2)^2;\; s^2\gtrsim s^1$ &$m_{{\rm KK},\, \mathbb{P}^2}$ &$m_s$  \\
				(II)&$s^0\gg (s^1)^3;\; s^1\gtrsim s^2$ &$m_{\text{KK},\, X}$ &$m_s$  \\
				(III)&$s^1(s^2)^2\gg s^0\gtrsim s^1s^2;\; s^2\gtrsim s^1\gg 1$ &$m_{{\rm KK},\, \mathbb{P}^2}$ &$M_{\rm Pl;\, 11}$  \\
				(IV)&$(s^1)^3 \gg s^0\gtrsim (s^1)^2;\; s^1\gtrsim s^2$ &$m_{\text{KK},\, X}$ &$M_{\rm Pl;\, 11}$  \\
				(i)& $s^0\gg s^1, s^2\sim 1$ &$m_s$ &$m_s$  \\
				(ii)& $s^0\sim s^1(s^2)^2,\, s^1\ll s^2$ & $m_{{\rm KK},\, \mathbb{P}^2}$& $M_{\rm Pl;\, 10}$\\
				(iii)& $s^0\sim (s^1)^4,\, s^1\gg s^2$ & $m_{\text{KK},\, X}$& $M_{\rm Pl;\, 10}$\\
				(iv)& $(s^2)^2\gtrsim s^0\gtrsim s^2; s^1\sim 1$ & $m_{{\rm KK},\, \mathbb{P}^2}$& $M_{\rm Pl;\, 8}$\\
				(v)& $s^2\gtrsim s^0\gg 1; s^1\sim 1$ & $m_{\text{M-th}}$& $M_{\rm Pl;\, 8}$\\
				(vi)& $s^0\sim 1$& $m_{\text{M-th}}$& $M_{\rm Pl;\, 5}$\\
				(V)& Otherwise&$m_{\text{M-th}}$ &$M_{\rm Pl;\, 11}$
			\end{tabular}
		\end{center}
		\caption{Classification of the different asymptotic regions of the heterotic string theory on $\mathbb{P}^{(1,1,1,6,9)}[18]$ in terms of the saxion limit and leading tower $m_{\rm t}$ and species scale $\Lambda_{\rm{sp}}$. In uppercase Roman numerals are identified those regions having the same dimensionality as the space of infinite distance limits, while those measure-zero limit sets are numbered in lowercase. In the interfaces (actually intersections) between different regions, the leading towers and species cut-offs scale in the same way.\label{tab:limits hetCY}}
	\end{table}
	
	%%%%%%%%%%%%
	%%%%%%%%%%%%
	
	As it was also the case for toroidal compactifications (see Section \ref{s:maxsugra}), the pattern holds even stepwise, if we decompose the limit into several steps associated to the different towers that we encounter before reaching the species scale.
	Consider for example the limit given by $\mathbf{s}(\lambda)\sim(e^{4\sigma},e^{\sigma},e^{2\sigma})$, which sits in the interior of region (III). Along that limit, $m_{\rm t}=m_{\text{KK},\, \mathbb{P}^2}\ll m_{{\rm KK},\, X}\ll m_{\rm KK,\, M-th}\ll\Lambda_{\rm sp}=M_{\rm Pl;\, 11}\ll M_{\rm Pl;\, 10}\ll M_{\rm Pl;\, 8}\sim m_{\rm s}\ll M_{\rm Pl;\, 5}$. 
	Hence, we first decompactify the $\mathbb{P}^2$ cycle, then the complete CY $X$ and finally the M-theory interval, until reaching the 11d theory. 
	When computing the product, one then finds
	\begin{equation}
		\vec{\zeta}_{\text{KK},\, \mathbb{P}^2}\cdot\vec{\mathcal{Z}}_{\rm Pl,\, 11}=\vec{\zeta}_{\text{KK},\, \mathbb{P}^2}\cdot\vec{\mathcal{Z}}_{\rm Pl,\, 10}=\vec{\zeta}_{\text{KK},\, \mathbb{P}^2}\cdot\vec{\mathcal{Z}}_{\rm Pl,\, 8}=\vec{\zeta}_{\text{KK},\, \mathbb{P}^2}\cdot\vec{\mathcal{Z}}_{\rm s}=\frac{1}{2}\;,
	\end{equation}
	whereas $\vec{\zeta}_{\text{KK},\, \mathbb{P}^2}\cdot\vec{\mathcal{Z}}_{\rm Pl,\, 5}=\frac{1}{6}$,\footnote{This is easy to understand by looking at Table \ref{tab:limits hetCY}, as $m_{{\rm KK}, \mathbb{P}^2}$ is the leading tower in some of the regions in which $\Lambda_{\rm sp}$ is given by $M_{\rm Pl; 8}$, $M_{\rm Pl; 10}$ and $M_{\rm Pl; 11}$, but never $M_{\rm Pl; 5}$. As a matter of fact, for the lower part of region (vi), where $s^1\gg s^2\gg s^0\sim 1$, $m_{{\rm KK}, \mathbb{P}^2}$ is heavier than $\Lambda_{\rm sp}=M_{\rm Pl;5}$.}. Hence, the pattern holds between the leading tower and any of the intermediate species scales obtained in each decompactification step until finally reaching the 11d theory. Moreover,
	\begin{equation}
		\vec{\zeta}_{\text{KK},\, \mathbb{P}^2}\cdot\vec{\mathcal{Z}}_{\rm Pl,\, 11}=\vec{\zeta}_{\text{KK},\, X}\cdot\vec{\mathcal{Z}}_{\rm Pl,\, 11}=\vec{\zeta}_{\text{KK, M-th}}\cdot\vec{\mathcal{Z}}_{\rm Pl,\, 11}=\frac{1}{2},
	\end{equation}
	while $\vec{\zeta}_{\rm osc}\cdot\vec{\mathcal{Z}}_{\rm Pl,\, 11}=\frac{1}{3}$,	so that the pattern also holds between the final (true) species scale and all the subleading towers that will play the role of the leading tower after each decompactification step of the process.	This is a generalization of the \emph{nested} decompactifications described in Appendix \ref{ap:generalities} for product manifolds.
	
	Notice that, unlike type IIA string theory on $\mathbb{P}^{(1,1,1,6,9)}[18]$ (see Section \ref{ss:ExampleII}), in this example there are no D2-branes that can become light yielding an effective decompactification to 6d F-theory on our (elliptically-fibered) Calabi--Yau manifold. Because of this, the limits with $s^0$ constant always have $m_{\rm M-th}$ as the leading tower, with $M_{\rm Pl;\, 5}$ being the species scale. This explains the difference between region (vi) in Figure \ref{fig:regions het 4dN=1} (see also Table \ref{tab:limits hetCY}) and Figure \ref{fig:asympt lim  IIAP11169}.
	
	For this example and those of Sections \ref{ss:ExampleII}, \ref{ss:ExampleIII}, \ref{sss:IIA/heterotic} and \ref{ss:heteroticCY3}, the internal manifolds are different Calabi--Yau threefolds rather than toroidal compactifications (or quotients thereof). While for the later case (or more generally when having product manifolds) there is a diagonal/boxed internal metric and the different cycles can be decompactified independently (see Appendix \ref{ap:generalities} for more on this), in the case of Calabi--Yau threefolds one finds for the volume some non-homogeneous expression $V_X=\frac{1}{6}\kappa_{abc}s^a s^b s^c$ in terms of the different K\"ahler saxions and intersection numbers, such that a complicated moduli space metric arises. As already described in some of the examples, for single-field asymptotic limits the internal manifold can be seen as a fibration, with the volume and metric simplifying considerably, as reviewed more generally in Section \ref{ss:preliminary}. Interestingly, as already stressed, the pattern not only holds along these limits, but also for asymptotic directions interpolating between them, along which different terms in the volume can compete, and thus they cannot be disregarded. In this subsection, we have only shown the realization of the pattern in a 4d $\mathcal{N}=1$ setup for one particular Calabi--Yau example, but we expect that the generalization to any CY should work analogously as we did for 4d $\mathcal{N}=2$ in Section \ref{s:8supercharges}. The geometrical features characterizing the infinite distance limits are the same, and only the interpretation and masses of the towers change, as explained above.

	\subsection{Type IIB/F-theory compactifications}
	\label{ss:IIBF}
	
	The next framework where the pattern can be checked is F-theory compactified to four dimensions. The compactification manifold $X$ is K\"ahler, with the axio-dilaton $\tau=C_0+ \text{i}\, e^{-\phi}$ being non-constant and undergoing monodromies around the 7-branes. Because of this, it is convenient to adopt a 10d Einstein-frame description from the type IIB point of view, with a metric of the form
	\begin{equation}
		\dd s^2=e^{2A}\dd s_4^2+\ell_s^2\dd s^2_X\, , \qquad \text{with }\, \, e^{2A}=\frac{\ell_s^2\Mpf^2}{4\pi V_X}\, .
	\end{equation}
	For our purposes of analyzing the infinite distance limits, it will be enough to work in the low-warping approximation  \cite{Denef:2008wq,Grimm:2010ks} for the 4d $\mathcal{N}=1$ EFT.
	
	Focusing in the K\"ahler sector, its moduli $\{v_a\}_{a=1}^{h^4}$ can be used to expand the Einstein frame K\"ahler form $J=v_a[D^a]$, with $[D^a]$ the Poincaré duals of a basis of divisors $D^a\in H_4(X,\mathbb{Z})$ and $v_a$ measures the volume of the  dual effective curves $\{C_a\}$. On the other hand, the K\"ahler sector can be parametrized by the following EFT saxions
	\begin{equation}
		\mathbf{s}=\frac{1}{2}J\wedge J=s^a[C_a]\in H^4(X,\mathbb{R})\quad \Longrightarrow \quad s^a=\int_{D^a}\mathbf{s}=\frac{1}{2}\kappa^{abc}v_bv_c\, ,
	\end{equation}
	where $\kappa_{abc}$ are the triple intersection numbers of $X$ and $[C_a]\in H^4(X,\mathbb{R})$ the Poincaré duals of the set of effective curves $C_a \in H_2(X,\mathbb{Z})$. After grouping the EFT saxions into (chiral) complex fields $t^a=a^a+{\rm i}s^a$, we can obtain the K\"ahler potential, as a function of the $s^a$, which reads (up to an irrelevant constant)
	\begin{equation}
		K_{\rm ks}=-2\log\int_X J\wedge J\wedge J=-2\log\kappa(\mathbf{v},\mathbf{v},\mathbf{v})\, .
	\end{equation}
	The volume of the manifold will be given by $V_X=\frac{1}{6}\kappa(\mathbf{v},\mathbf{v},\mathbf{v})$ in 10d Planck units. As an example, we consider $X$ to be the $n$-twisted $\mathbb{P}^{1}$ fibration over $\mathbb{P}^2$ described by a gauged linear sigma model \cite{Lanza:2021udy}.\footnote{\label{fn:gauged sigma model}Following \cite{Denef:2008wq}, the 2d sigma model is given by
		\begin{center}
			\begin{tabular}{c|ccccc|c}
				$U(1)_i$&$Q_1^i$&$Q_2^i$&$Q_3^i$&$Q_4^i$&$Q_5^i$&FI\\\hline
				$U(1)_1$&1&1&1&$-n$&0&$v_1>0$\\
				$U(1)_2$&0&0&0&1&1&$v_2>0$
			\end{tabular}
		\end{center}
		where $Q^i_j$ are the $U(1)_i$ charge of the fields $\{x_j\}_{j=1}^5$, and $n$ takes positive integer values, so that
		\begin{equation}
			\notag X=\left\{x\in\mathbb{C}^5:\;|x_1|^2+|x_2|^2+|x_3|^2-n|x_4|^2=v_1,\; |x_4|^2+|x_5|^2=v_2\right\}/U(1)^2\, ,
		\end{equation}
		where the volume of $X$ will depend on the FI parameters $v_1$ and $v_1$. Given $\mathcal{D}_I=X\cap\{x_I=0\}$ the toric divisors of $X$, we choose the following
		\begin{equation}
			\notag D^1=\mathcal{D}_1\simeq \mathcal{D}_2\simeq \mathcal{D}_3\, , \quad E=\mathcal{D}_4\, , \quad D^2 =\mathcal{D}_5\, , \qquad\text{with }E=D^2-n D^1\;,
		\end{equation}
		with the cone of effective divisors given by $\mathcal{C}_I=\langle\{D^1,E\}\rangle_{\mathbb{Z}}$.
		The intersection numbers are given by
		\begin{equation}\label{eq: vol p1p2}
			\notag \mathcal{I}=\kappa_{abc}D^aD^b D^c=(D^1)^2D^2+nD^1(D^2)^2+n^2(D^2)^3\, ,
		\end{equation}
		and the dual effective curves, such that $C_a\cdot D^b=\delta_a^b$, are $C_1\simeq D^1\cdot E\simeq D^1\cdot(D^2-nD^1)$ and $C_2\simeq D^1\cdot D^1$. We can identify $C_2$ with the $\mathbb{P}^1$ fibre and $C_1$ with its pushforward to the base through the $\mathcal{D}_4=E$ section. With this in mind we can expand the EFT saxions as 
		\begin{equation}\label{eq: sv rel}
			\notag \mathbf{s}=\frac{1}{2} J\wedge J=s^1[C_1]+s^2[C_2]\, , \qquad\text{with }\, \, \left\{
			\begin{array}{l}
				s^1=v_1v_2+\frac{1}{2}nv_2^2\\
				s^2=\frac{1}{2}(v_1+nv_2)^2
			\end{array}
			\right. .
		\end{equation}
		As $\mathcal{C}_I$ is generated by $D^1$ and $E$, we have that the saxionic cone is given by$\Delta=\{\mathbf{s}\in H^4(X,\mathbb{R}):s^1>0,\; s^2-ns^1>0\}$. Finally, we can invert \eqref{eq: sv rel} to obtain
		\begin{equation}
			\notag v_1=\sqrt{2(s^2-ns^1)},\qquad v_2=\frac{1}{n}\left(\sqrt{2s^2}-\sqrt{2(s^2-ns^1)}\right)\, ,
		\end{equation}
		which can be used to obtain the volume and K\"ahler potential.} The compact volume and the K\"ahler potential are given by: 
	\begin{subequations}\label{eq:volumeandKahlerpot}
		\begin{align}
			V_X&=\frac{\sqrt{2}}{3n}\left[(s^2)^{3/2}-(s^2-ns^1)^{3/2}\right]>0\, ,\label{eq: vol IIB}\\
			K_{\rm ks}&=-2\log\left[(s^2)^{3/2}-(s^2-ns^1)^{3/2}\right]+{\rm const}\, .
		\end{align}
	\end{subequations}
	At this point we have everything we need to compute the leading towers and species scales. As the 4d and 10d dilaton are the same and are decoupled from the K\"ahler sector, we will restrict ourselves to limits purely along $s^1$ and $s^2$. Thus, the leading towers will correspond to KK modes, whose masses are computed as follows
	\begin{equation}
		m_{{\rm KK},\, \ast}^2\sim\frac{e^{2A}}{\ell_s^2R_{\ast}^2}\sim\frac{M_{\rm Pl;\, 4}^2}{V_XR_\ast^2}\, .
	\end{equation}
	As usual, $R_\ast$ denotes here the scale (in string units) of the cycle decompactifying at the fastest rate. By inspection of \eqref{eq:volumeandKahlerpot} (see also footnote \ref{fn:gauged sigma model}), we conclude that the only decompactification options are of the whole $X$ or $D^1$ (note that in this latter case we decompactify for dimensions), since decompactifying $D^2$ or $E=D^2-nD^1$ at a pace equal or faster than $D^1$ results in full decompactification. Hence, we have the following two possibilities for the leading tower:
	\begin{equation}
		m_{{\rm KK},\, X}\sim \Mpf V_X^{-2/3}\, , \qquad m_{{\rm KK},\, D^1}\sim \Mpf V_X^{-1/2}v_1^{-1/2}\, .
	\end{equation}
	Analogously, taking into account that we can only decompactify to ten and eight dimensions implies that the only possibilities for the species scale are\footnote{Notice that we are not considering here any emergent string limit, since we fix the 4d dilaton to a constant value.}
	\begin{equation}
		M_{\rm Pl;\, 10}\sim \Mpf V_X^{-1/2}\, , \qquad M_{\rm Pl;\, 8}\sim \Mpf V_X^{-1/3} v_1^{-1/3}\, ,
	\end{equation}
	where all volumes are already measured in 10d Planck units. Similarly to what happened for the heterotic string example in Section \ref{ss:heteroticCY3}, the volume of the compact manifold $V_X$ has different terms that can dominate depending on the particular limit taken by the saxions (c.f. eq. \eqref{eq: vol IIB}), resulting in a non-diagonal moduli space metric. However, we can follow the same approach as in Section \ref{ss:heteroticCY3} above and classify the different asymptotic limits by their leading tower(s) and species scale. Subsequently, using \eqref{eq: patron check} we can compute $\vec{\zeta}_{\rm t}\cdot\vec{\mathcal{Z}_{\rm sp}}$ along different infinite distance trajectories. This leads to the following three limits:
	\begin{center}
		\begin{tabular}{rl|cc}
			&Limit&$m_{\rm t}$&$\LSP$\\\hline
			&$s^2\gtrsim(s^1)^2$&$m_{{\rm KK},\,D^1}$&$M_{\rm Pl;\,8}$\\
			%&$s^2\sim(s^1)^2$&$m_{{\rm KK},D^1}$&$M_{\rm Pl;8}\sim M_{\rm Pl;10}$\\
			&$(s^1)^2\gg s^2\gg ns^1$&$m_{{\rm KK},\, D^1}$&$M_{\rm Pl;\,10}$\\
			%$(\star)$&$s^2\sim \lambda s^1$, $(\lambda>n)$&$m_{{\rm KK},X}\sim m_{{\rm KK},D^1}$&$M_{\rm Pl;10}$\\
			&$s^2\sim n s^1$&$m_{{\rm KK},\, X}$&$M_{\rm Pl;\,10}$
		\end{tabular}
	\end{center}
	In all the above limits the pattern $\vec{\zeta}_{\rm t}\cdot\vec{\mathcal{Z}_{\rm sp}}=\frac{1}{d-2}=\frac{1}{2}$ is fulfilled. As it was the case in Section \ref{ss:heteroticCY3}, in the interfaces between the different asymptotic regions the $\vec{\zeta}_{\rm t}\cdot\vec{\mathcal{Z}}_{\rm sp}=\frac{1}{2}$ holds regardless of the chosen vector. While the pattern is always verified when considering the leading tower(s) and the species scale, when computing the product with the subleading towers the result is case-dependent. For the $(s^1)^2\gg s^2\gg ns^1$ limit, $\Lambda_{\rm sp}=M_{\rm Pl;\, 10}\gg m_{{\rm KK},\, X}\gg m_{{\rm KK},\, D^1}=m_{\rm t}$ and  $\vec{\mathcal{Z}}_{\rm Pl,\, 10}\cdot\vec{\zeta}_{{\rm KK},\, D^1}=\vec{\mathcal{Z}}_{\rm Pl,\, 10}\cdot\vec{\zeta}_{{\rm KK},\, X}=\frac{1}{2}$, while in the $s^2\gg (s^1)^2$ limits $\Lambda_{\rm sp}=M_{\rm Pl;\, 8}\gtrsim m_{{\rm KK},\, X}\gg m_{{\rm KK},\, D^1}=m_{\rm t}$ but $\vec{\mathcal{Z}}_{\rm Pl,\, 8}\cdot\vec{\zeta}_{{\rm KK},\, X}=\frac{4}{9}$. This is not surprising, as in the former case we are decompactifying to 10d, and the associated $\vec{\mathcal{Z}}_{\rm Pl,\, 10}$ and $\vec{\zeta}_{{\rm KK},\, X}$ trivially fulfill the pattern, while for the latter case the decompactification is to 8d, and we do not expect the KK tower decompactifying to ten dimensions to have the correct product here (note that indeed this was not the case either for toroidal decompactifications). 
	
	One could wonder whether the $\vec{\zeta}_{\rm t}\cdot\vec{\mathcal{Z}}_{\rm sp}=\frac{1}{2}$ pattern holds only asymptotically or if it is exact for finite distance points in moduli space (as it was the case for toroidal compactifications). This seems to depend on the limit taken, as $s^2\sim n s^1$ limits are exact (irrespective of the proportional factor between the saxions), while for example a trajectory of the form $(s^1,s^2)\sim(\lambda,\lambda^3)$ with $n=2$ has $\vec{\zeta}_{\rm t}\cdot\vec{\mathcal{Z}}_{\rm sp}=\vec{\zeta}_{\text{KK},\, D^1}\cdot\vec{\mathcal{Z}}_{\rm Pl,\, 8}=\frac{1}{2}+\frac{1}{6}\lambda^{-\sqrt{\frac{2}{5}}}+\mathcal{O}\left(\lambda^{-2\sqrt{\frac{2}{5}}}\right)$ (with corrections at all others being positive), so that along this limit $\vec{\zeta}_{\rm t}\cdot\vec{\mathcal{Z}}_{\rm sp}>\frac{1}{2}$, only being saturated for $\lambda\to\infty$.
	
	Notice that here, as well as in all previous examples, the corrections (if any) are non-negative, which suggests that a possible generalization  (if there exists one) of the pattern \eqref{eq:patternmass} to the interior of the moduli space could be
	\begin{equation}
		\frac{\vec\nabla M}{M} \cdot\frac{\vec\nabla \LSP}{\LSP}\geq \frac{1}{d-2}\quad \text{ over all }\mathcal{M},
	\end{equation}
	saturating the inequality asymptotically at the infinite distance limits. This is still well defined in the entire asymptotic regime (and not only in the strict infinite distance limit), as there are still light towers of states and we can identify $M=m_{\rm t}$. However, the notion of the lightest tower is not well defined anymore in the interior of the moduli space, so the mass of the tower should be replaced by some other quantity $M$ that approaches $M\to m_{\rm t}$ asymptotically and diverges if $\LSP$ develops a maximum. We hope to come back to this question in the future.
	
	\subsection{Type IIA on a Calabi--Yau orientifold}
	
	For our last 4d $\mathcal{N}=1$ example, we will consider type IIA string theory compactified on the projection of a Calabi--Yau threefold $X$ under a O6 orientifold $\iota:X\to X$, \cite{Grimm:2004ua, Hitchin:2000jd, Acharya:2002ag}. The moduli $X$ will be encoded in a (string frame) K\"ahler form $J$ and $(3,0)$-form $\Omega$ with normalization
	\begin{equation}\label{eq: norm IIA}
		\frac{\rm i}{8}\Omega\wedge\bar{\Omega}=\frac{1}{6}J\wedge J\wedge J\, ,
	\end{equation}
	and satisfying the projection condition $\iota^\ast J=-J$ and $\iota^\ast\Omega=\bar{\Omega}$. We will have two sets of chiral fields, $t^a=a^a+{\rm i}s^a$ and $\hat t^\alpha=\hat a^\alpha+{\rm i}\hat s^\alpha$, respectively parametrizing $B_2+{\rm i}J$ (in the string frame) and $C_3+{\rm i}e^{-\phi}{\rm Re}\,\Omega$ as
	\begin{subequations}
		\begin{align}
			B_2+{\rm i}J&=(a^a+{\rm i}s^a)[D_a^+]\, ,\\
			C_3+{\rm i}e^{-\phi}{\rm Re}\,\Omega&=(\hat a^\alpha+{\rm i}\hat s^\alpha)[\Sigma^-_\alpha]\, ,
		\end{align}
	\end{subequations}
	where $\phi$ is the 10d dimensional dilaton, and $[D_a^+]$ and $[\Sigma^-_\alpha]$ form basis for the odd 2-form and even 3-form cohomology classes $H_-^2(X,\mathbb{R})$ and $H_+^3(X,\mathbb{R})$.\footnote{Notice that Poincaré duality relates even/odd cycles with odd/even cohomology classes, as orientifold involution inverts orientation.} With this in mind we can obtain the string frame volume of $X$ and the Hitchin function 
	\begin{subequations}
		\begin{align}
			V_X(s)&=\frac{1}{6}\int_X J\wedge J\wedge J=\frac{1}{6}\kappa_{abc}s^a s^b s^c\, ,\\
			\mathcal{H}(\hat{s})&=\frac{\rm i}{8}\int_Xe^{-2\phi}\Omega\wedge\bar{\Omega}\, ,
		\end{align}
	\end{subequations}
	which, in the perturbative regime where the backreaction of both the fluxes and sources are neglected and the warping is approximately constant, can be related to each other as follows
	\begin{equation}
		e^{2\phi}=\frac{V_X(s)}{\mathcal{H}(\hat{s})}\, ,
	\end{equation}
	where we stick to the normalization in \eqref{eq: norm IIA}. This allows us to write the 10d string frame metric with a conformal factor before $\dd s^2_4$
	\begin{equation}
		e^{2A}=\frac{\ell_s^2\Mpf^2}{2\pi \mathcal{H}(\hat{s})}=\frac{\ell_s^2\Mpf^2e^{2\phi}}{2\pi V_X(s)}\, ,
	\end{equation}
	and a K\"ahler potential
	\begin{equation}
		K = -\log V_X-2\log\mathcal{H}(\hat{s})\, .
	\end{equation}
	Once we have introduced the required mathematical framework, we consider the case studied \cite{Cvetic:2001nr}, where $X=(\mathbb{T}^2\times\mathbb{T}^2\times\mathbb{T}^2)/(\mathbb{Z}_2\times\mathbb{Z}_2)$ and the orientifold involution is taken to be $\iota:(z_1,z_2,z_3)\mapsto (\bar z_1,\bar z_2,\bar z_3)$. Moreover, the complex-structure is fixed as $\tau_j={\rm i}\frac{R_{2j}}{R_{2j-i}}$, so that
	\begin{align}
		\Omega&=R_1R_3R_5\, \dd z_1\wedge \dd z_2\wedge\dd z_3\notag\\
		&=(R_1\dd y_1+{\rm i}R_2 \dd y_2)\wedge(R_3\dd y_3+{\rm i}R_4 \dd y_4)\wedge(R_5\dd y_5+{\rm i}R_6 \dd y_6)\, .
	\end{align}
	This allows to, for certain basis $[\Sigma_\alpha^-]$,\footnote{While not of importance for our discussion, we include said basis for completeness:
		\begin{equation*}
			[\Sigma^-_0]=4\dd y_1\wedge \dd y_3\wedge\dd y_5\, ,\quad[\Sigma^-_1]=-4\dd y_1\wedge \dd y_4\wedge\dd y_6\, ,\quad[\Sigma^-_2]=-4\dd y_2\wedge \dd y_3\wedge\dd y_6\, ,\qquad[\Sigma^-_0]=-4\dd y_2\wedge \dd y_4\wedge\dd y_5\, .
		\end{equation*}
	} identify the following saxions:
	\begin{align}\label{eq: tr1}
		\hat{s}^0&=\frac{1}{4}e^{-\phi}R_1R_3R_5\, ,\quad \hat{s}^1=\frac{1}{4}e^{-\phi}R_1R_4R_6\, , \notag\\
		\hat{s}^2&=\frac{1}{4}e^{-\phi}R_2R_3R_6\, ,\quad \hat{s}^3=\frac{1}{4}e^{-\phi}R_2R_4R_5\, ,
	\end{align}
	such that
	\begin{equation}
		\mathcal{H}(\hat{s})=8\sqrt{\hat{s}^0\hat{s}^1\hat{s}^2\hat{s}^3}\, .
	\end{equation}
	On the other hand, the K\"ahler saxions $s^a$ measure the volume of the $a$-th 2-torus in string frame, so that
	\begin{equation}\label{eq: tr2}
		s^1=R_1R_2\, , \qquad s^2=R_3R_4\, , \qquad s^3=R_5R_6\, .
	\end{equation}
	We are now in conditions to obtain the possible leading towers \cite{Lanza:2021udy}, which can either be emergent strings or KK modes:
	\begin{equation}
		m_{\rm s}\sim\frac{e^A}{\ell_s}\sim \Mpf e^{\phi}V_X^{-1/2},\qquad m_{{\rm KK},\, \ast}\sim\frac{e^A}{\ell_sR_{\ast}}\sim \Mpf e^{\phi}V_X^{-1/2}R_\ast^{-1}\, ,
	\end{equation}
	where $R_{\ast}$ is the characteristic scale of the decompactified volume. As for the species scale, we can either have the string scale, $m_{\rm s}$ or the higher-dimensional Planck mass
	\begin{equation}
		M_{{\rm Pl};\, d}\sim \Mpf e^{\frac{d-4}{d-2}\phi}V_X^{-1/2}\volume_{10-d}^{\frac{1}{d-2}}\, ,
	\end{equation}
	where $\volume_{10-d}$ denotes the remaining volume (in string units) after decompactifying to $d$ dimensions. To obtain this expression we have followed a procedure analogous to that of Section \ref{ss:heteroticCY3} in order to rewrite the volumes in the appropriate Planck units.
	
	We are finally ready to show that the condition $\vec{\zeta}_{\rm t}\cdot\vec{\mathcal{Z}}_{\rm sp}=\frac{1}{2}$ holds in every limit. To do so we will work in the $\{e^{\phi},R_1,\ldots,R_6\}$ basis, and proceed similarly as in Section \ref{s:MthT7}. Grouping $\vec{R}=(R_0 \equiv e^{\phi},R_1,\ldots,R_6)$ and $\vec{s}=(s^1,s^2,s^3,\hat{s}^0,\ldots,\hat{s}^3)$, one can consider general limits of the form $R_i = e^{\rho_i} \sim e^{\rho_i(0)\lambda}$ and $s^I = e^{\sigma^I} \sim e^{\sigma^I(0)\lambda}$ for large $\lambda$. With this in mind, we can use \eqref{eq: tr1} and \eqref{eq: tr2} to obtain the relation
	\begin{equation}
		\vec{\rho}=M\vec{\sigma},\qquad \text{with }\ \ M=\frac{1}{4}\left(
		\begin{array}{rrrrrrr}
			2 & 2 & 2 & -1 & -1 & -1 & -1 \\
			2 & 0 & 0 & 1 & 1 & -1 & -1 \\
			2 & 0 & 0 & -1 & -1 & 1 & 1 \\
			0 & 2 & 0 & 1 & -1 & 1 & -1 \\
			0 & 2 & 0 & -1 & 1 & -1 & 1 \\
			0 & 0 & 2 & 1 & -1 & -1 & 1 \\
			0 & 0 & 2 & -1 & 1 & 1 & -1 \\
		\end{array}
		\right)\, .
	\end{equation}
	It is easy to check that $M$ is indeed invertible, and that it maps $\mathbb{R}^7_{\geq 0}$ to $\mathbb{R}_{\leq 0}\times\mathbb{R}^6_{\geq 0}$ in a one-to-one way,\footnote{Notice that the weak coupling limit corresponds to $\phi= \rho_0\to-\infty$.} so that we are able to take any possible limit in the $\{ R_0,R_1,\ldots,R_6\}$ basis. Now, in the same way as in \eqref{eq: metric G2}, we obtain
	\begin{align}
		\mathsf{G}^{R_iR_j}&=\frac{\partial R_ i}{\partial  s^I}\frac{\partial R_ j}{\partial  s^J}\mathsf{G}^{s^Is^J}=2R_iR_j\delta^{IJ}M_{iI}M_{jJ}=\left(\delta_{ij}+\frac{1}{2}(\delta_{i0}+\delta_{j0})\right)R_iR_j\, ,
	\end{align}
	with no sum over $i$ or $j$. On the other hand, using the same notation as in Section \ref{s:MthT7}, we have
	\begin{subequations}\label{e:masses T7}
		\begin{align}
			\log\frac{m_{{\rm KK},\, n_{\max}}}{\Mpf}&=\log R_0-\sum_{i=1}^6\left(\frac{1}{2}+\frac{\mathbb{I}_{\rm max}(i)}{n_{\rm max}}\right)\log R_i\, ,\\
			\log\frac{m_{\rm s}}{\Mpf}&=\log R_0-\frac{1}{2}\sum_{i=1}^6\log R_i\, ,\\
			\log\frac{M_{{\rm Pl;}\, 4+n}}{\Mpf}&=\frac{n}{n+2}\left\{\log R_{0}-\sum_{i=1}^6\left(\frac{1}{2}+\frac{\mathbb{I}_{\rm dec}(i)}{n}\right)\log R_i\right\}\, .			
		\end{align}
	\end{subequations}
	Now, as it was the case in Section \ref{s:MthT7}, we find that $\vec{\zeta}_{{\rm KK},\, n_{\max}}\cdot\vec{\mathcal{Z}}_{{\rm Pl;}\, 4+n}=\frac{1}{d-2}=\frac{1}{2}$ iff  $\{\tilde{R}_i\}_{i=1}^{n_{\max}}\subseteq \{\hat{R}_i\}_{i=1}^{n}$. In this sense, any time $m_{\rm t}$ is a KK tower, the pattern is exactly (i.e. not just asymptotically) verified, also for subleading towers with $\{\tilde{R}_i\}_{i}\subseteq \{\hat{R}_i\}_{i=1}^{n}$, which following an argument similar than \eqref{e: sub lighter} can be checked to be lighter than the species scale. On the other hand, it is immediate from \eqref{e:masses T7} that the string mass is given by the formal limit $m_{\rm s}=\lim_{n\to\infty} M_{{\rm Pl;}\, 4+n}$, with the  $\mathbb{I}_{\rm dec}(i)$ dependence disappearing. Therefore, $\vec{\zeta}_{{\rm KK},\, n_{\max}}\cdot\vec{\mathcal{Z}}_{{\rm s}}=\frac{1}{d-2}=\frac{1}{2}$ for any choice of $\{\tilde{R}_i\}_{i=1}^{n_{\max}}$ (even in those limits for which the species scale is not given by the string mass). Finally, one can also check that $|\vec{\zeta}_{\rm s}|^2=\frac{1}{2}$, so that the pattern also holds in emergent string limits. We thus conclude that any decompactification or emergent string limit of this example fulfills the proposed pattern. This result should also be easily generalized to any other $(\mathbb{T}^2\times\mathbb{T}^2\times\mathbb{T}^2)/\Gamma$ compactification, with $\Gamma\in SU(3)$, see \cite{Blumenhagen:2005mu,Blumenhagen:2006ci,Marchesano:2007de} for reviews. As argued in Section \ref{s:MthT7} and Appendix \ref{ap:generalities}, the mass and behavior of the towers depend only on the scaling with the internal compact volume, which is expected to be independent from quotienting by finite order groups.
	
	\section{On the quest for a bottom-up rationale}
	\label{s:bottomup}

	In the previous sections we have provided significant evidence for the relation \eqref{eq:pattern} in string theory compactifications. This pattern provides a very sharp relation between the growth of the density of states and the rate at which they become light asymptotically (c.f. \eqref{eq:patternN}). The more dense the tower is, the slower the mass goes to zero. The next natural open question is whether this pattern is a lamppost effect of the string theory landscape or a general feature of Quantum Gravity. In order to address this question, we would need to provide some bottom-up explanation of the pattern, independent of string theory. Here, we will give the first steps to find such an argument. Despite not being able to provide a purely bottom-up rationale for the pattern, we will identify and motivate some sufficient conditions that allow the pattern to hold; with the hope that this can be useful for future research.
	
	In a nutshell, we argue that through the Emergence Proposal \cite{Palti:2019pca} one can rewrite the product $\vec{\zeta}_{\rm t}\cdot \vec{\mathcal{Z}}_{\rm sp}=\partial^{\phi^i}\log m_{\rm t}\partial_{\phi^i}\log \Lambda_{\rm sp}$ as a linear combination of the ratios of logarithmic derivatives of the species scales and leading tower, independent of the explicit expression of the moduli space metric. Then we argue that, upon requiring the exponential rate of the towers and cut-offs to be well-defined over the space of infinite distance limits, $\infty(\mathcal{M})$ (see Appendix \ref{ap:detailsbottomup} for details on its construction), we can divide said space into patches over which $\vec{\zeta}_{\rm t}\cdot \vec{\mathcal{Z}}_{\rm sp}$ is constant. Then, assuming that every time several leading towers become light at the same rate there exists an additional one formed by bound states of these, the transition from one patch to the other is such that $\vec{\zeta}_{\rm t}\cdot \vec{\mathcal{Z}}_{\rm sp}$ takes the same value on both, and as a result, over the whole space of infinite distance limits. Finally, for those cases in which there exists at least either one emergent string limit within the moduli space of the theory or one overall decompactification to a higher dimensional vacuum, the observed value is set to $\vec{\zeta}_{\rm t}\cdot \vec{\mathcal{Z}}_{\rm sp}=\frac{1}{d-2}$.
	
	\subsection{The Emergence Proposal}
	\label{ss:Emergence}
	
	A promising avenue to find a bottom-up rationale is by means of the Emergence Proposal \cite{Palti:2019pca,vanBeest:2021lhn}. This proposal claims that all the IR dynamics (i.e. the kinetic terms of all fields in an EFT for example) should emerge from integrating out the massive degrees of freedom \cite{Harlow:2015lma,Grimm:2018ohb}.\footnote{See \cite{Heidenreich:2017sim,Heidenreich:2018kpg} for a different but related version of the Emergence Proposal based on unification of the strong coupling scales.} The proposal provides a bottom-up explanation for some Swampland conjectures, as some of these constraints relating EFT data with new massive states simply arise as natural consequences of RG flow renormalization. It implies, in particular, that the asymptotic kinetic terms of the moduli emerge upon integrating out the towers of states becoming light (i.e. those towers with a characteristic mass scale below the species scale \cite{Palti:2019pca,Blumenhagen:2023abk}). This reverses the logic of the Distance Conjecture, as the infinite field distance itself emerges because of the existence of a tower of states becoming massless to start with (see \cite{Grimm:2018ohb,Corvilain:2018lgw,Heidenreich:2018kpg} for the original works and \cite{Castellano:2022bvr, Hamada:2021yxy,Marchesano:2022axe,Castellano:2023qhp,Blumenhagen:2023yws,Kawamura:2023cbd,Seo:2023xsb,Blumenhagen:2023tev} for multiple follow-ups).
	
	In the following, we are going to investigate whether the Emergence Proposal can help to provide a bottom-up rationale for our pattern. Consider an asymptotic trajectory resulting in an infinite tower of states whose mass $m_{\rm t}(\phi)$ is parametrized by the vacuum expectation value of some massless scalar $\phi$.  The one-loop corrections to the field metric $\mathsf{G}$ associated to $\phi$ coming from integrating out a tower of states in $d>4$ read \cite{Grimm:2018ohb,Hamada:2021yxy,Castellano:2022bvr} 
	\beq
	\label{metric}
	\mathsf{G}_{\phi\phi}=c^{(\phi)}\int_0^{\LSP} \rho(\mu)(\partial_\phi m_{\rm t}|_\mu)^2 \mu^{d-4}\dd \mu\, ,
	\eeq
	where $\rho(\mu)$ is the density of states per unit mass and $c^{(\phi)}$ is some numerical factor that may depend on the spacetime dimension $d$ and the nature of the tower. The species scale $\LSP$ associated to the tower is given in Planck units by
	\beq
	\label{Lsp}
	\LSP=(bN)^{-\frac{1}{d-2}}\, ,\quad \text{with}\quad N=\int_0^{\LSP} \rho(\mu)\dd \mu\, ,
	\eeq
	where $b$ is again some undetermined numerical constant. Note that the above relations imply the following integral equation for $\LSP$:
	\begin{equation}
		\label{Lsp def}
		\LSP=\left(b\int_0^{\LSP}\rho(\mu)\dd \mu\right)^{-\frac{1}{d-2}}\,,
	\end{equation}
	so that \eqref{Lsp} is not an identity but only valid for the $\LSP$ solution of \eqref{Lsp def}.
	
	It is well-known \cite{Corvilain:2018lgw, Castellano:2022bvr} that if we integrate out a tower of Kaluza-Klein-like or string-like modes with characteristic mass up to the species scale we recover the logarithmic behavior of the distance predicted by the Distance Conjecture:
	\beq
	\label{dist}
	\mathsf{G}_{\phi\phi}\sim a^{(\phi)}(\partial_\phi \log m_{\text{t}})^2\Longrightarrow \Delta_{\phi}=\int_{\phi_1}^{\phi_2} \dd \phi \sqrt{ \mathsf{G}_{\phi\phi}}\sim-\sqrt{a^{(\phi)}} \log \left|\frac{m_{\text{t}}(\phi_2)}{m_{\text{t}}(\phi_1)}\right|\, ,
	\eeq
	with $a^{(\phi)}>0$ some numerical factor fixed upon integrating \eqref{metric}. If the Emergence proposal holds, $(a^{(\phi)})^{-1/2}$ should reproduce precisely the exponential decay rate of the mass of the tower in terms of the distance, since the classical field metric should purely arise from quantum corrections of integrating out the tower in a dual sense. However, to reliably compute $a^{(\phi)}$, one has to properly integrate out all relevant states in a frame where they look non-perturbative, which is often out of scope (see though \cite{Blumenhagen:2023tev} for recent results). 
	From \eqref{dist} we obtain that
	\begin{equation}
		\partial^\phi\log m_{\rm t}=\mathsf{G}^{\phi\phi}\partial_\phi\log m_{\rm t}\sim\left(a^{(\phi)}\partial_\phi\log m_{\rm t}\right)^{-1}\;.
	\end{equation}
	The above expression is still valid for multi-field trajectories depending on several moduli $\{\phi^i\}_{i}$. Considering a diagonal metric\footnote{A priori, $\mathsf{G}_{\phi^i\phi^j}$ being non-degenerate, one can diagonalize the metric upon an appropriate choice of moduli (which amounts to a moduli space reparametrization) at every point in $\mathcal{M}$. However, in practice, this usually complicates the one-loop computations since the Feynman rules get more involved. We do not enter into this and assume it works by the Emergence Proposal.} for simplicity, one would have

	\begin{equation}\label{eq: patron}
		\vec{\zeta}_{\rm t}\cdot \vec{\mathcal{Z}}_{\rm sp}=\partial^{\phi^i}\log m_{\rm t}\partial_{\phi^i}\log \Lambda_{\rm sp}=\sum_{\phi^i} \frac{1}{a^{(\phi^i)}}\frac{\partial_{\phi^i}\log \Lambda_{\rm sp}}{\partial_{\phi^i}\log m_{\rm t}}\,,
	\end{equation}
	where again we take $\partial_{\phi^i}\log m_{\rm t}\neq 0$, so that $\mathsf{G}_{\phi^i\phi^i}\neq 0$ along this trajectory (otherwise we would just restrict to the components $\phi^i$ along which $m_{\rm t}$ depends on). Therefore, if $\Lambda_{\rm sp}$ and $m_{\rm t}$ present the same dependence (i.e. polynomial, logarithmic, etc.) with each $\phi^j$ \cite{Castellano:2021mmx, Castellano:2022bvr}, then \eqref{eq: patron} results in $\vec{\zeta}_{\rm t}\cdot \vec{\mathcal{Z}}_{\rm sp}$ being a constant.
	
	To summarize, we can use the Emergence proposal to argue for the exponential behavior of the tower mass and the species scale in terms of the distance in moduli space, which therefore implies that the product \eqref{eq:pattern} asymptotes to a constant. This argument, by itself, does not seem to be constraining enough to determine a universal value for the latter. Such constant seems to be a model-dependent result, and this is why it is (a priori) surprising that we always obtain the same value for every infinite distance limit in string compactifications.
	
	\subsection{Sufficient conditions to satisfy the pattern at every asymptotic limit}	
	\label{ss:continuity}
	
	The above reasoning implies that by the Emergence Proposal $\vec{\zeta}_{\rm t}\cdot \vec{\mathcal{Z}}_{\rm sp}$ approaches a constant asymptotic trajectory. Rather than finding a complete bottom-up explanation on why said constant is indeed $\frac{1}{d-2}$ for every infinite distance limit, we will settle for the weaker result of showing that, upon some general enough \emph{sufficient conditions}, $\vec{\zeta}_{\rm t}\cdot \vec{\mathcal{Z}}_{\rm sp}$ takes the same constant value over each connected components of the set of infinite distance points. This, together with the existence of a limit for which said value is $\frac{1}{d-2}$ (such as an emergent string limit or a homogeneous decompactification) would finally solve the problem. In order to do this, we must first understand how the scalar charge-to-mass ratios for the towers and species scale behave as we change the asymptotic trajectories along the moduli space.
	
	For this, consider a generic EFT endowed with a moduli space $(\mathcal{M},\mathsf{G})$, and a set of towers $m_I$ that are not necessarily defined over the whole moduli space. Following the Emergent String Conjecture \cite{Lee:2019wij} and \cite{Castellano:2022bvr}, associated to each tower we will have a density parameter $p_I$, such that
	\begin{equation}\label{eq:massspectrum}
		m_{I,\, n}\sim n^{1/p_I}m_{I,\, 0},\qquad \Lambda_{I}\sim M_{{\rm Pl};\, d}\, N_I^{-\frac{1}{d-2}}\sim N_I^{1/p_I}m_{I,\, 0}\;,
	\end{equation}
	where $\Lambda_{I}$ is the species scale computed by taking into account only the states of the $m_I$ tower. The $p_I\to\infty$ limit corresponds to the oscillator modes of an emergent string, while having finite and positive $p_I$ will be rather interpreted as a KK tower (we will not impose $p$ to be an integer).
	
	Consider now some infinite distance limit in which several towers of states may become light at possibly different rates. This provides a set of scales or cut-offs $\Lambda_J$ associated to each tower independently. However, the final result for the actual $\LSP$ must be computed by taking into account \emph{all} states at or below the cut-off, resulting in a  scale $\LSP$ which may be lower than the individual $\Lambda_J$ ones; unless we move along a direction in which a single tower $m_{I_0}$ completely dominates the density of states, so that $\Lambda_{I_0}\simeq \LSP$.
	
	This way, when considering a limit along which several towers scale at the same rate (in other words, they span some facet/edge/etc., in the exterior of the convex hull of the $\zeta$-vectors, ${\rm Hull}(\{\vec{\zeta}_I\}_I)$), the following two possibilities can happen \cite{Castellano:2022bvr} (see Figure \ref{fig:add mult} below):
	\begin{itemize}
		\item \textbf{Multiplicative species:} \cite{Castellano:2021mmx} The total number of species is given by the product of the number of species for each tower $N=\Pi_k N_k$, which occurs whenever we have light states with `mixed charges'. In other words, a (sub-)lattice of the space of charges $\vec n$ is populated. This is the case for decompactification limits with KK-towers having momentum components along several internal directions, or for the D0-D2 bound states in Section \ref{ss:ExampleII}. In that case, as shown in  \cite{Castellano:2021mmx,Castellano:2022bvr}, the species scale can be equivalently computed as follows
		\begin{equation}\label{eq:multspecies}
			\Lambda_{\rm sp}=M_{\text{Pl;}\, d}\left(\frac{M_{{\rm Pl;}\,d}}{m_{\rm eff}}\right)^{-\frac{p_{\rm eff}}{d-2+p_{\rm eff}}}\, ,
		\end{equation}		
		in terms of an effective mass and density parameter,
		\begin{equation}\label{e:eff tow}
			m_{\rm eff}=(m_1^{p_1} \ldots m_k^{p_k})^{1/p_{\rm eff}},\qquad p_{\rm eff}=\sum_{i=1}^kp_i\; ,
		\end{equation}
		where the spectrum of each individual tower behaves as $m_{i,\, n} = n^{1/p_i}\, m_i$. The scalar charge-to-mass vector $\vec{\zeta}_{\rm eff}$ of this effective tower is precisely located at the closest point to the origin of the facet spanned by the towers, i.e. it is perpendicular to the latter. This implies that the $\mathcal{Z}$-vector of the species scale is also perpendicular to said facet,
		\begin{align}\label{e: Z mult}
			\vec{\mathcal{Z}}_{\rm sp}&=\frac{1}{d-2+p_{\rm eff}}\sum_{i=1}^k(d-2+p_i)\vec{\mathcal{Z}}_I=\notag\\
			&=\frac{p_{\rm eff}}{d-2+p_{\rm eff}}\vec{\zeta}_{\rm eff}\perp{\rm Hull}(\{\vec{\zeta}_1, \ldots ,\vec{\zeta}_k\})\; ,
		\end{align}
		so it points precisely towards the direction along which the corresponding towers decay at the same rate. In other words, it lies at the interface separating regions of moduli space characterized by having different towers as the leading one.
		Note that if one of the contributing towers corresponds to an emergent string, $p=p_{\rm eff}\to\infty$ and $\vec{\mathcal{Z}}_{\rm eff}$ is simply given by the string scale vector.
		\item \textbf{Additive species:} In this scenario there exists no mixing between the different towers, so that the number of species becoming light is the sum of the individual states associated to each tower, $N=\sum_I N_I$. It  then follows that
		\begin{equation}\label{e:add}
			\vec{\mathcal{Z}}=\frac{1}{N}\sum_I N_I \vec{\mathcal{Z}}_I=\frac{1}{N}\sum_I N_I \frac{p_I }{d-2+p_I}\vec{\zeta}_I\;.
		\end{equation}
		If a single tower contributes dominantly to $\LSP$ (i.e. $N\sim N_I$), it sets alone the species scale as all other towers with $N\gg N_I$ can be neglected. %Note again that when a string tower is in the set of towers, it dominates and sets the species scale over the whole region.
		If, on the other hand, several towers contribute with $N_1\sim \ldots \sim N_k\sim N$, it means that we are moving along a direction where the corresponding individual species scales $\Lambda_I$ decay at the same rate, so that $\hat{T}\cdot \vec{\mathcal{Z}}=\hat{T}\cdot \vec{\mathcal{Z}}_1= \ldots =\hat{T}\cdot \vec{\mathcal{Z}}_k$, with $\hat{T}$ being the tangent vector of the asymptotic geodesic. This implies that $\vec{\mathcal{Z}}\perp {\rm Hull}(\{\vec{\mathcal{Z}}_1,\ldots,\vec{\mathcal{Z}}_k\})$, which in general does not force $\vec{\mathcal{Z}}\perp {\rm Hull}(\{\vec{\zeta}_1,\ldots,\vec{\zeta}_k\})$ unless $p_1=\ldots=p_k$ (when the two convex hulls are parallel to each other).
	\end{itemize}
	\begin{figure}[htb]
		\begin{center}
			\begin{subfigure}{0.5\textwidth}
				\includegraphics[width=0.9\textwidth]{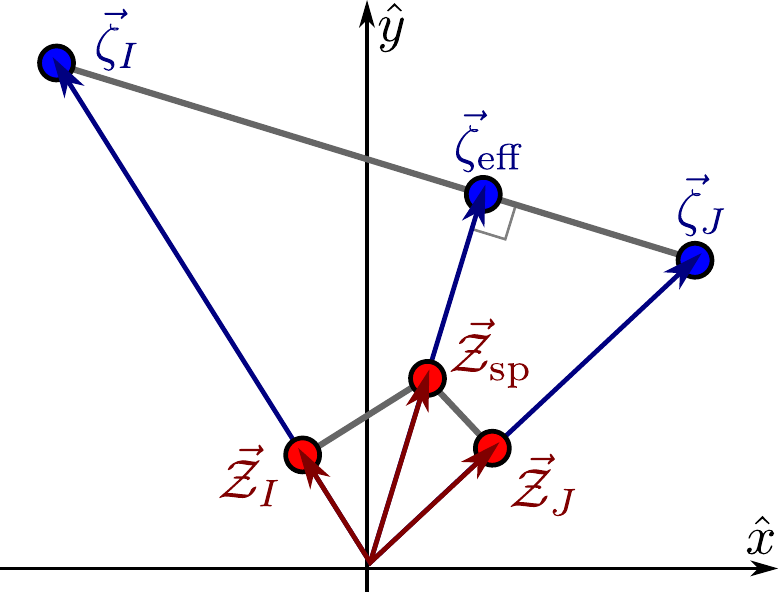}
				\caption{
					\label{fig:mult}}
			\end{subfigure}
			\begin{subfigure}{0.3\textwidth}
				\includegraphics[width=\textwidth]{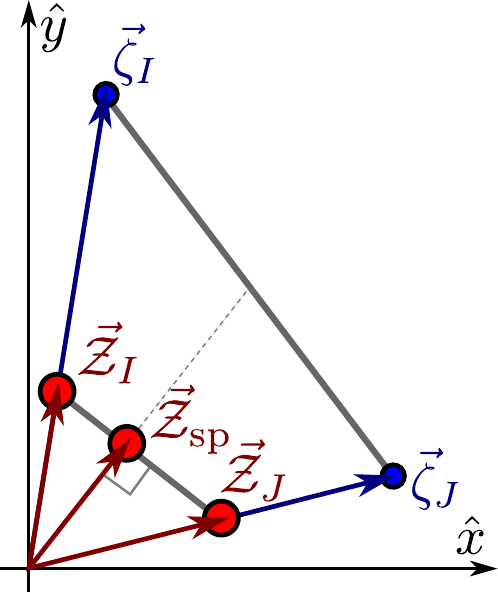}
				\caption{%\small Scalar charge-to-mass and species vectors.
					\label{fig:add}}
			\end{subfigure}
			\caption{\small Sketch of the two possibilities for the species scale $\Lambda_{\rm sp}$ behavior along limits for which two (or more) leading towers can become light at the same rate, spanning a facet of ${\rm Hull}(\{\vec{\zeta}_I\}_I)$. For simplicity, we just take two such towers, $\vec{\zeta}_I$ and $\vec{\zeta}_J$. When there is an effective tower $\vec{\zeta}_{\rm eff}$ of bound states with mixed charges, the associated \emph{multiplicative species} \textbf{(a)} dominates over the individual $\vec{\mathcal{Z}}_I$ and $\vec{\mathcal{Z}}_J$, and is perpendicular to the facet spanned by the individual towers, ${\rm Hull}(\{\vec{\zeta}_I,\vec{\zeta}_J\})$. On the contrary, if the effective towers are absent, the resulting \emph{additive species} \textbf{(b)} $\vec{\mathcal{Z}}_{\rm sp}$ is associated to the sum of the states given by each tower alone, being moreover perpendicular to ${\rm Hull}(\{\vec{\mathcal{Z}}_I,\vec{\mathcal{Z}}_J\})$ and only providing the actual cut-off when both individual species fall at the same rate. Note that in this case $\vec{\mathcal{Z}}_{\rm sp}$ is not expected to be perpendicular to ${\rm Hull}(\{\vec{\zeta}_I,\vec{\zeta}_J\})$, in general.}
			\label{fig:add mult}
		\end{center}
	\end{figure}

	With this in mind, we will argue for the following \emph{sufficient conditions} that allow the pattern \eqref{eq:pattern} to hold universally.
	
	\begin{condition}[Condition 1]	
		The exponential rates $\lambda_I=-\hat{T}\cdot\vec{\zeta}_I$ of the different towers $m_I$ are continuous over the asymptotic regions where they are defined. Furthermore, the product $\vec{\zeta}_{\rm t}\cdot\vec{\mathcal{Z}}_{\rm sp}$ must be well defined along any asymptotic direction.
		\label{A:s1}
	\end{condition}	
	Roughly speaking, this condition implies intuitively that the exponential rate of the leading tower is purely determined by the asymptotic direction approching some infinite distance point, regardless of the particular trajectory that we take as a representative of this direction.
	Furthermore, it also implies that whenever a given tower of states stops being the leading one, it is because there is a second tower that starts dominating instead, in such a way that both have the same exponential mass decay rate along some particular direction. This way, even if the identification of the leading tower changes, its exponential rate still behaves in a continuous fashion on the space of infinite distance limits $\infty(\mathcal{M})$.	
	
	To argue for this, we consider the different masses $m_I$ to be continuous along the regions in $\mathcal{M}$ over which they are defined, and define the exponential rates $\lambda_I$ as functions over the subsets of $\infty(\mathcal{M})$ of possible asymptotic geodesics within the $m_I$ domain of definition (see Appendix \ref{ap:detailsbottomup} for details). Thus, given an asymptotic direction with unit tangent vector $\hat{T}$, $\lambda_I(\hat{T})=\hat{T}\cdot\vec{\zeta}_I=-\hat{T}^a \partial_a \log m_I$. In order for $\lambda_I$ to be well-defined and continuous, we need to show that its value is independent of the asymptotic geodesic we take as representative of a direction $\hat{T}\in\infty(\mathcal{M})$. To do so, take two asymptotic geodesics $\gamma$ and $\gamma'$ with the same direction $\hat{T}$, separated by some distance $\Delta$. Suppose that along each of them the mass $m_I$ has an exponential rate $\lambda_I$ and $\lambda_I'$. Fixing $P\in\gamma$ and $P'\in\gamma'$ as well as two asymptotic points $Q\in\gamma$ and $Q'\in\gamma'$, such that $d(P,Q)=d(P',Q')$ and $d(Q,Q')=\Delta$, we have that
	\begin{equation}
		m_I(Q)\sim m(P)e^{-\lambda_Id(P,Q)},\qquad m_I(Q')\sim m(P')e^{-\lambda_I'd(P,Q)}\;.
	\end{equation}
	As $m_I(P)$ and $m_I(P')$ are fixed and independent of $d(P,Q)$, we obtain $m_I(Q)\sim m_I(Q')\exp\{(\lambda_I'-\lambda_I)d(P,Q)\}$, which if $\lambda_I\neq\lambda_I'$ results in parametrically different values taken by $m_I$ at points separated a fixed distance $\Delta$. This is indeed problematic, since it would result in
	\begin{equation}
		|\vec{\zeta}_I|^2=|\vec{\zeta}_I^{\,\parallel}|^2+|\vec{\zeta}_I^\perp|^2\sim|\nabla^\perp\log m_I|^2\sim\left[(\lambda'_I-\lambda_I)\frac{d(P,Q)}{\Delta}\right]^2\to\infty\;,
	\end{equation}
	as $d(P,Q)\to\infty$, where $\parallel$ and $\perp$ are the perpendicular and parallel components with respect to $\hat{T}$. If the scalar mass-to-charge ratio vectors of our towers are bounded not only from below but also from above (see e.g. \cite{Etheredge:2022opl}), then one must have $\lambda_I=\lambda_I'$. Thus, $\lambda_I$ should be well-defined along any asymptotic direction along which it exists in the first place, irrespective of the specific geodesic. The immediate consequence of this is that, since $\lambda_I=\hat{T}\, \cdot\, \vec{\zeta}_I=\hat{T}\, \cdot\, (\vec{\zeta}_I^{\,\parallel}+\vec{\zeta}_I^\perp)=\hat{T}\, \cdot\, \vec{\zeta}_I^{\,\parallel}$, the only possible difference in $\vec{\zeta}_I$ evaluated along parallel asymptotic geodesics must be in the components perpendicular to $\hat{T}$. This actually puts strong constraints on how the $\vec{\zeta}$-vectors may change or \emph{slide} as we move in moduli space. 
	It implies\footnote{Take some specific geodesic $\gamma_0$ towards some definite infinite distance limit $\hat{T}$, and consider geodesics parallel to it and separated a distance $\Delta$ along some non-compact perpendicular direction $\hat{N}$. In the plane spanned by these two, take the slightly different tangent vector $\hat{T}'$ (i.e. $\hat{T}\cdot\hat{T}'=1-\epsilon$ with $\epsilon\ll 1$). By continuity of $\lambda_I$, $\vec{\zeta}_I$ will have the same expression for $\Delta\to\infty$ and $\epsilon\to 0$, as in the slightly off direction the $\Delta\to\infty$ limit is realized. In other words, sliding occurs along regions when $m_I$ is not homogeneous, with its expression (and that of $\vec{\zeta}_I$) changing between points separated finite distances in moduli space. Moving further from the directions in which the sliding occurs only realizes the $\Delta\to\infty$ limit, doing so with `different velocities' in the moduli.} that the sliding of the $\vec{\zeta}_I$ vectors only occurs along measure-zero subsets of the space of infinite distance trajectories, as it will be further explored in \cite{taxonomyTBA}.
	This is indeed what happens in the 9d moduli spaces with sixteen supercharges (Section \ref{ss:het s1}) and 4d CY compactifications (Sections \ref{ss:heteroticCY3} and \ref{ss:IIBF}) where the volume is not well approximated asymptotically by a single monomial. Even if the masses of the towers have complicated expressions in terms of the moduli, the sliding of the $\vec{\zeta}_I$ vectors occurs as a function of the impact parameter and not as a function of the asymptotic direction. Therefore, $\{\vec{\zeta}_I\}_I$ will take fixed values in almost all (i.e. except for a measure-zero subset) the asymptotic regions where they are defined, only `jumping' when traversing those loci where the sliding occurs.
	
	Analogously, a similar argument can be run for the species scale vectors $\vec{\mathcal{Z}}_J$, such that they take a fixed asymptotic expression over their asymptotic domains except for perhaps measure-zero loci over which sliding could occur, which is accompanied by a jumping in the asymptotic $\vec{\mathcal{Z}}_J$ expression. This way, the exponential rate functions $\alpha_J(\hat{T})=\hat{T}\cdot\vec{\mathcal{Z}}_J=-\hat{T}^a \partial_a \log \Lambda_J$ are also well-defined. Note that, as states from subleading towers may also contribute to $\Lambda_{\rm sp}$, even if well-defined, in principle we cannot expect the exponential rate of the species scale to be continuous over $\infty(\mathcal{M})$, as some of the towers setting the species might cease to exist when exiting their domain. For example, a subleading string tower can be setting the species scale, but if we move away from the asymptotic domain over which it is defined, other (perhaps KK) towers will now set $\LSP$, with the subsequent change in the exponential rate. This is indeed what happens in Section \ref{ss:het s1} when moving parallel to the type I$'$ self-dual line, along which the type I strings are obstructed.
	
	Finally, in order to discuss the pattern, we also need that the product $\vec{\zeta}_{\rm t}\cdot\vec{\mathcal{Z}}_{\rm sp}$ is well defined along any asymptotic direction $\hat{T}\in \infty(\mathcal{M})$. This is a stronger condition than requiring it for each vector independently, and we have no particular motivation to justify it. But if it is not satisfied, such that it can take different values depending on the trajectory representing $\hat T$, then it does not make sense to talk about the pattern along that direction anymore. In general, the well-definition of the $\vec{\zeta}_{\rm t}\cdot\vec{\mathcal{Z}}_{\rm sp}$ product will be guaranteed by \Cref{A:s2} in the border between regions along which different towers/species dominate, but this is not necessarily the case when there is sliding if $\vec{\zeta}_{\rm t}$ and $\vec{\mathcal{Z}}_{\rm sp}$ change in an arbitrary different way (see Appendix \ref{app:derivation}).
	
	\begin{condition}[Condition 2]
		For every infinite distance limit $\hat{T}$ along which two or more towers decay at the same rate, there must exist bound states involving all leading towers, such that the species scale must be given by that corresponding to the case of multiplicative species in \eqref{eq:multspecies}.
		\label{A:s2}
	\end{condition}
	Asymptotic directions  along which two ore more towers decay at the same rate correspond precisely to the interfaces/intersections between the aforementioned domains of fixed vectors $\vec{\zeta}_I$. As we cross these interfaces, the leading vector $\vec{\zeta}_{\rm t}$ jumps since the identification of the tower that decays at the fastest rate changes. Hence, along the interface, we have two or more leading towers sharing the same exponential rate $\lambda_{\rm t}=\hat{T}\cdot\vec{\zeta}_{1}=\ldots=\hat{T}\cdot\vec{\zeta}_{k}$, where we fix $k$ to be the number of towers decaying at the same rate along said direction. These towers span a lattice of charges $(n_1,\ldots,n_k)$, where $n_i$ is the tower level in \eqref{eq:massspectrum} associated to each tower. If a (sub-)lattice of these charges is populated by states that contribute to the species scale, we realize the scenario of multiplicative species above and the species scale is given by \eqref{eq:multspecies}. As already explained, this species scale decays faster than the individual ones associated to each tower alone, and it is moreover perpendicular to the hull spanned by the leading towers of states ${\rm Hull}(\{\vec{\zeta}_1,\ldots,\vec{\zeta}_k\})$. This implies that
	\beq
	\vec{\zeta}_1\cdot \vec{\mathcal{Z}}_{\rm sp}=\ldots= \vec{\zeta}_k\cdot \vec{\mathcal{Z}}_{\rm sp}= \vec{\zeta}_{\rm eff}\cdot \vec{\mathcal{Z}}_{\rm sp}\, ,
	\eeq
	so that the product of the species scale with any of the vectors spanning ${\rm Hull}(\{\vec{\zeta}_1, \ldots ,\vec{\zeta}_k\})$ is the same, and therefore the result remains constant as we cross the interfaces where the identification of the leading tower changes. As a consequence, the different facets of the convex hull of the species scale vectors indicate the regions in moduli space where a given tower plays the role of the leading one.
	
	Contrarily, if the towers do not form bound states and the total number of species is simply given by the sum of the species of each tower (as in the scenario of additive species above, \eqref{e:add}), then the product $\vec{\zeta}_{\rm t}\cdot \vec{\mathcal{Z}}_{\rm sp}$ would generically change upon crossing the interfaces. Notice that, in such a case, we get independent towers of states becoming light at the same rate along the interface, so we would naively\footnote{This is the naive expectation assuming that we indeed get two or more completely independent towers, but more work would be required to determine the type of theory arising at infinite distance.} recover two massless gravitons asymptotically, which goes against Swampland expectations \cite{Bedroya:2023tch,Kim:2019ths}.
	
	We can further motivate the case of multiplicative species as follows. First, notice that, whenever we can identify the tower levels $n_k$ with some sort of gauge charges, the Completeness Hypothesis \cite{Banks:2010zn,Polchinski:2003bq} would require the existence of multiparticle states for each possible value of the charges. Hence, requiring the existence of a (sub-)lattice/tower of bound states resembles a strong version of the Completeness Hypothesis,  where one demands the existence of bound states or enough long-lived resonances (rather than just multiparticle states) so that they can contribute to lower the species scale cut-off. In fact, whenever $n_k$ correspond to gauge charges under massless $U(1)$ gauge fields, Condition \ref{A:s2} reduces to the (sub-)lattice or tower Weak Gravity Conjecture \cite{Heidenreich:2015nta,Heidenreich:2016aqi,Montero:2016tif,Andriolo:2018lvp} (both versions would suffice to get the behavior of multiplicative species).\footnote{We only require the existence of the (sub-)lattice for charges associated to the leading towers and not the subleading ones, since our pattern \eqref{eq:pattern} is only guaranteed to hold for the former. This fits nicely with recent results of \cite{Cota:2022yjw,Cota:2022maf} which show that the (sub-)lattice WGC is satisfied for gauge fields whose charge states decay at the fastest rate asymptotically, but not for certain subleading ones.} More generally, Condition \ref{A:s2} is a generalization of the tower Weak Gravity Conjecture even to the case in which $n_k$ are not conserved gauge charges as there is no associated massless $U(1)$ gauge field (e.g., in certain limits of toroidal orbifolds or Calabi--Yau compactifications). Even in those cases, by analogy, we expect that Condition \ref{A:s2} can be formulated as a strong version of the Completeness Hypothesis that constraints how close we can get to restore a global symmetry asymptotically.
	
	Equivalently, we could have formulated Condition \ref{A:s2} as a condition on the leading tower of states as we move along a direction along which the species scale of different domains decay at the same rate. Along that direction, the species of the leading towers at each side behave as additive species. Hence, for the pattern to hold, there must exist a new additional tower of states that signals a different type of limit along the interface. For instance, this is what happens at the interface of the different dual frames in the moduli space of M-theory on $K3$ in Figure \ref{fig:CHMthy7d}. At the interface between the region described by type 10d IIB and 11d M-theory, the leading tower is actually given by KK modes that signal a decompactification to 9d type IIA. The consequence of this conditions is that the generators of the convex hull of the scalar charge-to-mass ratio of the towers lie precisely at the interfaces where the identification of the species scale changes and the theory is better described by a different dual frame. This way, the different dual frames are determined by the facets of the convex hull of the towers.
	
	\begin{condition}[Condition 3]
		For every connected component of $\infty(\mathcal{M})$ there exists at least one direction associated to an emergent string limit or the homogeneous decompactification of an internal cycle to a higher dimensional vacuum.\label{A:s3}
	\end{condition}
	This condition parallels the Emergent String Conjecture \cite{Lee:2019wij} but it is a weaker statement, since it only demands the existence of at least one limit that has the interpretation of being a string perturbative limit or a decompactification to a higher dimensional vacuum, such that the species scale is simply given by the higher dimensional Planck scale. As explained in Section \ref{s:pattern}, in these specific cases where the species scale is purely set by the leading tower (more so for the emergent string limit, where $m_{\rm t}=\LSP=m_{\rm s}$) we know $\vec{\zeta}_{\rm t}\cdot\vec{\mathcal{Z}}_{\rm sp}$ to take the value $\frac{1}{d-2}$.
	
	With the above conditions in mind, we are now endowed with the tools to give an explanation of the pattern (see Appendix \ref{app:derivation} for more details on this derivation). First, \Cref{A:s1} allows us to divide the space of asymptotic limits into different regions, such that $\vec{\zeta}_{\rm t}\cdot\vec{\mathcal{Z}}_{\rm sp}$ remains constant in the interior of each region. The intersections of the regions are given either because we have several co-leading towers or species decaying at the same rate, or because they correspond to some \emph{sliding loci} where the $\vec{\zeta}_{\rm t}$ or $\vec{\mathcal{Z}}_{\rm sp}$ vectors jump as a function of the asymptotic direction (since they change continuously as a function of the impact parameter) as described above. In the former case, by \Cref{A:s2} we will have that the leading species is located perpendicular to the convex hull of the towers. Hence, the only difference between the scalar charge-to-mass vectors of the towers scaling at the same rate is in the components perpendicular to $\hat{T}\propto \vec{\mathcal{Z}}_{\rm sp}$ ($\hat{T}\propto \vec{\zeta}_{\rm t}$), which  implies that $\vec{\zeta}_{\rm t}\cdot\vec{\mathcal{Z}}_{\rm sp}$ takes the same value at both sides, as well as at the intersection. The same is implied at the \emph{sliding loci} by \Cref{A:s1}. Therefore, for every connected component of $\infty(\mathcal{M})$, any point can be reached from any other by crossing a finite number of intersections/sliding loci, so that $\vec{\zeta}_{\rm t}\cdot\vec{\mathcal{Z}}_{\rm sp}$ takes the same value over all of said component of $\infty(\mathcal{M})$. Now, in order to fix its value, it is enough to know $\vec{\zeta}_{\rm t}\cdot\vec{\mathcal{Z}}_{\rm sp}$ for one limit. This is where \Cref{A:s3} enters into play, finally setting $\vec{\zeta}_{\rm t}\cdot\vec{\mathcal{Z}}_{\rm sp}=\frac{1}{d-2}$.
	
	\subsection{Relation to Emergent String Conjecture}

	To conclude, we want to comment on the relation of the pattern \eqref{eq:pattern} with the Emergent String Conjecture (ESC) \cite{Lee:2019wij}, since they are clearly linked and the informed reader might be wondering to what extent one follows from the other. For the non-experts, let us recall that the ESC implies that every infinite distance limit should either correspond to a decompactification to higher dimensions or to a string perturbative limit where a critical string becomes tensionless.
	
	In Section \ref{ss:continuity} we identified some sufficient conditions that allow the pattern to hold universally in the moduli space, so that we can compare them now with the ESC. \Cref{A:s1} does not follow from the ESC, since it is rather a condition on the asymptotic structure of the towers and how the tower and species vectors can change as we move in moduli space. \Cref{A:s3} clearly follows from the ESC, although it is a weaker statement. The interesting link, though, is associated to \Cref{A:s2}, which is the most important feature underlying the pattern. A priori, it is not obvious whether the ESC implies  \Cref{A:s2}, or why the latter requirement is stronger, as we explain in the following. Consider some decompactification limit in which we have several Kaluza-Klein towers so that several directions open up asymptotically. If all these towers are truly KK towers from the perspective of the \emph{same} duality frame, then it is guaranteed that we will populate the lattice of KK momenta and satisfy  \Cref{A:s2}. This is because for very large momenta, one can use WKB approximation to compute the eigenvalues of the laplacian of the internal space, and they should scale as $m_{\rm KK}\sim n^{1/p_{\rm eff}}m_0$ where $p_{\rm eff}$ is equal to the total number of decompactifying dimensions. Note that this has precisely the structure of the \emph{effective tower} (see \eqref{e:eff tow}) in the multiplicative species scenario discussed in Section \ref{ss:continuity}, so  \Cref{A:s2} holds. However, the ESC a priori does not require that the limits associated to each tower can be interpreted as decompactification limits from the perspective of the same dual frame. For instance, consider the case in which we take a limit along which a KK tower decay at the same rate than a tower of winding modes. In that case, even if both towers signal a decompactification limit in some dual frame, they do not do so within the same duality frame and therefore the total number of species is actually additive (let us denote this as a case of \emph{non-compatible} decompactification limits). Hence, if we only had these two separate towers, we would not get a lattice of bound states and  \Cref{A:s2} (and consequently the pattern) would not hold. However, in practice, whenever this scenario occurs in string theory, we always get additionally a tower of string oscillator modes precisely along the direction where the KK and winding modes decay at the same rate, so that we realize an emergent string limit (rather than decompactifying two extra dimensions) and the pattern again holds. This seems to be always the case even in more complicated string theory examples where we are not simply considering circle decompactifications and we do not have winding modes of a perturbative string. Instead, we may have towers of particles coming from wrapped branes. But even in those instances, the rich network of string dualities always allow us to identify some critical string (in some dual frame) becoming tensionless along the interface between the different  \emph{non-compatible} decompactification limits. We want to remark that this is indeed crucial for the pattern to hold, and from a bottom-up perspective, it imposes a non-trivial constraint on how the different infinite distance limits glue together in the moduli space (or more precisely in $\infty(\mathcal{M})$).
	
	Therefore, if we interpret the ESC as the milder claim that the leading tower must be either a KK tower in some dual frame, or an emergent string, then it does not immediately imply \Cref{A:s2} and it is, therefore, a weaker condition that the pattern. The above scenario of \emph{non-compatible} decompactification limits would still be consistent with this mild version of the ESC even if we did not have the string becoming tensionless at the interface, since the leading tower would still be a KK tower (although the total number of species would be additive). However, if we interpret or refine the ESC as the claim that there must be either a single dual frame where all the leading towers can be seen as KK towers or we get an emergent string yielding the leading tower, then it implies \Cref{A:s2}. In this latter case, the pattern would essentially follow from this stronger version of the ESC, mod some subtleties related to the sliding of the $\zeta$-vectors  that are addressed in \Cref{A:s1}. This would be interesting, as the pattern would then hint a new avenue to try to provide a bottom-up explanation for the ESC, which so far is only motivated by string theory examples. If we were able to provide a bottom-up rationale for the pattern (possibly based on black hole physics or entropy arguments), then we could use it to argue for the ESC and show that a consistent quantum gravity theory necessarily requires from the existence of perturbative strings and extra dimensions.

	\section{Conclusions}
	\label{s:conclusions}
	
	In this paper (and its companion \cite{patternPRL}), we point out an interesting pattern that it is satisfied in all (up to now) known examples of infinite distance limits in the moduli space of string theory compactifications, regardless of the level of supersymmetry or the type of compactification manifold. This pattern is a sharp relation between the asymptotic value of the variation rate (in moduli space) of the species scale cut-off and the mass of the leading tower of states, given by \eqref{eq:pattern}. We check that it holds for multi-field geodesic trajectories where several moduli are taken to infinity, and even if the species scale cut-off is not only determined by the leading tower of states but captures information of other subleading towers. 
	
	At the very least, this pattern is a common thread underlying all known string theory examples that have been explored so far, and it make manifest a very constrained structure behind the large casuistics of different types of infinite distance limits and how they can fit together in the moduli space. We suspect, though, that the universality of the pattern is rooted in a deeper underlying quantum gravity principle, rather than being just a lamppost effect of known string constructions. Hence, the most important goal for the future is to search for a purely bottom-up rationale that could explain the pattern independently of string theory. Promising avenues include thinking of black holes or entropy bounds, since the pattern relates the number of species (which itself provides the entropy of the smallest semiclassical black hole) with the mass of the tower. Alternatively, we can also think of the number of species as a measure of the density of states in an Einstein gravity theory, so that the less dense the tower is, the fastest can become light according to the pattern. Another avenue would be to use S-matrix bootstrap techniques, since the species scale cut-off can be understood as the scale at which a semiclassical Einstein gravity description breaks down and higher derivative terms start dominating over the tree-level Einstein term. It would be fantastic if one could argue for a relation between this scale and e.g., the scale of the first massive spin-2 field of a KK tower.
	
	Finding a bottom-up rationale for the pattern would have profound consequences for the Swampland program, since it implies a more precise formulation of the Distance conjecture that constrains the nature of the tower and implies a sharp bound on how fast it becomes light. If the pattern holds, then it automatically implies a lower bound on the exponential rate of the tower given by $\lambda_{\text{t}} \geq \frac{1}{\sqrt{d-2}}$, which supports the bound proposed by \cite{Etheredge:2022opl} and it is closely related to the Emergent String Conjecture \cite{Lee:2019wij}. Furthermore, it provides a clear recipe to determine the species scale cut-off upon knowledge of the leading tower of states along different directions. It would also be interesting to explore how it extends to the interior of the moduli space, where the notion of a leading tower of states is no longer well defined. 
	
	In this work, we also identified some sufficient conditions that the towers of states and the asymptotic geometry of the moduli space must satisfy to allow for the pattern to hold. Interestingly, the most important condition resembles a sort of (sub-)lattice WGC where the role of the gauge charges is played by the levels of the tower. This condition also follows from a strong interpretation of the Emergent String Conjecture. Hence, many ideas in the Swampland program get interconnected and can be re-derived from this simple equation relating the variation of the species scale and the leading tower of states becoming light asymptotically. We hope to come back to these ideas in the future.
	
	\section*{Acknowledgements}
	
	We are very thankful to Jos\'e Calder\'on-Infante, Muldrow Etheredge, Ben Heidenreich, \'Alvaro Herr\'aez, Luis Iba\~nez, Jacob McNamara, Miguel Montero, David Prieto, Tom Rudelius, Cumrun Vafa and Timo Weigand for illuminating discussions. A.C. and I.R. also wish to acknowledge the hospitality of the Department of Theoretical Physics at CERN during the early stages of this work. I.R. and I.V. would like to thank Simons Center for Geometry and Physics, Stony Brook University for hospitality and support during the middle stages of this work. The authors acknowledge the support of the Spanish Agencia Estatal de Investigacion through the grant ``IFT Centro de Excelencia Severo Ochoa'' CEX2020-001007-S and the grant PID2021-123017NB-I00, funded by MCIN/AEI/10.13039/ 501100011033 and by ERDF A way of making Europe. The work of A.C. is supported by the Spanish FPI grant No. PRE2019-089790 and by the Spanish Science and Innovation Ministry through a grant for postgraduate students in the Residencia de Estudiantes del CSIC. The work of I.R. is supported by the Spanish FPI grant No. PRE2020-094163. The work of I.V. is also partly supported by the grant RYC2019-028512-I from the MCI (Spain) and the ERC Starting Grant QGuide-101042568 - StG 2021.
	
	%\newpage

	\appendix
	
	\section{Generalities on charge-to-mass and species vectors}
	\label{ap:generalities}
	
	In this appendix we present a derivation of the general formulae associated to the computation of the relevant charge-to-mass and species vectors that arise upon compactifying a $D$-dimensional gravitational theory on some closed manifold of real dimension $n\in \mathbb{N}$. Later on, we generalize the analysis to the case in which we consider the compact space to be of the product form $\mathcal{X}'_n=\mathcal{X}_{n_1}\times \ldots \times\mathcal{X}_{n_N}$, with $n_i$ denoting the dimensionality of the corresponding submanifold. With such information we revisit the pattern \eqref{eq:pattern}, thus checking it explicitly in the cases at hand. Therefore, the discussion here can be seen as complementary to the material presented in Sections \ref{s:pattern} and \ref{s:maxsugra}.
	
	\subsubsection*{Compactification on an $n$-dimensional cycle}
	
	Let us start by studying the kind of charge-to-mass vectors that typically appear in string-motivated EFTs. In order to be as general as possible, we consider a $D$-dimensional theory compactified down to $d=D-n$ spacetime dimensions. We denote $\mathcal{V}_n$ the overall volume modulus associated to the internal compact manifold, $\mathcal{X}_n$, measured in $D$-dimensional Planck units. Suppose that we focus on a sector of the theory described by the following simple action \cite{Etheredge:2022opl}
	\begin{equation}\label{eq:Ddim}
		S_{D} \supseteq \int \dd^{D}x\, \sqrt{-g_D}\,  \left[  \frac{1}{2\kappa_{D}^2}\mathcal{R}_{D} - \frac{1}{2}\left(\partial \hat \phi \right)^2 \right]\, ,
	\end{equation}
	where $\hat{\phi}$ is some generic canonically normalized modulus. Note that one may also think of $\hat \phi$ as parametrizing some fixed (asymptotically) geodesic trajectory in a multi-moduli setup. Upon compactification on the $n$-fold $\mathcal{X}_n$, one arrives at
	\begin{equation}\label{eq:ddim}
		S_{d} \supseteq \int \dd^{d}x\, \sqrt{-g_d}\,  \left[ \frac{1}{2\kappa_{d}^2} \left(\mathcal{R}_{d} - \frac{d+n-2}{n (d-2)} \left(\partial \log \mathcal{V}_n \right)^2 \right)- \frac{1}{2} \left(\partial \hat \phi \right)^2 \right]\, ,
	\end{equation}
	where we have retained only the scalar-tensor sector of the lower dimensional theory, ignoring possible extra fields arising in the dimensional reduction process.\footnote{To obtain \eqref{eq:ddim} in such form one needs to perform a Weyl rescaling of the $d$-dimensional metric as follows $g_{\mu \nu} \to g_{\mu \nu} \mathcal{V}_n^{-\frac{2}{d-2}}$.} One can then define a canonically normalized volume modulus
	\begin{equation}\label{eq:canonicalvolume}
		\hat \rho = \frac{1}{\kappa_d}\sqrt{\frac{d+n-2}{n(d-2)}} \log \mathcal{V}_n\, ,
	\end{equation}
	which indeed controls the overall Kaluza-Klein scale associated to the compact internal space
	\begin{equation}\label{eq:KKscale}
		m_{\text{KK},\, n} \sim  M_{\text{Pl};\, d}\, e^{-\kappa_d \sqrt{\frac{d+n-2}{n (d-2)}} \hat \rho}\, .
	\end{equation}
	As customary, this tower of states becomes exponentially light when taking the decompactification limit $\hat \rho \to \infty$. In terms of scalar charge-to-mass vectors one would then write
	\begin{equation}\label{eq:kkcharge2mass}
		\vec{\zeta}_{\text{KK},\, n} = \left( 0 , \sqrt{\frac{d+n-2}{n (d-2)}} \right)\, ,
	\end{equation}
	where the first (last) entry corresponds to the normalized modulus $\hat \phi$ ($\hat \rho$).
	
	Let us also assume that the scalar $\hat \phi (x)$ is non-compact, and that the higher dimensional theory satisfies the Distance Conjecture \cite{Ooguri:2006in}. Therefore, there should exist an infinite tower of particles with mass behaving asymptotically as follows
	\begin{equation}\label{eq:SDCDdim}
		m_{\text{tower}} \sim  M_{\text{Pl};\, D}\, e^{-\kappa_D \lambda_D \hat \phi}\, ,
	\end{equation}
	where $\lambda_D$ is nothing but the $D$-dimensional scalar charge-to-mass ratio along the positive $\hat \phi$-direction. If such tower of particles is inherited by the lower-dimensional theory, they would present a mass which in Planck units depends on both $\hat \phi$ and the volume modulus $\hat \rho$ through the relation
	\begin{equation}\label{eq:SDCddim}
		m_{\text{tower}} \sim  M_{\text{Pl};\, d}\, \exp\left\{-\kappa_d \lambda_D \hat \phi - \kappa_d \sqrt{\frac{n}{(d+n-2)(d-2)}} \hat \rho\right\}\, ,
	\end{equation}
	where the second term in the exponent arises just from the ratio $M_{\text{Pl};\, D}/M_{\text{Pl};\, d}$. Again, in terms of scalar charge-to-mass vectors one obtains
	\begin{equation}\label{eq:chargetomasstower}
		\vec{\zeta}_{\text{t}} = \left( \lambda_D , \sqrt{\frac{n}{(d+n-2)(d-2)}} \right)\, .
	\end{equation}
	Note that if $\hat{\phi}$ denotes the $D$-dimensional dilaton in some string theory, then $\lambda_D=\frac{1}{\sqrt{D-2}}=\frac{1}{\sqrt{d+n-2}}$ \cite{Etheredge:2022opl, vandeHeisteeg:2023ubh}, whilst if it corresponds to a volume modulus from a higher compactification (i.e. from $D'=D+n'$ to $D$ spacetime dimensions), then $\lambda_D=\sqrt{\frac{D+n'-2}{n'(D-2)}}=\sqrt{\frac{d+n+n'-2}{n'(d+n-2)}}$. Remarkably, this also encompasses the case in which one of the moduli corresponds to some dilatonic field, since upon taking the limit $n'\to \infty$ the first entry of the scalar charge-to-mass vector becomes $\frac{1}{\sqrt{D-2}}$ \cite{Etheredge:2022opl,Castellano:2021mmx}.
	
	For the species scale, on the other hand, we will distinguish between two possibilities, as predicted by the Emergent String Conjecture \cite{Lee:2019wij}. First of all, if the limit corresponds to an emergent critical string, the QG cut-off will be given by the string scale (up to logarithmic corrections) since the set of light states will be dominated by an exponentially large number of string excitation modes. Because of this,
	\begin{equation} \label{eq:stringmassdependence}
		\LSP \sim m_{\rm string}\sim M_{\text{Pl};\, D}\, \exp\left\{-\kappa_D\frac{1}{D-2} \hat \phi\right\}\, .
	\end{equation}
	Hence $\vec{\mathcal{Z}}_{\rm osc}=\vec{\zeta}_{\rm osc}$, so that in this limit
	\begin{equation}\label{eq:generalstringlimit}
		\vec{\zeta}_{\text{t}}\cdot\vec{\mathcal{Z}}_{\rm sp}=|\vec{\zeta}_{\rm osc}|^2=\frac{1}{d-2}\, ,
	\end{equation}
	thus fulfilling \eqref{eq:pattern}. Notice that \eqref{eq:generalstringlimit} above is also verified when $\vec{\zeta}_{\text{t}}=\vec{\zeta}_{\text{KK},\, n}$ (see Figure \ref{sfig:KKstring}), since for an emergent string limit the KK tower falls at the same rate as the string mass \cite{Lee:2019wij}. Otherwise, one could retrieve a critical string in $d<10$.
	
	The second possibility would correspond to explore some decompactification limit, namely when the tower from \eqref{eq:SDCDdim} is of Kaluza-Klein nature (in some duality frame). In such a case, one would have three different species vectors: those which are parallel to the original $\zeta$-vectors and a new one arising as an effective combination thereof \cite{Calderon-Infante:2023ler}. For the former, one can write 
	\begin{equation}\label{eq:Zvectorgeneral}
		\vec{\mathcal{Z}}_{{\rm KK},\, n'}=\frac{n'}{d+n'-2}\vec{\zeta}_{{\rm KK},\, n'}\, , \qquad \vec{\mathcal{Z}}_{{\rm KK},\, n}=\frac{n}{d+n-2}\vec{\zeta}_{{\rm KK},\, n}\, ,
	\end{equation}
	where $\vec{\zeta}_{{\rm KK},\, n}$ is given by \eqref{eq:kkcharge2mass} above and with
	\begin{equation}
		\vec{\zeta}_{{\rm KK},\, n'} = \left( \sqrt{\frac{d+n+n'-2}{n' (d+n-2)}} , \sqrt{\frac{n}{(d+n-2)(d-2)}} \right)\, ,
	\end{equation}
	thus satisfying $|\vec{\zeta}_{{\rm KK},\, n'}|^2=\frac{d+n'-2}{n' (d-2)}$. Therefore, whenever we explore an asymptotic direction parallel to one of these two, the species scale will be parametrically controlled by the Planck scale of the $(d+n')$-dimensional (resp. $(d+n)$) theory. As an example, upon taking the limit $\hat\phi, \hat\rho \to \infty$ along the $\vec{\zeta}_{{\rm KK},\, n'}$\,-direction one finds
	\begin{align} \label{eq:kkspeciesvector}
		\LSP \sim M_{{\rm Pl};\, d+n'}\sim M_{\text{Pl};\, d}\, \left(\frac{m_{{\rm KK},\, n'}}{M_{\text{Pl};\, d}}\right)^{\frac{n'}{d+n'-2}}\, ,
	\end{align}
	with $m_{{\rm KK},\, n'}$ denoting the mass scale of the corresponding KK-like tower. For intermediate directions, however, the dominant species vector is that obtained by combining the previous ones as follows \cite{Calderon-Infante:2023ler}
	\begin{equation}\label{eq:effectiveKKspeciesvector}
		\vec{\mathcal{Z}}_{{\rm KK},\, n+n'} = \frac{1}{d+n+n'-2}\left( n'\, \vec{\zeta}_{{\rm KK},\, n'} + n\, \vec{\zeta}_{{\rm KK},\, n}  \right)\, ,
	\end{equation}
	which is indeed controlled by the Planck scale of the $(d+n+n')$-dimensional parent theory, see Figure \ref{sfig:twoKK}. With this we can now check if the pattern \eqref{eq:pattern} is satisfied. Once again, for the directions determined by any of the three species vectors one easily verifies that $\vec{\zeta}_{\text{t}} \cdot \vec{\mathcal{Z}}_{\text{sp}}=\frac{1}{d-2}$. In particular, when probing the $\vec{\mathcal{Z}}_{{\rm KK},\, n+n'}\,$-direction what one effectively does is decompactifying both cycles at the same rate, such that the total KK mass yields a charge-to-mass vector of the form
	\begin{equation} \label{eq:effectivezeta}
		\vec{\zeta}_{{\rm KK},\, n+n'} = \frac{1}{n+n'}\left( n'\, \vec{\zeta}_{{\rm KK},\, n'} + n\, \vec{\zeta}_{{\rm KK},\, n}  \right)\, ,
	\end{equation}
	which happens to lie at the point closest to the origin within the polytope generated by $\vec{\zeta}_{{\rm KK},\, n'}$ and $\vec{\zeta}_{{\rm KK},\, n}$, see Figure \ref{sfig:twoKK}.
	
	For intermediate cases, given that the species scale is determined by $\vec{\mathcal{Z}}_{{\rm KK},\, n+n'}$ together with the fact that $\vec{\zeta}_{{\rm KK},\, n+n'}$ is orthogonal to the line joining the two $\zeta$-vectors, one finds that \eqref{eq:pattern} still holds for any asymptotically light tower.
	
	\subsection*{Generalization to `nested' compactifications}
	\label{ss:nestedcompactifications}	
	
	The previous analysis can be easily generalized to the case in which our $D$-dimensional theory is compactified down to $d=D-n$ on an $n$-dimensional manifold given by the Cartesian product $\mathcal{X}_n=\mathcal{X}_{n_1}\times \ldots \times\mathcal{X}_{n_N}$, with $n=\sum_{i=1}^N n_i$. This can be alternatively seen as a step-by-step (or `nested') compactification
	\begin{equation}\label{eq:compactchain}
		\notag D=d+\sum_{i=1}^N n_i\to d+\sum_{i=2}^N n_i\to \ldots \to d+n_N\to d\, ,
	\end{equation}
	where the order of the compactification chain is unimportant and only amounts to a certain rotation of the associated scalar charge-to-mass vectors, hence not affecting neither their length nor the angles subtended between them. With this in mind, one finds that the KK tower obtained from the decompactifying any $\mathcal{X}_{n_i} \subset \mathcal{X}_n$ is given by
	\begin{equation}\label{eq: gen eq}
		\zeta^j_{{\rm KK},\, n_i}=\left\{
		\begin{array}{ll}
			0& \qquad \text{if }j<i\\
			\sqrt{\frac{d+\sum_{l=i}^Nn_l-2}{n_i(d+\sum_{l=i+1}^Nn_l-2)}}& \qquad \text{if }i=j\\
			\sqrt{\frac{n_j}{(d+\sum_{l=j}^Nn_l-2)(d+\sum_{l=j+1}^Nn_l-2)}}& \qquad \text{if }j>i
		\end{array}
		\right.
	\end{equation} 
	Notice that this also encompasses the case in which one of the moduli corresponds to some $D$-dimensional dilaton, upon setting $n_0\to \infty$, so that the zero-th entry of the scalar charge-to-mass vector becomes $\frac{1}{\sqrt{D-2}}$.
	
	On the other hand, given a subset $\{\vec{\zeta}_{{\rm KK},\, m_j}\}_{j=1}^M\subseteq \{\vec{\zeta}_{{\rm KK},\, n_i}\}_{i=1}^N$, one can show that
	\begin{equation}\label{eq: mult decompact}
		\vec{\zeta}_{{\rm KK},\, \sum_j m_j}=\frac{1}{\sum_{j=1}^M m_j}\sum_{j=1}^M m_j\, \vec{\zeta}_{{\rm KK},\, m_j}\, ,
	\end{equation}
	corresponds to the `effective' KK tower associated to the joint decompactification of $\mathcal{X}_{m_1}\times \ldots \times \mathcal{X}_{m_M}$, where the volume of each of the cycles grows at the same rate. Incidentally, it can be seen to coincide with the point of the polytope spanned by $\{\vec{\zeta}_{{\rm KK},\, m_j}\}_{j=1}^M$ located closest to the origin.
	
	Taking infinite distance limits, the easiest possibility would be an emergent string limit, for which $\vec{\zeta}_{\text{t}}\cdot\vec{\mathcal{Z}}_{\rm sp}=|\vec{\zeta}_{\rm osc}|^2=\frac{1}{d-2}$ is trivially fulfilled. The other option would correspond to explore a decompactification limit from $d$ to $d+\sum_{j=1}^M m_j$ dimensions, with $\{m_j\}_{j=1}^M\subseteq\{n_i\}_{i=1}^N$, where we allow the possibility of a dilaton-like direction by setting $m_0\equiv \infty$. In this case the species scale will be parametrically given by the Planck scale of the ($d+\sum_{j=1}^M m_j$)-dimensional theory,\footnote{If $m_0\equiv \infty$ then the species scale is again given by the string scale.} so
	\begin{align} \label{eq:kkspeciesvectorgeneral}
		\LSP&\sim M_{{\rm pl},\, d+\sum_{j=1}^M m_j}\sim M_{\text{Pl};\, d}\exp\left\{-\kappa_d\frac{\sum_{j=1}^M m_j}{(d+\sum_{j=1}^M m_j-2)(d-2)}\hat{\rho}\right\}\notag\\
		&\sim M_{\text{Pl};\, d}\left(\frac{m_{{\rm KK},\, \sum_{j=1}^M m_j}}{M_{{\rm Pl},d}}\right)^{\frac{\sum_{j=1}^M m_j}{d+\sum_{j=1}^M m_j-2}},
	\end{align}
	where $\hat{\rho}$ is the normalized modulus denoting the volume being decompactified. As a result
	\begin{equation}
		\vec{\mathcal{Z}}_{\rm sp}=\frac{\sum_{j=1}^M m_j}{d+\sum_{j=1}^M m_j-2}\vec{\zeta}_{{\rm KK},\, \sum_{j=1}^M m_j}=\frac{1}{d+\sum_{j=1}^M m_j-2}\sum_{j=1}^M m_j\, \vec{\zeta}_{{\rm KK},\, m_j}\, ,
	\end{equation}
	where \eqref{eq: mult decompact} is used. Now, for the leading tower, we have two possibilities. First of all, we might be moving in the joint compactification direction, so $\vec{\zeta}_{\text{t}}=\vec{\zeta}_{{\rm KK},\, \sum_{j=1}^M m_j}$, and thus
	\begin{equation}
		\vec{\zeta}_{\text{t}}\cdot\vec{\mathcal{Z}}_{\rm sp}=\frac{\sum_{j=1}^M m_j}{d+\sum_{j=1}^M m_j-2} |\vec{\zeta}_{{\rm KK},\, \sum_{j=1}^M m_j}|^2=\frac{1}{d-2}\, .
	\end{equation}
	The other possibility is that we move in some other direction, where while still decompactifying  $\mathcal{X}_{m_1}\times \ldots \times \mathcal{X}_{m_M}$, not all cycles do so at the same speed. Then we will have a leading tower $\vec{\zeta}_{\text{t}}=\vec{\zeta}_{{\rm KK},{m_{i_0}}}\in \{\vec{\zeta}_{{\rm KK},m_j}\}_{j=1}^M$, so that
	\begin{align}
		\vec{\zeta}_{\text{t}}\cdot\vec{\mathcal{Z}}_{\rm sp}&= \frac{1}{d+\sum_{j=1}^M m_j-2}\sum_{j=1}^M m_j\, \vec{\zeta}_{{\rm KK},\, {m_{i_0}}}\cdot\vec{\zeta}_{{\rm KK},\, m_j}\notag\\
		&=\frac{1}{d+\sum_{j=1}^M m_j-2}\left[m_{i_0}|\, \vec{\zeta}_{{\rm KK},\,{m_{i_0}}}|^2+\sum_{j\neq i_0}m_j\, \vec{\zeta}_{{\rm KK},\,{m_{i_0}}}\cdot\vec{\zeta}_{{\rm KK},\, m_j}\right]\notag\\
		&=\frac{1}{d+\sum_{j=1}^M m_j-2}\frac{d+\sum_{j=1}^Mm_{j}-2}{d-2}=\frac{1}{d-2}\, ,
	\end{align}
	where for the last sum in the second line we have used \eqref{eq: gen eq}. The generalization of this, for which several (but not all) of the cycles decompactify the fastest at the same pace is straightforward, as $\vec{\zeta}_{\text{t}}$ will be a convex combination of KK vectors (actually determined by the closest point to the origin of the polytope generated by the latter). Indeed, this follows from the fact that all possible $\vec{\zeta}_{\rm t}$ are located in the polytope spanned by the $\vec{\zeta}_{\text{KK},\, m_j}$ vectors corresponding to dimensions being decompactified, to which $\vec{\mathcal{Z}}_{\rm sp}$ is perpendicular, by construction.
	
	\section{Details on the Hypermultiplet metric}
	\label{ap:hypermetric}
	
	The material presented in this appendix is complementary to the discussion in Section \ref{ss:hypers}, where the fate of the pattern within certain heavily quantum-corrected moduli spaces was analyzed. Here we provide more details regarding the relevant non-perturbative corrections, as well as their contribution to the exact hypermultiplet metric. Section \ref{ss:exactmetric} briefly summarizes the procedure employed in \cite{Alexandrov:2014sya} to obtain the aforementioned line element, upon using the twistorial formulation of quaternionic-K\"ahler spaces. Section \ref{ss:SL2Z} reviews the duality properties of the hypermultiplet moduli space arising from type II compactifications on CY threefolds \cite{Bohm:1999uk, Robles-Llana:2007bbv}, both at the classical and quantum levels. Finally, in Section \ref{ss:detailshyper} we use these considerations to argue how the pattern survives at the quantum level in a highly non-trivial way.
	
	\subsection{The Exact Metric and the contact potential}
	\label{ss:exactmetric}
	
	The exact hypermultiplet metric for type IIA string theory compactified on a CY$_3$ has been recently computed exactly to all orders in $g_s$ incorporating the contributions of \emph{mutually local} D2-brane instantons in \cite{Alexandrov:2014sya}. The strategy followed in that work was to exploit the twistorial description of quaternionic-K\"ahler manifolds (see e.g., \cite{Alexandrov:2008ds,Alexandrov:2010qdt}), combined with certain symmetries which are also expected to be preserved at the quantum level. In the following we will briefly review such computation in order to explicitly show the very non-trivial metric one arrives at, which is strongly corrected both at the perturbative and non-perturbative level, thus putting naively in danger any conclusion drawn from the tree-level metric displayed in \eqref{eq:classicalhypermetric}.
	
	The crucial ingredient to obtain the hypermultiplet metric is the so-called \emph{contact potential} $\chi^{\rm IIA}$, which is a real-valued function defined over a twistor space $\mathcal{Z}$ constructed as a $\mathbb{P}^1$-bundle over the moduli space $\mathcal{M}_{\rm HM}$. It moreover has a connection given by the $SU(2)$ part, $\vec{p}=\left( p^+, p^-, p^3\right)$, of the Levi-Civita connection on $\mathcal{M}_{\rm HM}$, which in turn determines the holomorphic contact structure associated to $\mathcal{Z}$ (see e.g., the review \cite{Alexandrov:2011va}). Therefore, one may define a holomorphic 1-form as follows
	\begin{align}\label{eq:holomorphic1form}
		\mathcal{X}=-4 \text{i} \chi^{\text{IIA}} Dt\, ,
	\end{align}
	where $t$ is a complex coordinate on $\mathbb{P}^1$ and $Dt= \dd t +p^+-\text{i} p^3t+p^- t^2$. Now, in order to obtain the metric on $\mathcal{M}_{\rm HM}$ one first computes the contact potential $\chi^{\rm IIA}$ including all D-instanton corrections, which reads \cite{Alexandrov:2009zh}
	\begin{align}\label{eq:chiIIAtwistor}
		\chi^{\rm IIA} =\, \frac{\mathcal{R}^2}{2} e^{-K_{\text{cs}}} + \frac{\chi_{E}(X_3)}{96\pi} -\frac{\text{i} \mathcal{R}}{16 \pi^2} \sum_{\gamma} \Omega(\gamma) \left( Z_{\gamma} \mathcal{J}_{\gamma}^{(1, +)} + \bar{Z}_{\gamma} \mathcal{J}_{\gamma}^{(1, -)}\right)\, ,
	\end{align}
	where $\mathcal{R}=e^{-\phi} \mathcal{V}_{A_0}/2$ is the mirror dual of the ten-dimensional IIB dilaton, $K_{\text{cs}}$ is the complex structure K\"ahler potential, and $Z_{\gamma}(z)=q_I z^I-p^I \mathcal{F}_I$ denotes the central charge function of a D2-instanton with (integral) charges $\gamma=\left( q_I, p^I\right)$. Their degeneracy is captured by the Donaldson-Thomas invariants $\Omega(\gamma)$, which count (in a BPS indexed way) the relevant instantons within the class $[\gamma] \in H_3(X_3, \mathbb{Z})$ \cite{Alexandrov:2013yva}.\footnote{The Donaldson-Thomas invariants $\Omega(\gamma)$ can be related, upon using Mirror Symmetry, to the genus-0 Gopakumar-Vafa invariants in the type IIB dual description \cite{Gopakumar:1998ii, Gopakumar:1998jq}.} We have also defined the twistorial integrals \cite{Alexandrov:2014sya}
	\begin{align}\label{eq:Jintegral}
		\mathcal{J}_{\gamma}^{(1, \pm)}= \pm \int_{\ell_{\gamma}} \frac{\dd t}{t^{1 \pm 1}}\, \log \left( 1-\sigma_{\gamma} e^{-2\pi \text{i} \Theta_{\gamma}(t)}\right)\, ,
	\end{align}
	where $\ell_{\gamma}$ is a BPS ray on $\mathbb{P}^1$, $\sigma_{\gamma}$ is a sign function that we will take to be $+1$ in the following, and $\Theta_{\gamma}(t)$ are functions defined over the twistor space $\mathcal{Z}$ which, in the case of mutually local instantons, are given by
	\begin{align}\label{eq:Thetafn}
		\Theta_{\gamma}(t) = q_I \zeta ^I -p_I \tilde{\zeta}_I + \mathcal{R} \left( t^{-1} Z_{\gamma}-t \bar{Z}_{\gamma}\right)\, .
	\end{align}
	As a next step, one needs to determine the $SU(2)$ connection $\vec{p}$ as functions on the base $\mathcal{M}_{\text{HM}}$ and the complex coordinate $t \in \mathbb{P}^1$, from which one extracts the triplet of quaternionic 2-forms $\vec{\omega}$ as follows 
	\begin{align}
		\vec{\omega}=-2 \left(\dd \vec{p} + \frac{1}{2} \vec{p} \times \vec{p} \right)\, .
	\end{align}
	The advantage of knowing $\vec{\omega}$ is that these are defined by the almost complex structures $\vec{J}$ characterizing the quaternionic-K\"ahler manifold $\mathcal{M}_{\text{HM}}$ as well as by its metric. Therefore, upon specifying e.g., $J^3$ by providing a basis of holomorphic 1-forms on $\mathcal{M}_{\text{HM}}$, one may retrieve the metric via the relation $g(X,Y)=\omega^3 (X, J^3 Y)$, for all $X,Y \in T\mathcal{M}_{\text{HM}}$. Once all this has been done, one arrives at the quantum-corrected line element (we henceforth set all magnetic charges $p^I=0$, which can be achieved via some symplectic rotation) \cite{Alexandrov:2014sya}:
	\begin{align}\label{eq:quantumhypermetric}
		\nonumber \dd s^2_{\text{HM}} &= \frac{1}{2\left(\chi^{\text{IIA}}\right)^2} \left( 1-\frac{\chi^{\text{IIA}}}{\mathcal{R}^2 U}\right)(\dd \chi^{\text{IIA}})^2 + \frac{1}{2 \left(\chi^{\text{IIA}}\right)^2\left( 1-\frac{\chi^{\text{IIA}}}{\mathcal{R}^2 U}\right)} \left( \dd \varrho - \tilde{\zeta}_J \dd\zeta^J+\zeta^J \dd\tilde{\zeta}_J + \mathcal{H} \right)^2\\
		\nonumber  &+\frac{\mathcal{R}^2}{2 \left(\chi^{\text{IIA}}\right)^2} \left| z^I \mathcal{Y}_I\right|^2 + \frac{1}{2 \chi^{\text{IIA}} U} \left| \mathcal{Y}_I M^{IJ} \bar{v}_J - \frac{\text{i} \mathcal{R}}{2\pi} \sum_{\gamma} \Omega_{\gamma} \mathcal{W_{\gamma} \dd Z_{\gamma}}\right|^2\\
		\nonumber &-\frac{1}{2 \chi^{\text{IIA}}} M^{IJ} \left( \mathcal{Y}_I + \frac{\text{i} \mathcal{R}}{2\pi} \sum_{\gamma} \Omega_{\gamma}\, q_I \mathcal{J}_{\gamma}^{(2, +)} \left( \dd Z_{\gamma}-U^{-1} Z_{\gamma}\, \partial e^{-K_{\text{cs}}}\right)\right)\\
		\nonumber &\times \left( \bar{\mathcal{Y}}_J - \frac{\text{i} \mathcal{R}}{2\pi} \sum_{\gamma'} \Omega_{\gamma'}\, q'_J \mathcal{J}_{\gamma'}^{(2, -)} \left( \dd \bar{Z}_{\gamma'}-U^{-1} \bar{Z}_{\gamma'}\, \bar{\partial} e^{-K_{\text{cs}}}\right)\right)\\
		\nonumber &+\frac{\mathcal{R}^2\, e^{-K_{\text{cs}}}}{2 \chi^{\text{IIA}}} \Bigg( \mathsf{G}_{i \bar j} \dd z^i \dd z^{\bar j} - \frac{1}{\left( 2\pi U\right)^2} \left|  \sum_{\gamma} \Omega_{\gamma} \mathcal{W}_{\gamma} Z_{\gamma} \right|^2 \left| \partial K_{\text{cs}}\right|^2\\
		&+\frac{e^{K_{\text{cs}}}}{2\pi} \sum_{\gamma} \Omega_{\gamma} \mathcal{J}_{\gamma}^{(2)}\left| \dd Z_{\gamma}-U^{-1} Z_{\gamma}\, \partial e^{-K_{\text{cs}}}\right|^2\Bigg)\, ,
	\end{align}
	where $\mathcal{Y}_I$ is a (1,0)-form adapted to $J^3$ which reads
	\begin{align}\label{eq:holomorphic1formJ3}
		\mathcal{Y}_I= \dd\tilde{\zeta}_I -\mathcal{F}_{IK} \dd\zeta^K - \frac{1}{8\pi^2} \sum_{\gamma} \Omega_{\gamma} q_I \dd \mathcal{J}_{\gamma}^{(1)}\, ,
	\end{align}
	whilst $U$ denotes some real function that is defined as follows\footnote{Note that the quantity $U$ defined in \eqref{eq:Udef} can be intuitively thought of as an instanton corrected version of the complex structure K\"ahler potential.}
	\begin{align}\label{eq:Udef}
		U= e^{-K_{\text{cs}}} - \frac{1}{2\pi} \sum_{\gamma} \Omega_{\gamma} \left| Z_{\gamma}\right|^2\mathcal{J}_{\gamma}^{(2)} + v_I M^{IJ} \bar{v}_J\, ,
	\end{align}
	with the matrix $M^{IJ}$ being the inverse of $M_{IJ}=-2\text{Im}\, \mathcal{F}_{IJ} - \sum_{\gamma} \Omega_{\gamma} \mathcal{J}_{\gamma}^{(2)} q_I q_J$, and the vector $v_I$ is given by
	\begin{align}
		v_I=\frac{1}{4\pi} \sum_{\gamma} \Omega_{\gamma} q_I \left( Z_{\gamma} \mathcal{J}_{\gamma}^{(2, +)} + \bar{Z}_{\gamma} \mathcal{J}_{\gamma}^{(2, -)}\right)\, .
	\end{align}
	We have also introduced the quantities $\mathcal{W}_{\gamma}=\bar{Z}_{\gamma} \mathcal{J}_{\gamma}^{(2)}- \mathcal{J}_{\gamma}^{(2, +)}v_I M^{IJ} q_J$ and $\mathcal{H}$, the latter being a 1-form generalizing the K\"ahler connection on the complex structure moduli space (see \cite{Alexandrov:2014sya} for details); as well as the following twistorial integrals (c.f. \eqref{eq:Jintegral}) 
	\begin{align}\label{eq:JintegralII}
		\mathcal{J}_{\gamma}^{(2, \pm)}&= \pm \int_{\ell_{\gamma}} \frac{\dd t}{t^{1 \pm 1}}\, \frac{1}{\sigma_{\gamma} e^{-2\pi \text{i} \Theta_{\gamma}(t)}-1}\, , \qquad \mathcal{J}_{\gamma}^{(2)}= \int_{\ell_{\gamma}} \frac{\dd t}{t}\, \frac{1}{\sigma_{\gamma} e^{-2\pi \text{i} \Theta_{\gamma}(t)}-1}\, , \notag\\
		\mathcal{J}_{\gamma}^{(1)}&= \int_{\ell_{\gamma}} \frac{\dd t}{t}\, \log\left(1-\sigma_{\gamma} e^{-2\pi \text{i} \Theta_{\gamma}(t)} \right)\, ,
	\end{align}
	which may be rewritten in terms of Bessel functions, thus capturing the exponentially suppressed behavior --- at large central charge --- associated to D-instanton effects.
	
	Several comments are in order. First, notice how cumbersome the quantum-corrected metric becomes when compared with its classical analogue in \eqref{eq:classicalhypermetric}. %Of course, it is possible to retrieve the latter upon setting all the $q_I$, and consequently all $Z_{\gamma}$, to zero.
	Particularly interesting are the corrections to the metric components associated to the non-compact scalars, namely the 4d dilaton and the complex structure moduli. Regarding the former, it is the contact potential $\chi^{\rm IIA}$ which may be taken to parametrize the quantum hypermultiplet moduli space.\footnote{In fact, the real function $\chi^{\rm IIA}$ can be physically identified with the quantum-exact four-dimensional dilaton $\varphi_4$ \cite{deWit:2006gn}, and it plays a role similar to a would-be K\"ahler potential \cite{deWit:1999fp,deWit:2001brd}.} As for the latter, we clearly see that the classical piece $\mathsf{G}_{i \bar j} \dd z^i \dd z^{\bar j}$ receives strong instanton corrections which can even overcome the tree-level contribution \cite{Marchesano:2019ifh}. Moreover, there also appear cross-terms of the form $(\dd \chi^{\text{IIA}} \dd z^i + \text{c.c.})$, which arise from the 1-form $\dd \mathcal{J}_{\gamma}^{(1)}$ inside $\mathcal{Y}_I$ in \eqref{eq:holomorphic1formJ3} above. Hence, a direct evaluation of the pattern \eqref{eq:pattern} at infinite distance points within $\mathcal{M}_{\text{HM}}$ in principle requires from the use of the full lime element \eqref{eq:quantumhypermetric}, which can become rather involved depending on the limit of interest. Therefore it is highly non-trivial for the inner product $\vec{\zeta}_{\text{t}} \cdot \vec{\mathcal{Z}}_{\text{sp}}$ to verify \eqref{eq:pattern} at any infinite distance boundary, even if it does so already at the classical level.

	\subsubsection{The contact potential $\chi^{\rm IIA}$}
	\label{sss:chiIIA}
	
	Before moving on, let us have a closer look at the contact potential to get a grasp on its physical meaning. This will also provide us with some useful formulae that will be used several times in the following.
	
	Therefore, we start from the twistorial expression for $\chi^{\rm IIA}$, as shown in eq. \eqref{eq:chiIIAtwistor}, which may be written as follows \cite{Alexandrov:2008gh}
	\begin{equation}\label{eq:fullcontactpotential}
		\chi^{\rm IIA}= \chi^{\rm IIA}_{\rm class} + \chi^{\rm IIA}_{\rm quant}\, .
	\end{equation}
	The first term corresponds to the classical piece
	\begin{equation}\label{eq:classicalcontactpot}
		\chi^{\rm IIA}_{\rm class} = \frac{\mathcal{R}^2}{2} e^{-K_{\text{cs}}}\, ,
	\end{equation}
	such that $\chi^{\rm IIA}_{\rm class}$ matches with $e^{-2\varphi_4}$, as one can easily check upon using eqs. \eqref{eq:CSmetric} and \eqref{eq:slagvolumes}. On the other hand, for the quantum corrected piece, $\chi^{\rm IIA}_{\rm quant}$ --- in the case of mutually local instantons arising from D2-branes\footnote{This set of instantons is mapped by Mirror Symmetry to D($-1$) and D1-instantons wrapping holomorphic 0- and 2-cycles within the CY threefold, respectively \cite{Robles-Llana:2007bbv}.} wrapping sLag representatives of the 3-cycle classes $[A_I]$ --- one finds \cite{Robles-Llana:2007bbv,Alexandrov:2014sya,Cortes:2021vdm} (see also the review \cite{Alexandrov:2011va}) 
	\begin{equation}\label{eq:quantumchi}
		\chi^{\rm IIA}_{\rm quant} = \frac{\chi_{E}(X_3)}{96\pi} + \frac{\mathcal{R}}{2\pi^2} \sum_{\gamma} \Omega (\gamma) \sum_{m=1}^{\infty} \frac{|k_I z^I|}{m} \cos \left( 2\pi m k_I \zeta^I\right) K_1 \left( 4\pi m \mathcal{R}|k_I z^I|\right)\, ,
	\end{equation}
	where the term proportional to the Euler characteristic of the threefold, $\chi_{E}(X_3)=2\left( h^{1,1}(X_3)-h^{2,1}(X_3)\right)$, comes from a one-loop $g_s$-correction, whilst the second piece arises from the non-perturbative D2-brane instantons. To actually see how \eqref{eq:quantumchi} arises from eq. \eqref{eq:chiIIAtwistor} above, one needs to substitute the definition of the quantities $\mathcal{J}_{\gamma}^{(1, \pm)}$ (c.f. eq. \eqref{eq:Jintegral}), then expand the logarithm around $\Theta_{\gamma}=0$ and finally rewrite the integrals in terms of the modified Bessel function upon using the following identity
	\begin{align}\label{eq:Besselintegral}
		\int_{0}^{\infty} \frac{\dd y}{y}\, \left(ay + \frac{b}{y}\right) e^{-\left(ay+b/y\right)/2} = 4 \sqrt{ab} K_1 \left( \sqrt{ab}\right)\, .
	\end{align}

	Notice that the contribution to \eqref{eq:quantumchi} associated to the D2-instantons is controlled by their BPS central charge, which coincides (up to order one factors) with the corresponding 4d action
	\begin{equation}\label{eq:D2instantonaction}
		S_{m,\, k_I} = 4\pi m \mathcal{R} |k_I z^I| + 2\pi {\rm i} m k_I \zeta^I\, ,
	\end{equation}
	where $k_I=(k_0, \mathbf{k})$ denote the (quantized) instanton charges. The axionic vevs $\zeta^I$ measure the oscillatory part of the corrections, whereas the non-compact scalars $(z^I, \mathcal{R})$ determine their `size' through the modified Bessel function $K_1(y)$.
	
	\subsection{$SL(2,\mathbb{Z})$ Duality}
	\label{ss:SL2Z}
	
	Here we provide some details regarding the $SL(2,\mathbb{Z})$ invariance that the type IIA hypermultiplet metric inherits from its dual type IIB compactification via Mirror Symmetry. This will moreover highlight the effect that the D2-brane instanton corrections have on certain (classical) infinite distance singularities $\mathcal{M}_{\rm HM}^{\rm IIA}$ in the LCS limit studied in Section \ref{ss:hypers} (see also Section \ref{ss:detailshyper} below).
	
	\subsubsection{The classical metric}
	
	Let us first exhibit the duality of the theory at the classical level. The tree-level metric was shown in \eqref{eq:classicalhypermetric} above, and we repeat it here for the comfort of the reader:
	\begin{align}\label{eq:classicalhypermetricII}
		h_{p q}\, \dd q^p \dd q^q &= \left( \dd \varphi_4\right)^2 + \mathsf{G}_{i \bar j} \dd z^i \dd z^{\bar j} + \frac{e^{4\varphi_4}}{4} \left( \dd \varrho - \left( \tilde{\zeta}_J \dd\zeta^J-\zeta^J \dd\tilde{\zeta}_J \right)\right)^2 \notag\\
		& -\frac{e^{2\varphi_4}}{2} \left( \text{Im}\, \mathcal{B}\right)^{-1\ IJ} \left( \dd\tilde{\zeta}_I -\mathcal{B}_{IK} \dd\zeta^K\right) \left( \dd\tilde{\zeta}_J -\bar{\mathcal{B}}_{JL} \dd\zeta^L\right)\, ,
	\end{align}
	where the different fields describing the hypermultiplet sector of type IIA on the threefold $X_3$ were discussed around \eqref{eq:CSmoduli}. In order to uncover the $SL(2,\mathbb{Z})$ invariance of the action at tree-level, it is useful to switch to the type IIB mirror description, where the symmetry is manifest, and then map the duality transformations back to the original type IIA setup. Regarding the first step,  we will simply state here the relevant identifications, whilst referring the reader interested in the details to the original references \cite{Candelas:1990rm, Aspinwall:1993nu}. These read \cite{Bohm:1999uk}: 
	\begin{align}\label{eq:mirrormap}
		\zeta^0 &= \tau_1\, , \quad \zeta^i = c^i-\tau_1 b^i\, , \quad z^i=b^i+\text{i}t^i\, , \quad \mathcal{R}=\frac{\tau_2}{2}\, , \notag\\
		\tilde{\zeta}_0 &= c^0-\frac{1}{2} \rho_j b^j + \frac{1}{2} \kappa_{ijk} c^i b^j b^k - \frac{1}{6} \tau_1\, \kappa_{ijk} b^i b^j b^k\, , \quad \tilde{\zeta}_i=\rho_i - \kappa_{ijk} c^j b^k + \frac{1}{2} \tau_1\, \kappa_{ijk} b^j b^k\, , \notag\\
		\varrho &= 2b^0 + \tau_1 c^0 + \rho_j \left( c^j - \tau_1 b^j\right)\, ,
	\end{align}
	where $\tau=\tau_1 + \text{i} \tau_2= C_0 + \text{i}\, e^{-\phi_{\rm{IIB}}}$ is the type IIB axio-dilaton, $\vartheta^i \equiv b^i+ \text{i} t^i$ denote the (complexified) K\"ahler moduli of the mirror threefold $Y_3$, $\left( c^i, \rho_i\right)$ arise as period integrals of the RR and 2-form and 4-form fields $\left(C_2, C_4\right)$ over integral bases of $H^2(Y_3)$ and $H^4(Y_3)$, respectively; and finally $\left(b^0, c^0 \right)$ are scalar fields dual to the four-dimensional components of the 2-forms $C_2$ and $B_2$. We stress that the complex structure moduli $z^i$ appearing in the mirror map above should be taken as the `flat' (inhomogeneous)  coordinates associated to the expansion of the prepotential around the LCS point \cite{Hori:2003ic}. Therefore, upon applying such map to the line element displayed in \eqref{eq:classicalhypermetricII} one obtains \cite{Ferrara:1989ik}:
	\begin{align}\label{classicalhypermetric}
		\nonumber h_{pq} \dd q^p \dd q^q &= (\dd \varphi_4)^2 + \mathsf{G}_{i \bar j} \dd \vartheta^i \dd \bar \vartheta^j + \frac{1}{24} e^{2\varphi_4}\cK (\dd C_0)^2\\
		\nonumber  &+\frac{1}{6}e^{2\varphi_4}\cK \mathsf{G}_{i \bar j} \left(\dd c^i-C_0 \dd b^i\right) \left(\dd c^j-C_0 \dd b^j\right)\\
		&+\frac{3}{8\cK}e^{2\varphi_4}\mathsf{G}^{i \bar j}\left(\dd \rho_i - \kappa_{ikl} c^k \dd b^l\right)\left(\dd \rho_j - \kappa_{jmn} c^m \dd b^n\right)\\
		\nonumber &+\frac{3}{2\cK}e^{2\varphi_4}\left(\dd c^0-\frac{1}{2}(\rho_i\dd b^i -b^i \dd \rho_i)\right)^2\\
		\nonumber&+\frac{1}{2}e^{4\varphi_4}\left(\dd b^0+C_0\dd c^0 +c^i \dd \rho_i+\frac{1}{2}C_0(\rho_i\dd b^i- b^i \dd \rho_i)-\frac{1}{4} \kappa_{ijk}c^i c^j\dd  b^k\right)^2\, .
	\end{align}
	Now, as already mentioned, the 4d theory inherits from the 10d supergravity a continuous $SL(2,\mathbb{R})$ symmetry which is broken down to a discrete $SL(2,\mathbb{Z})$ subgroup by non-perturbative (i.e. instanton) effects. The action of any such element $g \in SL(2,\mathbb{Z})$ on the type IIB coordinates reads as \cite{Gunther:1998sc, Bohm:1999uk}
	\begin{align}\label{eq:SdualitytransIIB}
		\tau \rightarrow \frac{a\tau + b}{c\tau+d}\,,\qquad
		t^i \rightarrow |c\tau+d| t^i \,,\qquad
		\begin{pmatrix}
			c^i\\b^i
		\end{pmatrix}
		\rightarrow
		\begin{pmatrix}
			a \quad  b\\c \quad  d
		\end{pmatrix}
		\begin{pmatrix}
			c^i\\b^i
		\end{pmatrix}\, , 
	\end{align}
	where we have only displayed the transformations that are most relevant for our purposes here.\footnote{\label{fnote:SL2Zcontactpotential}Notice that under \eqref{eq:SdualitytransIIB}, the 4d dilaton transforms non-trivially, namely $e^{-2\varphi_4} \rightarrow \frac{e^{-2\varphi_4}}{|c\tau+d|}$.} One can then easily check that these are already enough so as to prove the invariance of the first two rows in \eqref{classicalhypermetric} under $SL(2,\mathbb{Z})$.
	
	Finally, it is now straightforward to translate the S-duality transformations \eqref{eq:SdualitytransIIB} into a set of analogous ones in the type IIA mirror dual compactification upon using the mirror map \eqref{eq:mirrormap}. This leads to \cite{Bohm:1999uk}
	\begin{align}\label{eq:SdualitytransIIA}
		&\Xi \rightarrow \frac{a\, \Xi + b}{c\, \Xi+d}\,,\qquad
		\text{Im}\, z^i \rightarrow |c\, \Xi+d|\, \text{Im}\, z^i \,,\qquad \notag\\
		&\begin{pmatrix}
			\zeta^i + \zeta^0\, \text{Re}\, z^i\\ \text{Re}\, z^i
		\end{pmatrix}
		\rightarrow
		\begin{pmatrix}
			a \quad  b\\c \quad  d
		\end{pmatrix}
		\begin{pmatrix}
			\zeta^i + \zeta^0\, \text{Re}\, z^i\\ \text{Re}\, z^i
		\end{pmatrix}\, , 
	\end{align}
	where we have defined the complex field $\Xi= \zeta^0 + 2 \text{i} \mathcal{R}$. Note that this is again sufficient to show the invariance of the metric components in \eqref{eq:classicalhypermetricII} associated to the 4d dilaton, the complex structure and the $\zeta^I$ coordinates.
	
	\subsubsection{Quantum corrections}
	
	One can go beyond the previous tree-level analysis and study $SL(2,\mathbb{Z})$ duality once quantum corrections have been taken into account. Following the discussion of Section \ref{ss:exactmetric}, we will only consider the effect of `electric' D2-brane instantons, i.e. those wrapping the $A_I$\,-cycles introduced in \eqref{eq:symplecticpairing}.  
	
	Recall that the quantum hypermultiplet metric can be effectively encoded into the contact (or tensor) potential, $\chi^{\rm IIA}$, which reads (see Section \ref{sss:chiIIA})
	\begin{align}\label{eq:apchiIIA}
		\chi^{\rm IIA} &=\, \frac{\mathcal{R}^2}{2} \frac{{\rm i} \int\Omega \wedge \bar \Omega }{|Z^0|^2} + \frac{\chi_{E}(X_3)}{96\pi} \notag\\
		&+ \frac{\mathcal{R}}{2\pi^2} \sum_{\gamma} \Omega (\gamma) \sum_{m=1}^{\infty} \frac{|k_I z^I|}{m} \cos \left( 2\pi m k_I \zeta^I\right) K_1 \left( 4\pi m \mathcal{R}|k_I z^I|\right) \, ,
	\end{align}
	where the first, second and third terms correspond to the classical, one-loop and D2-instanton contributions, respectively. Now, instead of trying to show how the exact hypermultiplet metric \eqref{eq:quantumhypermetric} still respects $SL(2,\mathbb{Z})$ duality, we will concentrate on rewriting the above expression in a way which manifestly reflects the symmetry. This will allow us to relate certain non-perturbative corrections to classically-derived terms, thus providing more evidence in favour of our argumentation in Section \ref{ss:detailshyper} below. 
	
	Let us start by extracting a common $\sqrt{\mathcal{R}}$ factor from each of the three terms in \eqref{eq:apchiIIA}, yielding
	\begin{align}\label{eq:modinvariantchi}
		\frac{\chi^{\rm IIA}}{\sqrt{\mathcal{R}}} &=\, \frac{\mathcal{R}^{3/2}}{2} \frac{{\rm i} \int\Omega \wedge \bar \Omega }{|Z^0|^2} + \frac{\chi_{E}(X_3)}{96\pi} \mathcal{R}^{-1/2} \notag\\
		&+ \frac{\mathcal{R}^{1/2}}{2\pi^2} \sum_{\gamma} \Omega (\gamma) \sum_{m=1}^{\infty} \frac{|k_I z^I|}{m} \cos \left( 2\pi m k_I \zeta^I\right) K_1 \left( 4\pi m \mathcal{R}|k_I z^I|\right) \, .
	\end{align}
	Therefore, given that the contact potential transforms under $SL(2, \mathbb{Z})$ precisely the same way as $\sqrt{\mathcal{R}}$ does (see footnote \ref{fnote:SL2Zcontactpotential}), we can now concentrate on finding a modular invariant expression for the RHS of \eqref{eq:modinvariantchi}. To do so, we first expand the classical term around the LCS, as follows
	\begin{equation}\label{eq:chiclassical}
		\begin{aligned}
			\frac{\mathcal{R}^{3/2}}{2} \frac{{\rm i} \int\Omega \wedge \bar \Omega }{|Z^0|^2} =&\,  4 \mathcal{R}^{3/2} \bigg[ \frac{1}{3!} \kappa_{ijk} v^i v^j v^k + \frac{\zeta(3) \chi_{E}(X_3)}{4(2 \pi)^3}\\
			&+ \frac{1}{2(2 \pi)^3} \sum_{\textbf{k}>0} n_{\textbf{k}}\, \text{Re}\, \left \lbrace \text{Li}_3 \left( e^{2\pi \text{i}k_i z^i} \right) + 2\pi k_i v^i \text{Li}_2 \left( e^{2\pi \text{i}k_i z^i} \right)\right \rbrace \bigg ]\, ,
		\end{aligned}
	\end{equation}
	where $\zeta(x)$ denotes the Riemann zeta function, $\text{Li}_k (x)= \sum_{j=1}^{\infty} \frac{x^j}{j^k}$ is the polylogarithm function and we have defined $v^i \equiv \text{Im}\, z^i$ in the above expression. The physical interpretation of each term is clear: the first piece corresponds to the classical volume term of the mirror dual type IIB compactification on $Y_3$, whilst the second and third ones arise as perturbative and non-perturbative $\alpha'$-corrections that modify the former away from the large volume point. The integers $n_{\textbf{k}}$ denote the genus-zero Gopakumar-Vafa invariants that `count' the multiplicity of holomorphic 2-cycles in a given class $[k_i \gamma^i] \in H^+_2(Y_3, \mathbb{Z})$.
	
	Next, we divide the instanton piece in \eqref{eq:modinvariantchi} into two different terms, namely we separate the contributions associated to D2-branes wrapped on the SYZ cycle from those wrapping the remaining $A_I$\,-cycles. The reason for doing so will become clear in the following. This leads to
	\begin{align}\label{eq:chiD2}
		\frac{\chi^{\rm IIA}_{\text{D2}}}{\sqrt{\mathcal{R}}} =\, &\frac{\mathcal{R}^{1/2} \chi_{E}(X_3)}{8\pi^2} \sum_{k_0, m \neq 0} \left| \frac{k_0}{m}\right| e^{ 2\pi i m k_0 \zeta^0} K_1 \left( 4\pi \mathcal{R}|m k_0|\right)\notag\\
		&+ \frac{\mathcal{R}^{1/2}}{4\pi^2} \sum_{\textbf{k}>0} n_{\textbf{k}} \sum_{m\neq0, k_0 \in \mathbb{Z}} \frac{|k_I z^I|}{|m|} e^{ 2\pi i m k_I \zeta^I} K_1 \left( 4\pi m \mathcal{R}|k_I z^I|\right)\, ,
	\end{align}
	where we have substituted the Donaldson-Thomas invariants $\Omega(\gamma)$ by $\chi_{E}(X_3)/2$ and $n_{\textbf{k}}$ for $\gamma = \left(k_0 \neq 0, \textbf{k}=0 \right)$ and $\gamma = \left(k_0 \in \mathbb{Z}, \textbf{k} > 0 \right)$, respectively. 
	
	With this, we are finally ready to rewrite \eqref{eq:modinvariantchi} in a manifestly modular invariant way. Notice that the first term in eq. \eqref{eq:chiclassical} is left unchanged under the set of transformations in \eqref{eq:SdualitytransIIA}, reflecting the fact that the tree-level hypermultiplet metric at LCS/Large Volume is modular invariant. Consider now the terms which are proportional to the Euler characteristic of the threefold, $\chi_{E}(X_3)$, which read 
	\begin{align}\label{eq:chiEuler}
		\frac{\chi^{\rm IIA}_{\chi_{E}}}{\sqrt{\mathcal{R}}} =\, &\frac{\chi_{E}(X_3)}{2(2\pi)^3} \bigg[ 2\mathcal{R}^{3/2} \zeta(3) + \frac{\pi^2}{6} \mathcal{R}^{-1/2} + 4\pi \mathcal{R}^{1/2} \sum_{k_0, m > 0} \left| \frac{k_0}{m}\right| e^{ 2\pi i m k_0 \zeta^0} K_1 \left( 4\pi \mathcal{R}|m k_0|\right) \bigg]\, .
	\end{align}
	Upon performing a Poisson resummation\footnote{The Poisson resummation identity reads as follows \begin{equation}
			\sum_{n \in \mathbb{Z}} F(x+na) = \frac{1}{a} \tilde{F} \left(\frac{2\pi n}{a} \right) e^{2\pi i n x/a}\, ,
		\end{equation}
		with $\tilde{F} (\omega)=\int_{-\infty}^{\infty} \dd x\, F(x) e^{-i \omega x}$ the Fourier transform of $F(x)$.} on the integer $k_0$ in eq. \eqref{eq:chiEuler} above, one arrives at \cite{Robles-Llana:2007bbv}
	\begin{align}\label{eq:chiEulerII}
		\frac{\chi^{\rm IIA}_{\chi_{E}}}{\sqrt{\mathcal{R}}}\, =\, \frac{\chi_{E}(X_3)}{2(2\pi)^3} \sum_{m, n \in \mathbb{Z}\setminus \lbrace (0,0) \rbrace} \frac{\mathcal{R}^{3/2}}{|m\, \Xi +n|^3}\, ,
	\end{align}
	which is indeed modular invariant. Note that in order to obtain such expression one needs to substitute $\zeta(k)=\sum_{n>1} n^{-k}$ for $k=2,3$.
	
	Finally, we group together those terms containing sums over Gopakumar-Vafa invariants, such that, after performing again a Poisson resummation over the unconstrained integer $k_0$, one finds \cite{Robles-Llana:2007bbv}
	\begin{align}\label{eq:chiGV}
		\frac{\chi^{\rm IIA}_{\text{GV}}}{\sqrt{\mathcal{R}}}\, =\, \frac{1}{(2\pi)^3} \sum_{\textbf{k}\neq 0} n_{\textbf{k}} \sum_{m, n \in \mathbb{Z}\setminus \lbrace (0,0) \rbrace} \frac{\mathcal{R}^{3/2}}{|m\, \Xi +n|^3} \left( 1+2\pi |m\, \Xi +n| k_i v^{i}\right) e^{-S_{m, n}}\, ,
	\end{align}
	where $S_{m, n}= 2 \pi k_i \left( |m\, \Xi +n| v^i + i m \left( \zeta^i+\zeta^0 \text{Re}\, z^i \right) -in \text{Re}\, z^i\right)$. This last term can be seen to be the mirror dual of the quantum corrections arising from Euclidean type IIB $(p,q)$-strings, and it tells us that the exponentially suppressed terms within the complex structure K\"ahler potential --- close to the LCS point --- are related by $SL(2, \mathbb{Z})$ duality to certain D2-brane instanton contributions. This fact will indeed play a key role later on in Section \ref{ss:detailshyper}.
	
	\subsection{The evaluation of the Pattern within $\mathcal{M}_{\rm HM}$}
	\label{ss:detailshyper}
	
	In Section \ref{sss:classivalvsquantum} in the main text, we were interested in evaluating the pattern \eqref{eq:pattern} for certain trajectories lying entirely within the hypermultiplet moduli space $\mathcal{M}_{\rm HM}$. Such infinite distance paths were of the form
	\begin{equation}\label{eq:generictrajII}
		\text{Im}\, z^i \sim \sigma^{e^1}\, , \qquad e^{-\varphi_4}\sim \sigma^{e^2}\, , \qquad \sigma \to \infty\, ,
	\end{equation}
	with $e^1, e^2 \geq 0$, thus including both the weak coupling and large complex structure (LCS) points. Classically, i.e. without taking into account D-instanton corrections, both kind of limits were shown to fulfill the pattern. Quantum-mechanically, however, one expects large instanton contributions to modify the computation, at least in some cases. The purpose of this subsection is to put all the machinery previously described into work in order to prove that eq. \eqref{eq:pattern} still holds even after taking into account all relevant quantum effects, as advertised in Section \ref{sss:instantons}. We analyze each of these limits in turn.	
	
	\subsubsection*{Weak Coupling Point}
	
	In this case, since the singularity that is being approached is at weak string coupling, we do not expect neither perturbative nor non-perturbative effects to become important, and the classical analysis from Section \ref{sss:classivalvsquantum} should be reliable. This can be readily confirmed upon looking at the behavior of the sum in eq. \eqref{eq:quantumchi}, since for $\mathcal{R} \to \infty$ and $z^I$ finite one finds
	\begin{equation}\label{eq:asympotic Bessel infinity}
		K_1 \left( 4\pi m \mathcal{R}|k_I z^I|\right) \sim \sqrt{\frac{1}{8m \mathcal{R}|k_I z^I|}}\, e^{-4\pi m \mathcal{R}|k_I z^I|} %\left(1 \, +\, \mathcal{O}(1/\mathcal{R}|k_I z^I|) \right)
		\, ,
	\end{equation}
	such that $\chi^{\rm IIA}_{\rm quant}= \text{const.}\, + \mathcal{O}\left( e^{-\mathcal{R}|k_I z^I|}\right) \ll \chi^{\rm IIA}_{\rm class}$ asymptotically. Similarly, the moduli space metric deviates from the tree-level one by additional terms which at leading order behave as follows (c.f. \eqref{eq:quantumhypermetric}) 
	\begin{equation}
		\delta  \dd s_{\rm HM}^2 =  \delta  \dd s_{\rm HM}^2\rvert_{\text{1-loop}} + \delta  \dd s_{\rm HM}^2\rvert_{\text{D-inst}}\, \sim\, \frac{\chi_{E}(X_3)}{\chi^{\rm IIA}}\, +\, \sum_{\gamma} \Omega_{\gamma}\, e^{-S_{m,\, k_I}}\, ,
	\end{equation}
	and thus it is enough to use the classical approximation \eqref{eq:classicalhypermetricII}. Therefore, we conclude that the calculations performed after \eqref{eq:fundstringmass} remain valid, and the pattern is still verified.
	
	Let us also say a few words about the S-dual limit, since it will play a crucial role in what follows. As we mention in the main text, the weak coupling singularity here discussed translates into a physically equivalent one at both strong coupling and LCS, namely $\left( \mathcal{R}' \sim \sigma^{-1} , \text{Im}\, z^{i\, '} \sim \sigma \right)$. Notice that $\mathcal{R}'\, \text{Im}\, z^{j\, '} \to \text{const.}$ , which means, in particular, that the tree-level piece of $\chi^{\rm IIA}$ dominates over the quantum corrections, i.e. the D2-brane instanton contributions decouple.\footnote{This is not completely true, since the instanton sum can still lead to additional \emph{finite distance} degenerations, which are the S-dual versions of the conifold locus.} Hence, one can again safely use the classical metric \eqref{eq:classicalhypermetricII} to compute the inner products between the relevant charge-to-mass and species vectors. These are associated to the D4-string, with tension
	\begin{equation}\label{eq:D4SYZSdual}
		\left(\frac{T_{\text{D4}}}{\Mpf^2} \right)= \frac{2 \mathcal{R}'}{\left(\chi^{\rm IIA}\right)'} = \frac{1}{\chi^{\rm IIA}} \sim \frac{1}{\sigma^{2}}\, ,
	\end{equation}
	and the KK scale:
	\begin{equation}\label{eq:KKSYZSdual}
		\left(\frac{m_{\text{KK},\, B^0}}{\Mpf} \right)^2 \sim \frac{1}{\text{Im}\, z^{i\, '} \left(\chi^{\rm IIA}\right)'} \sim  \frac{1}{\text{Im}\, z^{i}\, \chi^{\rm IIA}} \sim \frac{1}{\sigma^2}\, ,
	\end{equation}
	where in order to arrive at the second equalities we have used the S-duality transformation rules (see eq. \eqref{eq:SdualitytransIIA}). 
	
	\subsubsection*{LCS Point}
	
	A slightly different story holds for the second kind of limit, namely that corresponding classically to large complex structure at fixed 4d dilaton
	\begin{equation}\label{eq:LCSfixeddilaton}
		z^j ={\rm i} \xi^j \sigma\, , \qquad \varphi_4= \text{const.}\, , \qquad \sigma \to \infty\, .
	\end{equation}
	This limit is indeed the mirror dual to the one explored in \cite{Marchesano:2019ifh,Baume:2019sry}. In terms of the relevant coordinates controlling the behavior of the contact potential, such trajectories are of the form $(z^j (\sigma), \mathcal{R} (\sigma)) \sim \left( {\rm i} \sigma, \sigma^{-3/2}\right)$, which means that for small enough instanton charges $k_I$, the correction term controlled by the Bessel function in \eqref{eq:quantumchi} will behave as 
	\begin{equation}\label{eq:strongorrections}
		K_1 \left( 4\pi m \mathcal{R}|k_I z^I|\right) \sim\frac{1}{4\pi m \mathcal{R}|k_I z^I|}\, .
	\end{equation}
	More precisely, the charges must be such that
	\begin{equation}\label{eq:instantoncondition}
		4\pi m \mathcal{R}|k_0 + k_i z^i| \ll 1\, ,
	\end{equation}
	for the associated D2-instantons to contribute significantly to the tensor potential $\chi^{\rm IIA}_{\text{quant}}$. As already noted in \cite{Marchesano:2019ifh}, this parallels the behavior of the exponentially light towers of D3-brane bound states appearing in the mirror dual vector multiplet moduli space \cite{Grimm:2018ohb}.
	
	To see what is the upshot of including such quantum corrections to the hypermultiplet metric along the limit specified by \eqref{eq:LCSfixeddilaton} one can follow the same strategy as in \cite{Baume:2019sry}, and exploit the $SL(2, \mathbb{Z})$ duality of the theory. This allows us to translate the aforementioned limit into a simpler one where we can readily identify the relevant asymptotic physics. Indeed, after performing the duality we end up exploring the following `classical' limit
	\begin{equation}\label{eq:smallCSsmalldilaton}
		\text{Im}\, z^{j\, '} \sim \sigma^{-1/2}\, , \qquad \mathcal{R}' = \frac{e^{-\phi\, '} \mathcal{V}_{A_0}'}{2} \sim \sigma^{3/2}\, , \qquad e^{2\varphi_4'} \sim \sigma^{-3/2}\, ,
	\end{equation}
	where one should think of $z^{i\, '}= \frac{1}{2\pi {\rm i}} \log x^i$ as flat complex structure variables defined close to the LCS point ($x^i \to 0$), see below. Notice that this is nothing but the mirror dual of the F1 limit studied in \cite{Baume:2019sry}. There, the relevant quantum corrections to the classical type IIB hypermultiplet metric are induced by $\alpha'$ and worldsheet instantons, whilst D-brane contributions decouple. Importantly, here such `corrections' are already captured by the \emph{exact} complex structure metric \eqref{eq:CSmetric}, thus simplifying the analysis enormously.
	
	Therefore, recall that away from the LCS point, the periods of the holomorphic $(3,0)$-form $\Omega$ receive corrections from their flat values, namely \cite{B_hm_2000, Hori:2003ic}
	\begin{equation}\label{eq:analyticCSvariables}
		z^{j\, '} = \frac{1}{2\pi {\rm i}} \log x^j + \mathcal{O}(x^i)\, ,
	\end{equation}
	such that upon increasing $x^i$ towards one, the logarithmic approximation for $z^{i\, '}$ stops being valid and the polynomial corrections clearly dominate. Hence, instead of reaching a point where $\text{Im}\, z^{i\, '} \to 0$ asymptotically, what happens is that the complex structure variables generically approach some constant $\mathcal{O}(1)$ value (see e.g., \cite{Blumenhagen:2018nts,Joshi:2019nzi,Alvarez-Garcia:2021mzv}). %(see \cite{Blumenhagen:2018nts,Joshi:2019nzi,Alvarez-Garcia:2021mzv} for one-parameter examples where this behavior has been observed). 
	This does not prevent, on the other hand, the $\mathcal{R}$ coordinate from keep flowing towards weak coupling, such that a more accurate parametrization of the asymptotic trajectory would be the following:
	\begin{equation}\label{eq:smallCSsmalldilatoncorrected}
		\text{Im}\, z^{j\, '} = \text{const.}\, , \qquad \mathcal{R}' \sim \sigma^{3/2}\, , \qquad e^{2\varphi_4'} \sim \sigma^{-3}\, .
	\end{equation}
	Notice that this belongs to the family of geodesics in \eqref{eq:generictrajII} with $\mathbf{e}=(0, 3/2)$. Hence, our previous analysis for the weak coupling singularity around \eqref{eq:asympotic Bessel infinity} applies here and we conclude that the pattern still holds.
	
	From the original perspective, though, a direct evaluation of the scalar product \eqref{eq:pattern} seems to be rather involved, since the metric receives strong corrections that deviate from the simple block diagonal form displayed in \eqref{eq:classicalhypermetricII} above (c.f. eq. \eqref{eq:quantumhypermetric}). However, let us stress again that we do not need to do this, as we already know what is the S-dual limit of \eqref{eq:smallCSsmalldilatoncorrected}: It corresponds to an infinite distance trajectory of the form $\left( \mathcal{R} \sim \sigma^{-3/2} , \text{Im}\, z^{i} \sim \sigma^{3/2} \right)$, thus located at strong coupling and LCS (see discussion around \eqref{eq:D4SYZSdual}). Incidentally, this nicely explains why the pattern was still verified along the classically obstructed limit \eqref{eq:LCSfixeddilaton}, since the products in eqs. \eqref{eq:patternviolation} and \eqref{eq:patternviolationII} are formally identical to the ones that need to be computed along the present quantum corrected trajectory.
	
	\section{More on the mathematical structure of $\infty(\mathcal{M})$}
	\label{ap:detailsbottomup}
	In this appendix we will review the construction of the set of infinite distance limits $\infty(\mathcal{M})$ associated to some moduli space $(\mathcal{M},\mathsf{G})$, as well as its topology and structure. This can be then used to define functions over it, once we are given certain quantities defined on $\mathcal{M}$.
	
	Following \cite{Klingenberg+2011}, given a geodesically complete, simply connected $n$-manifold $\mathcal{M}$ with non-positive curvature, two geodesics $c$ and $c'$ defined on it are said to be \emph{positively (negatively) asymptotic} if there exists a constant $a_{cc'}\geq 0$ such that $d(c(\tau),c'(\tau))\leq a_{cc'}$ for any affine parametrization $\tau\geq 0\;(\leq 0)$. It can be shown that being asymptotic is an equivalence relation on the set of geodesics within $\mathcal{M}$, and that given a geodesic $c$ and a point $p\in \mathcal{M}$, then there exists exactly one geodesic $\tilde{c}$ such that $[c]_\pm=[\tilde{c}]_\pm$ and $\tilde{c}(0)=p$. Now, for such a manifold $\mathcal{M}$, we will refer as \emph{points of infinity}, $\infty(\mathcal{M})$, to the different classes of positive and negative asymptotic geodesics. We denote by $\infty(\mathcal{M})$ the set of infinite points of $\mathcal{M}$.
	
	We can relax our previous definition by simply requiring that, given an arbitrary point $p_0\in\mathcal{M}$, there exists $r_0>0$ such that the curvature is non-positive for all points $p\in \mathcal{M}$ with $d(p,p_0)>r_0$, thus allowing for positive curvature in the `bulk' of the moduli space. The simply connectedness property of $\mathcal{M}$ is needed in order to avoid situations such as $\mathcal{M}=\mathbb{R}\times\mathbb{S}^1$, where we can have geodesics $c_\theta,\; c_{\theta'}$ with different angles $\theta\neq\theta'$ with respect to the `vertical' direction that intersect an infinite number of times while having $d(c_\theta(\tau),c_{\theta'}(\tau))=2\sin\left(\frac{|\theta-\theta'|}{2}\right)\tau\to\infty$ as $\tau\to \infty$. In some cases, such as when the moduli space is given by a homogeneous space, as for example $\mathcal{M}=E_{n(n)}/K_n$ for M-theory compactified on a $n=11-d$ torus, one can quotient out the compact subspace and simply work with a submanifold of the complete $\mathcal{M}$ where all the remaining moduli are non-compact. In other cases, close to the infinite distance points a global shift symmetry on the axionic scalars is restored \cite{Corvilain:2018lgw} such that all relevant physical quantities (e.g., the metric $\mathsf{G}$, etc.) loose all dependence on them, and thus one can simply focus on the non-compact (saxionic) directions.
	
	It can be then shown that $\infty(\mathcal{\mathcal{M}})\simeq \mathbb{S}^{n-1}$, with the usual topology inherited from $\mathbb{R}^{n-1}$, since all the positive (negative) asymptotic geodesics of the same class have the same unit tangent vector $\hat{T}=\frac{\dot{c}(s)}{\|\dot{c}(s)\|}\in \mathbb{S}^{\dim M-1}$ as $s\to\infty\; (-\infty)$. In those cases where $\mathcal{M}$ is not geodesically complete, such as the K\"ahler cone in some 4d $\mathcal{N}=1$ examples from Section \ref{s:N=1}, $\infty(\mathcal{M})$ is built through geodesics that can be defined over $s\in [s_0,\infty)$ or $(-\infty,s_0]$, with $\infty(\mathcal{M})$ being homeomorphic to some subset of $\mathbb{S}^{n-1}$. The space $\text{cp}(\mathcal{M})=\mathcal{M}\cup\infty(\mathcal{M})$ it is known as \emph{compactification of $\mathcal{M}$}, inherits a natural topology from $\mathcal{M}$. On the other hand, when $\mathcal{M}$ is geodesically complete, one rather has $\text{cp}(\mathcal{M})\simeq D^{\dim M}$.
	
	With this in mind, we can consider an EFT with a moduli space $(\mathcal{M},\mathsf{G})$, and such that the set of relevant scales for tower masses are continuous (with continuous derivative as well) functions $\{m_I:\mathcal{U}_I\subseteq\mathcal{M}\to\mathbb{R}_{>0}\}_I$, with possibly $\mathcal{U}_I\neq \mathcal{M}$, as they might be defined only over some subdomain $\mathcal{U}_{I}$ of the moduli space. We can further define the subsets $\{\mathcal{U}^{\infty}_I\}_I$ formed by those infinite distance points that can be accessed via geodesics contained in $\mathcal{U}_I$. Note that as long as one geodesic $\gamma$ reaching infinity can be defined over $\mathcal{U}_I$, then $[\gamma]\in\mathcal{U}^\infty_I$.
	
	The topology inherited from $\mathcal{M}$ to $\infty(\mathcal{M})$ is defined in terms of limits of successions (i.e. a set $\mathcal{C}\subseteq\infty(\mathcal{M})$ is closed if every succession defined on it has a limit on $\mathcal{C}$). As we expect the boundaries of domains of definition $\mathcal{U}_I$ of the different towers to be more or less sharply defined (for example being self-dual lines of the theory), one could in principle move to infinite distance limits in a trajectory parallel to such border, so in principle one could expect sets $\{\mathcal{U}_I^\infty\}_I$ to be closed even if $\{\mathcal{U}_I\}_I$ are not. In any case, as it discussed in Section \ref{s:bottomup}, the sets used in the bottom-up argument for the pattern will be closed, as they are the inverse image of closed sets by continuous operations.
	
	Functions on $\infty(\mathcal{M})$ can be constructed from functions in $\mathcal{M}$ that take the same limiting constant value along any representative geodesics going to a given point in $\infty(\mathcal{M})$. In this sense, the exponential rates of the towers 
	\begin{equation}
		\begin{array}{rrcl}
			\lambda_I:&\mathcal{U}_I^{\infty}\subseteq\infty(\mathcal{M})&\longrightarrow&\mathbb{R}\\
			&\hat{T}&\longmapsto&\hat{T}\cdot\vec{\xi}_I=-\hat{T}^a \partial_a \log m_I\;,
		\end{array}
	\end{equation}
	and the analogous $\{\alpha_J:\mathcal{W}^\infty_J\subseteq\infty(\mathcal{M})\to\mathbb{R}\}_J$ for each species scale $\{\Lambda_J:\mathcal{W}_J\subseteq\mathcal{M}\to\mathbb{R}_{>0}\}_J$ are examples of functions defined in (subsets of) $\infty(\mathcal{M})$, once the well-definition requirement of Condition \ref{A:s1} is taken into account\footnote{Notice that, as we expect the domain of definition $\{\mathcal{U}_I\}_I$ of the different species to be sharply defined, at least in asymptotic regions of $\mathcal{M}$, the associated $\{\mathcal{U}_I^{\infty}\}_I$ (and as a result $\{\mathcal{W}_J^{\infty}\}_J$) will be closed subsets of $\infty(\mathcal{M})$, as one can in principle move towards infinite distance limits with geodesics parallel to the boundary of $\mathcal{U}_I$.\label{fn:closed}}. By construction, these functions are continuous, and continuous operations with them (such as addition, subtraction, or the maximum) will be so too.
	
	Note that, unlike the different towers, which in many cases only make sense to be defined in the asymptotic region of moduli space where they become light, we expect the species scale to be well defined all over  the moduli space, $\Lambda_{\rm sp}:\mathcal{M}\to\mathbb{R}_{>0}$, with $\Lambda_{\rm sp}\sim \max\{\Lambda_J\}_J$ as we approach the asymptotic regions. Something similar occurs with scalar potentials $V:\mathcal{M}\to \mathbb{R}$, where the de Sitter coefficient $\frac{\|\nabla V\|}{V}:\infty(\mathcal{M})\to \mathbb{R}_{\geq 0}$ is expected to be a well-defined quantity, see \cite{Calderon-Infante:2022nxb}. A natural question is whether continuous functions derived from $V(\phi^i)$ or $\Lambda_{\rm sp}$ can be analogously defined over ${\rm cp}(\mathcal{M})$, as that way topology/continuity arguments might shed some light on the existence of vacua in the bulk of moduli space, among other things.
	
	On the other hand, it is not immediate that $\vec{\zeta}_{\rm t}\cdot\vec{\mathcal{Z}}_{\rm sp}$ is a well-defined function over ${\infty}(\mathcal{M})$. However, the different string theory constructions thotoughly checked along this paper seem to suggest this is indeed the case, and as explained in Section \ref{s:bottomup}, it is one of the sufficient conditions required for the $\vec{\zeta}_{\rm t}\cdot\vec{\mathcal{Z}}_{\rm sp}=\frac{1}{d-2}$ pattern to hold.
	
	\subsection{Derivation of the Pattern from the sufficient conditions in Section \ref{ss:continuity} \label{app:derivation}}
	
	Here we provide more details regarding the derivation of the pattern \eqref{eq:pattern} upon assuming the sufficient conditions declared in Section \ref{ss:continuity}. First of all, \Cref{A:s1} allows us to define the functions
	\begin{equation}\label{e:maxmax}
		\tilde{\lambda}_I(\hat{T})=\max\{\hat{T}\cdot\vec{\zeta}_{K}\}_K-\lambda_I(\hat{T})\,,\qquad 
		\tilde{\alpha}_J(\hat{T})=\max\{\hat{T}\cdot\vec{\mathcal{Z}}_L\}_L-\alpha_J(\hat{T})\,,
	\end{equation}
	(there is some abuse of notation here, but we can set $\lambda_I(\hat{T}), \alpha_J(\hat{T})\equiv 0$ for the limits over which they are not defined) which yield $0$ over such limits along which $m_I$ or $\Lambda_J$ dominate.  It is easy to see that $\{\tilde{\lambda}_I\}_I$ are continuous. By the nuances explained in Section \ref{s:bottomup}, $\{\tilde{\alpha}_J\}_J$ is only piece-wise continuous, but this will be enough for our purposes.

	As they are (piece-wise)continuous over the space of infinite distance limits, the sets $\tilde{\lambda}_I^{-1}(\{0\})$ and $\tilde{\alpha}_J^{-1}(\{0\})$ will be closed\footnote{Note that, following footnote \ref{fn:closed}, the different domains over which one species might dominate will be closed, so that the argument is still valid even if $\tilde{\alpha}_J$ is only piece-wise continuous.} coverings of the latter, with a finer covering being $\mathcal{C}=\{\tilde{\lambda}_I^{-1}(\{0\})\}_I\cap\tilde{\alpha}_J^{-1}(\{0\})\}_{I,J}$ (note that indeed many of these will be empty sets). For a given element in $\mathcal{C}$, all its points, barring perhaps intersections with other $\tilde{\lambda}_I^{-1}(\{0\})\}_I\cap\tilde{\alpha}_J^{-1}(\{0\})$ or directions where sliding occurs, $\vec{\zeta}_{\rm t}\cdot\vec{\mathcal{Z}}_{\rm sp}$ will have a constant value (note again here that all the masses have fixed asymptotic expressions, which results in $\Lambda_{\rm sp}$ and $\mathsf{G}$ to do the same). As for the aforementioned regions where this might not be the case, we have:
	\begin{itemize}
		\item \textbf{Intersection between covering elements:} Given two covering sets in $\mathcal{C}$, along their intersection we have several leading towers (or species) becoming light at the same rate. As explained before, this happens when $\hat{T}\perp{\rm Hull}(\{\vec{\zeta}_1,\ldots,\vec{\zeta}_k\})$ (equiv. $\hat{T}\perp{\rm Hull}(\{\vec{\mathcal{Z}}_1,\ldots,\vec{\mathcal{Z}}_k\})$). Hence, by \Cref{A:s2} we will have that the leading species (resp. tower) is located perpendicular to the convex hull, being dominant at both sides of the intersection. As the only difference between the scalar charge-to-mass vectors of the towers (species) scaling at the same speed is in the components perpendicular to $\hat{T}\propto \vec{\mathcal{Z}}_{\rm sp}$ ($\hat{T}\propto \vec{\zeta}_{\rm t}$), this  implies that $\vec{\zeta}_{\rm t}\cdot\vec{\mathcal{Z}}_{\rm sp}$ takes the same value at both sides, as well as at the intersection.
		
		\item In the case in which there is \textbf{sliding}, things turn out to be trickier. The sliding of both $\vec{\zeta}_{\rm t}$ and $\vec{\mathcal{Z}}_{\rm sp}$ can cause the product $\vec{\zeta}_{\rm t}\cdot\vec{\mathcal{Z}}_{\rm sp}$ to be not continuous or even ill-defined over an asymptotic trajectory. To see this, we split into parallel and perpendicular components to $\hat{T}$ to find
		\begin{equation}
			(\vec{\zeta}_{\rm t}\cdot\vec{\mathcal{Z}}_{\rm sp})(\Delta)=\vec{\zeta}_{\rm t}^{\,\parallel}\cdot\vec{\mathcal{Z}}_{\rm sp}^\parallel+\vec{\zeta}_{\rm t}(\Delta)^\perp\cdot\vec{\mathcal{Z}}_{\rm sp}^\perp(\Delta)\,,
		\end{equation}
		where $\Delta$ symbolically parameterize the non-compact directions perpendicular to $\hat{T}$. If $\vec{\zeta}_{\rm t}(\Delta)^\perp\cdot\vec{\mathcal{Z}}_{\rm sp}^\perp(\Delta)$ is not constant for all values of $\Delta$, then the product is not well defined over asymptotic directions where the sliding occurs. Then it is immediate that for this to be the case $\vec{\zeta}_{\rm t}\cdot\vec{\mathcal{Z}}_{\rm sp}$ must be constant over the sliding region, and as a result the product takes the same value at both sides of the sliding loci. However, by \Cref{A:s1}, we expect $\vec{\zeta}_{\rm t}\cdot\vec{\mathcal{Z}}_{\rm sp}$ to be a well-defined quantity at any point of the set of infinite distance limits. Note that, in principle, there is no physical reason to demand this, but if this was not the case it would make no sense to study the pattern along directions where sliding happens.
		
	\end{itemize}
	In conclusion, we have that $\vec{\zeta}_{\rm t}\cdot\vec{\mathcal{Z}}_{\rm sp}$ is constant in the interior of different regions of the space of asymptotic limits. Furthermore, in the limits between these regions (be them associated to co-leading species or towers, or sliding loci), the product is the same as in the limiting trajectories. This means that in each of the connected components of $\infty(\mathcal{M})$, for which every point can be reached from any other by crossing a finite number of intersections/sliding loci (as the $\mathcal{C}$ covering is finite), the function $\vec{\zeta}_{\rm t}\cdot\vec{\mathcal{Z}}_{\rm sp}$ takes the same value. It is then enough to know said product along some limit. Finally \Cref{A:s3} can be used, as in emergent string limits or homogeneous decompactifications said value is known to be $\frac{1}{d-2}$ (see Appendix \ref{ap:generalities}), thus resulting in $\vec{\zeta}_{\rm t}\cdot\vec{\mathcal{Z}}_{\rm sp}=\frac{1}{d-2}$ for every asymptotic limit.

	\bibliography{ref.bib}
	\bibliographystyle{JHEP}
	
\end{document}